\documentclass[final,authoryear,5p,times,twocolumn]{elsarticle}

\usepackage{graphicx, color, amsthm, amssymb}
\usepackage{txfonts, longtable, dcolumn, lscape, multirow, lineno, pifont, afterpage}
\usepackage{natbib}
\usepackage{hyperref}
\bibpunct{(}{)}{;}{a}{}{,}
\journal{Icarus}

\usepackage{microtype}

\newcommand\eatpunct[1]{}

\begin{document}

\begin{frontmatter}

\title{Thermophysical modeling of main-belt asteroids from WISE thermal data}

\author[label1]{J.~Hanu\v s\corref{cor1}}
\cortext[cor1]{Corresponding author. Tel: +420 221912572. Fax: +420 221912577.}
\ead{hanus.home@gmail.com}

\author[label2]{M.~Delbo'}
\author[label1]{J.~\v Durech}
\author[label3]{V.~Al\' i-Lagoa}

\address[label1]{Astronomical Institute, Faculty of Mathematics and Physics, Charles University, V~Hole{\v s}ovi{\v c}k{\'a}ch 2, 18000 Prague, Czech Republic}
\address[label2]{Universit\'e C\^ote d'Azur, CNRS--Lagrange, Observatoire de la C\^ote d'Azur, CS 34229 -- F 06304 NICE Cedex 4, France}
\address[label3]{Max-Planck-Institut f\"ur extraterrestrische Physik, Giessenbachstra{\ss}e, Postfach 1312, 85741 Garching, Germany}

\begin{abstract}\boldmath
\textbf{By means of a varied-shape thermophysical model of Hanu\v s et al. (2015, Icarus, Volume 256) that takes into account asteroid shape and pole uncertainties, we analyze the thermal infrared data acquired by the NASA's Wide-field Infrared Survey Explorer of about 300 asteroids with derived convex shape models. We utilize publicly available convex shape models and rotation states as input for the thermophysical modeling. For more than one hundred asteroids, the thermophysical modeling gives us an acceptable fit to the thermal infrared data allowing us to report their thermophysical properties such as size, thermal inertia, surface roughness or visible geometric albedo. This work more than doubles the number of asteroids with determined thermophysical properties, especially the thermal inertia. In the remaining cases, the shape model and pole orientation uncertainties, specific rotation or thermophysical properties, poor thermal infrared data or their coverage prevent the determination of reliable thermophysical properties. Finally, we present the main results of the statistical study of derived thermophysical parameters within the whole population of main-belt asteroids and within few asteroid families. Our sizes based on TPM are, in average, consistent with the radiometric sizes reported by Mainzer et al. (2016, NASA PDS, Volume 247). The thermal inertia increases with decreasing size, but a large range of thermal inertia values is observed within the similar size ranges between D$\sim$10--100 km. We derived unexpectedly low thermal inertias ($<$20\,J\,m$^{-2}$\,s$^{-1/2}$\,K$^{-1}$) for several asteroids with sizes $10<D<50$~km, indicating a very fine and mature regolith on these small bodies. The thermal inertia values seem to be consistent within several collisional families, however, the statistical sample is in all cases rather small. The fast rotators with rotation period $P\lesssim4$ hours tend to have slightly larger thermal inertia values, so probably do not have a fine regolith on the surface. This could be explained, for example, by the loss of the fine regolith due to the centrifugal force, or by the ineffectiveness of the regolith production (e.g., by the thermal cracking mechanism of Delbo' et al. 2014, Nature, Issue 508).}
\unboldmath
\end{abstract}
 
\begin{keyword}
Asteroids \sep Asteroids, surfaces \sep Infrared observations \sep Photometry \sep Numeric
\end{keyword}

\end{frontmatter}


\section{Introduction}\label{introduction}

The recent availability of thermal infrared data obtained by the NASA's Wide-field Infrared Survey Explorer \citep[WISE,][]{Wright2010} opens exciting possibilities of determining surface characteristics of thousands of minor bodies of our solar system \citep{Mainzer2011a}. This characterisation can be performed by the analysis of WISE data by thermophysical models (hereafter TPM, see Sect.~\ref{sec:TPM}). WISE observations changed asteroid thermophysical modeling from being limited by the availability and accuracy of thermal infrared data to being limited by the availability of the a priori information required by the TPMs, namely the spin and shape solutions \citep{Koren2015}. This is why early TPM analyses of WISE data focused on a small number of objects: of near-Earth asteroids (NEAs) (341843) 2008 EV5 \citep{AliLagoa2014} and (29075) 1950 DA \citep{Rozitis2014b}, and four main-belt asteroids (MBAs), four NEAs and 1 Trojan \citep{Hanus2015a}.

The main object characteristics that one aims to determine are thermal inertia $\Gamma$, Bond albedo $A$ (or geometric visible albedo $p_\mathrm{V}$), surface roughness $\overline\theta$ and volume-equivalent diameter (i.e., the diameter of a sphere with the same volume as of the asteroid shape model). The shape can be elongated and in general quite different compared to a sphere. 
Surface roughness at a scale bigger than the typical diurnal heat propagation distance (few mm to few cm) causes a surface to emit thermal radiation in a non Lambertian way \citep{Lagerros1998, Rozitis2011, Delbo2015}. In particular, the absorbed solar flux is preferentially radiated back to the sun, a phenomenon that is called thermal infrared beaming. 
Thermal inertia is defined as a function of the density of the surface regolith $\rho$, thermal conductivity $\kappa$, and heat capacity $C$: $\Gamma= (\rho\kappa C)^{1/2}$, and measures the resistance of a material to temperature change, and thus controls the temperature distribution of the surface of an atmosphere-less body. 
Non-zero thermal inertia breaks the symmetry of the temperature distribution on asteroids. So, $\Gamma$ directly controls the strength of the Yarkovsky effect, which is the rate of change in the semi-major axis of the orbit of an asteroid ($\mathrm{d}a/\mathrm{d}t$) due to the recoil force of the thermal photons \citep[see, e.g.,][]{Bottke2006, Vokrouhlicky2015}. 

In the case of asteroids, values between almost zero or below 10 to almost 1000 J\,m$^{-2}$\,s$^{-1/2}$\,K$^{-1}$ have been derived (see Table~\ref{tab:literature}). The lowest values are typical for very large asteroids \citep[$D>100$ km,][]{MullerPHD}, large Trojans \citep{Muller2010, Horner2012} and large trans-Neptunian objects (TNOs) \citep{Lellouch2013}. These low thermal inertia values have been interpreted as due to very fine and mature regolith or even fluffy surfaces with extremely high porosities \citep[e.g.,][]{Vernazza2012, Lellouch2013}. Most $D>100-200$ km MBAs have a thermal inertia of the order of few tens of J\,m$^{-2}$\,s$^{-1/2}$\,K$^{-1}$. On the other hand, much smaller NEAs (sizes from several hundred meters to few kilometers) have thermal inertia values of the order of several hundreds \citep{Delbo2015}. However, there are almost no thermal inertia determinations for MBAs in a size range of 10--100 km. Our current study fills this gap. 

The findings concerning different thermal inertia values between small and large asteroids were later confirmed by the work of \citet{Gundlach2013}: in particular, asteroids with sizes smaller than 100 km in diameter were found to be covered by relatively coarse regolith grains with typical particle sizes in the millimeter to centimeter regime, whereas large asteroids (with diameters bigger than 100 km) possess very fine regolith with grain sizes between 10 and 100 microns. Modeling by \citet{Rozitis2014b} suggested a lunar-like thermal inertia characteristic of fine surface regolith on a 1-km NEA. Presence of cohesion forces could prevent the escape of the fine particles driven by the solar wind pressure and the centrifugal force from the surface. So, thermal inertia correlates with the regolith grain size \citep{Gundlach2013}. Particularly, objects covered with a very fine regolith (for instance, grain sizes between 10 and 100 microns on asteroids larger than 100 km) have typical values of thermal inertia of the order of 10 J\,m$^{-2}$\,s$^{-1/2}$\,K$^{-1}$ (see the compilation in Table~\ref{tab:literature} for several examples). On the other hand, coarse regolith grains with typical particle sizes of millimeters to centimeters implies thermal inertia values of several hundred J\,m$^{-2}$\,s$^{-1/2}$\,K$^{-1}$ (typical for NEAs).

The database of the WISE thermal infrared asteroid observations, with their unprecedented photometric accuracy (often better than 1\%) not achievable by current ground-based telescopes and with no contamination by the Earth's atmosphere, can be analyzed by means of a TPM in order to derive thermal inertias for several hundreds of asteroids with known shape models. Typically, shape models are based on radar imaging or on inversion of photometric lightcurves. The lightcurve-based shape models are stored in the public Database of Asteroid Models from Inversion Techniques \citep[DAMIT\footnote{\url{http://astro.troja.mff.cuni.cz/projects/asteroids3D}},][]{Durech2010}.

Classically, a TPM is used with an a priori knowledge of the shape and the rotational state of the asteroid. However, the high precision of WISE data introduces a new challenge: as it was already noticed, the shape model plays a crucial role in the derivation of the asteroid physical parameters \citep{AliLagoa2014, Emery2014, Rozitis2014}. This motivated our recent study \citep{Hanus2015a}, where we introduced a {\em varied shape } TPM scheme (VS-TPM) that takes into account asteroid shape and pole uncertainties, and where we demonstrated its reliability on nine asteroids. Here we apply the VS-TPM method to all main-belt asteroids with lightcurve-based shape models and sufficient amount of thermal infrared data in WISE filters W3 and W4 (see Sect.~\ref{sec:thermal_data}).

We describe the thermal infrared fluxes obtained by the WISE satellite in Sect.~\ref{sec:thermal_data} and the shape models and the optical lightcurves used for their determination in Sect.~\ref{sec:photometry}. The VS-TPM is described in Sect.~\ref{sec:TPM} and applied to three hundred asteroids in Sect.~\ref{sec:results}. 
In Sect.~\ref{sec:interpretation}, we present the main findings of the statistical study of thermophysical parameters within the whole population of MBAs and within few asteroid families. We conclude our work in Sect.~\ref{sec:conclusions}.

\section{Data}\label{sec:data}

\subsection{Thermal infrared fluxes}\label{sec:thermal_data}

We make use of the data acquired by the WISE satellite \citep{Wright2010}, in particular the results of the NEOWISE project dedicated to the solar system bodies \citep[see, e.g., ][]{Mainzer2011a}. The thermal infrared data were downloaded from the WISE All-Sky Single Exposure L1b Working Database via the IRSA/IPAC archive \footnote{\url{http://irsa.ipac.caltech.edu/Missions/wise.html}} and processed in the same way as data used in our previous studies focused on asteroid (341843) 2008 EV$_5$ \citep{AliLagoa2014} and nine asteroids \citep{Hanus2015a}. Bellow, we briefly summarize our procedure, additional details can be found in papers mentioned above. 

We consider only thermal data from filters W3 and W4 (isophotal wavelengths at 12 and 22 $\mu$m) from the fully cryogenic phase of the mission, because these data are thermal-emission dominated, whilst the fluxes in filters W1 and W2 (isophotal wavelength at 3.4 and 4.6 $\mu$m) usually at least partially consist of reflected sunlight for typical main-belt objects.

Our selection criteria are based on a combination of criteria from \citet{Mainzer2011b}, \citet{Masiero2011}, and \citet{Grav2012a}. We obtained the reported observation tracklets from the Minor Planet Center (MPC) and used them for a cone search radius of $1''$ around the MPC ephemeris of the object when querying the IRSA/IPAC catalogs. We only consider data with artifact flags p, P, and 0, quality flags A, B, and C, and data with a magnitude error bars smaller than 0.25 mag. Moreover, we require the IRSA/IPAC modified Julian date to be within four seconds of the time specified by the MPC and that the data are not partially saturated. A positive match from the WISE Source Catalog within $6.5''$ around the tracklet indicates that there is an inertial source at a distance smaller than the point-spread function width of the W1 band. We consider that these data are contaminated if the inertial source fluxes are greater than 5\% of the asteroid flux and we remove them. We implement the correction to the red and blue calibrator discrepancy in W3 and W4 filters \citep{Cutri2012}.

We selected only datasets where we had at least 5 points in both W3 and W4 filters. Similarly as in \citet{Hanus2015a}, we increased the nominal error bars of the fluxes by factors 1.4 and 1.3 for the W3 and W4 data, respectively. To be more specific, we studied the consistency of the error bars within two WISE measurements of the same source in frames obtained 11 seconds apart from each other. Such double measurements were allowed due to a 10\% field overlap between two subsequent frames. Because 11 seconds is not enough time for rotation to explain the differences in the observed fluxes, the uncertainties are clearly underestimated and thus we should consider them. To account for that, we decided to enlarge the uncertainties the way they roughly followed the normal distribution. Similar results were inferred by Nathan Myrhvold \citep[][and private communication]{Myhrvold2017}. He analyzed all the double detections, while our subset contained about 400 of such detections.

We also utilized the thermal infrared fluxes obtained by the IRAS satellite \citep{Tedesco2002} as well. These data in four different filters (I1, I2, I3, I4, isophotal wavelengths 12, 25, 60 and 100 $\mu$m) were downloaded from the The Supplemental IRAS Minor Planet Survey \citep[SIMPS,][]{Tedesco2004}. We rejected fluxes in filter I4 due to their generally poor quality. The precision of the IRAS fluxes is usually about 10 times lower than of the WISE fluxes, which essentially means that the TPM fits will be governed by the WISE data.

For each asteroid, we downloaded the photometric data and shape solution from the DAMIT database. In several cases, we also obtained new lightcurve data from the Asteroid Lightcurve Data Exchange Format database \citep[ALCDEF\footnote{\url{http://alcdef.org}};][]{Warner2011d}.
For those asteroids that have new lightcurve data compared to the DAMIT version, we generated revised spin state and shape solutions using the lightcurve inversion \citep{Kaasalainen2001a,Kaasalainen2001b}. All these represent our "nominal" shape models ($\sim300$).
The references to the shape model publications are reported in Table~\ref{tab:TI}.

\subsection{Shape models and disk-integrated photometry}\label{sec:photometry}

A shape model is used as an \textit{a priori} information in thermophysical modeling of the infrared fluxes of asteroids described in the previous section. Our sample mostly consists of shape models already published and available in the public DAMIT database ($\sim$300). All shape models from the DAMIT database were derived by the lightcurve inversion method of \citet{Kaasalainen2001a} and \citet{Kaasalainen2001b} and are usually represented by a convex polyhedron with $\sim$1\,000 triangular facets. Sidereal rotation period, the pole orientation, the shape model as well as used lightcurves are all available in the DAMIT database. 

The overall shape model quality, which is dependent on the amount, type and variety of the photometric data used for the model determination, differs within our sample. Only recently, a quality flag $QF$ that reflects each shape model reliability was introduced in \citet{Hanus2018a}. Such a measure, available for each shape model, gives us an idea how well the shape model should represent the true shape of the asteroid (the high-quality shape models have $QF=3$, while rather coarse shape models have $QF=1$). Typical uncertainties in the pole orientation are $\sim$5--30 degrees \citep[e.g.,][]{Hanus2011}, where the lower values correspond to solutions based on large datasets of dense lightcurves ($QF=3$), whilst the larger values are typical for solutions mostly based on sparse measurements from astrometric surveys ($QF=$ 1--1.5). Often, the restricted geometry of observations due to asteroid's low inclination of its orbit does not allow us to break the symmetry of the inversion method \citep{Kaasalainen2006}, so two pole solutions with similar ecliptic latitudes $\beta$ and difference in ecliptic longitude $\lambda$ of $\sim180^{\circ}$ are typically available. This ill-posedness of the inversion methods is often called the pole ambiguity. The shape models within the ambiguous solutions are rather similar although differently oriented with respect to the observer. 

All shape models are based on dense-in-time and/or sparse-in-time photometry. Generally, combined datasets are used for the shape modeling. The dense data are typically obtained during one night and well sample the rotation period. On the other hand, the sparse data contain a few hundred individual calibrated measurements during $\sim$15 years and are obtained by several astrometric surveys. Sparse data sample various observing geometries, which helps to constrain the pole orientation despite their lower photometric quality. 

For 12 asteroids, new photometric dense lightcurves became available since their shape models were published. Such data allowed us to better constrain their shape models, and consequently decrease the uncertainty in the pole orientation. We applied the lightcurve inversion method to the updated photometric datasets and derived shape models by following the same procedures as in \citet{Hanus2011,Hanus2016a}. These revised shape models are then used in the TPM.

Moreover, for five asteroids, we derived their shape models for the first time. We present rotation states of these models and used photometry in Tables~\ref{tab:TI} and ~\ref{tab:observations}. We utilize these new models in the TPM as well.

\section{Thermophysical modeling (VS-TPM)}\label{sec:TPM}

In the classical TPM analysis a single shape model and spin solution per asteroid is used as an \textit{a priori} and fixed information and model parameters ($D$, $A$, $\Gamma$, $\overline\theta$) are varied until best fit between the calculated and the observed thermal infrared fluxes is obtained. However, this approach does not take into account the uncertainty of the shape and spin state solution. Here we use the varied-shape TPM (VS-TPM) introduced by \citet{Hanus2015a}. The VS-TPM consists of applying the classical TPM approach repeatedly using different (varied, bootstrapped) shape models of the same asteroids in order to map the uncertainty of the shape model and its rotation state on the values of the physical parameters. The varied shape models are generated by the lightcurve inversion method from bootstrapped photometric datasets. This implies that we need to possess the disk-integrated optical data that were used for the shape model determination. These data were downloaded from the DAMIT database.

The VS-TPM scheme consists of three steps:
 \begin{itemize}
  \item[1.] We bootstrap \citep{Press1986} the original photometric data (i.e., those downloaded from DAMIT or the updated dataset) in such way that we keep the original number of dense lightcurves in the dataset. We independently bootstrap also the sparse data: we randomly choose individual measurements within each lightcurve until we get a sparse lightcurve with the original number of individual measurements.
  \item[2.] We apply the lightcurve inversion method to each bootstrapped dataset and spin state solution and derive the shape model. We create 19 bootstrapped/varied shape models for each pole solution. We also add the nominal shape and spin pole solution, totaling to 20 varied shapes (and spin states) per object and per pole solution. Each shape model has its own spin solution that is usually consistent  within few degrees within the corresponding varied shapes.
  \item[3.] For each varied shape model and its rotational state, we perform the TPM analysis scheme the same way as for the nominal shape model (i.e., shape from the DAMIT). We run the VS-TPM for each pole solution individually.
 \end{itemize}

The CPU requirements are rather high: a TPM analysis scheme (the most time-consuming procedure) for one particular bootstrapped/varied shape model runs on a single CPU for about a day. We usually have two pole solutions for each asteroid, which results then in 40 individual TPM runs. We applied VS-TPM to about 300 asteroids. Additional details concerning the VS-TPM can be found in \citet{Hanus2015a}.

\paragraph[]{\eatpunct} We use a thermophysical model implementation of \citet{Delbo2007a, Delbo2004} that is based on TPM developed by \citet{Lagerros1996, Lagerros1997, Lagerros1998, Spencer1989, Spencer1990, Emery1998}. A TPM allows thermal infrared fluxes to be calculated at different wavelengths and at a number of epochs taking into account the shape of an asteroid, its spatial orientation, and a number of physical parameters such as the size of the body, the albedo $A$, the macroscopic surface roughness $\overline \theta$ (Hapke's mean surface slope), and the thermal inertia $\Gamma$. The values of the parameters are determined by minimizing the difference between the observed fluxes $f_i$ and the modeled fluxes $s^2F_i$, where we consider the scale factor $s$ for the asteroid size, and $i$ corresponds to individual observations. To find the optimal set of parameter values, we minimize the metric

\begin{equation}\label{chi2}
\chi^2 = \sum\frac{(s^2F_i-f_i)^2}{\sigma^2_i},
\end{equation}
where $\sigma_i$ represent the errors of fluxes $f_i$.

\paragraph[]{\eatpunct} The shape is represented either by a convex polyhedron with triangular facets (all models from the DAMIT database), or by a set of surfaces and normals (the so called Gaussian image). The latter representation is a direct output of the lightcurve inversion method \citep[see][]{Kaasalainen2001a} and its use in the TPM allows us to save a significant amount of computational time. According to our tests, both representations produce similar fluxes and could be considered equivalent. The shape model has usually a meaningless value of the initial size, and the parameter $s$, adjusted in the TPM fit to the data, is a factor that either scales linearly all vectors of the vertices of the polyhedron, or quadratically all the surfaces of the Gaussian image. The representation by the polyhedron gives us the {\em volume equivalent diameter} $D_\mathrm{V}$, while the representation by the Gaussian image leads to the {\em surface equivalent diameter} $D_\mathrm{S}$ (i.e. the diameter of a sphere with the same surface as the shape model scaled in size). However, we always transform $D_\mathrm{S}$ to $D_\mathrm{V}$ and present this quantity throughout this work. We use convex shape models so we do not need to take into account topographic shadowing effects and the heating due to the light reflected and emitted by facets on other facets. 

The usage of a convex shape model is motivated by the scarcity of available shape models with local surface features (based on radar measurements or spacecraft imaging). The ground-truth shape models and in-situ measurements of several asteroids that have been visited by spacecrafts offer a possible validation of the methods based on Earth-based measurements and convex shape models. For instance, \citet{ORourke2012} used a concave shape model of main-belt asteroid (21)~Lutetia based on Rosetta fly-by data \citep{Carry2010} as an input for a TPM and derived consistent (though more constrained) thermophysical properties as \citet{Muller2006} with a convex shape model of \citet{Torppa2003} as an input. Moreover, \citet{Rozitis2014} showed a good consistency between TPM results obtained for lightcurve- and radar-based shape models of NEA (1620)~Geographos. Similar consistency was also confirmed between the TPM results for near-Earth asteroid (101955)~Bennu \citep{Muller2012, Emery2014}. Furthermore, \citet{Rozitis2013} claim that large concavities are not always resulting in large self-heating effects for NEAs, it depends on the observational geometry and the aspects of the body being sampled in the disc-integrated data. Then, for MBAs, because they are colder and hence self-heating effects are much weaker (flux $\propto T^4$), we do not expect that neglecting this effect should have any measurable systematic impact on the sample of derived thermal inertias. It could, however, affect particular cases but this will have to wait for more ground-truth knowledge of MBAs with significant concavities to make a more quantitative argument. Therefore, we expect that the typical concavities as observed by the few spacecraft missions do not significantly affect the TPM results. We note that the largest asteroids usually lack the low-scale surface features, whilst the presence of concavities and their size increase with decreasing size (with some peculiar exceptions such as some asteroids rotating close to the disruption limit that are rather spherical). 

\paragraph[]{\eatpunct} The effect of roughness on the thermal infrared flux is accounted for by adding a spherical-section crater to each surface element of the shape, in which shadowing of crater facets on other facets and mutual heating are taken into account. The crater with an opening angle $\gamma_{\mathrm{c}}$ and the crater areal density with respect to the flat part of the surface element $\rho_{\mathrm{c}}$ can be varied from 0 to 90$^{\circ}$ and from 0 to 1, respectively, to cover different strength of the roughness. We calculate the TPM scheme for a set of ten roughness models, whose parameters are given in Table~\ref{tab:rough}. The correspondence between the Hapke's mean surface slope $\overline \theta$ and the adopted values of $\gamma_{\mathrm{c}}$ and $\rho_{\mathrm{c}}$ is also given in Table~\ref{tab:rough}. We compute $\overline \theta$ following the definition of \citet{Hapke1984}:

\begin{equation}\label{eq:thetabar}
 \tan \overline \theta = \frac{2}{\pi} \int^{\pi/2}_{0} \tan \theta \, a(\theta) \, \mathrm{d}\theta
\end{equation}
where $a(\theta)$ is a function describing the distribution of the tilts $\theta$ in the crater and in the flat surface element to which the crater belongs, weighted by the fractional area covered by the crater compared to that of the flat surface. In Eq.~(\ref{eq:thetabar}), $\theta$ is the angle of a given facet from horizontal, and $a(\theta)$ is the distribution of surface slopes \citep[see also][]{Emery1998}. The mean surface slope does not depend on the illumination, but it is an intrinsic property of the surface.

\begin{table}
\caption{\label{tab:rough}Ten different values of surface roughness used in the TPM. The table gives the opening angle $\gamma_{\mathrm{c}}$, the crater areal density $\rho_{\mathrm{c}}$, the Hapke's mean surface slope $\overline \theta$, and our designation.}
\centerline{
\begin{tabular}{cccc} \hline
$\gamma_{\mathrm{c}}$ & $\rho_{\mathrm{c}}$ & $\overline \theta$ & Designation \\ \hline\hline
0   & 0.0 &  0.0 & No roughness \\
30  & 0.3 &  3.9 & Low roughness \\
40  & 0.7 & 12.6 & Medium roughness \\
41  & 0.9 & 16.5 & Medium roughness \\
50  & 0.5 & 12.0 & Medium roughness \\
60  & 0.9 & 26.7 & High roughness \\
70  & 0.7 & 27.3 & High roughness \\
90  & 0.5 & 38.8 & High roughness \\
90  & 0.7 & 50.1 & Extreme roughness \\
90  & 0.9 & 58.7 & Extreme roughness \\ \hline
\end{tabular}
}
\end{table}


\paragraph[]{\eatpunct} Instead of explicitly calculating the heat diffusion within craters \citep{Delbo2009}, the analytical approximation of \citet{Lagerros1998} is used. The Lagerros crater approximation is applicable if the thermal infrared fluxes were obtained at lower phase angles ($\lesssim30-40^\circ$), which is the case of the WISE data of most MBAs. According to our tests, the differences in the fluxes produced by these two models are usually smaller than 1\%.

\paragraph[]{\eatpunct} We apply the color correction to model fluxes the same way as in \citet{AliLagoa2014, Hanus2015a}.

\paragraph[]{\eatpunct} We run the TPM model for different values of the thermal inertia $\Gamma \in (0, 2500)$  J\,m$^{-2}$\,s$^{-1/2}$\,K$^{-1}$. For each value of the surface roughness, we run the TPM for the thermal inertia $\Gamma=2500$ J\,m$^{-2}$\,s$^{-1/2}$\,K$^{-1}$ and the Bond albedo $A=0.08$, and get the first size estimate $D$. Before each following TPM run (while keeping the same surface roughness), we first compute the new value of the Bond albedo $A$ from the equation \citep[see, e.g.,][]{Harris2002}

\begin{equation}\label{eq:D}
D (\mathrm{km}) = \frac{1329}{\sqrt{p_{\mathrm{V}}}}\,10^{-0.2H},
\end{equation}
where we use diameter $D$ determined in the previous TPM run, and where the visible geometric albedo $p_{\mathrm{V}}$ can be expressed via $A = q\,p_{\mathrm{v}}$, where $q=0.290+0.684G$ is the phase integral \citep{Bowell1989}. We utilize the values of absolute magnitudes $H$ and slopes $G$ from the Asteroid absolute magnitude and slope catalog\footnote{\texttt{http://wiki.helsinki.fi/display/PSR/Asteroid+absolu\-te+magnitude+and+slope}} \citep[AAMS,][]{Muinonen2010, Oszkiewicz2011}. We then run the TPM model with decreasing values of $\Gamma$ until $\Gamma=0$. Following each step, we always recompute the $A$ value. The same procedure is performed for ten different values of the surface roughness.

\paragraph[]{\eatpunct} We set emissivity to $\epsilon=0.9$ as this is appropriate for objects with surfaces that emit a substantial portion of their thermal-infrared radiation shortwards of 8 $\mu$m \citep{Lim2005a}. 

\paragraph[]{\eatpunct} We do not account for the uncertainty in the $H$~and $G$ values. According to our tests (TPM with different $H$ values), a change of $\pm0.5$ mag in $H$ is compensated by the change of the Bond albedo $A$ (or $p_{\mathrm{V}}$). However, the size remains similar (see Eq.~\ref{eq:D}). Moreover, the thermal inertia is not sensitive on $H$ either.

\paragraph[]{\eatpunct} Due to the uncertainty in the rotation period, we have the typical uncertainty in the rotation phase of the asteroid of about 10$^{\circ}$. Therefore, we treat the initial rotation phase $\phi_0$ as a free parameter of the model to be adjusted by the best fitting procedure. We run the TPM scheme with different $\phi_0$ values from the expected interval (usually $\pm10^{\circ}$) given by the uncertainty in the rotation period with a step of 2$^{\circ}$.

\paragraph[]{\eatpunct} After scanning the parameter space of thermal inertia, surface roughness, initial rotational phase and Bond albedo, we find the solution with the lowest $\chi^2$ value. 

\paragraph[]{\eatpunct} To determine the uncertainties of thermal inertia $\Gamma$, diameter $D$ and Bond albedo $A$ for a TPM solution with a single shape model (i.e., the model from DAMIT), we utilize the standard statistical tools based on $\chi^2$ values. This approach for the uncertainty determination described, for example, in \citet{Press1986} has been commonly used \citep[see, e.g., ][]{AliLagoa2014, Emery2014, Hanus2015a}. First, we find all solutions within the 1$\sigma$ confidence interval, i.e., solutions with $\chi^2<(\chi^2_{\mathrm{min}}+\sqrt{2\nu})$, where we consider $\sqrt{2\nu}\sim\sigma$ and $\nu$ is the effective number of degrees of freedom. The range of possible solutions gives us then the upper and lower bounds of the derived parameter uncertainties. However, this approach gives reliable uncertainties if the $\chi^2$ values of the best-fitting solutions are comparable to the number of degrees of freedom (i.e., if the reduced $\chi^2\sim 1$), which is often not the case. Unfortunately, the contribution of the uncertainty of the shape model and the rotation state to the uncertainties of derived thermophysical properties is completely ignored here.

\paragraph[]{\eatpunct} To overcome the difficulties with the unrealistic uncertainties when the $\chi^2$ values are high, we estimate the uncertainties of the searched parameters by an empirical approach: we accept all solutions with $\chi^2<(\chi^2_{\mathrm{min}}+\chi^2_{\mathrm{min}} \sqrt{2\nu})$. This follows the procedure applied in \citet{Hanus2015a}.

\paragraph[]{\eatpunct} We prefer to use reduced chi-square values $\chi^2_{\mathrm{red}} = \chi^2 / \nu$, which are more illustrative than the ``non-reduced'' $\chi^2$. 

\paragraph[]{\eatpunct} To estimate the uncertainties of derived thermophysical properties ($\Gamma$, $D$ or $A$) based on the VS-TPM, we use an empirical method. We only consider the best-fitting solution for each varied shape. We find the smallest range within 14 solutions (corresponds to $\sim$68\% of the total number of 20). For example, each varied shape gives us a thermal inertia value. We sort these values and find the lowest range given by 14 thermal inertia values and compute the mean value. This usually rejects the most extreme values of $\Gamma$. Values of the thermophysical parameters with their uncertainties estimated by this method are included in Table~\ref{tab:TI}. 

\section{Results and discussion}\label{sec:results}

\subsection{VS-TPM of three hundred asteroids}\label{sec:TPM_modeling}

\begin{figure*}[!htbp]
\begin{center}
\resizebox{\hsize}{!}{\includegraphics{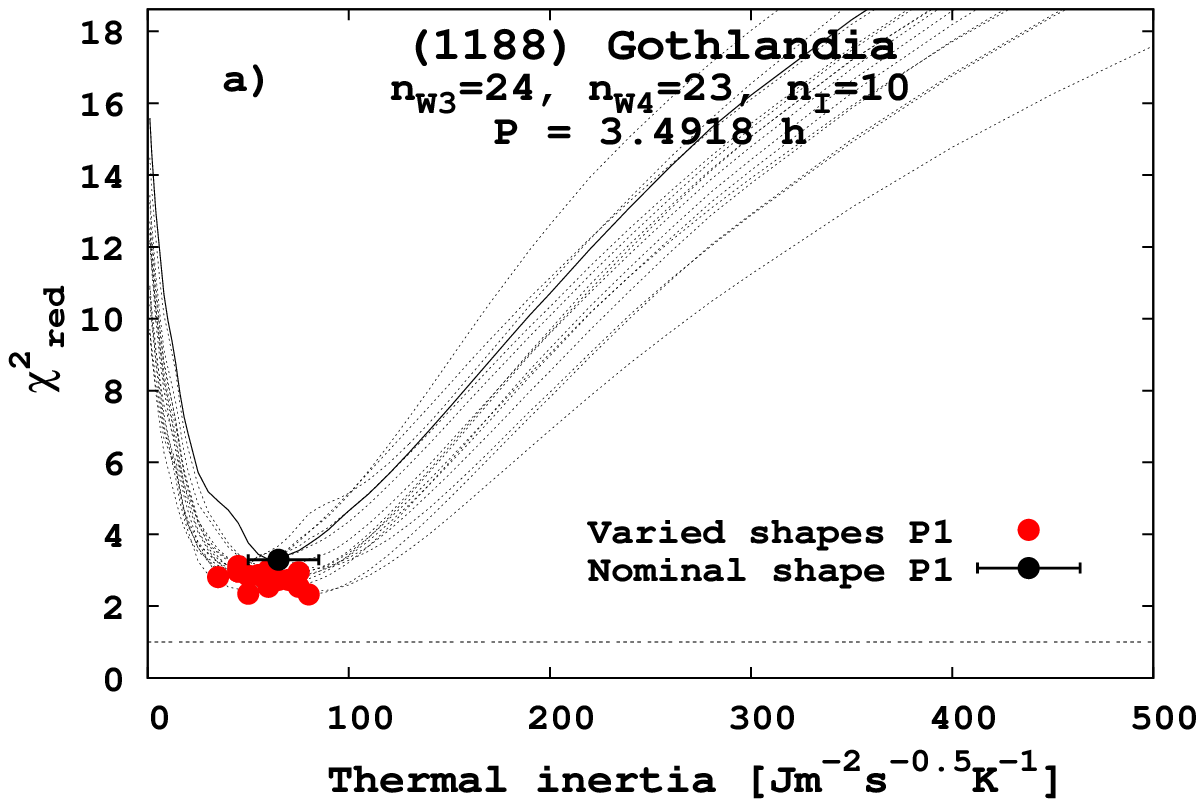}\includegraphics{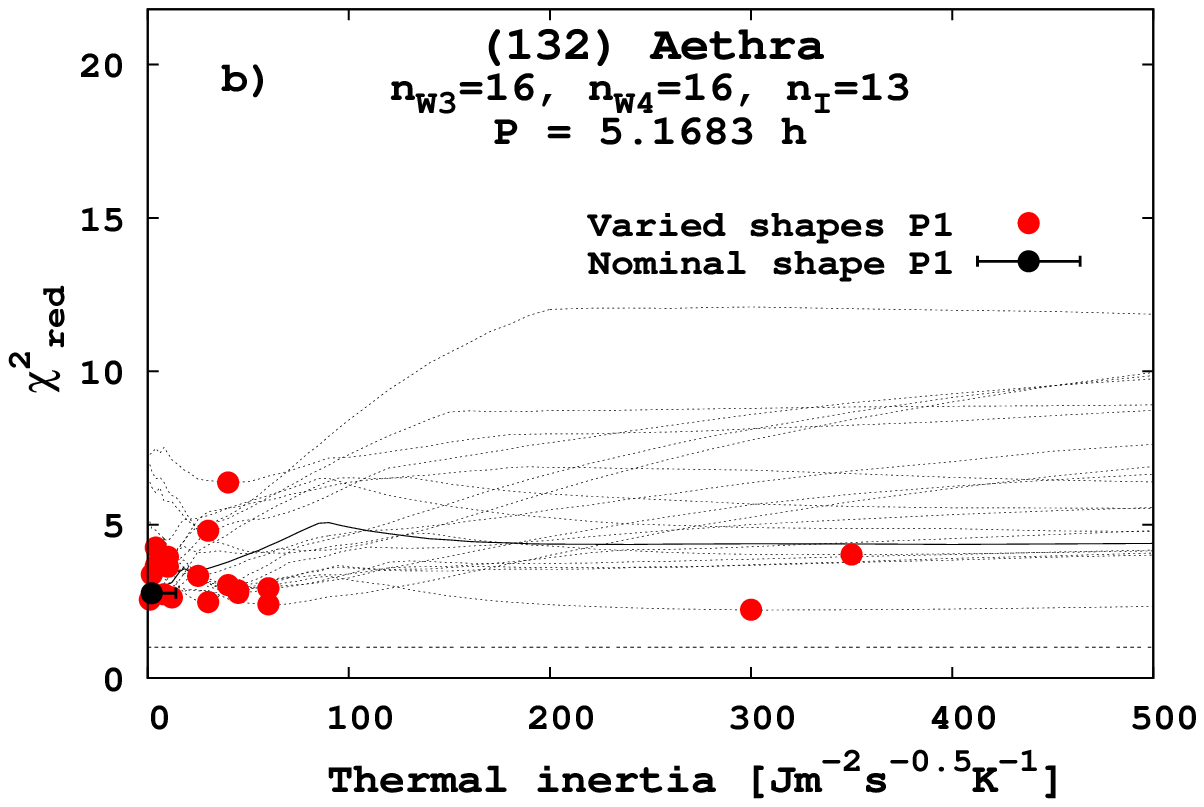}}\\
\resizebox{\hsize}{!}{\includegraphics{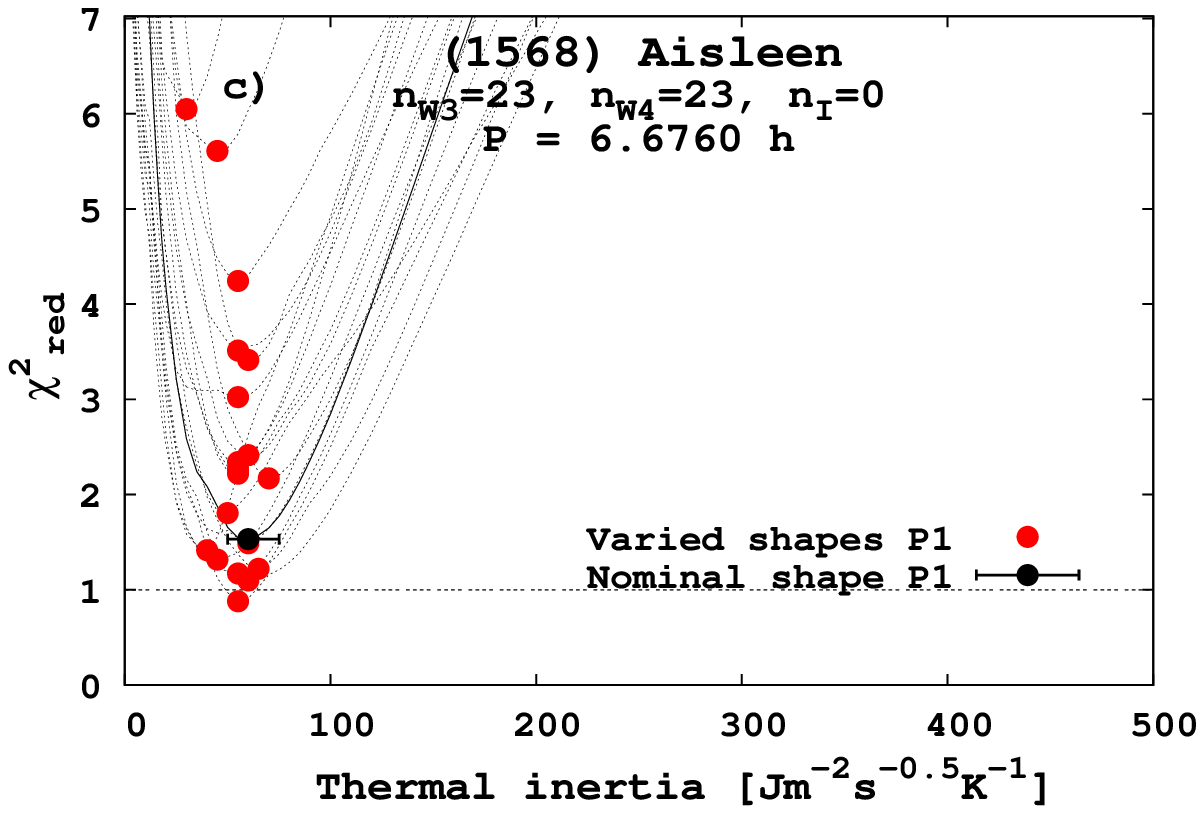}\includegraphics{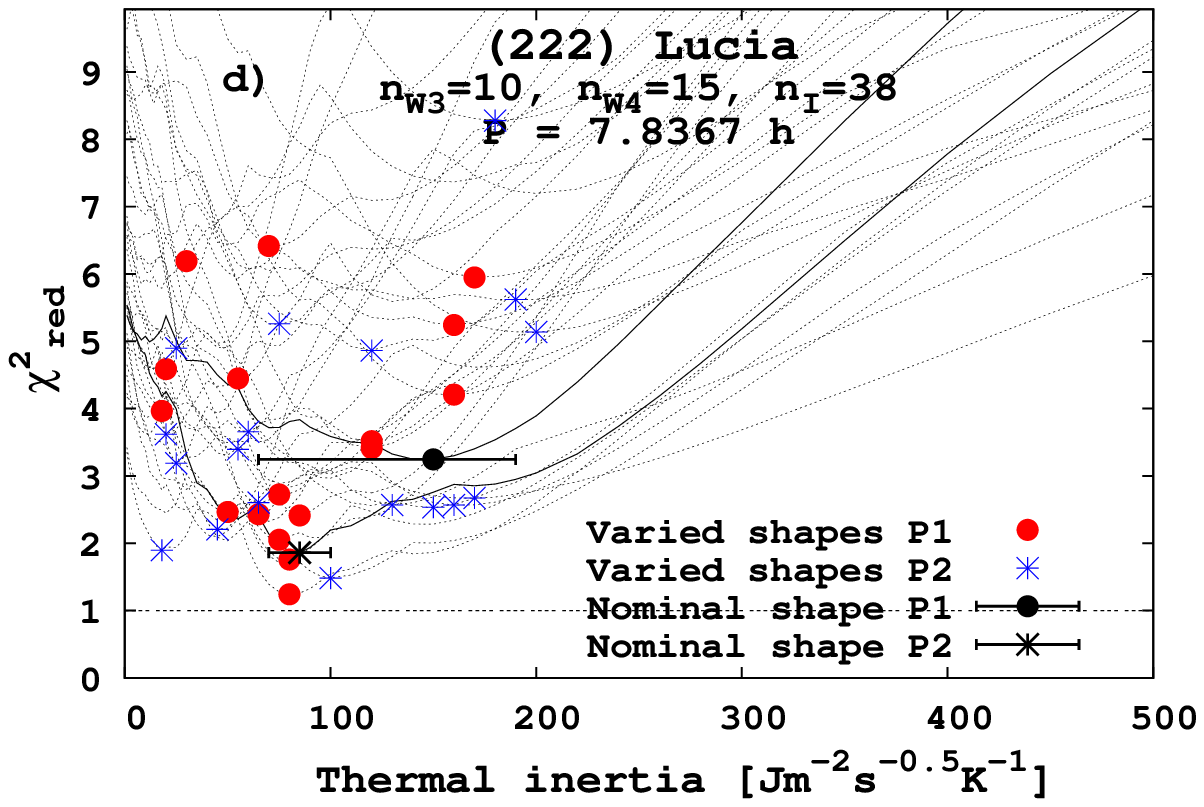}}\\
\resizebox{\hsize}{!}{\includegraphics{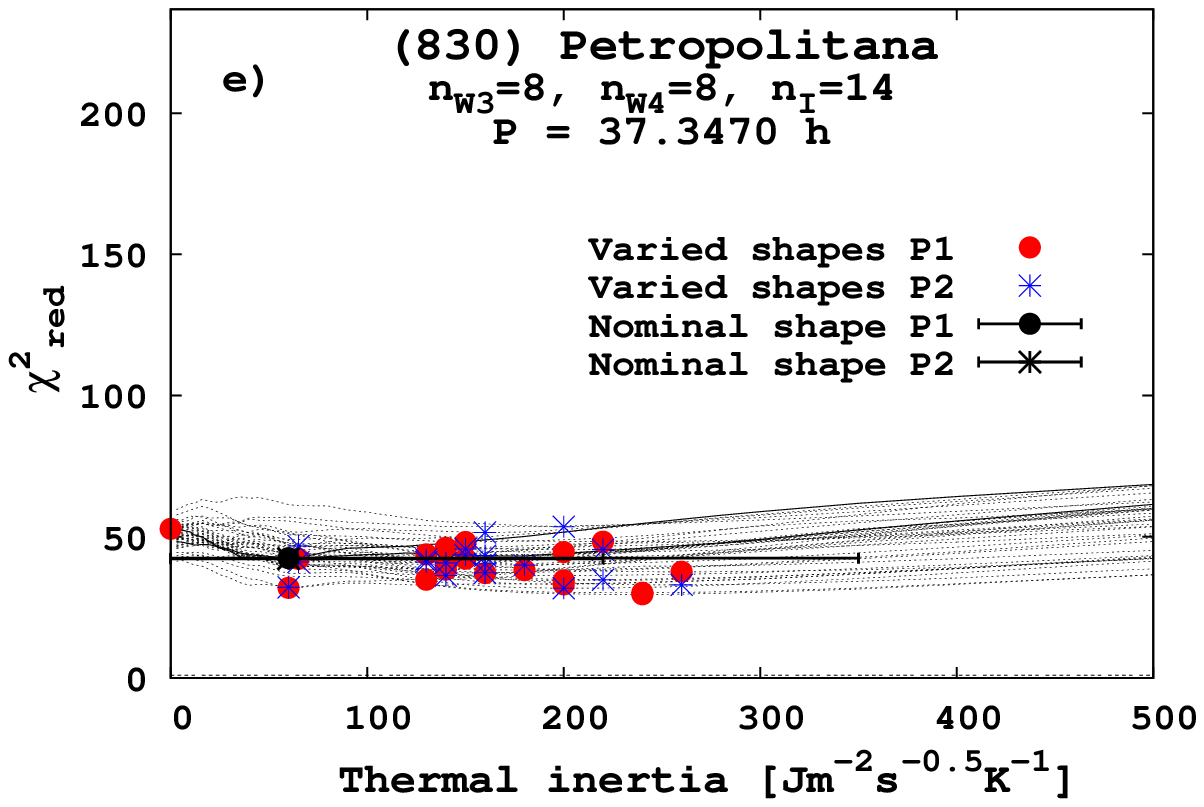}\includegraphics{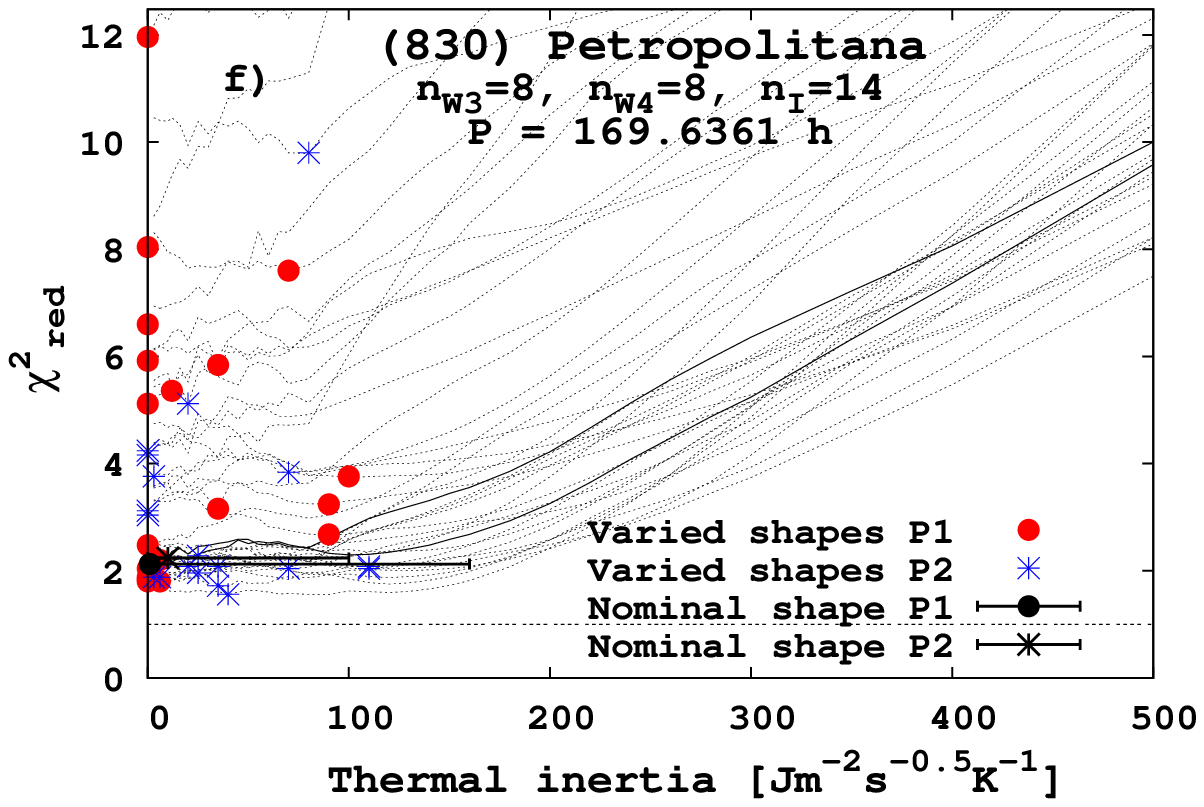}}\\
\end{center}
\caption{\label{fig:TPM_examples}Typical VS-TPM fits in the thermal inertia parameter space. We illustrate here five (a-e) qualitative results discussed in the text. Each plot also contains the number of thermal infrared measurements in WISE W3 and W4 filters and in all IRAS filters, and the rotation period. Case f) corresponds to the corrected/revised shape and spin solution of asteroid (830)~Petropolitana. P1 and P2 corresponds to ambiguous pole solutions.}
\end{figure*}

We applied the VS-TPM to $\sim280$ asteroids with shape models and sufficient amount of thermal infrared data from WISE. We obtained five qualitative VS-TPM results, which are illustrated in Fig.~\ref{fig:TPM_examples}:
 \begin{itemize}
  \item[a)] TPM solutions for all/most varied shapes are similar, they have reasonably low $\chi^2_\mathrm{red}$ values and a prominent minimum in $\Gamma$ (Fig.~\ref{fig:TPM_examples}a).
  \item[b)] TPM solutions for all/most varied shapes are similar, they have reasonably low $\chi^2_\mathrm{red}$ values, however, without a prominent minimum in $\Gamma$. Thermophysical properties cannot be constrained (Fig.~\ref{fig:TPM_examples}b).
  \item[c)] TPM solutions for all/most varied shapes exhibit similar trends (i.e., a qualitatively similar shape of the minimum) with a prominent minimum in $\Gamma$, but the range in the $\chi^2_\mathrm{red}$ is large ($\sim$5--10). Hence, the quality of the TPM fits depends on the varied shapes. However, the thermophysical parameters are consistent and well constrained across the varied shapes solutions (Fig.~\ref{fig:TPM_examples}c).
  \item[d)] TPM solutions for varied shapes are different, which suggests that the shape model uncertainty is important. Reliable thermophysical properties cannot be derived (Fig.~\ref{fig:TPM_examples}d).
  \item[e)] All TPM solutions have large $\chi^2_\mathrm{red}$ values, so the thermophysical properties are not reliable (Fig.~\ref{fig:TPM_examples}e). 
 \end{itemize}

Reliable thermophysical properties can be obtained in cases a) and c). Therefore, we only present these results in Table~\ref{tab:TI}). These cases represent $\sim$120 out of the initial $\sim$280 asteroids, for which we applied the VS-TPM. Most of the remaining cases fall to the b) ($\sim50$) and d) ($\sim100$) categories and only a few to the e) category.

Case b) usually happens if the thermal infrared dataset is rather small and/or with large uncertainties, the rotation period is too long ($\gtrsim15$ h) or the geometry of observation is close to pole-on. 

The category d) represents asteroids, for which the TPM fits are dependent on the varied shapes, so on the uncertainty in the shape model. Future shape model refinement by additional optical photometry would be necessary for obtaining a useful TPM solution.

The category e) correspond to cases likely affected by systematic effects. These could include incorrect rotation state (e.g., rotation period) or the shape, or incorrect fluxes (e.g., there could be a substantial offset of an individual flux measurement or we made an error in the data processing). As an example, we show asteroid (830)~Petropolitana in Fig.~\ref{fig:TPM_examples}e. The shape model presented in \citet{Hanus2016a} is likely wrong, because the shape and spin state solution was searched near the rotation period of $\sim$39 hours available at the CdR\&CdL\footnote{\url{http://obswww.unige.ch/~behrend/page\_cou.html}} database. However, the true period seems to be much longer ($\sim$169 h, personal communication with Dagmara Oszkiewicz and recent unpublished observations of Brian Skiff). We replaced the incorrect solution in DAMIT by the revised one and repeated the VS-TPM scheme. We obtained a significantly better fit to the thermal infrared data, however, the minimum in $\Gamma$ is rather broad, so thermal inertia cannot be constrained (Fig.~\ref{fig:TPM_examples}f).

In our figures of $\Gamma$ vs. $\chi^2_{\mathrm{red}}$ (see Fig.~\ref{fig:TPM_examples} and figures in the Appendix), we show the TPM solution with a nominal shape model as an input and for all (19) the varied shapes.

The standard uncertainty values based on the $\chi^2$ statistics are reliable only for a sub-sample of our studied asteroids. For the remaining asteroids, the $\chi^2_{\mathrm{red}}$ values are higher than $\sim1$ (see column 15 in Table~\ref{tab:TI}). We utilized here the semi-empirical approach for the uncertainty estimation described in Sect.~\ref{sec:TPM}.


The $\chi^2_{\mathrm{red}}$ values of $\sim2-5$ for the best-fitting set of parameters indicate that we are not fitting the thermal fluxes by our model very well. However, if the fit has a prominent minimum in $\Gamma$ and if the solution is consistent within the varied shapes, we consider such solution as reliable. The rather worse fit to the thermal data could be explained by two main reasons:
(i)~The tabulated uncertainties of the WISE data may not correspond to the 1$\sigma$ standard errors. We already increased the error bars based on the comparison of measurements taken in consecutive exposures \citep[see discussion in][Fig.~8]{Hanus2015a}, but there could be other biases that remain unaccounted for. Moreover, possible systematic errors in the WISE fluxes could be present, however, we do not have control on them other than that clear outliers could be identified.
(ii)~Model uncertainties (convex shape, pole orientation, systematics of the TPM) dominate over the flux uncertainties. 

For most asteroids, the VS-TPM can produce a better fit than the classical TPM with the published shape model. All varied shape models, although different in the goodness of thermal IR data fit, are indistinguishable in terms of reproducing the visible photometry. On the other hand, our method is not a true optimization, it rather maps the uncertainties in the shape and the pole orientation and their influence on the TPM results. 

Selecting 14 out of 20 varied-shape solutions for the estimation of fitted parameters and their uncertainties allowed us to reject potential unrealistic solutions that could originate from specific (peculiar) bootstrapped photometric datasets.

Another intriguing result is the fact that we obtained TPM fits with a large range of $\chi^2_{\mathrm{red}}$ values for a number of studied asteroids, however, with similar appearance of the minimum in $\Gamma$ (Fig.~\ref{fig:TPM_examples}c). This suggests a substantially different quality of the fits within the varied shape models on one side, and a stable TPM solution in $\Gamma$ on the other side. Some of the shape models produce $\chi^2_{\mathrm{red}}$ values larger by a factor up to 5--10 than the best fitting varied shape model. We illustrate such TPM fits for asteroid (1568)~Aisleen in Fig.~\ref{fig:TPM_examples}c. It is clear that the shape model (together with the pole orientation) is an important limiting factor for the quality of the fit, and so for a number of varied shapes we do not fit the thermal data well. On the other hand, we suspect that the quality of the TPM fit could be improved in the future when more realistic shape models will become available. The results of investigation of the stability of the TPM solution with respect to the shape model variations raise the confidence that for most asteroids presented in this study, the thermophysical solution is stable, and so derived thermophysical properties are realistic.




\section{Interpretation of derived thermophysical properties}\label{sec:interpretation}

\subsection{Sizes}

\begin{figure}
\begin{center}
\resizebox{\hsize}{!}{\includegraphics{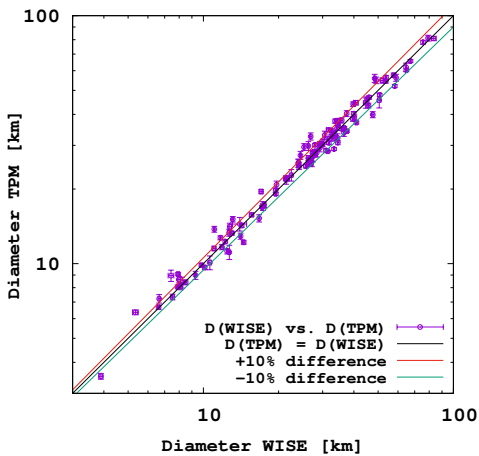}}\\
\end{center}
\caption{\label{img:WISE_D}Comparison between our sizes derived by the VS-TPM and radiometric sizes based on NEATM from \citet{Mainzer2016}. Both methods use the same thermal infrared datasets.}
\end{figure}

\begin{figure*}
\begin{center}
\resizebox{\hsize}{!}{\includegraphics{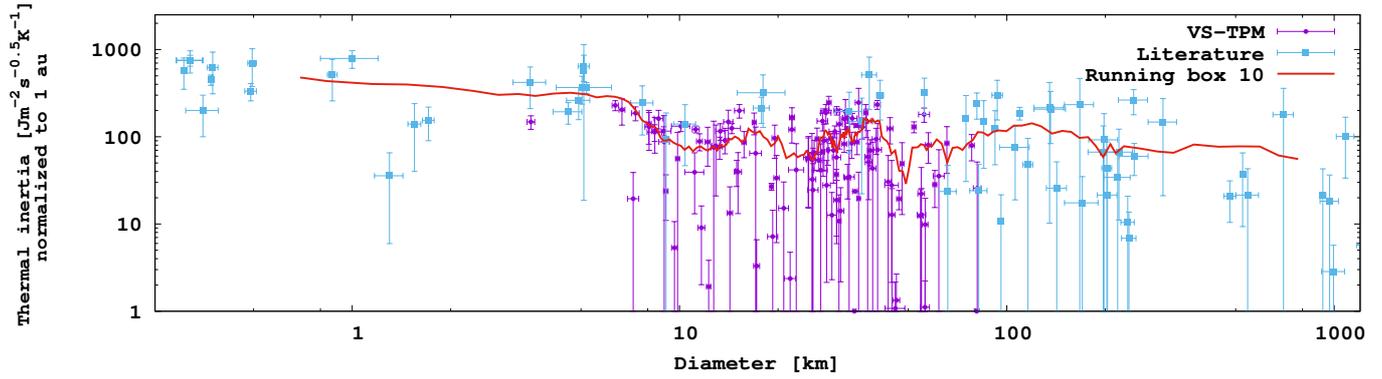}}\\
\end{center}
\caption{\label{img:gamma}Dependence of the thermal inertia $\Gamma$ on asteroid diameter $D$. We included our estimates (circles) and adopted literature values (squares).}
\end{figure*}

\begin{figure*}
\begin{center}
\resizebox{\hsize}{!}{\includegraphics{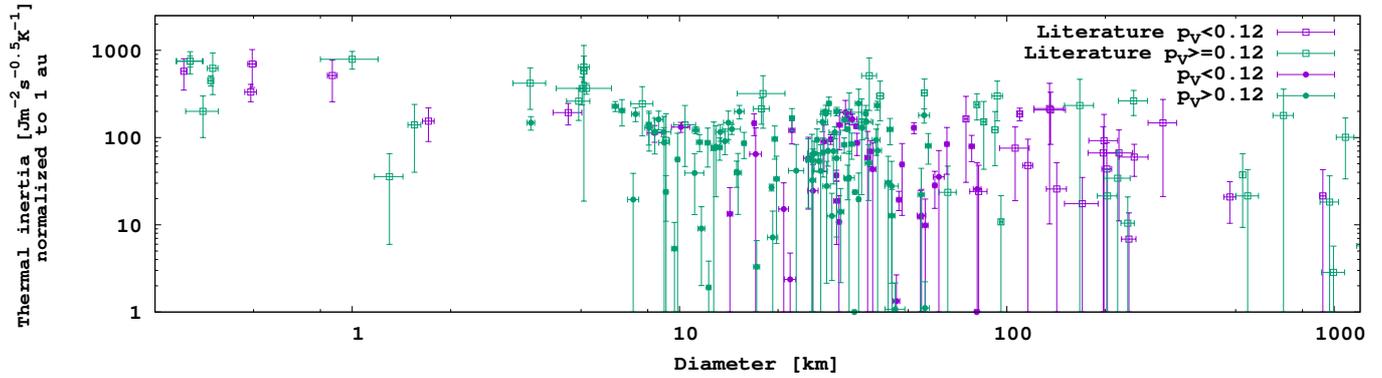}}\\
\end{center}
\caption{\label{img:albedo}Dependence of the thermal inertia $\Gamma$ on asteroid diameter $D$ for the low ($p_\mathrm{V}<0.12$) and high ($p_\mathrm{V}>0.12$) albedo objects. We included our estimates (circles) and adopted literature values (squares).}
\end{figure*}



\begin{figure*}
\begin{center}
\resizebox{\hsize}{!}{\includegraphics{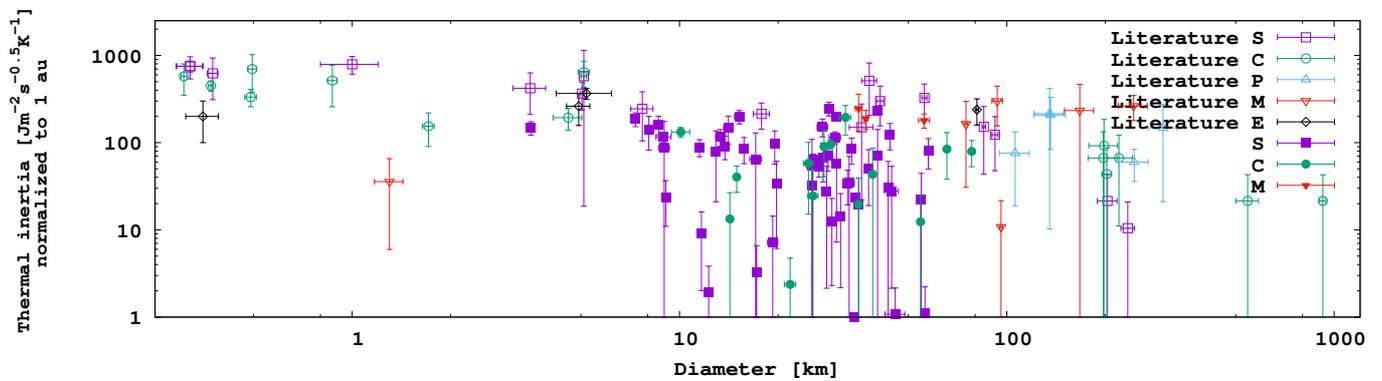}}\\
\end{center}
\caption{\label{img:taxonomy}Dependence of the thermal inertia $\Gamma$ on asteroid diameter $D$ for asteroids from S- and C-complexes, and P, M, and E types. We included our estimates and adopted literature values.}
\end{figure*}

\begin{figure*}
\begin{center}
\resizebox{\hsize}{!}{\includegraphics{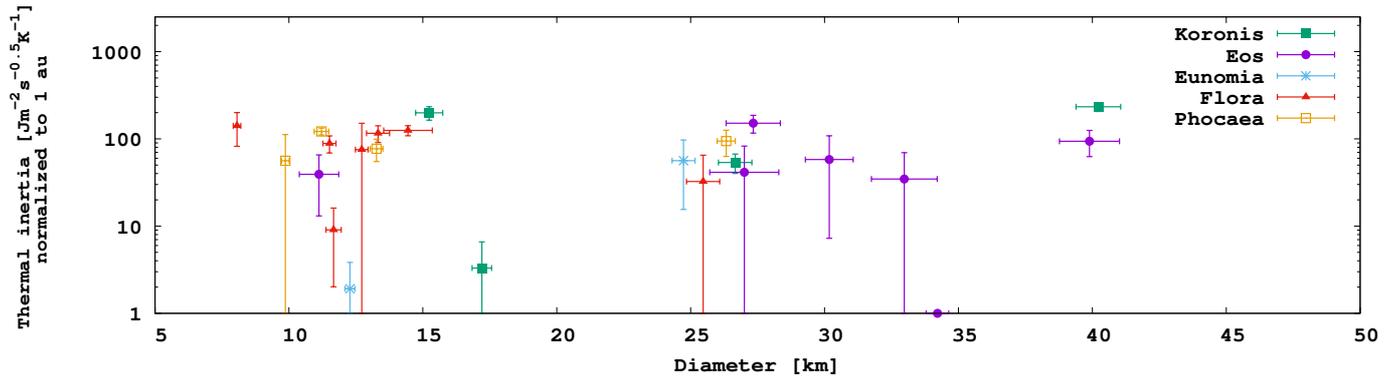}}\\
\end{center}
\caption{\label{img:families}Dependence of the thermal inertia $\Gamma$ on asteroid diameter $D$ for members of five collisional families -- Flora, Koronis, Eos, Eunomia and Phocaea.}
\end{figure*}

\begin{figure*}
\begin{center}
\resizebox{\hsize}{!}{\includegraphics{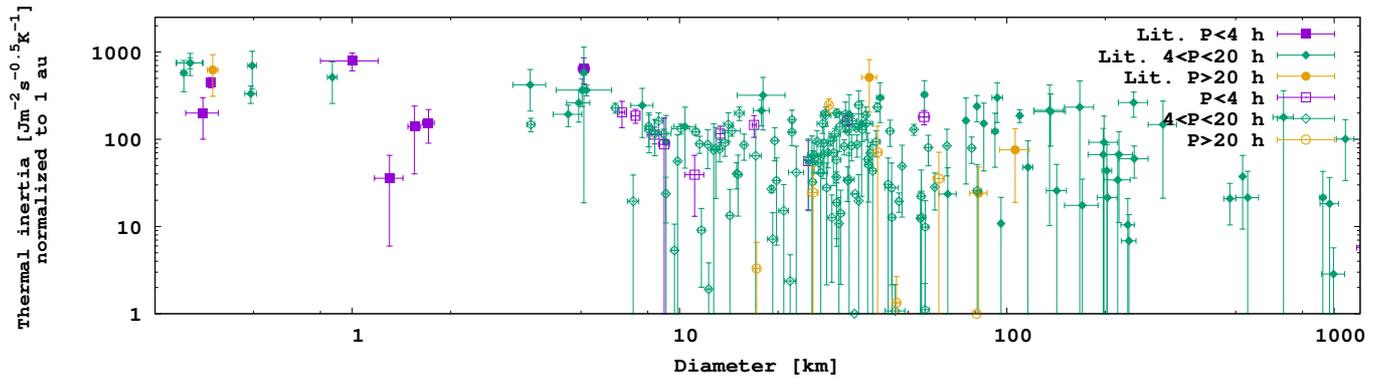}}\\
\end{center}
\caption{\label{img:rotators}Dependence of the thermal inertia $\Gamma$ on asteroid diameter $D$ for the fast ($P<4$ h), intermediate ($4<P<20$ h), and slow ($P>20$ h) rotators. We also included the literature values.}
\end{figure*}

\begin{figure*}
\begin{center}
\resizebox{\hsize}{!}{\includegraphics{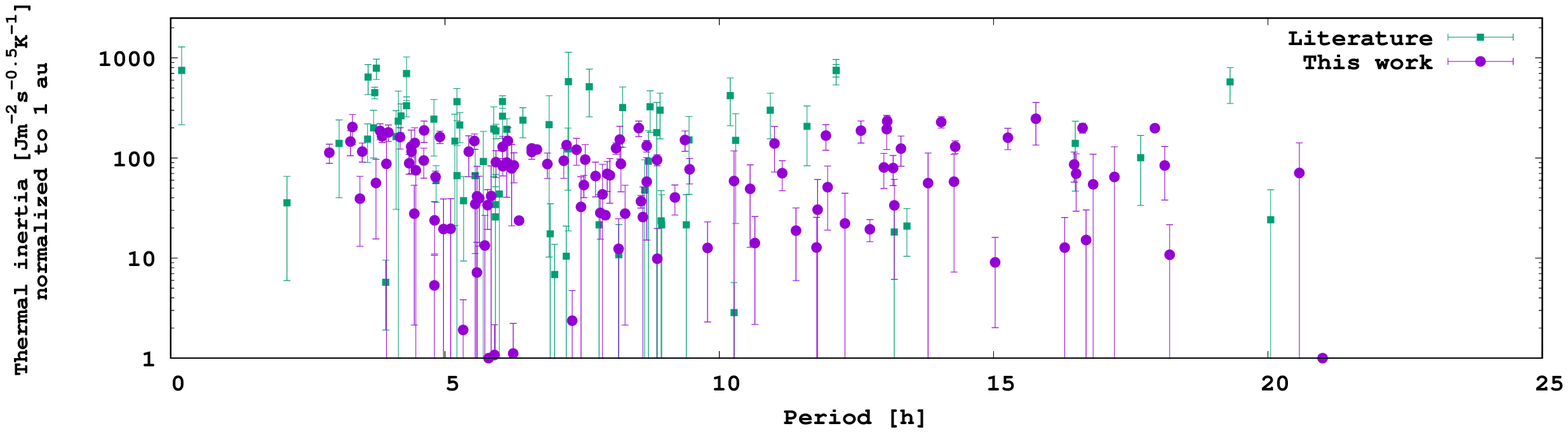}}\\
\end{center}
\caption{\label{img:rotators2}Dependence of the thermal inertia $\Gamma$ on asteroid rotation period $P$. We also included the literature values.}
\end{figure*}

We compared in Fig.~\ref{img:WISE_D} our diameters computed by the VS-TPM with the radiometric diameters reported by \citet{Mainzer2016}. Both diameters are based on the WISE thermal infrared data, so their differences of even 10\% has to originate in the different models utilized (i.e., of the TPM and NEATM). It is obvious that the spherical shape model used in the NEATM is a crude approximation, especially for elongated objects, because their projected sizes strongly depends on the geometry of the observation. If we study the asteroid populations from a global perspective, where we do not need accurate values for individual objects, the radiometric sizes still represent the best choice. Indeed, utilizing a large statistical sample (e.g., for the size frequency distributions of different populations used for the comparisons to the results of numerical models) averages the role of the non-spherical shape, and so these size samples are reliable \citep[the reliability of the sizes based on NEATM model and thermal data is discussed by][]{Usui2014}. The size comparison in Fig.~\ref{img:WISE_D} indicates that our sizes are consistent with the radiometric sizes of \citet{Mainzer2016}.

Our VS-TPM results only represent asteroids with sizes between 5 and 100 km. For larger asteroids, WISE data are usually affected by saturation, and for smaller asteroids, we do not have sets of lightcurves good enough to successfully obtain the shape models. It may be possible to correct for partial saturation, but we chose not to do it, because it has not been characterized for asteroids in particular, which would complicate the TPM interpretation. To increase our statistical sample on asteroids with larger and smaller sizes, we searched the literature for reported values of thermal inertia, size, and albedo determined by other authors (mostly by different TPM implementations) and present them together with the references in Table~\ref{tab:literature}. 

\subsection{Thermal inertia values}

We investigate the relationship between asteroid diameters and thermal inertias for all asteroids in our sample ($\sim$120 our determinations and 60 previously published values), but also for subgroups with respect to the albedo, taxonomy class, family membership or sidereal rotational period. It is not straightforward to compare thermal inertia values obtained at different heliocentric distances $r_\mathrm{hel}$, because $\Gamma$ is a function of the temperature $T$, hence $r_\mathrm{hel}$ \citep[see, e.g.,][]{Keihm1984, Muller2010, Delbo2015}

\begin{equation}\label{eq:TIhel}
 \Gamma \propto T^{3/2} \propto r^{-3/4}_\mathrm{hel}.
\end{equation}

We normalized the resulting thermal inertia values of all asteroids to $r_\mathrm{hel}=1$~au. We note that this model assumes that all the observations are obtained at similar $r_\mathrm{hel}$. However, for some asteroids, observations at two distinct epochs, thus $r_\mathrm{hel}$, are available. We provide the average value of $r_\mathrm{hel}$ in Table~\ref{tab:TI}, which introduces only a small inaccuracy, because the differences in $r_\mathrm{hel}$ between the two epochs are usually rather small. Most of the asteroids in our sample are from the main belt and were observed by WISE at typical heliocentric distances of $r_\mathrm{hel}\sim2-3.5$ au, which implies only a small correction of the thermal inertia. In the following, we use the corrected thermal inertia values, while Table~\ref{tab:TI} provides the original values of $\Gamma$ together with the heliocentric distance $r_\mathrm{hel}$. 

In Fig.~\ref{img:gamma}, we compare the $D$ vs. $\Gamma$ dependence of our whole sample. We also show the running box in $\Gamma$ of ten values to illustrate potential trend with size. Values of $\Gamma$ for most of our asteroids are between 0 and 200 J\,m$^{-2}$\,s$^{-1/2}$\,K$^{-1}$ and have large uncertainties. Considering all the $\Gamma$ values available, the trend of the increasing thermal inertia with decreasing size suggested in \citet{Delbo2007a} seems evident: the largest objects with $D>500$ km have $\Gamma\lesssim20$ J\,m$^{-2}$\,s$^{-1/2}$\,K$^{-1}$, typical thermal inertia is then growing with decreasing size, is $\sim$100 J\,m$^{-2}$\,s$^{-1/2}$\,K$^{-1}$ for objects between 10 and 100 km, and is reaching its maximum of several hundreds for kilometer-sized objects. It is also possible that in the size range between 5 and 100 km, covered by our new thermal inertia determinations, $\Gamma$ does not exhibit a significant trend in $D$ vs. $\Gamma$. This would indicate a sudden increase of $\Gamma$ values for sizes $<$10 km. The dispersion of the thermal inertia within similar size ranges is rather high, which could imply various grain sizes of the surface regolith. However, this conclusion is not very robust because of the large error bars affecting the values of the thermal inertia.

The $\Gamma$ values obtained from the literature seem to be, in general, larger than values derived here. This is mostly because both sources sample different populations. The adopted thermal inertia values correspond either to NEAs ($D<5$ km) or to large asteroids ($D>100$ km) and TNOs. For NEAs, we expect higher $\Gamma$ values than in our sample due to the $\Gamma$ vs. size inverse dependence. For TNOs, we suspect that the simple dependence following Eq.~(\ref{eq:TIhel}) might not be a reliable approximation, because their surface temperature is significantly lower than for main-belt asteroids. For $D$ in the 5--100 km range, the few literature values of $\Gamma$ tend to be larger than our values. However, the literature values of $\Gamma$ have rather large uncertainties and we also suffer by the low number statistics. Moreover, the adopted $\Gamma$ values in the size range of 5--100~km were mostly derived based on the IRAS thermal measurements \citep{Delbo2009} that usually have uncertainties $>$10\%. Another possible reason is that the Gamma values from the literature are not based on VS-TPM. So, the classical TPM uses one of the shapes (the nominal), which might not be the best one to interpret the thermal IR data. Therefore, we have some doubts about their reliability.

There is a group of objects with diameters $<$80 km with very low ($\Gamma<20$ J\,m$^{-2}$\,s$^{-1/2}$\,K$^{-1}$) thermal inertias that is observed for the first time. These objects should be covered by a layer of fine and mature regolith. Such fine regolith is a product of thermal disintegration and/or micrometeorite bombardment of larger rocks \citep{Delbo2014}. In both cases, the asteroid in the size range of several tens of kilometers needs millions of years to build up a fine regolith layer and this timescale is comparable or even larger than the collisional lifetime. In this sense, asteroids with low $\Gamma$ could be those who were lucky enough to avoid recent collisional event that would remove the fine-grained material from the surface (i.e., essentially increase $\Gamma$). 

In Fig.~\ref{img:albedo}, we plot the $D$ vs. $\Gamma$ dependence for the low ($p_\mathrm{V}<0.12$) and high ($p_\mathrm{V}>0.12$) albedo objects. We chose the value of $p_\mathrm{V}=0.12$, because it corresponds to the border between S-complex (high albedo) and C-complex (low albedo) taxonomy class asteroids \citep[see also the Supp. Materials of][]{Delbo2017}. The only obvious correlation in this plot is the selection effect of the lightcurve inversion method -- the majority of photometric data for smaller objects ($D\lesssim$20 km) is strongly biased towards the inner main belt and objects with higher albedo. We note that mid- and larger sized low-albedo objects are less likely to be in our sample because they saturate.

By using the SMASS II \citep{Bus2002} and the Tholen \citep{Tholen1984,Tholen1989} taxonomy, we assigned the taxonomic classification (if available) to the asteroids in our sample. For the purpose of our study, we distinguished S- and C-complexes (we included S, Q, Sa, Sq, Sr, Sk, Sl into the S-complex and B, C, Cb, Ch, Cg, Cgh into the C-complex) and split the X-complex into P, M, and E types according to their albedos ($p_\mathrm{V}<0.10$ for P, $0.10<p_\mathrm{V}<0.30$ for M, and $p_\mathrm{V}>0.30$ for E types). For few asteroids with unknown taxonomy classification, we adopted the taxonomic assignment from \citet{DeMeo2013}, which is based on the colors from the Sloan Digital Sky Survey. However, this approach is not always reliable for individual objects, so we also checked in these cases the values of the geometric visible albedo $p_\mathrm{V}$. In Fig.~\ref{img:taxonomy}, we plot the $D$ vs. $\Gamma$ dependence with respect to the most represented taxonomic types (S- and C-complexes, P, M, and E types). 
Thermal inertia values from the literature for M-types already seem, in average, larger than for S- and C-types. Our few additional solutions further supports this behavior. Moreover, E-types seem to have, in average, larger $\Gamma$ values as well, although the statistical sample is rather small and dominated by small ($D<10$ km) objects. 

We have five asteroid families that are represented in our sample by at least two members -- Flora, Koronis, Eos, Eunomia and Phocaea. Most of the convex shape models of asteroids that belong to these asteroid families were already studied by \citet{Hanus2013c}, from where we also adopted the membership revision, which is an essential procedure. Indeed, the initial family membership assignment is adopted from \citet{Nesvorny2015}, who used the hierarchical clustering method \citep[HCM, see e.g.,][]{Zappala1990,Zappala1994}. However, such family lists are contaminated by interlopers and the membership of each individual object should be carefully checked, for example, by considering the taxonomic type, the albedo or the color. In Fig.~\ref{img:families}, we plot the $D$ vs. $\Gamma$ dependence for asteroids that belong to these five asteroid families, where we excluded interlopers reported in \citet{Hanus2013c}. Thermal inertia values within these asteroid families (e.g., Flora, Eos, Eunomia) are rather consistent, we only have few values of thermal inertia $<$20 J\,m$^{-2}$\,s$^{-1/2}$\,K$^{-1}$, which could suggest different regolith grain size on the surface even for objects of common origin (i.e., same age and composition as is usually expected for family members). Alternatively, these objects could not be real members of the family or derived $\Gamma$ values are wrong. The small $\Gamma$ values of some family members could be an important link to non-catastrophic collisions that could refresh the surface and to the processes that could create the regolith \citep{Delbo2014}. Noticeable differences in thermal inertia values are present in the Koronis family. Two asteroids have thermal inertia values $\sim$100 J\,m$^{-2}$\,s$^{-1/2}$\,K$^{-1}$, while the other two $\sim$0--20 J\,m$^{-2}$\,s$^{-1/2}$\,K$^{-1}$. We suspect that some of the asteroids are not true members of the Koronis family. Specifically, asteroids (167)~Urda and (311)~Claudia are borderline cases based on their positions in the proper semi-major axis vs. size plot \citep{Hanus2013c}, and asteroid (1742)~Schaifers has quite a low albedo (0.11) compared to the typical albedos of the Koronis family members (average value is 0.22, the range is about $\pm0.1$). In this sense, asteroid (1618)~Dawn represents the only truly reliable member of the Koronis family. Asteroid (311)~Claudia has a consistent thermal inertia value to that of asteroid Dawn, which might support its membership in the Koronis family.

We show the VS-TPM fits in the thermal inertia parameter space for individual asteroids from Eos, Flora, Koronis and Phocaea collisional families in Figs.~\ref{img:TPM_Eos},~\ref{img:TPM_Flora},~\ref{img:TPM_Koronis} and ~\ref{img:TPM_Phocaea}. Thermal inertia values are rather consistent within the individual families, however, some minor differences are noticeable. Unfortunately, improvements in the thermal inertia determinations (i.e., lowering the uncertainties) and/or enlargement of the statistical sample are necessary. 

For fast rotating asteroids, it could be difficult to retain the very fine regolith grains on their surface because of the centrifugal force, and thus higher thermal inertias could be preferred. Another mechanism to consider is the thermal fatigue -- fast rotators should not experience large temperature differences during the day and night, and so the thermal cracking mechanism of \citet{Delbo2014} should not be that efficient as for intermediate rotators. To investigate the potential fingerprints of these scenarios, we show the $D$ vs. $\Gamma$ dependence for the fast ($P<4$ h, objects close to the disruption limit), intermediate ($4<P<20$ h, objects with a common rotation), and slow ($P>20$ h) rotators in Fig.~\ref{img:rotators} and the dependence of the thermal inertia $\Gamma$ on asteroid rotation period $P$ in Fig.~\ref{img:rotators2}. There are only two out of 17 fast rotating asteroids, which ranges of $\Gamma$ cover low values ($<$40 J\,m$^{-2}$\,s$^{-1/2}$\,K$^{-1}$). On the other hand, values of $\Gamma>40$ J\,m$^{-2}$\,s$^{-1/2}$\,K$^{-1}$ are still possible for these objects. Therefore, all these fast rotating asteroids could have higher thermal inertias. Unfortunately, our $\Gamma$ values are not constrained to a necessary resolution to draw a more reliable conclusions. So, the $D$ vs. $\Gamma$ dependence in Fig.~\ref{img:rotators} does not show any significant correlation.

\section{Conclusions}\label{sec:conclusions}


We performed thermophysical modeling of three hundred asteroids using the VS-TPM analysis, which produced acceptable fits (i.e., a reasonable minimum in the thermal inertia) to the thermal data for 122 asteroids. We report their thermophysical properties such as size, thermal inertia, surface roughness and geometric visible albedo (Table~\ref{tab:TI}). This work increased the number of asteroids with determined thermophysical properties, especially thermal inertias by about a factor of three. We attempted VS-TPM for $\sim$280 asteroids in total, but for $\sim$150  of these bodies, the shape model and pole orientation uncertainties, specific rotation or thermophysical properties, poor thermal infrared data or their coverage prevented the determination of reliable thermophysical properties. 

Derived sizes and geometric visible albedos are usually well constrained and have their uncertainties smaller than 10\% and 30\%, respectively. Moreover, our sizes (and albedos) are consistent with the radiometric sizes based on the NEATM from \citet{Mainzer2016} as is illustrated in Fig.~\ref{img:WISE_D}. Both sizes are based on the same thermal infrared dataset. On the other hand, the values of thermal inertia are significantly less constrained, which makes their interpretation difficult. Although some insight into the physical properties of main-belt asteroids can be made, unfortunately, improvements in the thermal inertia determinations (i.e., lowering the uncertainties) are still desired. This could be driven mostly by improvements in the shape models by utilizing additional photometric data for the shape modeling or by improvements of the TPM model, where, for instance, both optical and thermal infrared data could be utilized \citep[CITPM,][]{Durech2017c}. Also, new space-based thermal infrared measurements such as those obtained by the WISE satellite would greatly help. 

There is no doubt that the shape model and pole orientation uncertainty plays an important role for the thermophysical modeling and is the main reason for the low number of well constrained solutions and the large uncertainties of thermal inertia values. Specifically, the VS-TPM shows, in many cases, strong dependence of the thermophysical fit on the individual varied shape models. Often, the $\chi^2$ values of fits with some varied shape models are extremely large, so these solutions cannot be even accepted, but other varied shape models (of the same asteroid) fit the thermal infrared data reasonably well. Also, we have cases with inconsistent TPM fits with similar $\chi^2$ values within the varied shapes.

The VS-TPM allowed us to remove the pole ambiguity for seven asteroids. In all these cases, the TPM fits within the varied shapes that corresponded to the same original pole solution were clearly better than for the second (ambiguous) pole solution. We label these asteroids in Table~\ref{tab:TI}.

We confirmed the correlation between the size and the thermal inertia proposed by \citet{Delbo2007a}, however, the range of the thermal inertia for similar sizes is large and usually varies between zero and few hundreds (in the size range of 10--100 km). In general, larger objects have lower thermal inertia values. 

Surprisingly, we derived very low ($<$20 J\,m$^{-2}$\,s$^{-1/2}$\,K$^{-1}$) thermal inertias for several asteroids ($\sim$10) with various sizes. Asteroids with such properties that suggest a mature regolith on the surface are reported for the first time. We note that the uncertainties of thermal inertias for most of these asteroids are rather large and cover values even up to 100 J\,m$^{-2}$\,s$^{-1/2}$\,K$^{-1}$. Further confirmation of the low $\Gamma$ values by utilizing additional optical and/or thermal infrared data is desired. Thermal inertia values within several asteroid families are rather consistent with no obvious trends. However, we still have only few members in each studied family, which makes any interpretation difficult due to the low number statistics. 

The fast rotators with $P\lesssim4$ hours seem to have slightly larger thermal inertia values, so do not likely have a fine regolith on the surface. This could be explained, for example, by the loss of the fine regolith due to the centrifugal force, or by the ineffectiveness of the regolith production \citep[e.g., by the thermal cracking mechanism of][]{Delbo2014}.

Our current work represents a characterization effort to provide a context to whatever we learn about individual objects from missions and extensive studies of single objects. We provide thermal inertias for a set of MBAs in a size range that had almost no previous information. Still, this work shows that increasing the number of known thermal inertia values by a factor of three to $\sim$200 is not enough to push forward our understanding of the physical properties of asteroids. Therefore, we likely need dedicated studies (e.g., collisional families) that spend a few years on lightcurve observations, and maybe more missions such as the WISE satellite. The Near-Earth Object Camera \citep[NEOCam,][]{Mainzer2015} could be the next major driver for the new knowledge concerning the thermophysical properties of asteroids. 

Regarding future improvements on the TPM, it should happen based on studying the very few targets with ground-truth knowledge of physical properties and rich (thermal) data sets \citep[e.g.,][]{Rozitis2017}. Or maybe with thermal infrared data from spacecraft missions \citep[OSIRIS-REx data,][]{Lauretta2015}.

\section*{Acknowledgements}
The computations have been done on the ``Mesocentre'' computers, hosted by the Observatoire de la C\^{o}te d'Azur.

JH was supported by the grant 17-00774S of the Czech Science Foundation and JD by the grant 15-04816S of the Czech Science Foundation.

VAL and MD acknowledge support from the NEOShield-2 project, which has received funding from the European Union’s Horizon 2020 research and innovation programme under grant agreement no. 640351.

VAL: The research leading to these results has received funding from the European Union’s Horizon 2020 Research and Innovation Programme, under Grant Agreement no 687378.

This publication uses data products from NEOWISE, a project of the Jet Propulsion Laboratory/California Institute of Technology, funded by the Planetary Science Division of the NASA. We made use of the NASA/IPAC Infrared Science Archive, which is operated by the Jet Propulsion Laboratory, California Institute of Technology, under contract with the NASA.

\section*{References}
\bibliography{mybib}
\bibliographystyle{model2-names}

\clearpage
\appendix

\section{Tables}

\onecolumn
\scriptsize{
\begin{longtable}{r@{\,\,\,}l ccccccc}
\caption{\label{tab:literature}Compilation of previously published values of thermophysical properties (size $D$, thermal inertia $\Gamma$, geometric visible albedo $p_\mathrm{V}$), taxonomical class, heliocentric distance $r_{\mathrm{h}}$ of the object when the thermal infrared data used for the thermal inertia determination were obtained, rotation period $P$, and references to the publications. The taxonomical classes are either adopted directly from the publications, or based on the SMASS II \citep{Bus2002} or the Tholen \citep{Tholen1984,Tholen1989} taxonomy.}\\
\hline
 \multicolumn{2}{c} {Asteroid} & \multicolumn{1}{c} {$D$} & \multicolumn{1}{c} {$\Gamma$} & \multicolumn{1}{c} {$p_\mathrm{V}$} & \multicolumn{1}{c} {Tax} & \multicolumn{1}{c} {$r_{\mathrm{h}}$} & \multicolumn{1}{c} {$P$} & \multicolumn{1}{c} {Reference} \\
 \multicolumn{2}{c} {} & \multicolumn{1}{c} {[km]} & \multicolumn{1}{c} {[J\,m$^{-2}$\,s$^{-1/2}$\,K$^{-1}$]} &  &  & \multicolumn{1}{c} {[au]} & \multicolumn{1}{c} {[h]} & \\ \hline\hline
 
\endfirsthead
\caption{continued.}\\
 \multicolumn{2}{c} {Asteroid} & \multicolumn{1}{c} {$D$} & \multicolumn{1}{c} {$\Gamma$} & \multicolumn{1}{c} {$p_\mathrm{V}$} & \multicolumn{1}{c} {Tax} & \multicolumn{1}{c} {$r_{\mathrm{h}}$} & \multicolumn{1}{c} {$P$} & \multicolumn{1}{c} {Reference} \\
 \multicolumn{2}{c} {} & \multicolumn{1}{c} {[km]} & \multicolumn{1}{c} {[J\,m$^{-2}$\,s$^{-1/2}$\,K$^{-1}$]} &  &  & \multicolumn{1}{c} {[au]} & \multicolumn{1}{c} {[h]} & \\ \hline\hline

\endhead
\hline
\endfoot
\hline
1       &  Ceres            &  923$\pm$20     &  10$\pm$10    &  0.113$\pm$0.005  &  C     &  2.8   &  8.95    &  \citet{Muller1998}\\
2       &  Pallas           &  544$\pm$43     &  10$\pm$10    &  0.14$\pm$0.02    &  B     &  2.8   &  7.81    &  \citet{Muller1998}\\
3       &  Juno             &  234$\pm$11     &  5$\pm$5      &  0.24$\pm$0.03    &  S     &  2.7   &  7.21    &  \citet{Muller1998}\\
4       &  Vesta            &  525$\pm$1      &  20$\pm$15    &  0.34$\pm$0.01    &  V     &  2.3   &  5.34    &  \citet{Leyrat2012}\\
16      &  Psyche           &  244$\pm$25     &  125$\pm$40   &  0.12$\pm$0.02    &  M     &  2.7   &  4.20    &  \citet{Matter2013}\\
21      &  Lutetia          &  96$\pm$1       &  5$\pm$5      &  0.20$\pm$0.01    &  M     &  2.8   &  8.17    &  \citet{ORourke2012}\\
22      &  Kalliope         &  167$\pm$17     &  125$\pm$125  &  0.17$\pm$0.03    &  M     &  2.3   &  4.15    &  \citet{Marchis2012b}\\
32      &  Pomona           &  85$\pm$1       &  70$\pm$50    &  0.23$\pm$0.01    &  S     &  2.8   &  9.45    &  \citet{Delbo2009}\\
\hline
\end{longtable}
}
\twocolumn

\onecolumn
\scriptsize{
\begin{landscape}
\begin{longtable}{r@{\,\,\,}l ccccc ccc cccc ccc cc cc c}
\caption{\label{tab:TI}Thermophysical characteristics of asteroids derived by the VS-TPM. We provide the asteroid number and name, the ecliptic coordinates $\lambda$ and $\beta$ of the pole solutions, the sidereal rotational period $P$, the reference to the shape model, quality flag of the shape model $QF$, the number of IRAS $N_\mathrm{I}$, WISE in filters W3 $N_\mathrm{W3}$ and W4 $N_\mathrm{W4}$ thermal IR measurements, volume equivalent diameter $D$, thermal inertia $\Gamma$ in J\,m$^{-2}$\,s$^{-1/2}$\,K$^{-1}$ units, visual geometric albedo $p_\mathrm{V}$, Hapke's mean surface slope $\overline \theta$, reduced chi-square of the best fit $\chi^2_\mathrm{red}$, heliocentric distance $r_\mathrm{hel}$ of the thermal infrared observations, absolute magnitude $H$ and slope $G$ \citep[AAMS,][]{Muinonen2010, Oszkiewicz2011}, taxonomical class, and membership to a collisional family. For the taxonomy, we show the Tholen \citep[T1,][]{Tholen1984,Tholen1989} and SMASS II \citep[T2,][]{Bus2002} classes.}\\
\hline 
\multicolumn{2}{c} {Asteroid} & \multicolumn{1}{c} {$\lambda$} & \multicolumn{1}{c} {$\beta$} & \multicolumn{1}{c} {$P$} & \multicolumn{1}{c} {Reference} & \multicolumn{1}{c} {$QF$} & \multicolumn{1}{c} {$N_\mathrm{I}$} & \multicolumn{1}{c} {$N_\mathrm{W3}$} & \multicolumn{1}{c} {$N_\mathrm{W4}$} & \multicolumn{1}{c} {$D$} & \multicolumn{1}{c} {$D_\mathrm{WISE}$} & \multicolumn{1}{c} {$\Gamma$} & \multicolumn{1}{c} {$p_\mathrm{V}$} & \multicolumn{1}{c} {$\overline \theta$} & \multicolumn{1}{c} {$\chi^2_\mathrm{red}$} & \multicolumn{1}{c} {$r_\mathrm{hel}$} & \multicolumn{1}{c} {$H$} & \multicolumn{1}{c} {$G$} & \multicolumn{1}{c} {T1} & \multicolumn{1}{c} {T2} & \multicolumn{1}{c} {Family} \\
\multicolumn{2}{l} { } & [deg] & [deg] & [h] &  &  &  &  &  & \multicolumn{1}{c} {[km]} & \multicolumn{1}{c} {[km]} & \multicolumn{1}{c} {[SI]} &  &  &  & [au] & \multicolumn{1}{c} {[mag]} &  &  &  &  \\ \hline\hline

\endfirsthead
\caption{continued.}\\

\multicolumn{2}{c} {Asteroid} & \multicolumn{1}{c} {$\lambda$} & \multicolumn{1}{c} {$\beta$} & \multicolumn{1}{c} {$P$} & \multicolumn{1}{c} {Reference} & \multicolumn{1}{c} {$QF$} & \multicolumn{1}{c} {$N_\mathrm{I}$} & \multicolumn{1}{c} {$N_\mathrm{W3}$} & \multicolumn{1}{c} {$N_\mathrm{W4}$} & \multicolumn{1}{c} {$D$} & \multicolumn{1}{c} {$D_\mathrm{WISE}$} & \multicolumn{1}{c} {$\Gamma$} & \multicolumn{1}{c} {$p_\mathrm{V}$} & \multicolumn{1}{c} {$\overline \theta$} & \multicolumn{1}{c} {$\chi^2_\mathrm{red}$} & \multicolumn{1}{c} {$r_\mathrm{hel}$} & \multicolumn{1}{c} {$H$} & \multicolumn{1}{c} {$G$} & \multicolumn{1}{c} {T1} & \multicolumn{1}{c} {T2} & \multicolumn{1}{c} {Family} \\
\multicolumn{2}{l} { } & [deg] & [deg] & [h] &  &  &  &  &  & \multicolumn{1}{c} {[km]} & \multicolumn{1}{c} {[km]} & \multicolumn{1}{c} {[SI]} &  &  &  & [au] & \multicolumn{1}{c} {[mag]} &  &  &  &  \\ \hline\hline

\endhead
\hline
\endfoot
\hline
73     &  Klytia        &  266  &  68   &  8.28307   &  \citet{Hanus2011}       &  3  &  24  &  9   &  13  &  44.1$\pm$2.1  &  44.6$\pm$0.9       &  13$\pm$12   &  0.20$\pm$0.02  &  26.7  &  1.1  &  2.8  &  9.1   &  0.40  &  S      &        &                    \\
       &                &  44   &  83   &  8.28307   &  \citet{Hanus2011}       &  3  &  24  &  9   &  13  &  45.4$\pm$1.3  &  44.6$\pm$0.9       &  15$\pm$14   &  0.18$\pm$0.02  &  12.0  &  1.3  &  2.8  &        &        &         &        &                    \\
82     &  Alkmene       &  164  &$-$28  &  13.0008   &  \citet{Hanus2011}       &  2  &  12  &  10  &  10  &  58.6$\pm$1.2  &  57.6$\pm$0.7       &  32$\pm$12   &  0.25$\pm$0.03  &  38.8  &  1.9  &  3.3  &  8.2   &  0.23  &  S      &  Sq    &                    \\
       &                &  349  &$-$33  &  13.0008   &  \citet{Hanus2011}       &  2  &  12  &  10  &  10  &  57.9$\pm$1.3  &  57.6$\pm$0.7       &  29$\pm$15   &  0.26$\pm$0.03  &  38.8  &  2.3  &  3.3  &        &        &         &        &                    \\
99     &  Dike          &  233  &  49   &  18.1191   &  \citet{Hanus2016a}      &  3  &  3   &  7   &  8   &  66.5$\pm$0.9  &  67.4$\pm$0.4       &  35$\pm$19   &  0.07$\pm$0.03  &  55.4  &  2.2  &  3.2  &  9.3   &  0.11  &  C      &  Xk    &                    \\
125    &  Liberatrix    &  95   &  68   &  3.96820   &  \citet{Durech2007}      &  3  &  12  &  14  &  14  &  51.1$\pm$2.1  &  48.4$\pm$0.5       &  80$\pm$15   &  0.18$\pm$0.02  &  38.8  &  2.1  &  2.9  &  8.8   &  0.26  &  M      &  X     &                    \\
       &                &  280  &  74   &  3.96820   &  \citet{Durech2007}      &  3  &  12  &  14  &  14  &  50.1$\pm$1.3  &  48.4$\pm$0.5       &  60$\pm$15   &  0.19$\pm$0.02  &  26.7  &  1.6  &  2.9  &        &        &         &        &                    \\
\hline
\end{longtable}
\end{landscape}
}
\twocolumn

\onecolumn
\begin{longtable}{r@{\,\,\,}l lccc}
\caption{\label{tab:observations}Lightcurve observations used for model revisions and new determinations. We provide the number of dense lightcurves $N_\mathrm{lc}$ and individual sparse-in-time measurements $N_\mathrm{sp}$. The sparse data were obtained by two astrometric surveys -- U.S. Naval observatory in Flagstaff (USNO, IAU code 689) and Catalina Sky Survey \citep[CSS, IAU code 703,][]{Larson2003} or comes from the Lowell database.}\\
\hline
 \multicolumn{2}{c} {Asteroid} & Date & $N_\mathrm{lc}$ & $N_\mathrm{sp}$ & Reference/Observatory \\ \hline\hline

\endfirsthead
\caption{continued.}\\

\hline
 \multicolumn{2}{c} {Asteroid} & Date & $N_\mathrm{lc}$ & $N_\mathrm{sp}$ & Reference/Observatory \\
\hline\hline
\endhead
\hline
\endfoot
  155 &          Scylla & 1998 07 --  2009 01 &    & 110 & USNO-Flagstaff\\
      &                 & 1999 10 --  2010 05 &    & 108 & CSS \\
      &                 & 2008 11 --  2008 12 &  4 &     & \citet{Owings2009} \\
      &                 & 2008 11 --  2008 12 &  7 &     & \citet{Pilcher2009b} \\
      &                 & 2014 02 --  2014 02 &  2 &     & \citet{Stephens2014b} \\
  430 &          Hybris & 1998 11 --  2009 03 &    & 105 & USNO-Flagstaff\\
      &                 & 2005 06 --  2010 06 &    &  54 & CSS \\
  749 &        Malzovia & 1998 08 --  2008 10 &    & 121 & USNO-Flagstaff\\
\hline
\end{longtable}






\clearpage
\section{Figures}

\begin{figure*}[!htbp]
\begin{center}
\resizebox{0.8\hsize}{!}{\includegraphics{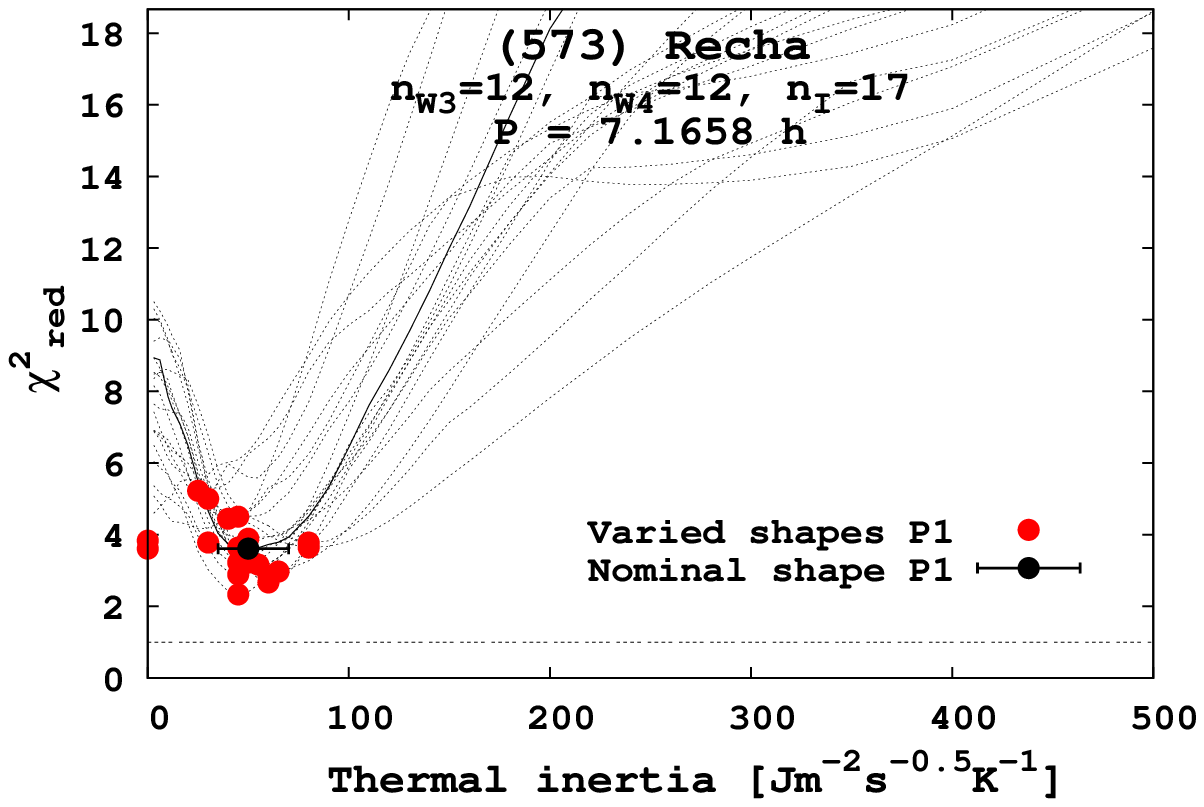}\includegraphics{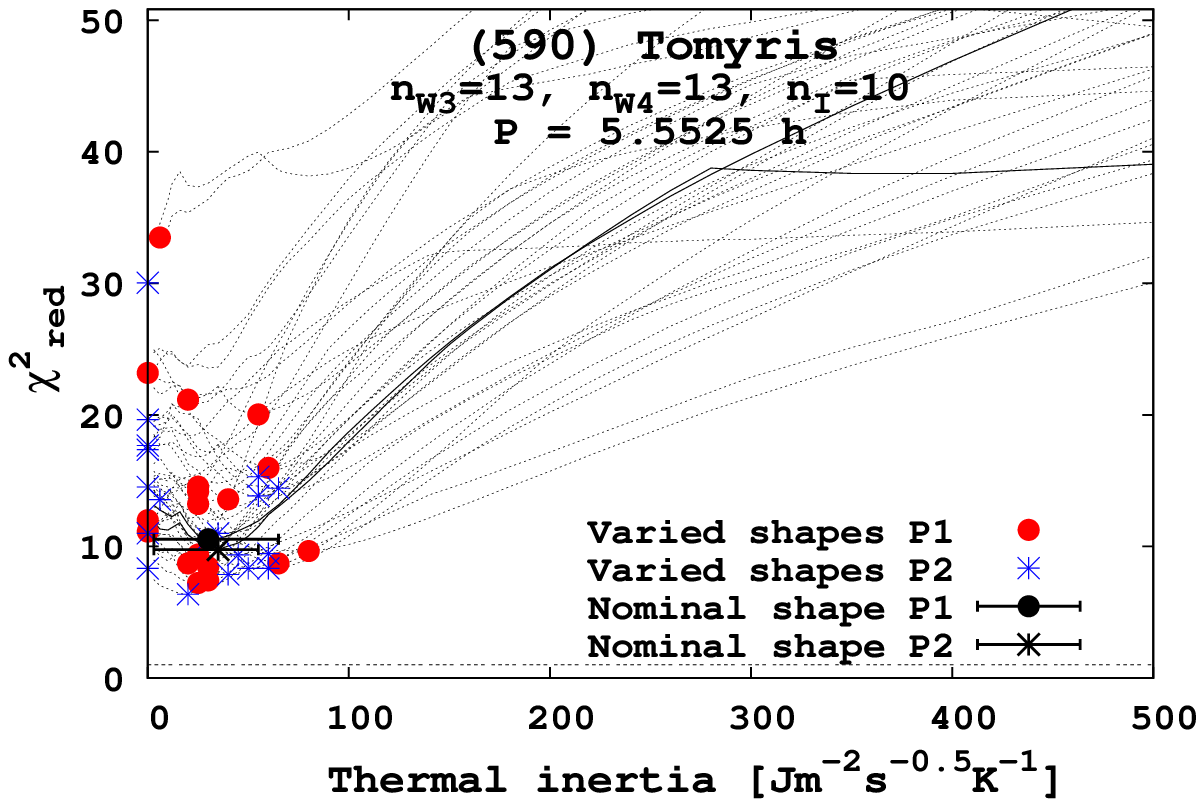}}\\
\resizebox{0.8\hsize}{!}{\includegraphics{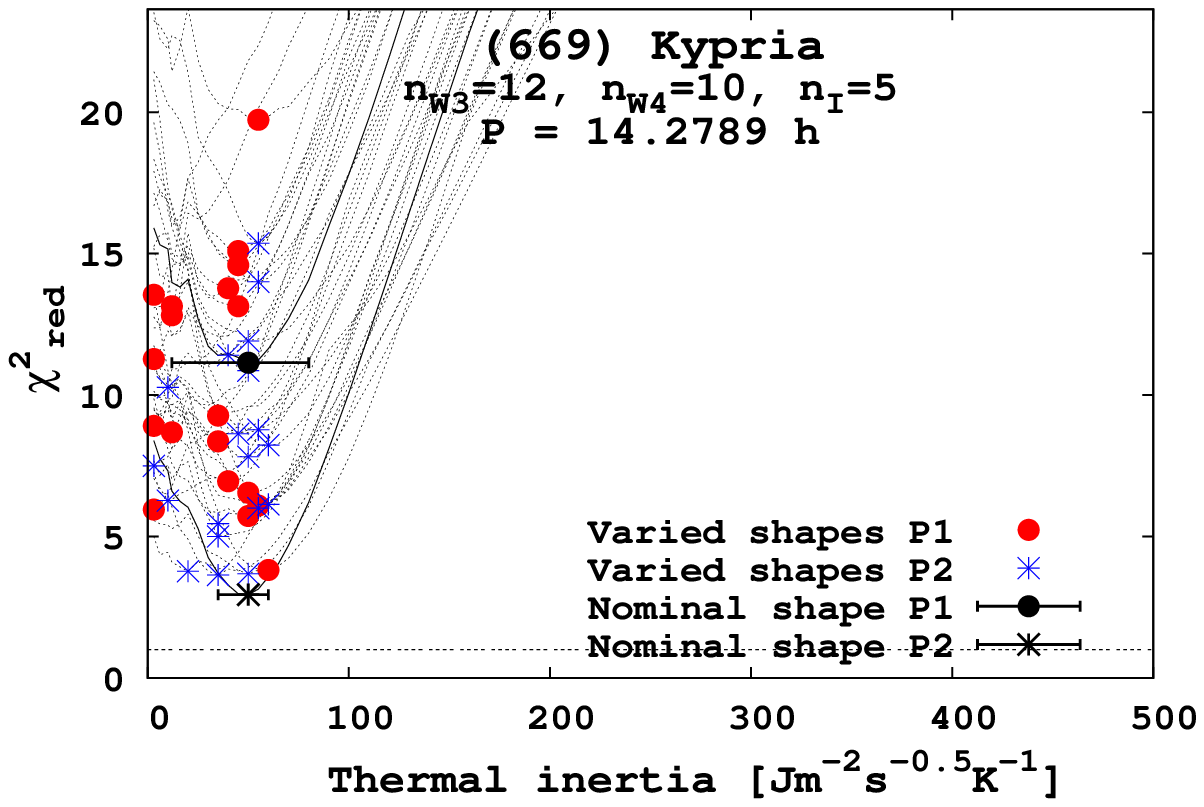}\includegraphics{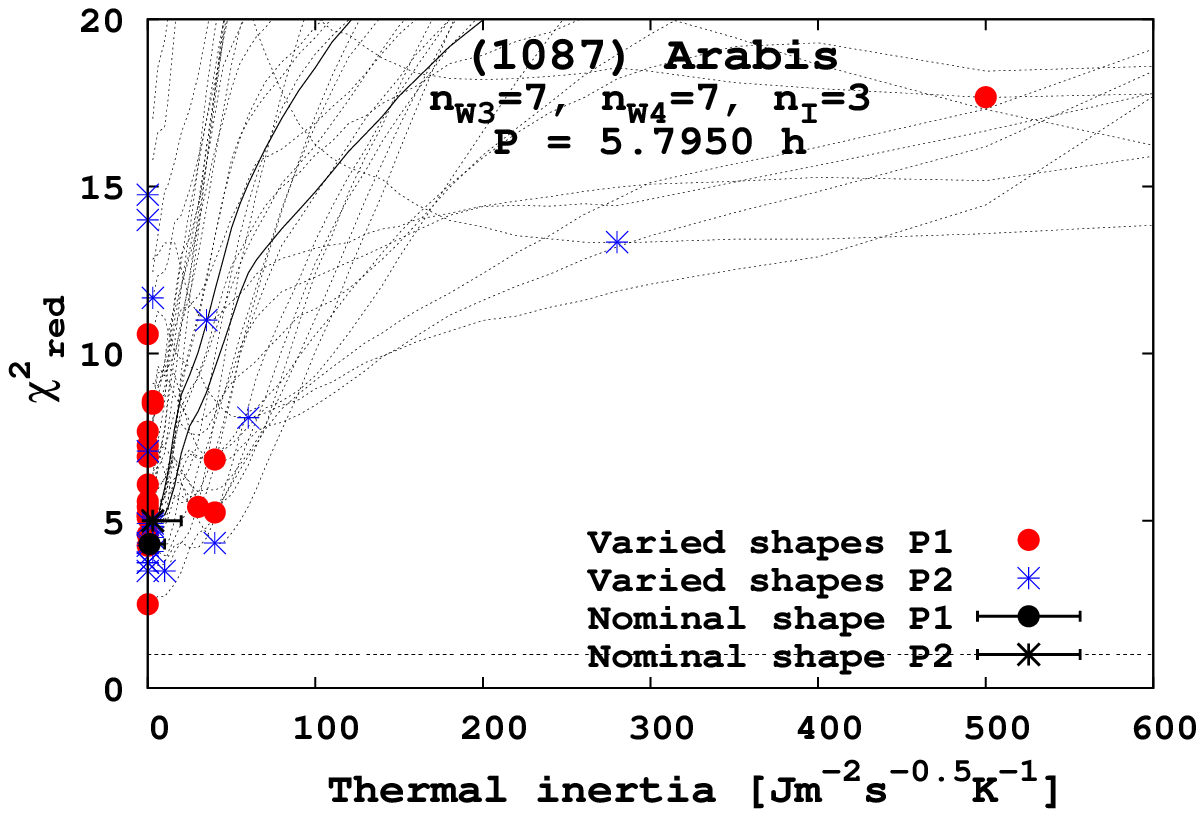}}\\
\resizebox{0.8\hsize}{!}{\includegraphics{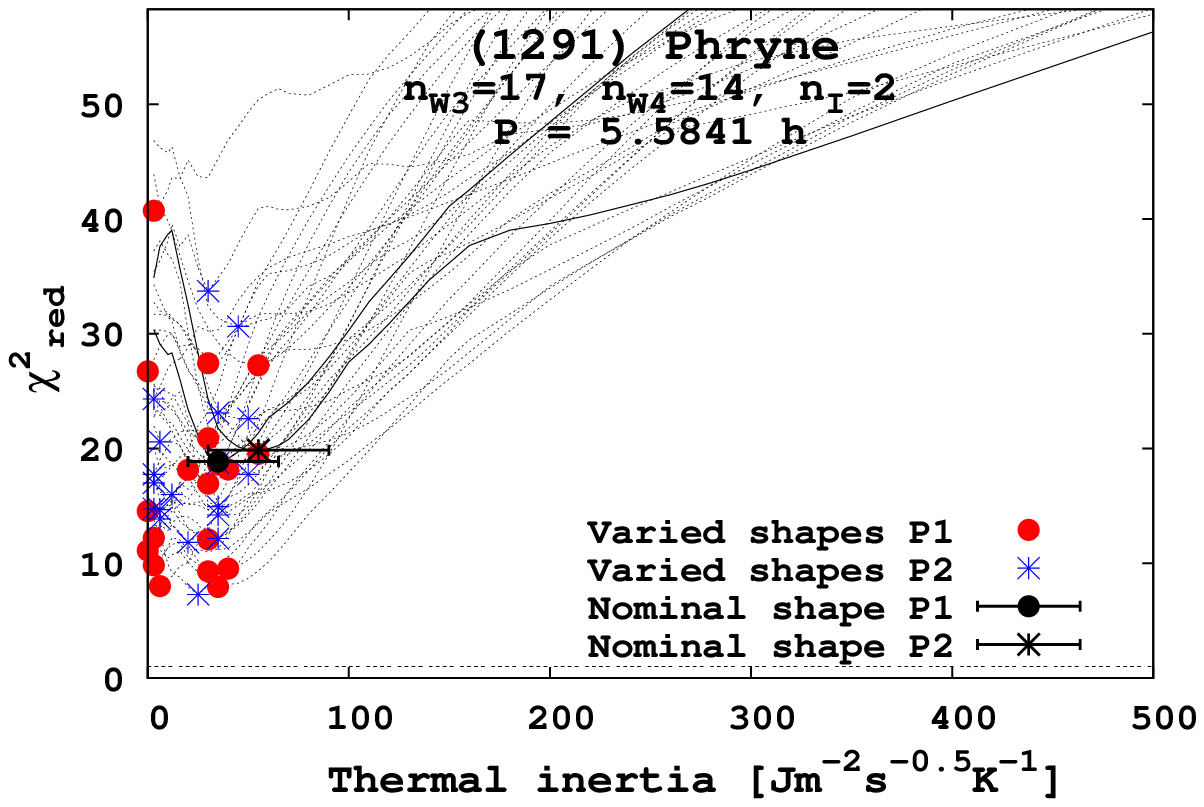}\includegraphics{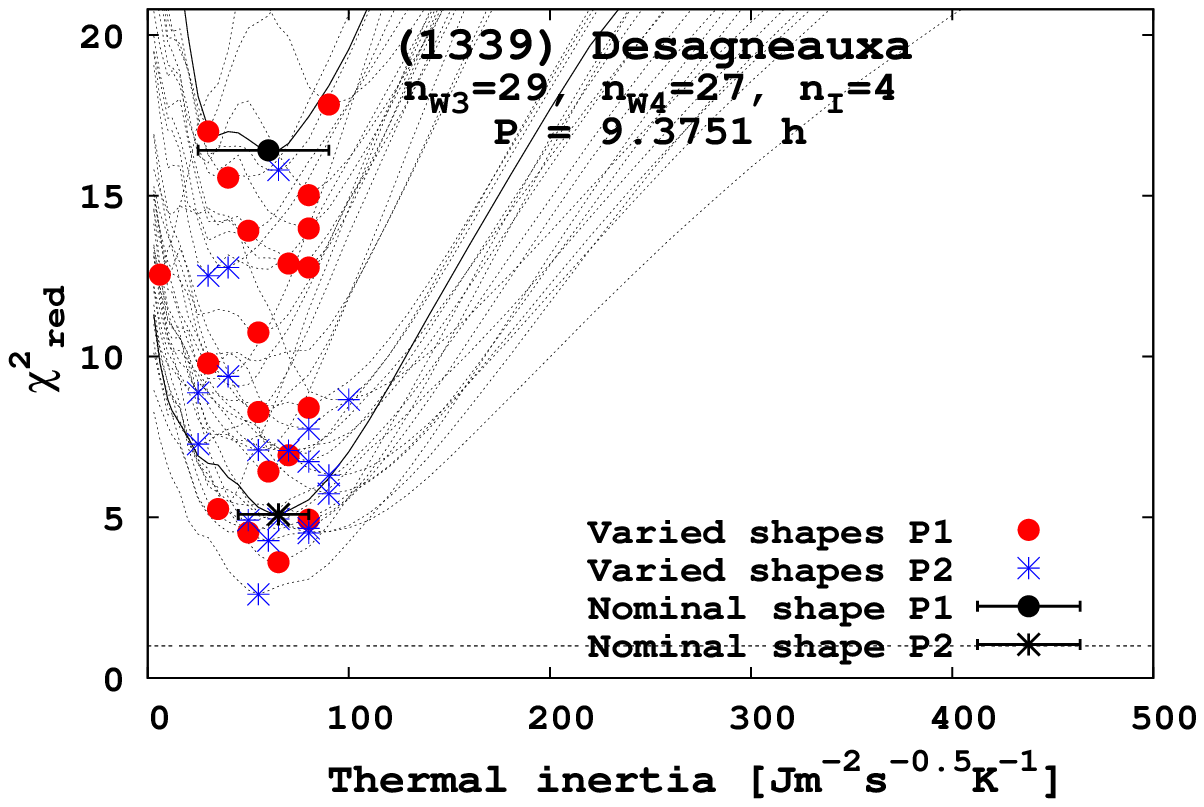}}\\
\resizebox{0.4\hsize}{!}{\includegraphics{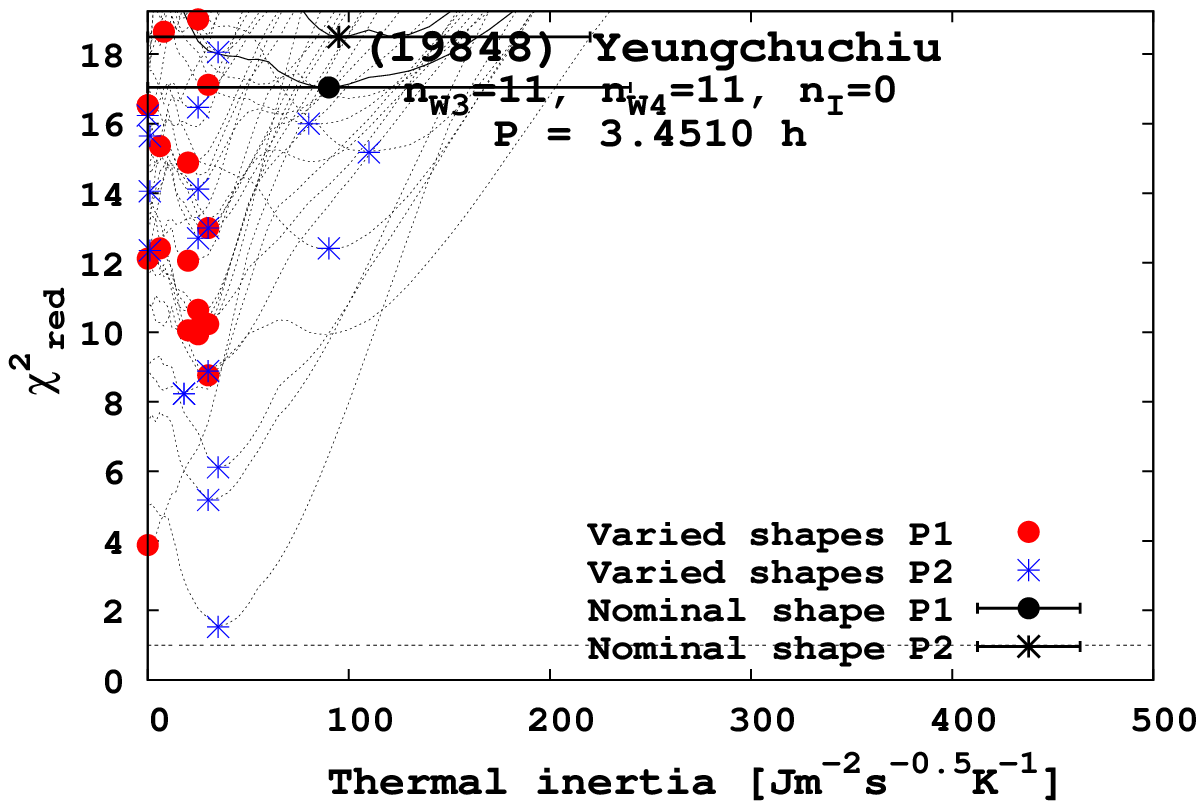}}\\
\end{center}
\caption{\label{img:TPM_Eos}VS-TPM fits in the thermal inertia parameter space for asteroids from Eos collisional family. Each plot also contains the number of thermal infrared measurements in WISE W3 and W4 filters and in all four IRAS filters, and the rotation period.}
\end{figure*}

\begin{figure*}[!htbp]
\begin{center}
\resizebox{0.8\hsize}{!}{\includegraphics{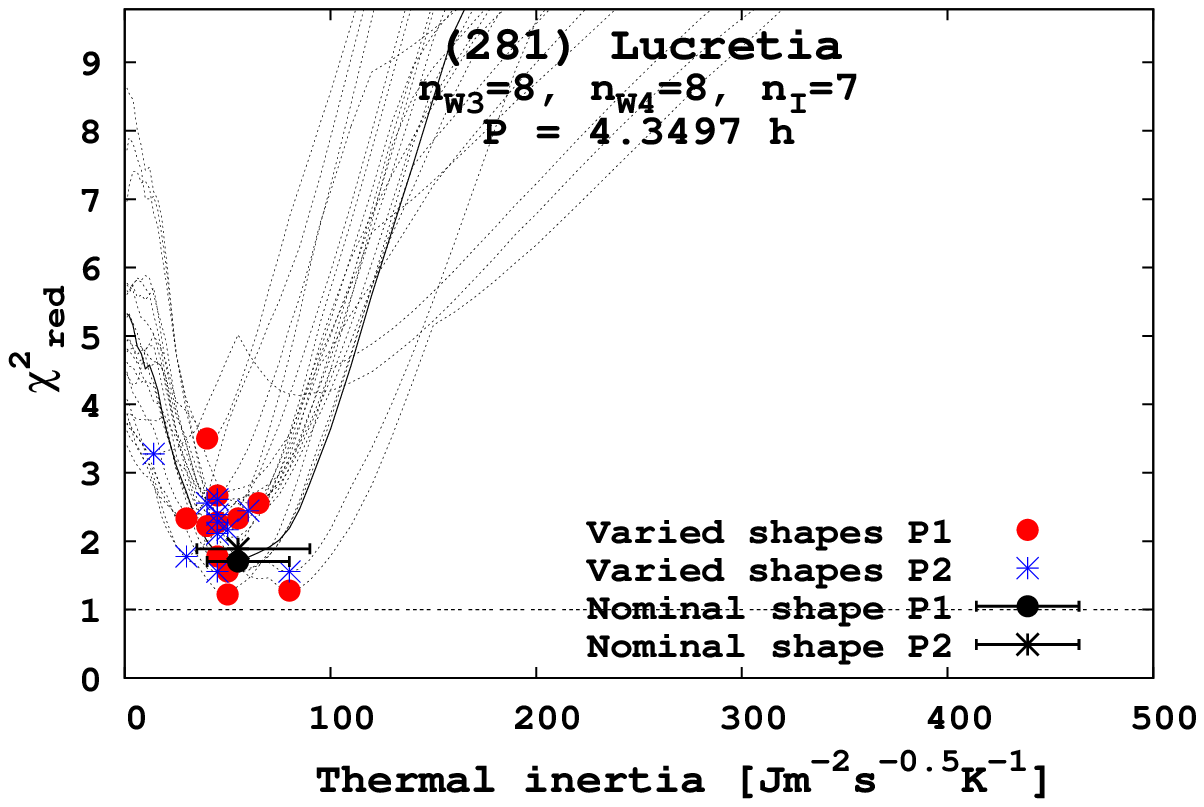}\includegraphics{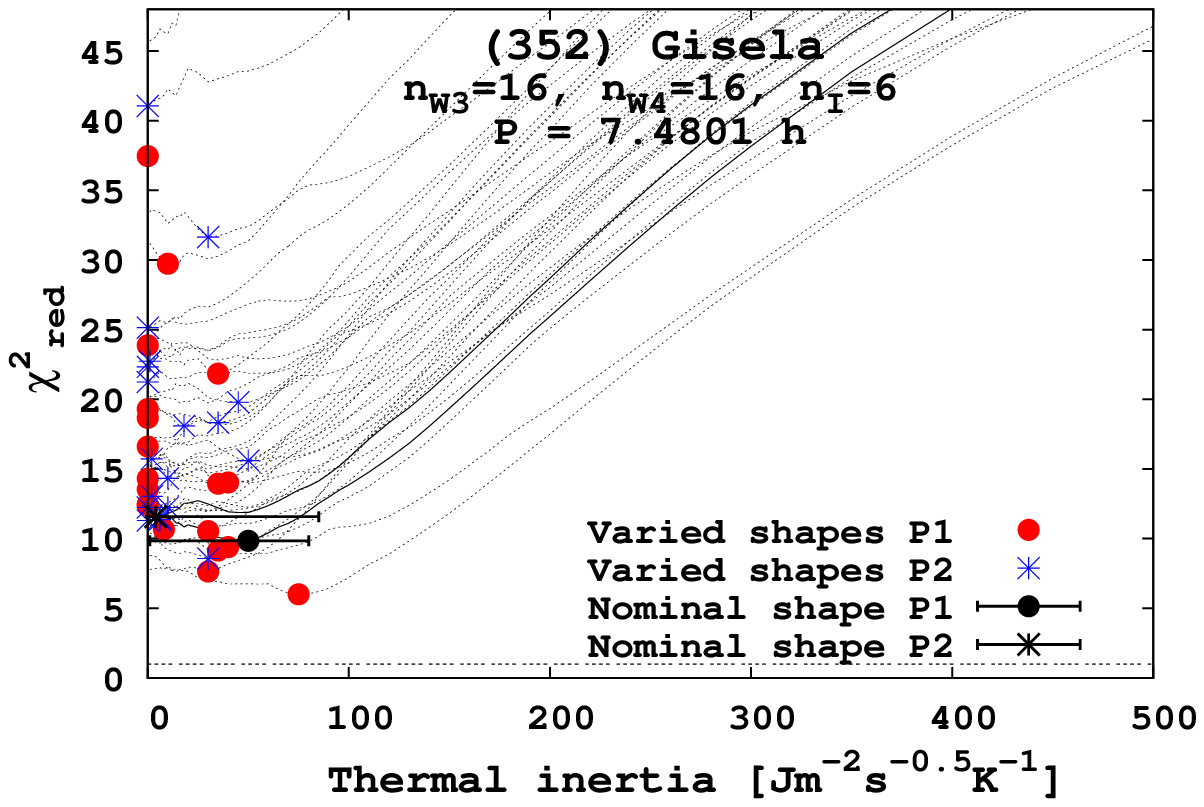}}\\
\resizebox{0.8\hsize}{!}{\includegraphics{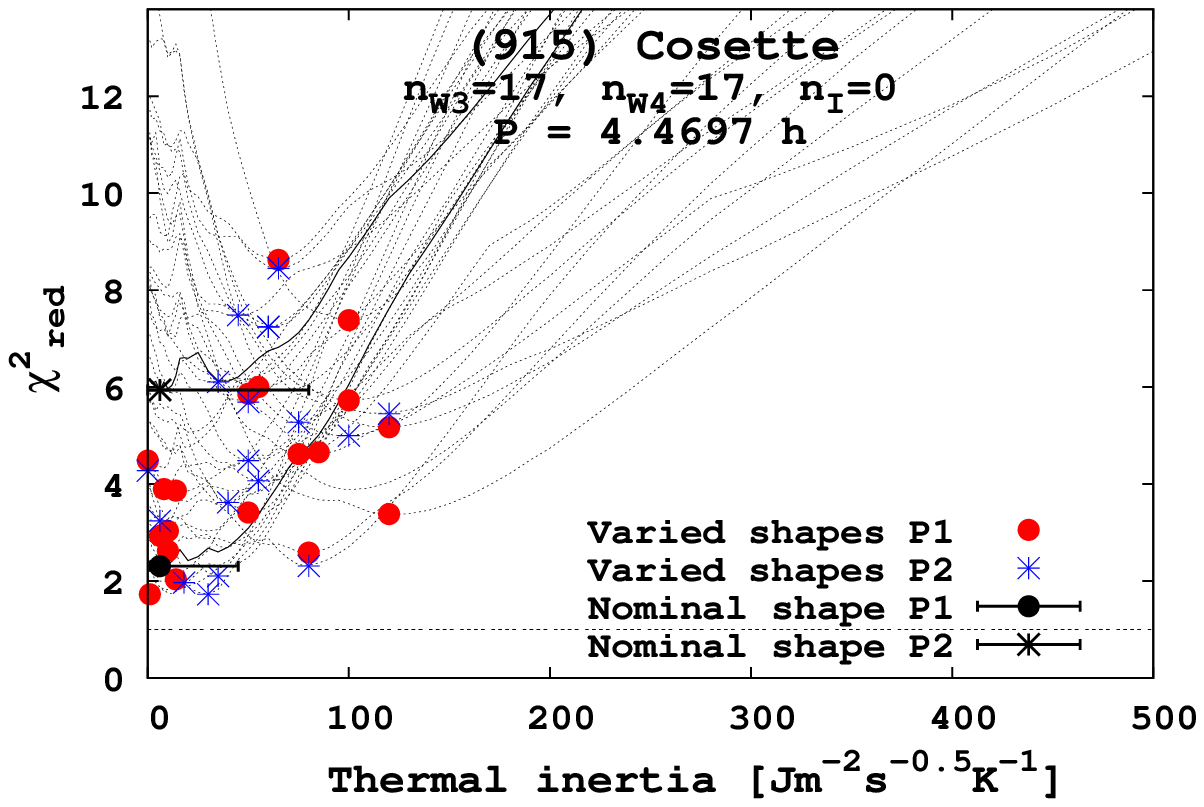}\includegraphics{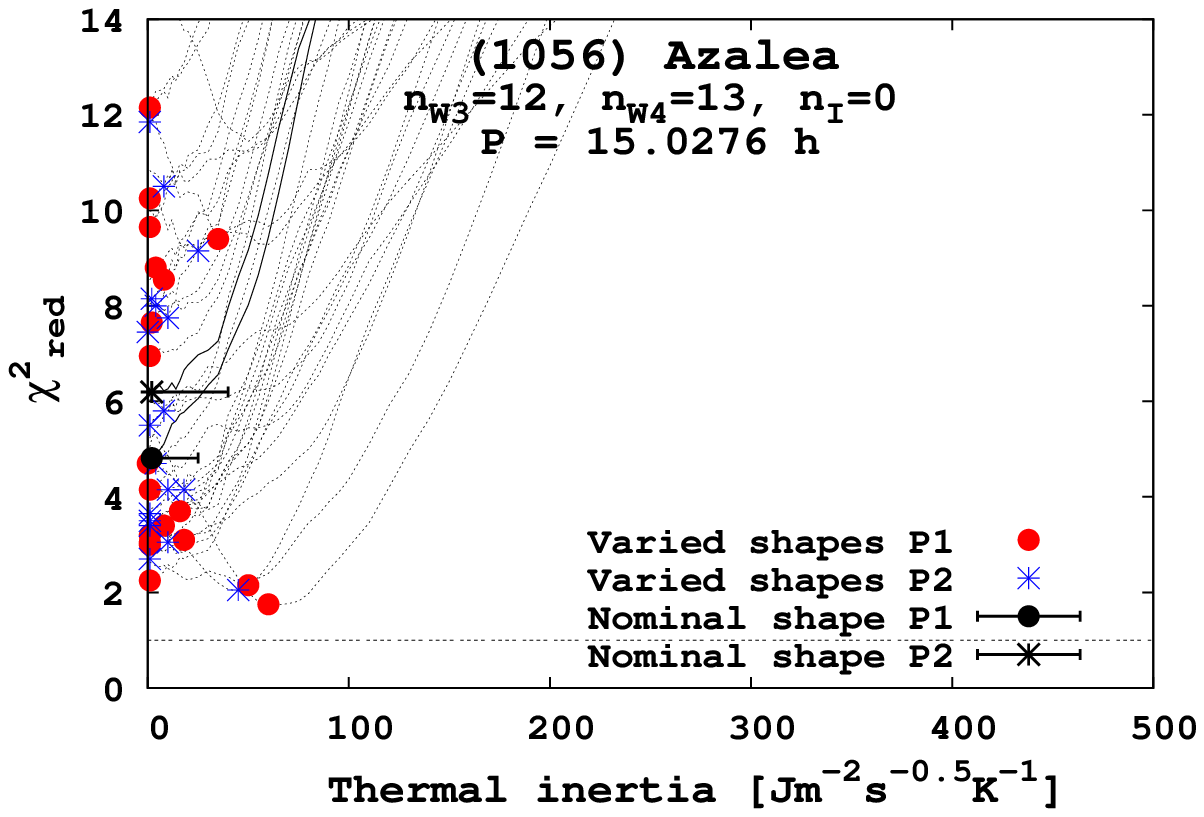}}\\
\resizebox{0.8\hsize}{!}{\includegraphics{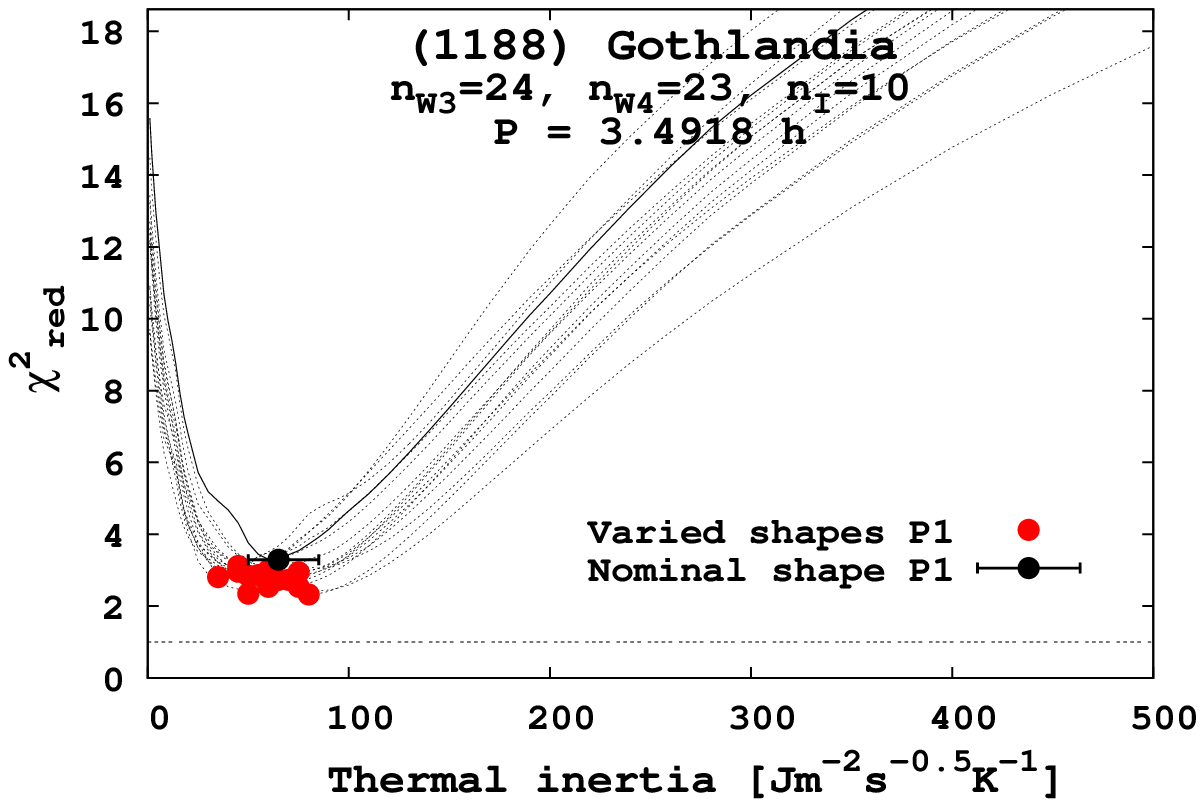}\includegraphics{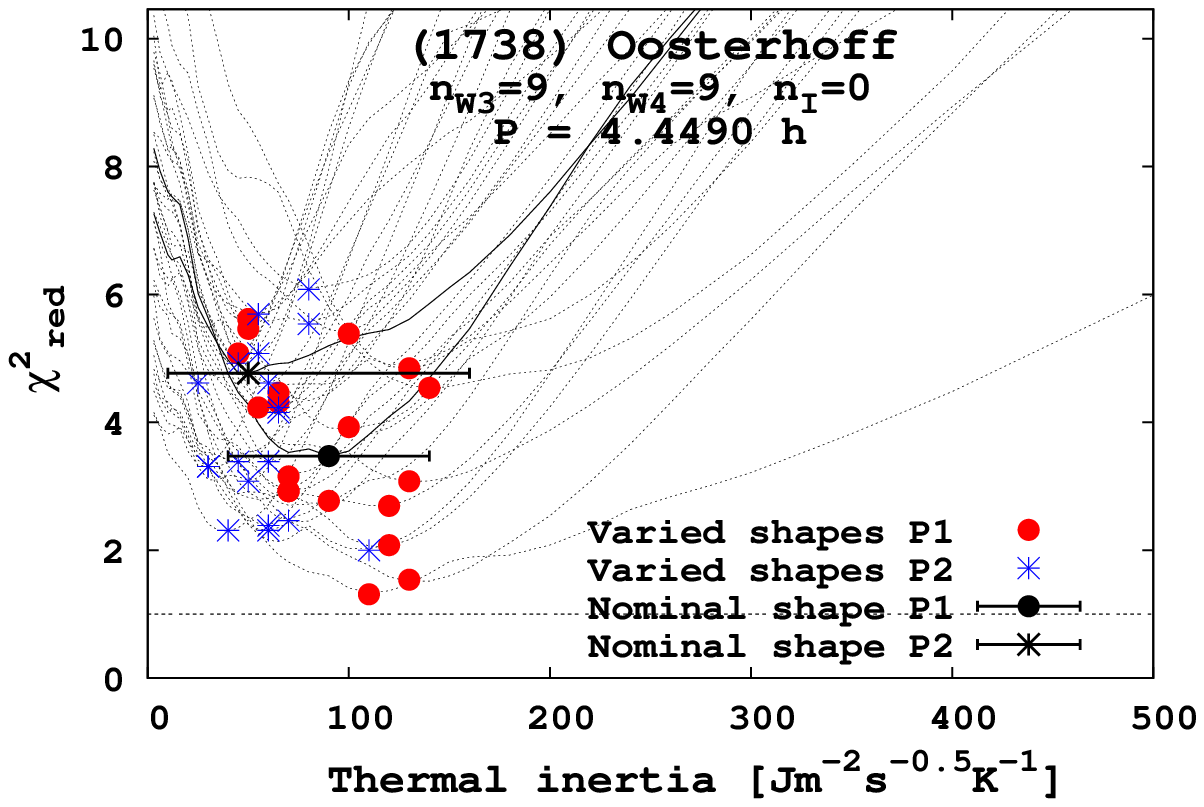}}\\
\end{center}
\caption{\label{img:TPM_Flora}VS-TPM fits in the thermal inertia parameter space for asteroids from Flora collisional family. Each plot also contains the number of thermal infrared measurements in WISE W3 and W4 filters and in all four IRAS filters, and the rotation period.}
\end{figure*}

\begin{figure*}[!htbp]
\begin{center}
\resizebox{0.8\hsize}{!}{\includegraphics{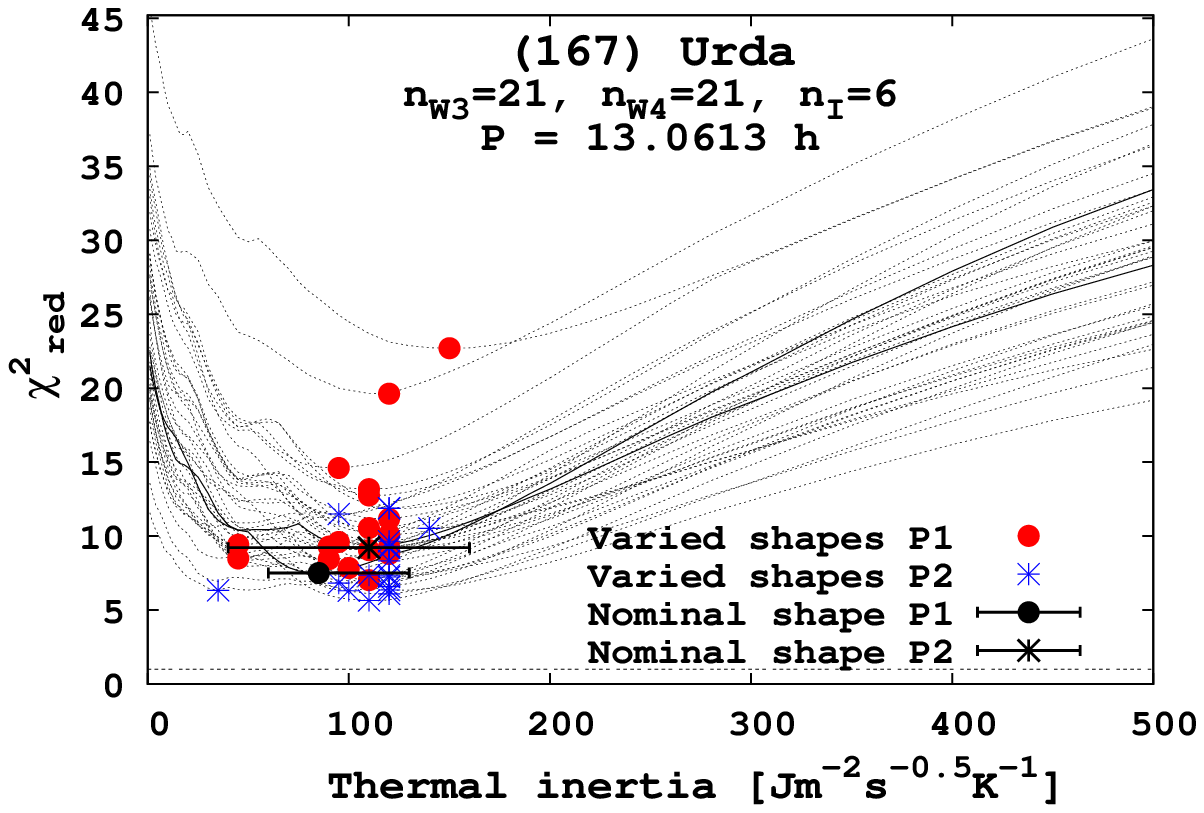}\includegraphics{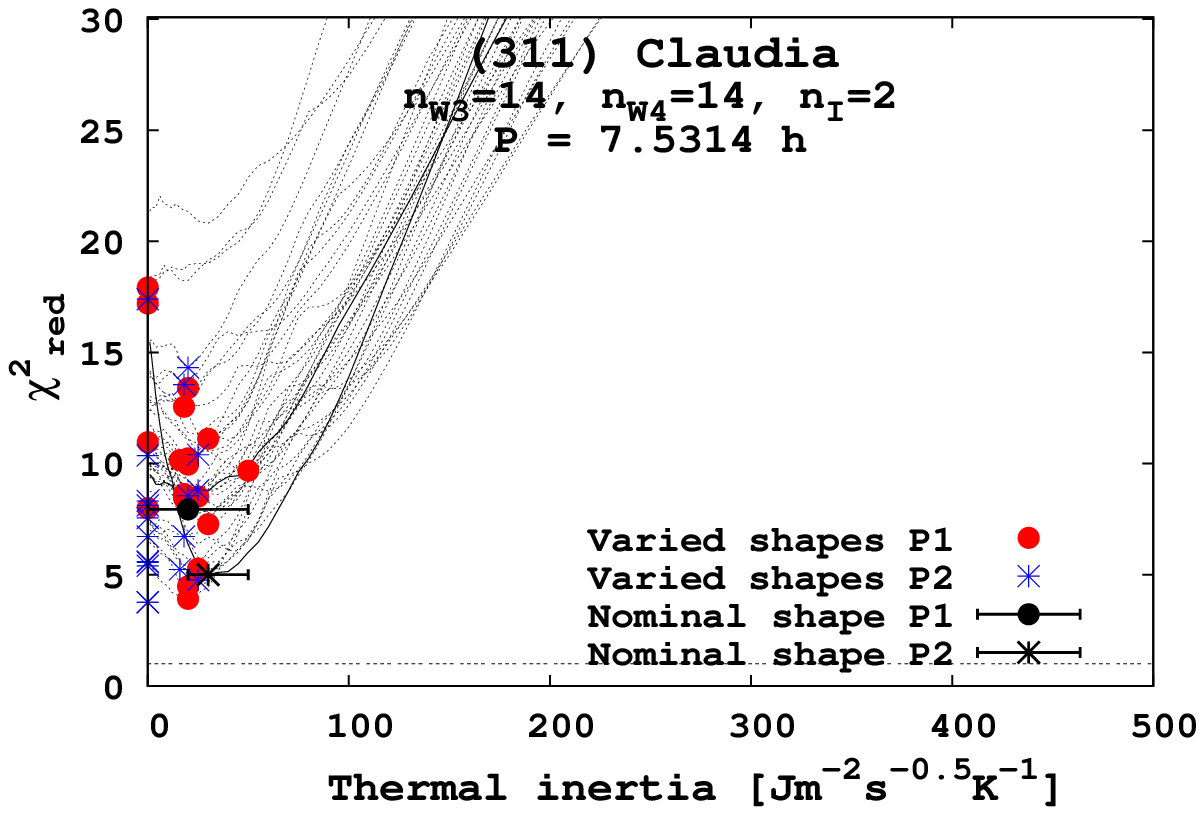}}\\
\resizebox{0.8\hsize}{!}{\includegraphics{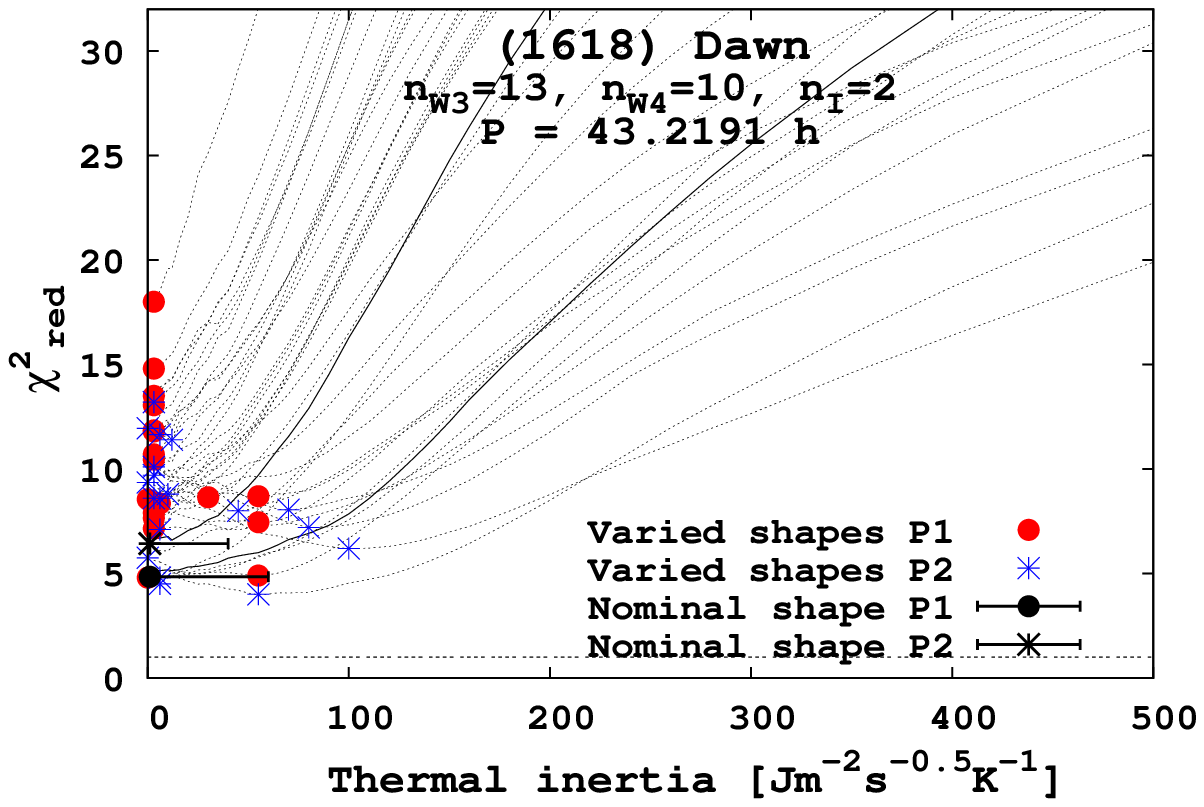}\includegraphics{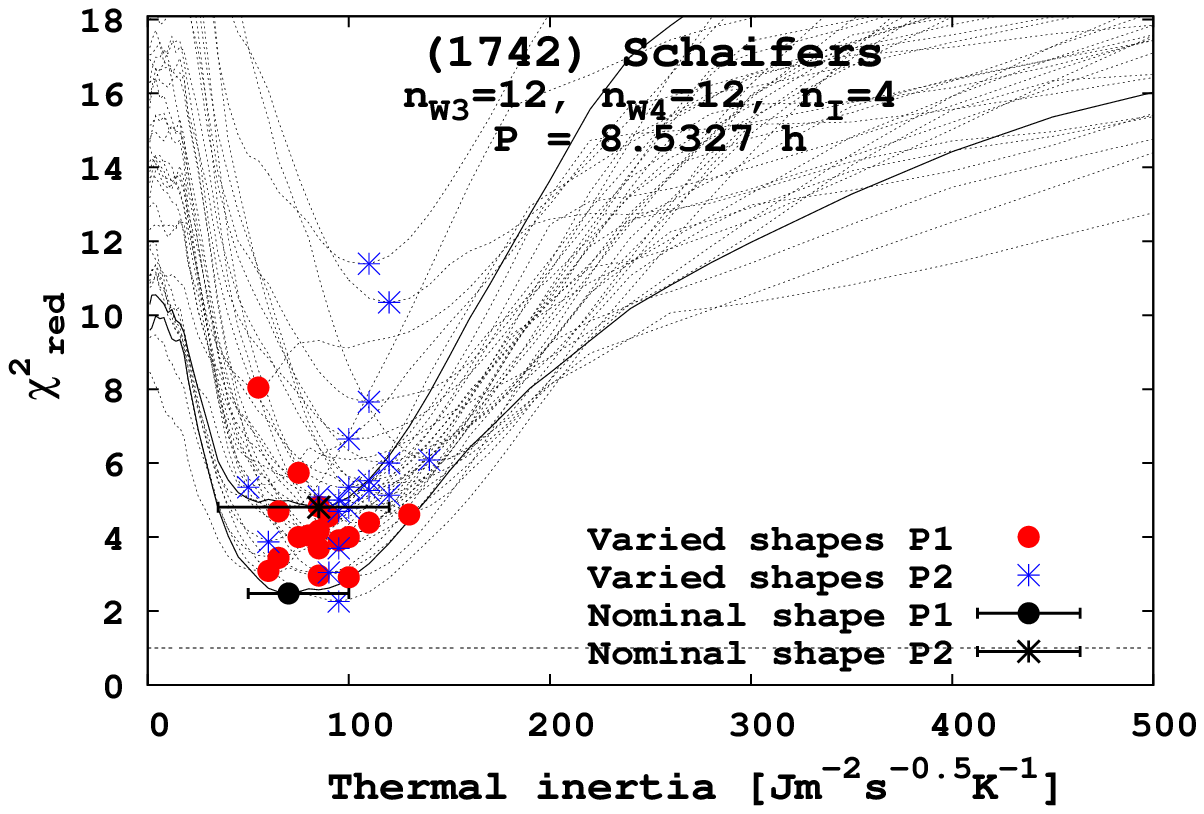}}\\
\end{center}
\caption{\label{img:TPM_Koronis}VS-TPM fits in the thermal inertia parameter space for asteroids from Koronis collisional family. Each plot also contains the number of thermal infrared measurements in WISE W3 and W4 filters and in all four IRAS filters, and the rotation period.}
\end{figure*}

\begin{figure*}[!htbp]
\begin{center}
\resizebox{0.8\hsize}{!}{\includegraphics{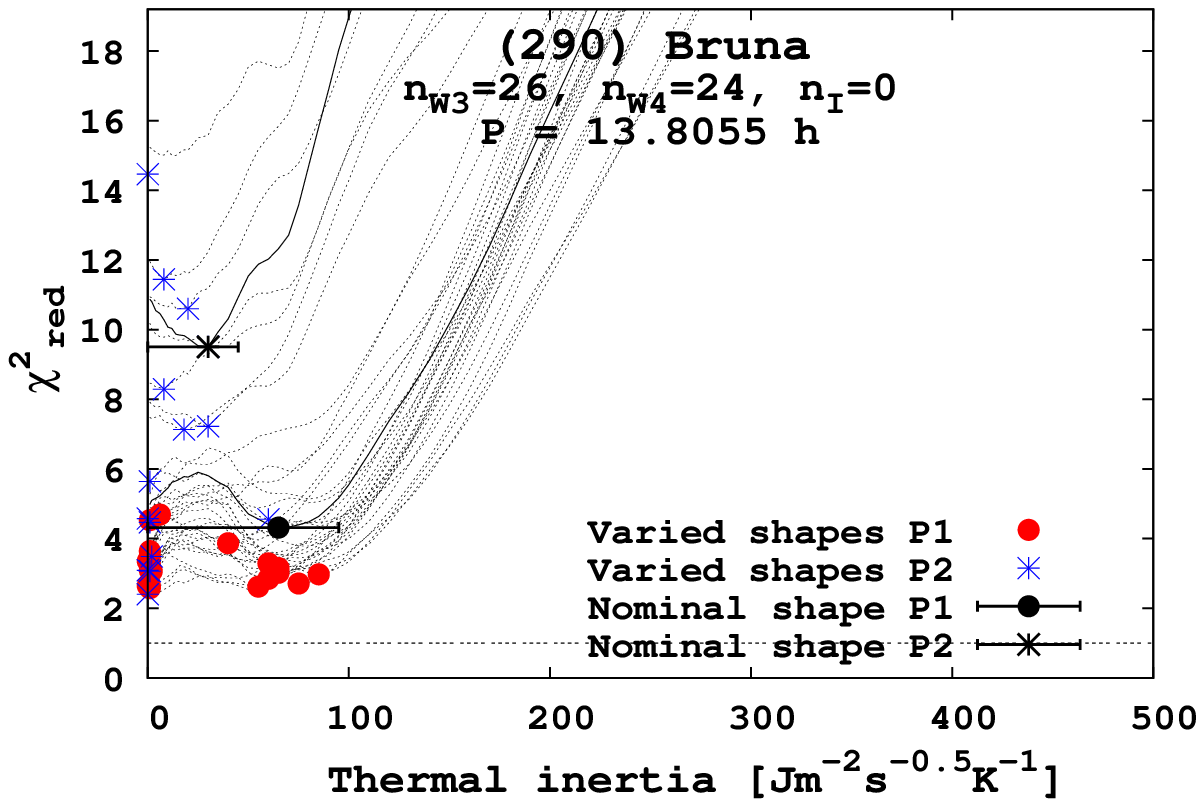}\includegraphics{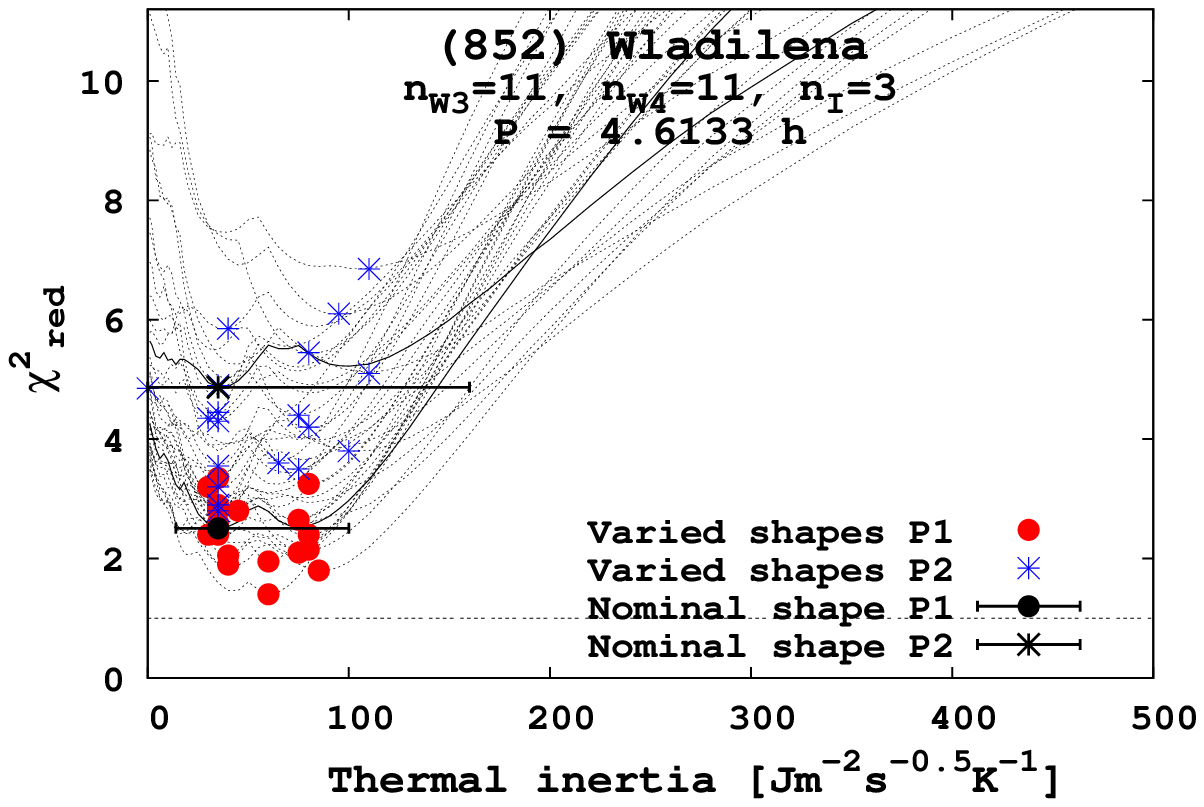}}\\
\resizebox{0.8\hsize}{!}{\includegraphics{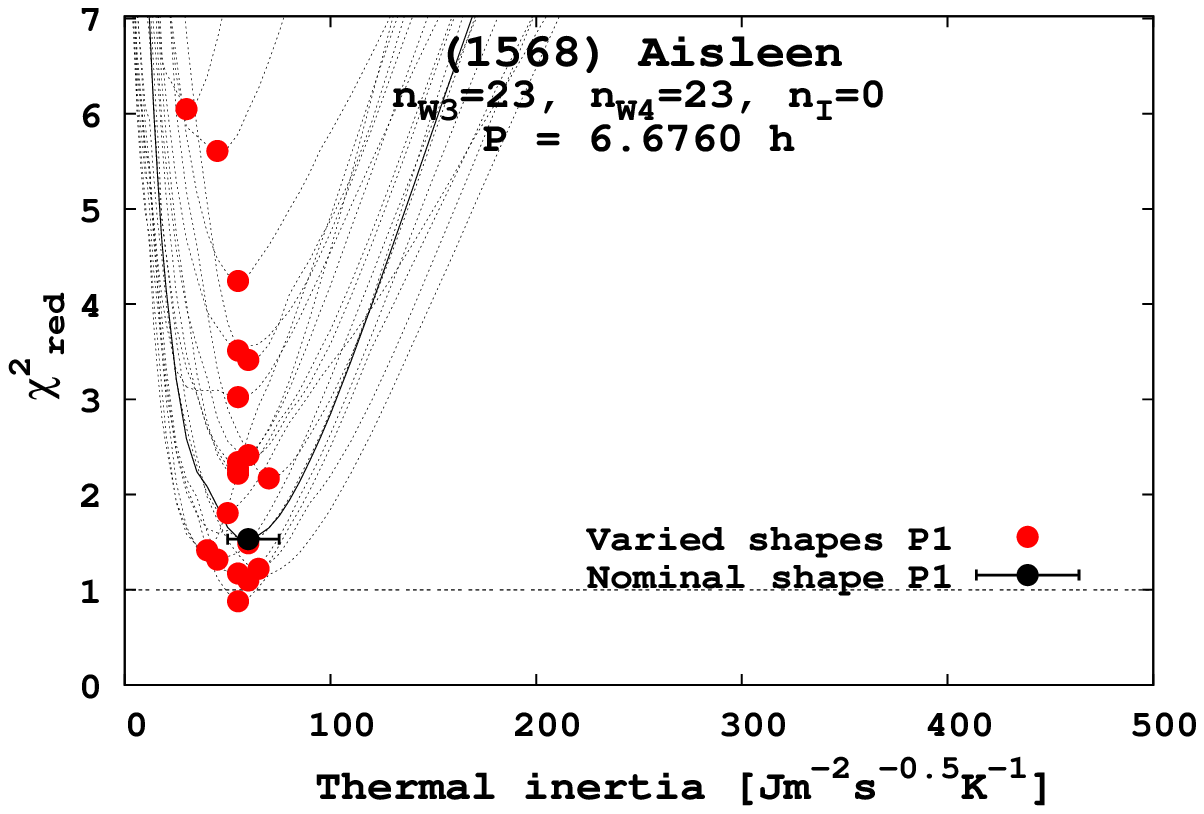}\includegraphics{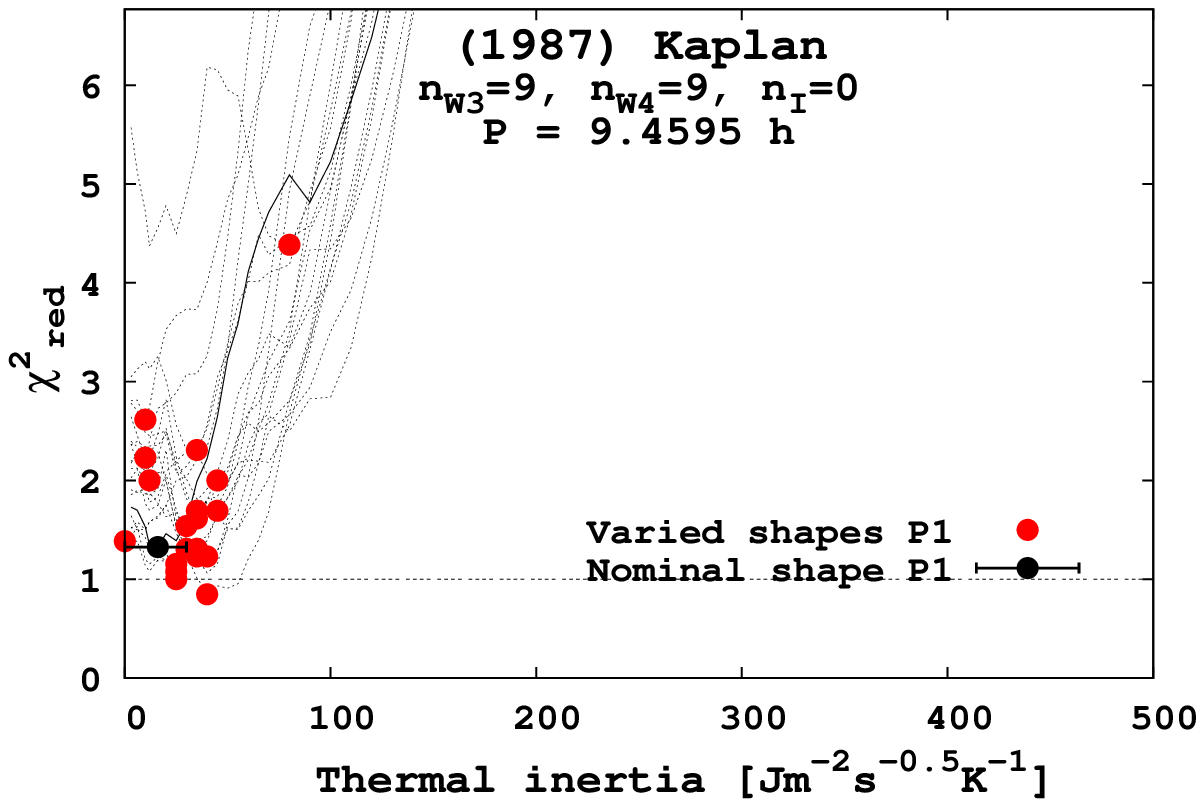}}\\
\end{center}
\caption{\label{img:TPM_Phocaea}VS-TPM fits in the thermal inertia parameter space for asteroids from Phocaea collisional family. Each plot also contains the number of thermal infrared measurements in WISE W3 and W4 filters and in all four IRAS filters, and the rotation period.}
\end{figure*}

\begin{figure*}[!htbp]
\begin{center}
\resizebox{0.8\hsize}{!}{\includegraphics{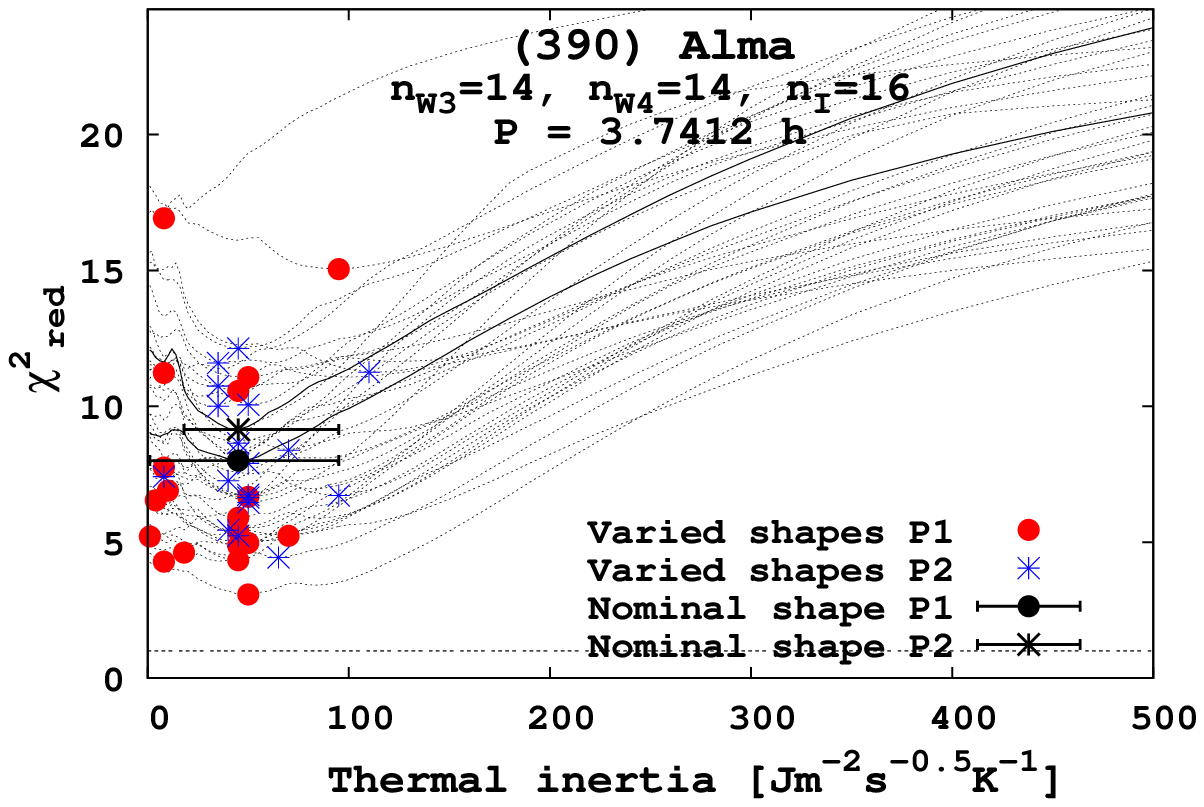}\includegraphics{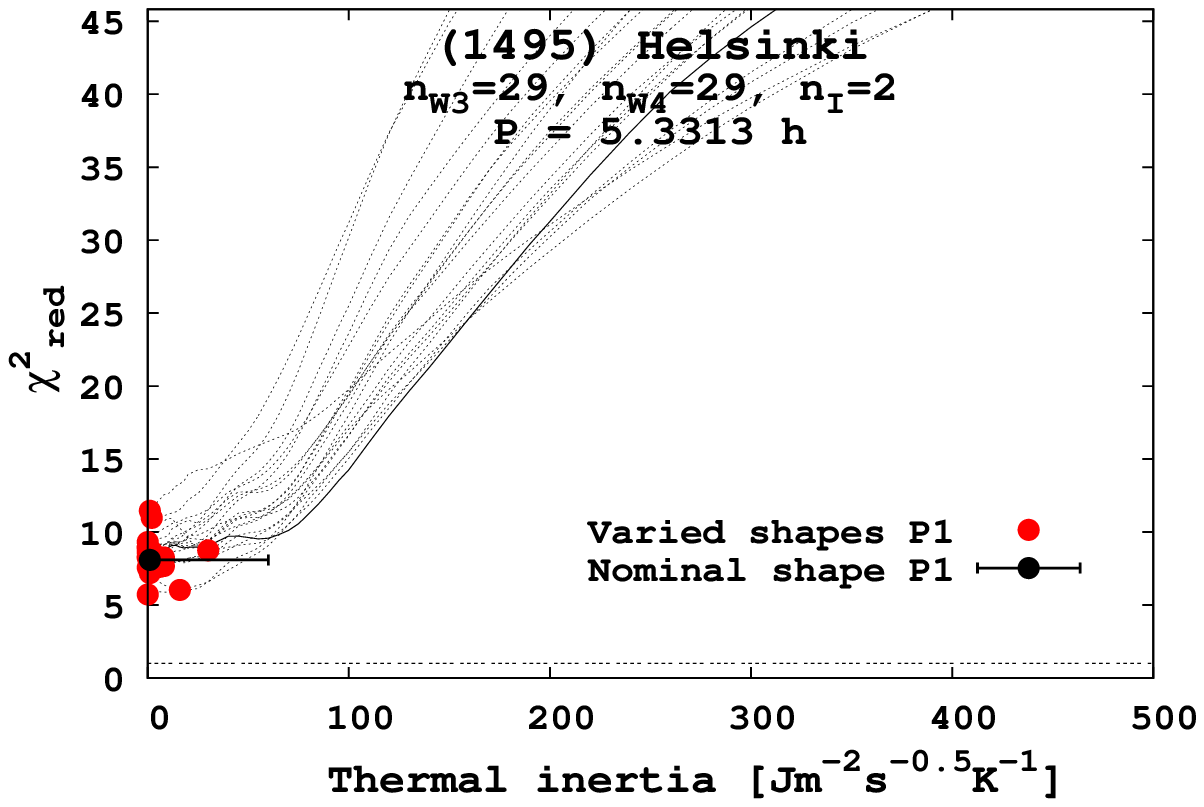}}\\
\end{center}
\caption{\label{img:TPM_Eunomia}VS-TPM fits in the thermal inertia parameter space for asteroids from Eunomia collisional family. Each plot also contains the number of thermal infrared measurements in WISE W3 and W4 filters and in all four IRAS filters, and the rotation period.}
\end{figure*}

\begin{figure*}[!htbp]
\begin{center}
\resizebox{0.8\hsize}{!}{\includegraphics{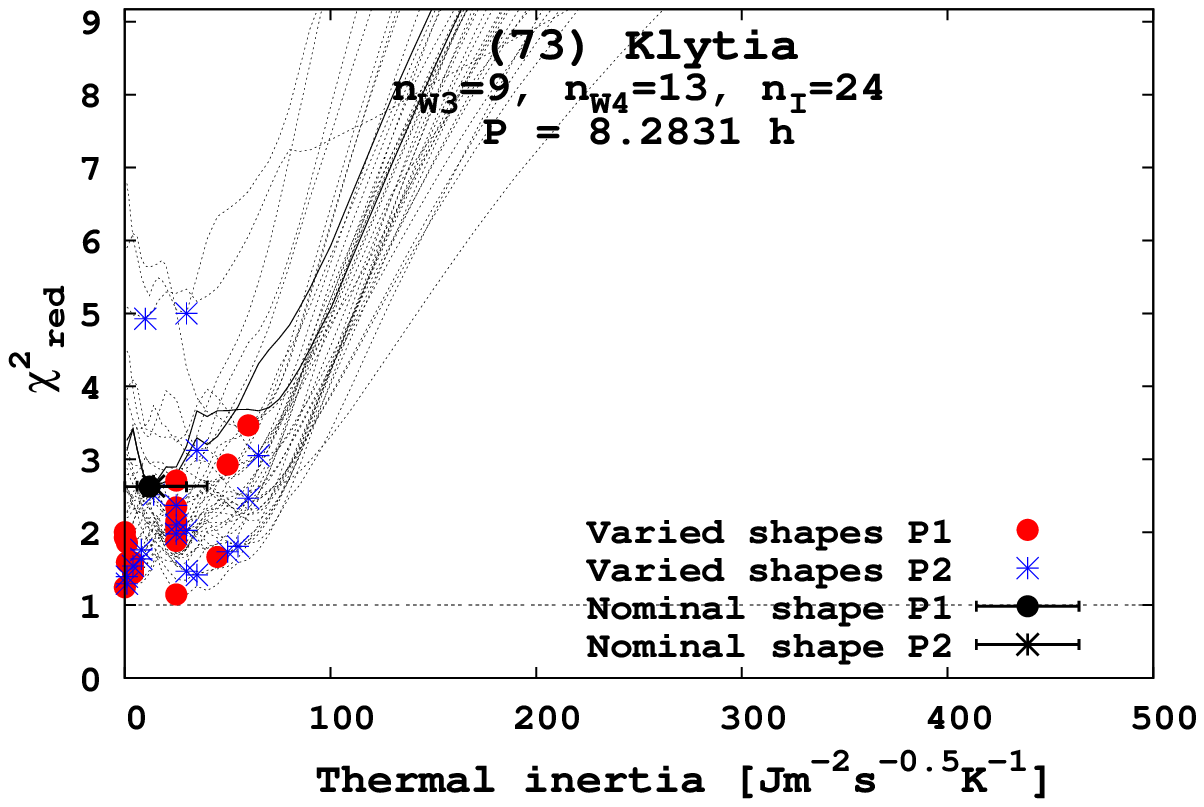}\includegraphics{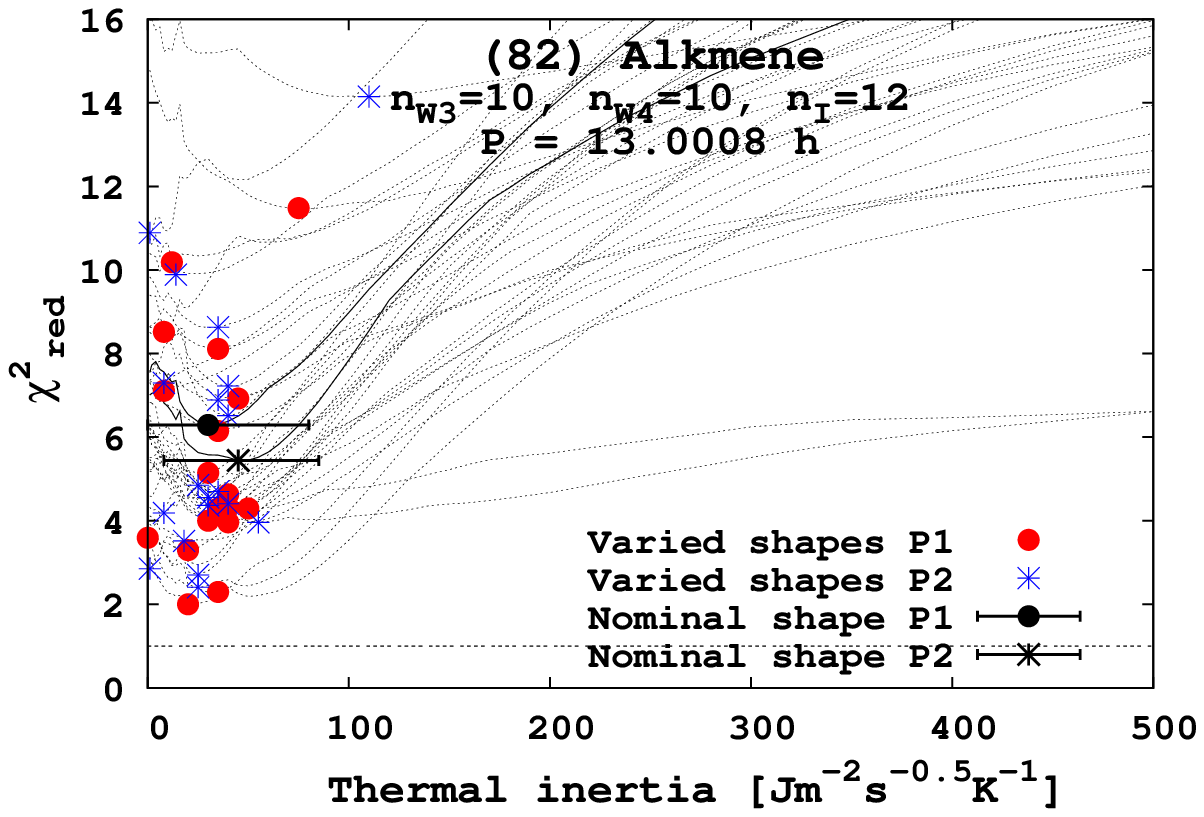}}\\
\resizebox{0.8\hsize}{!}{\includegraphics{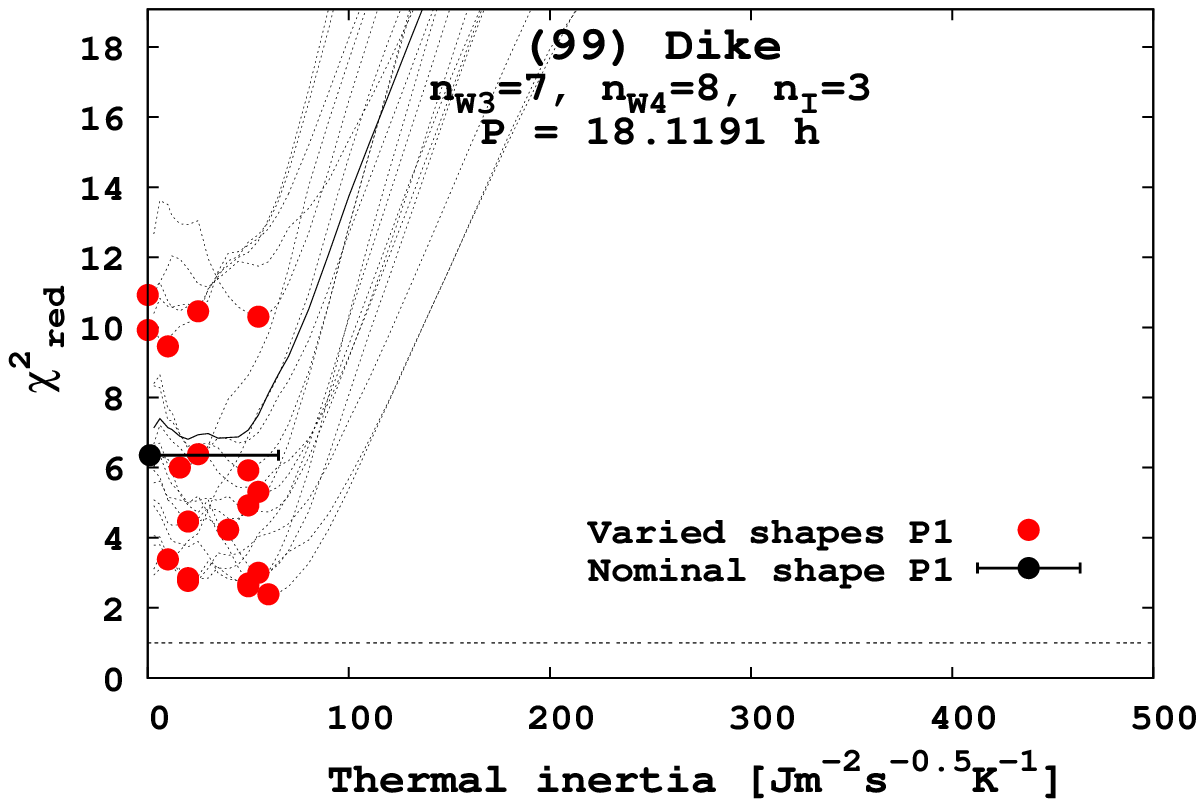}\includegraphics{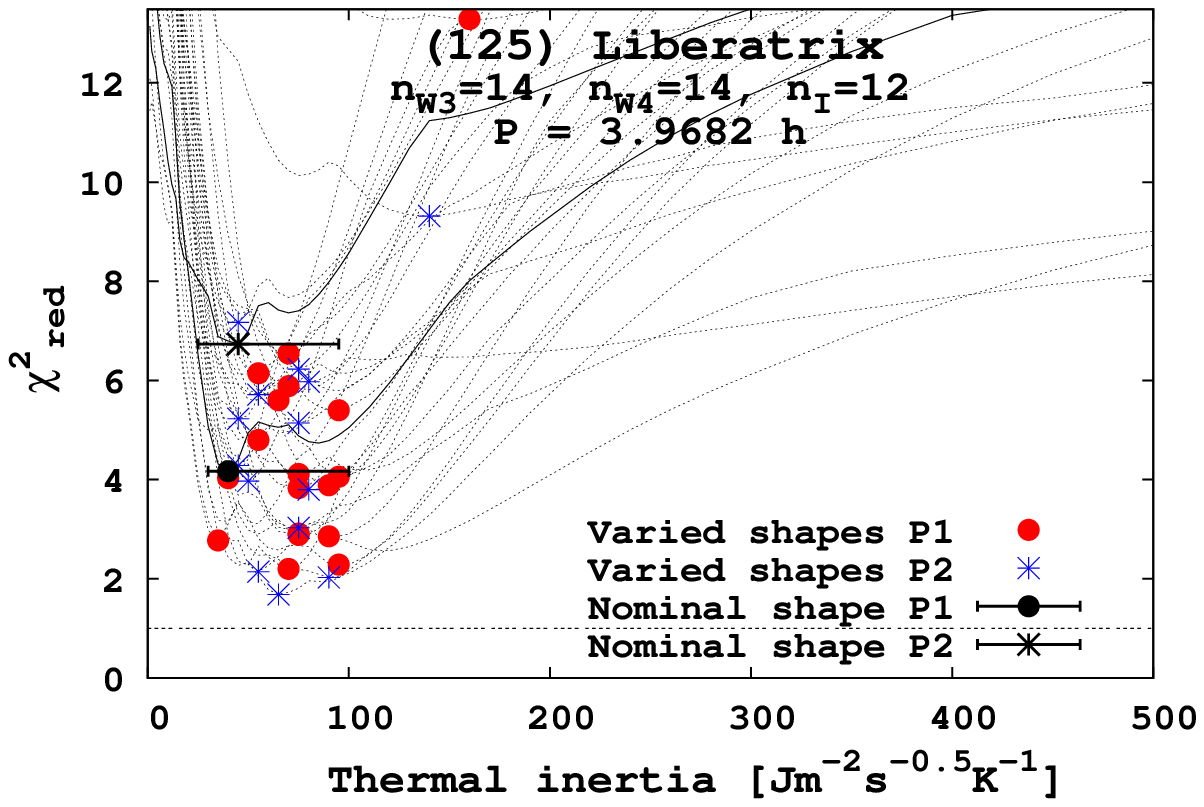}}\\
\resizebox{0.8\hsize}{!}{\includegraphics{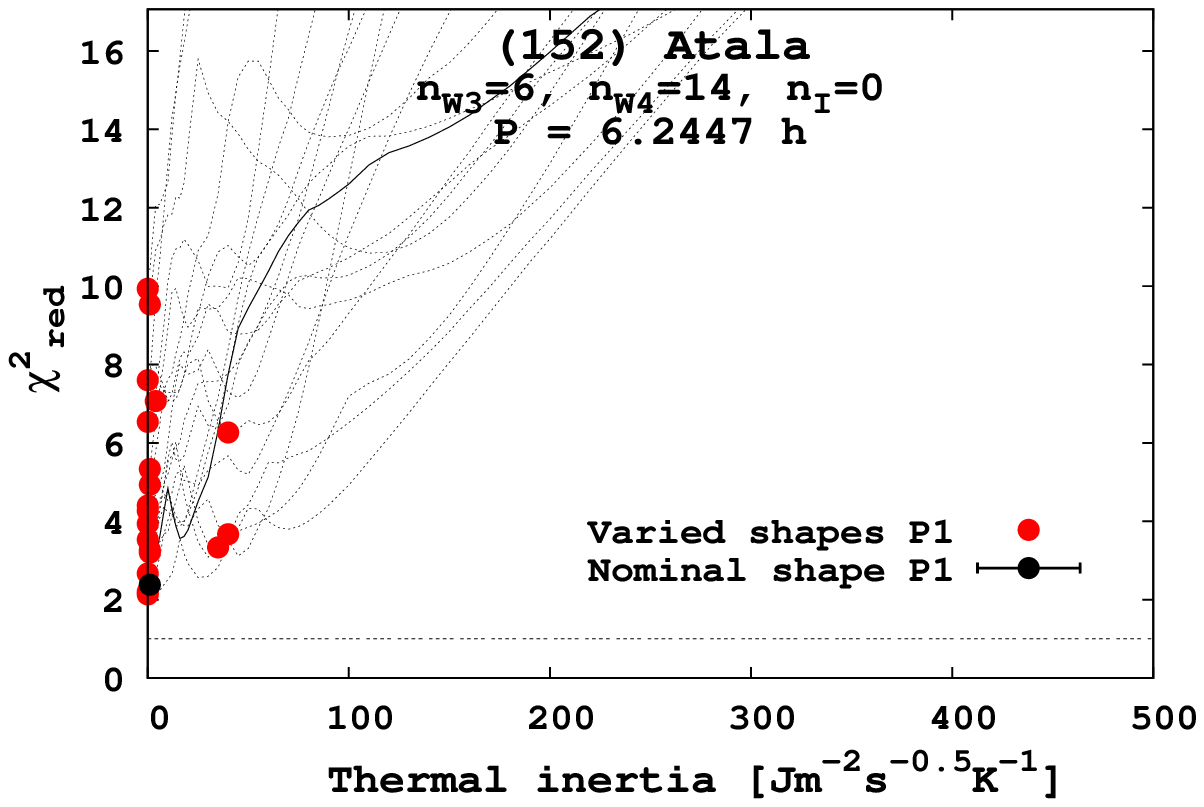}\includegraphics{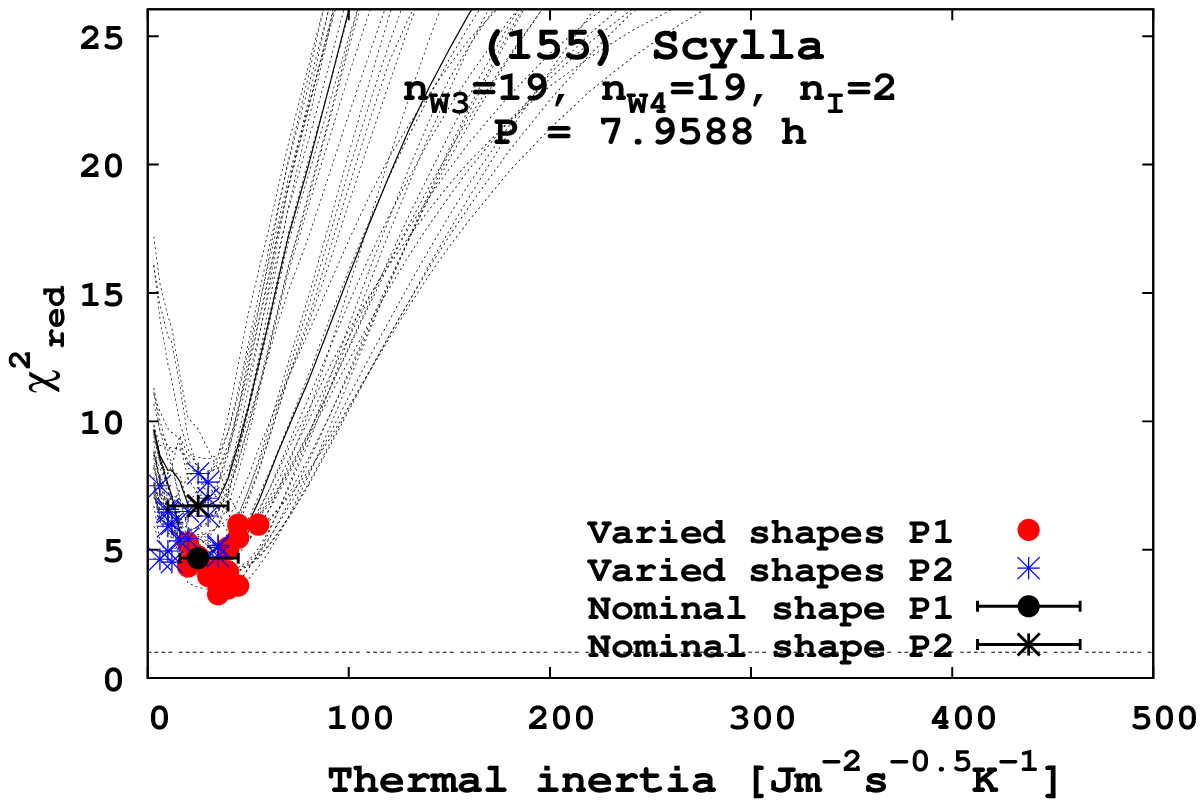}}\\
\resizebox{0.8\hsize}{!}{\includegraphics{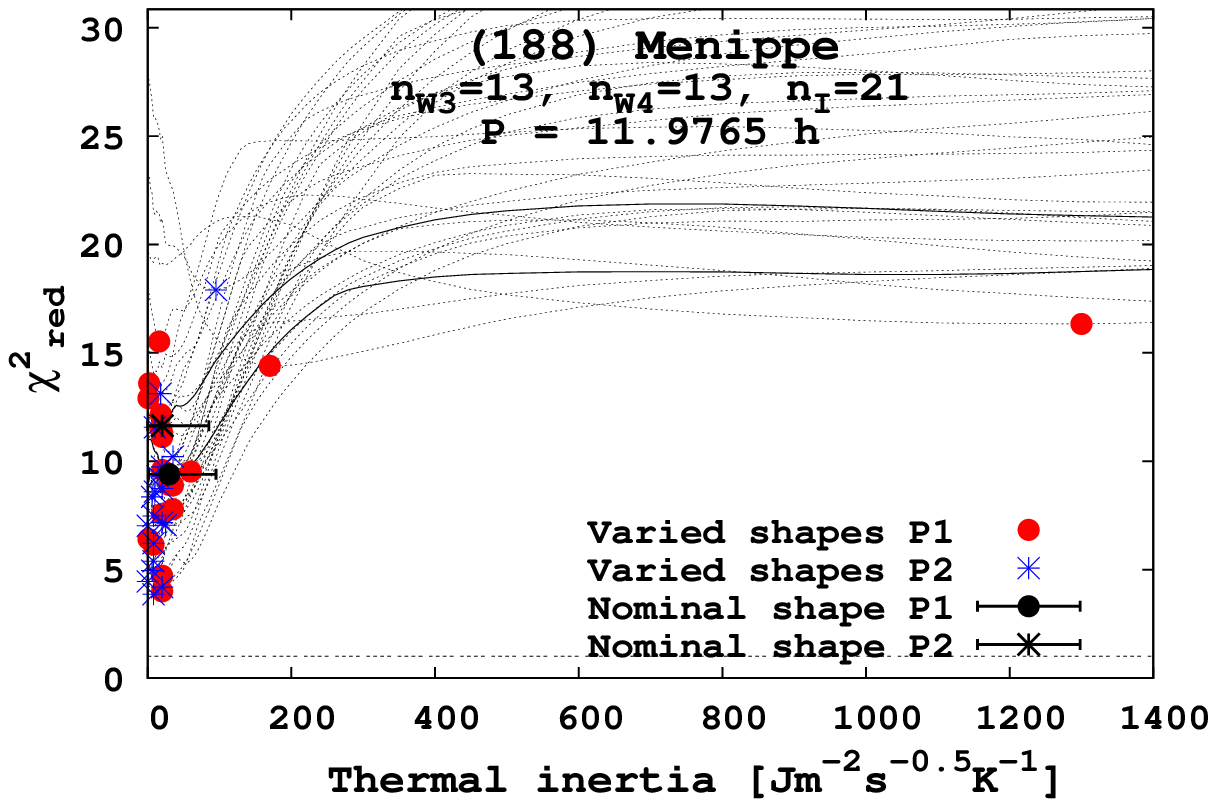}\includegraphics{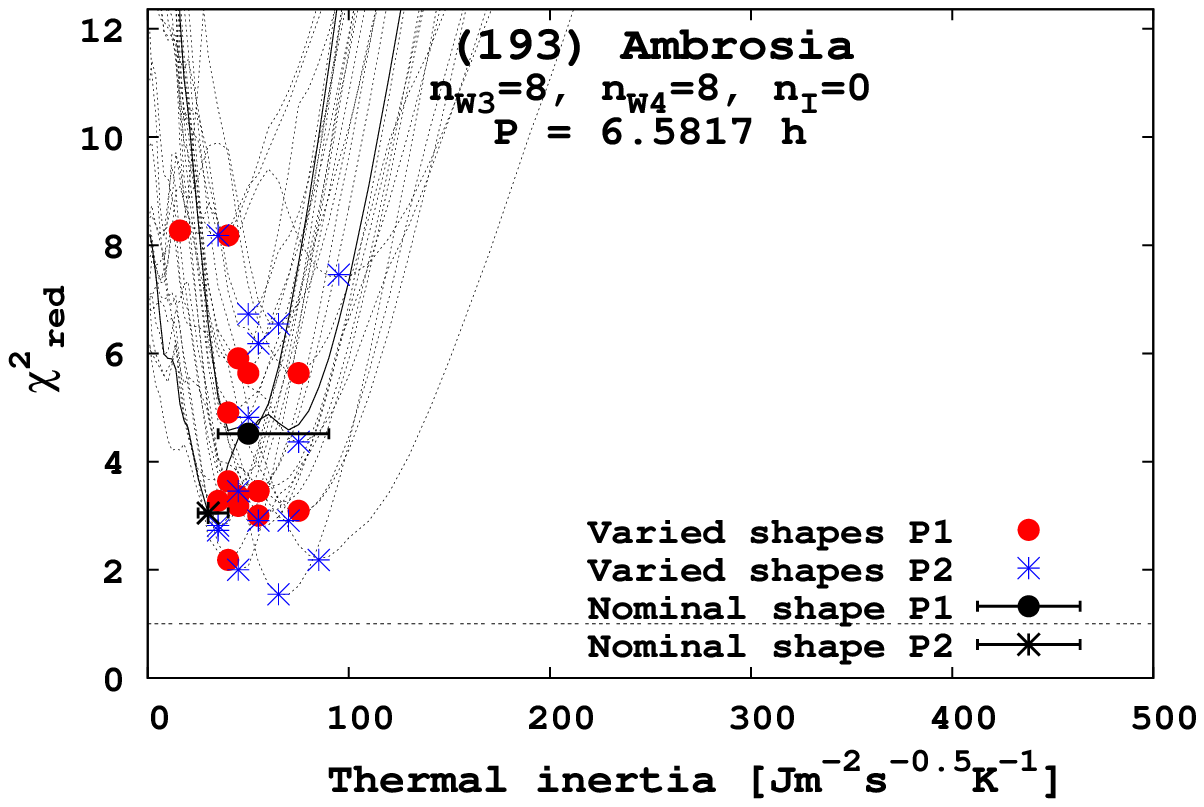}}\\
\end{center}
\caption{VS-TPM fits in the thermal inertia parameter space for eight asteroids. Each plot also contains the number of thermal infrared measurements in WISE W3 and W4 filters and in all four IRAS filters, and the rotation period.}
\end{figure*}

\begin{figure*}[!htbp]
\begin{center}
\resizebox{0.8\hsize}{!}{\includegraphics{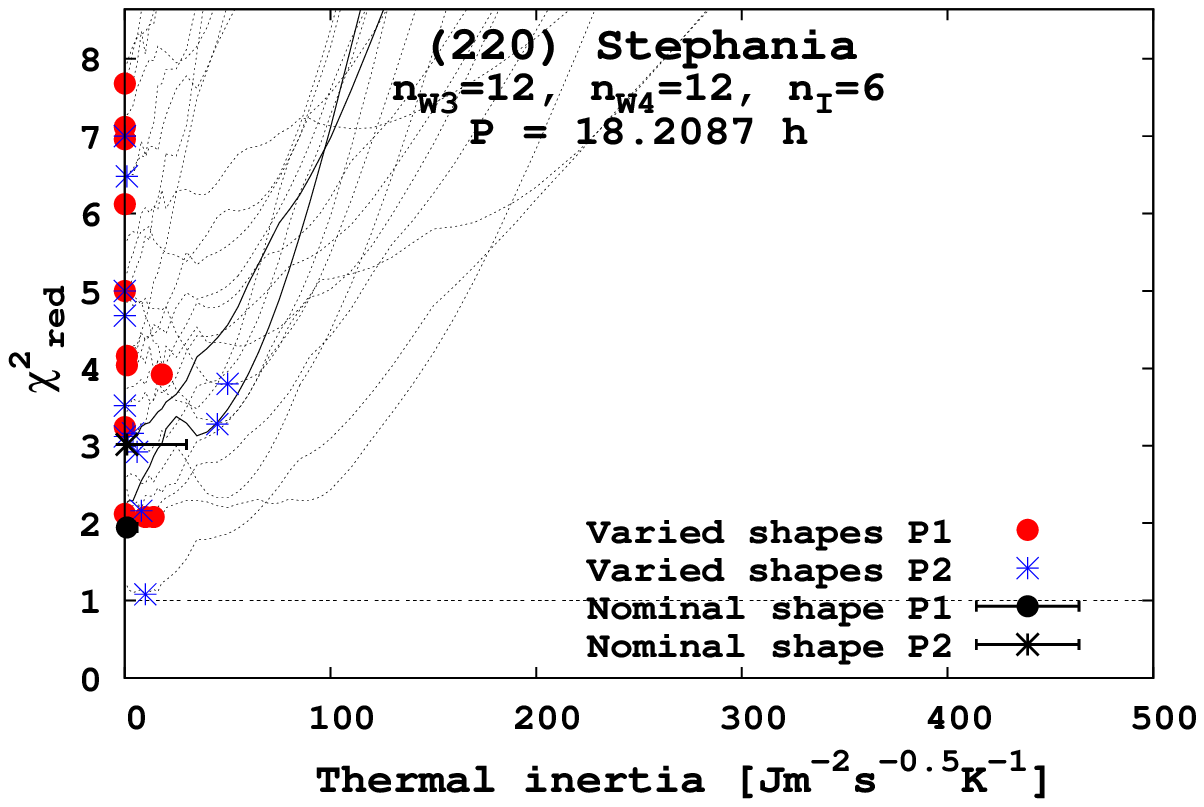}\includegraphics{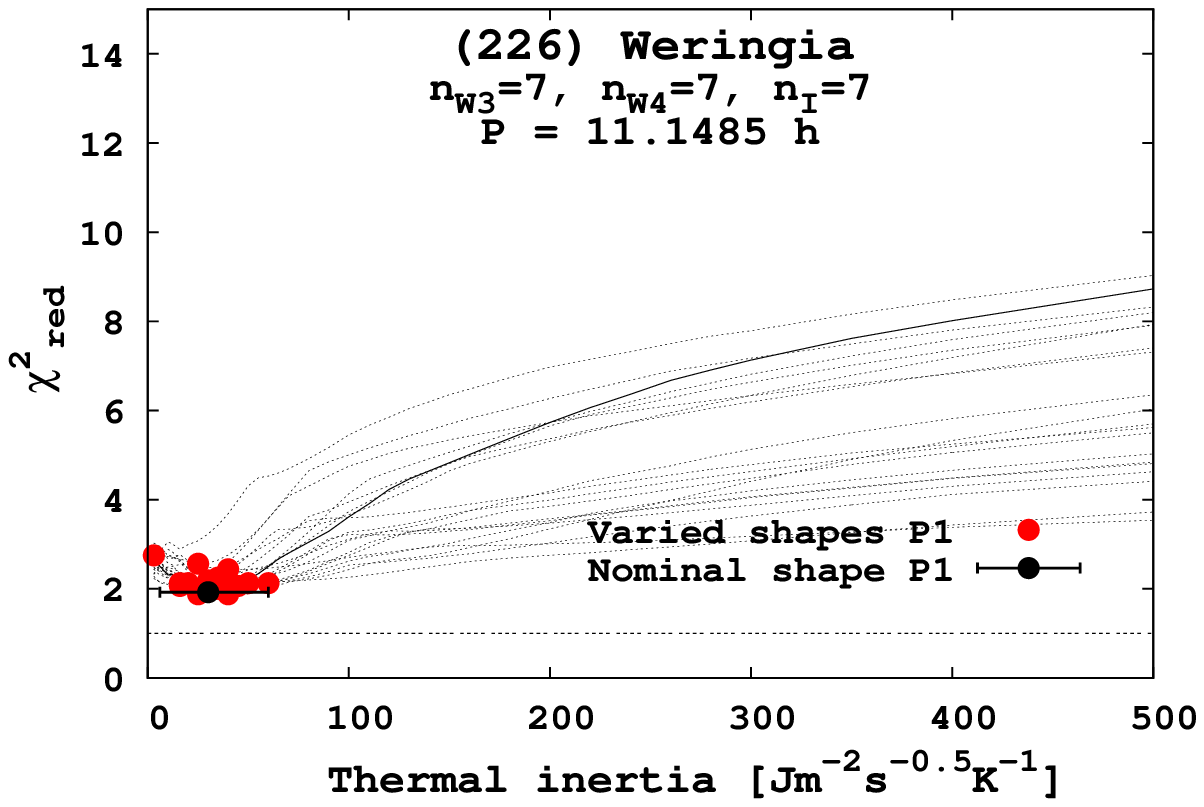}}\\
\resizebox{0.8\hsize}{!}{\includegraphics{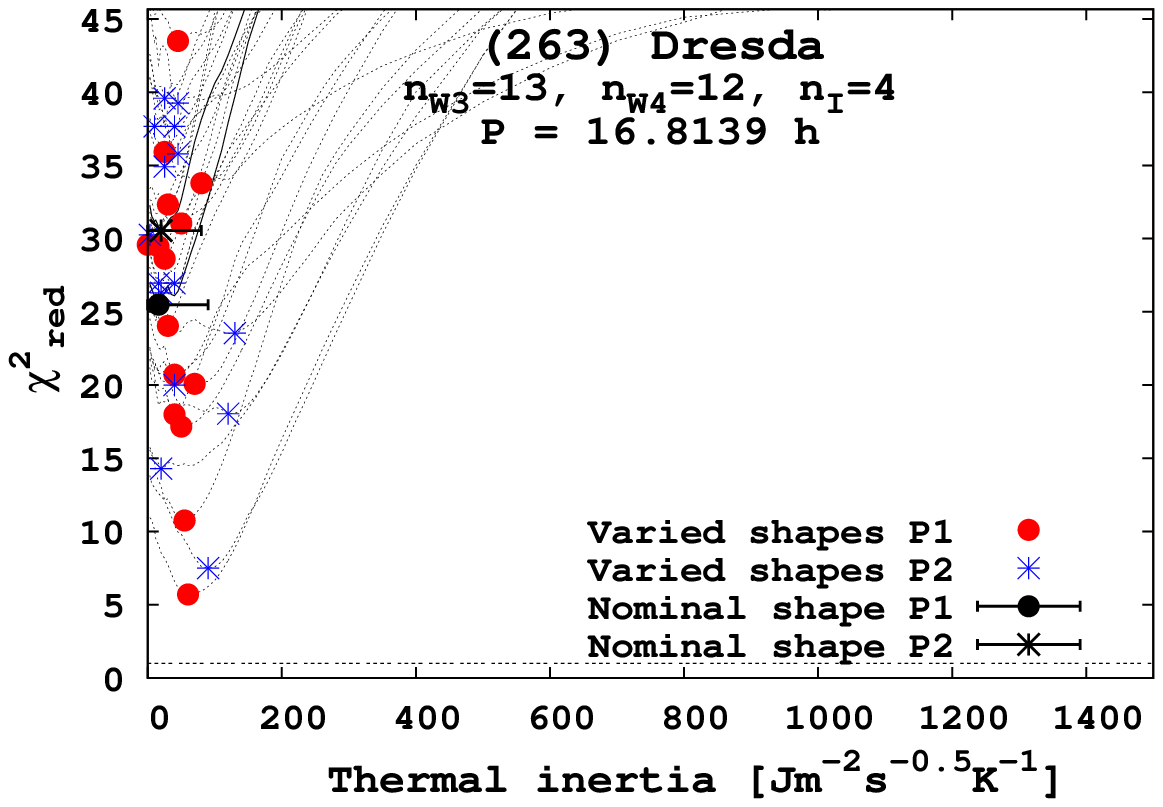}\includegraphics{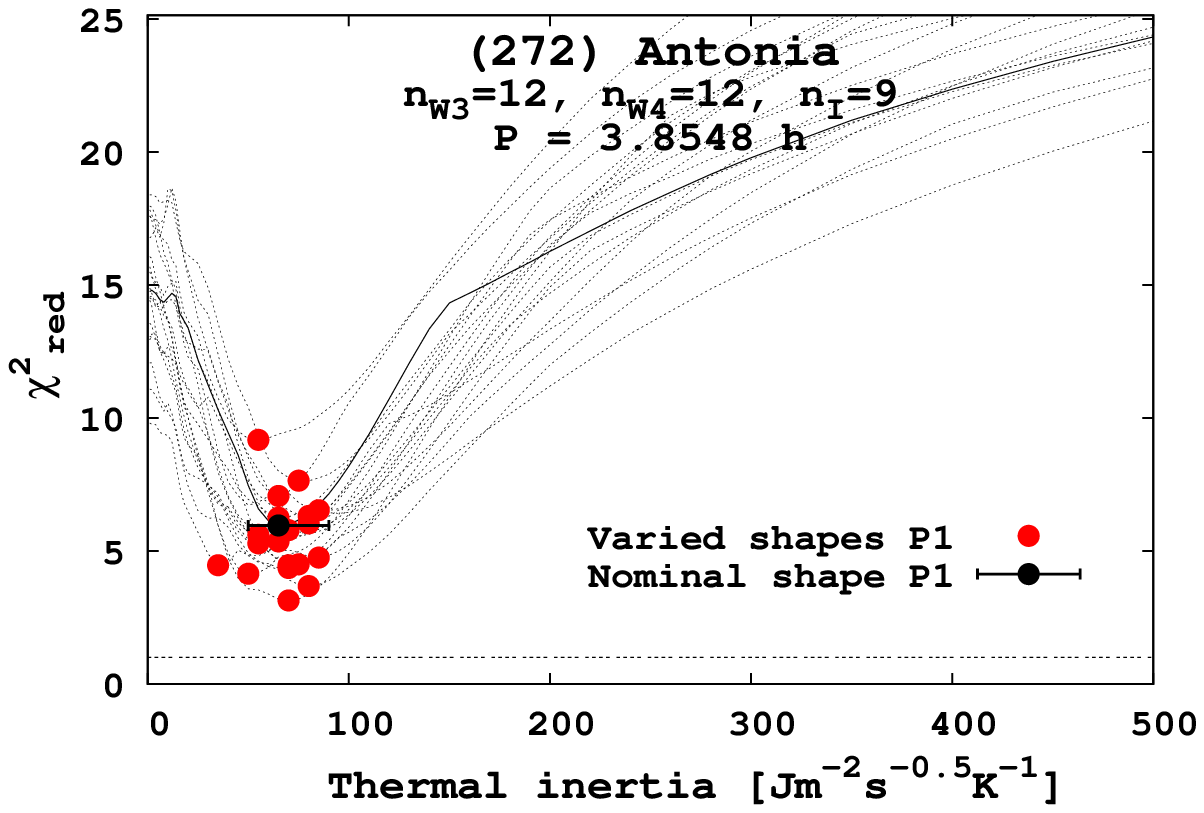}}\\
\resizebox{0.8\hsize}{!}{\includegraphics{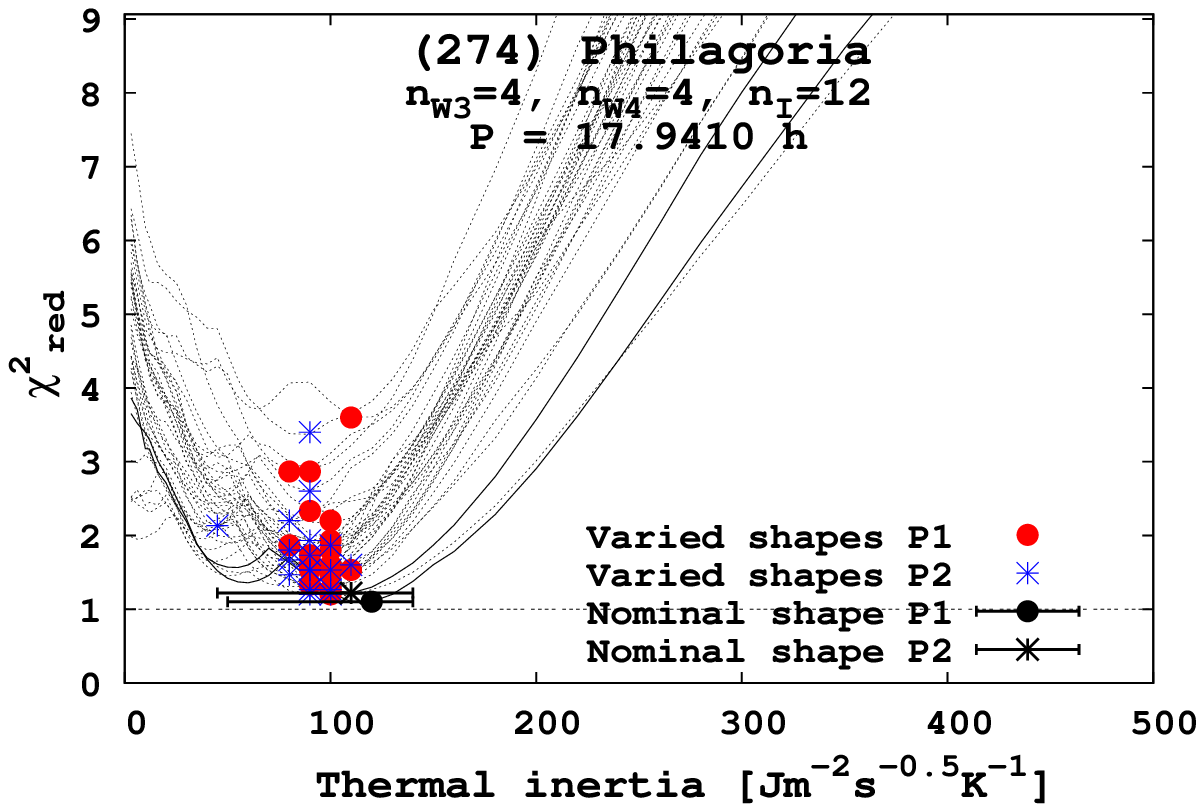}\includegraphics{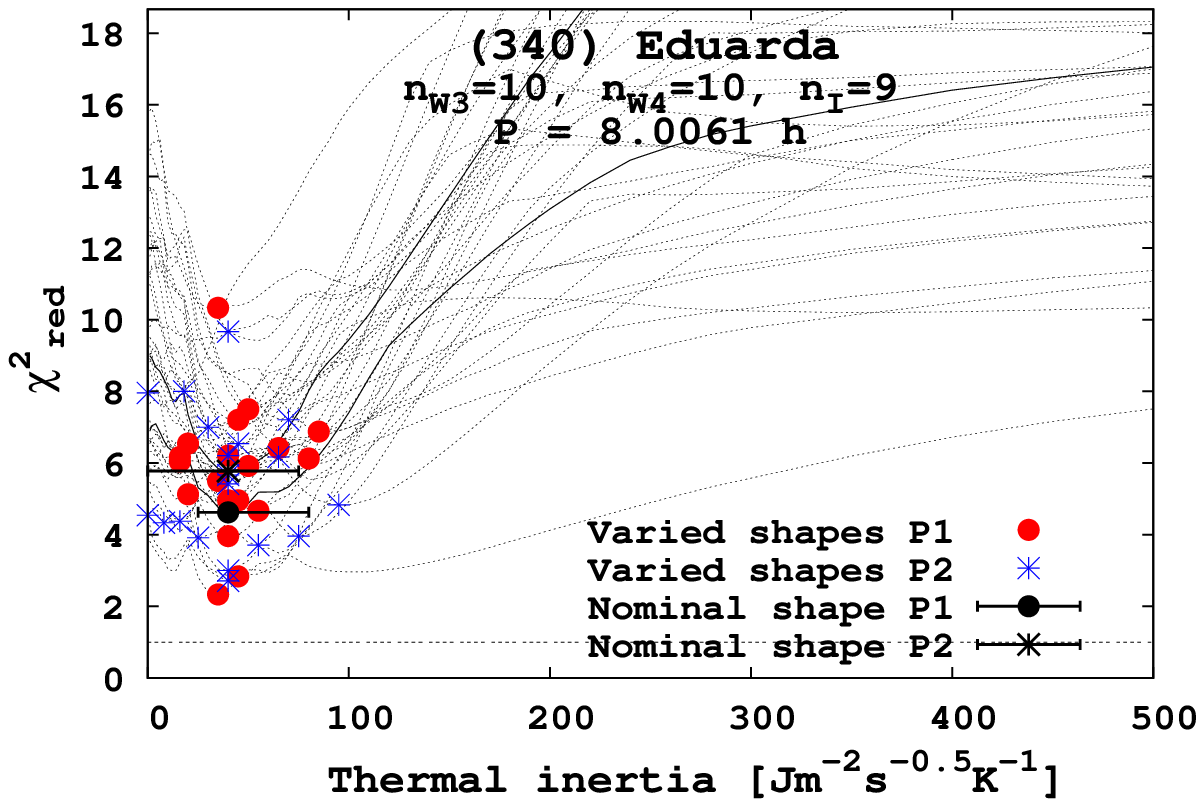}}\\
\resizebox{0.8\hsize}{!}{\includegraphics{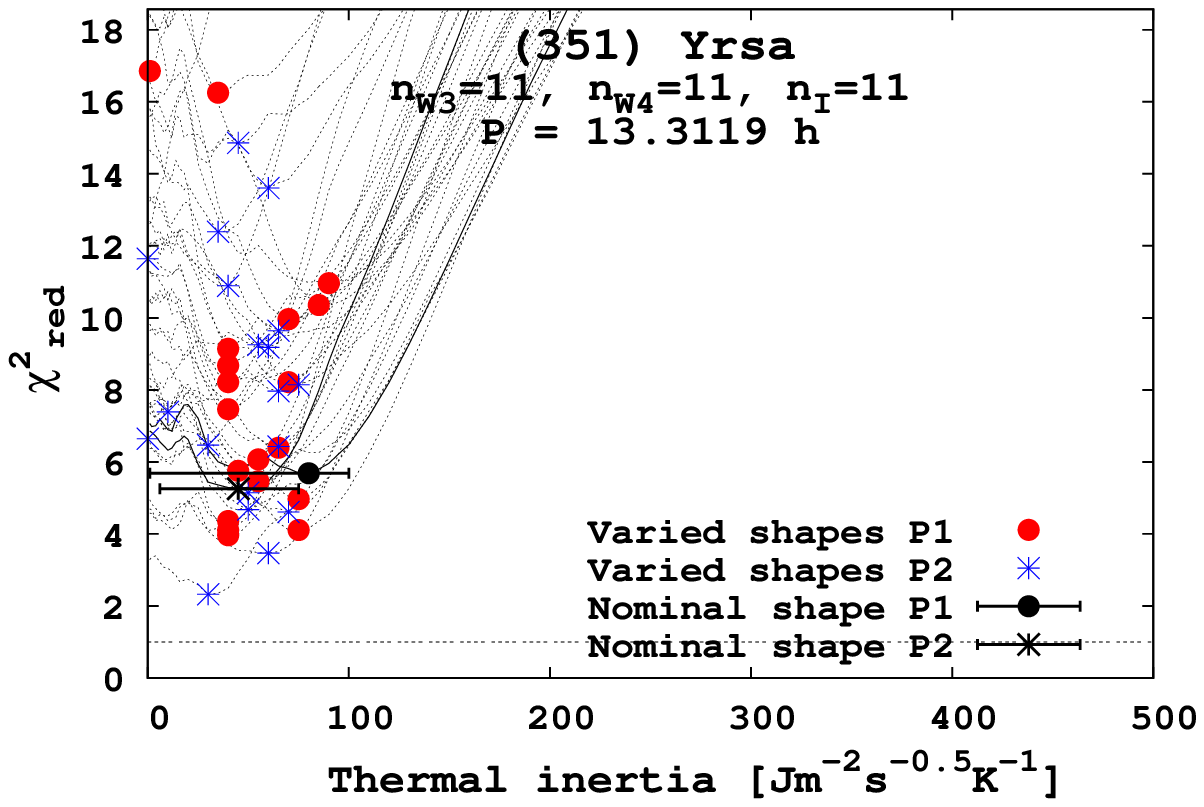}\includegraphics{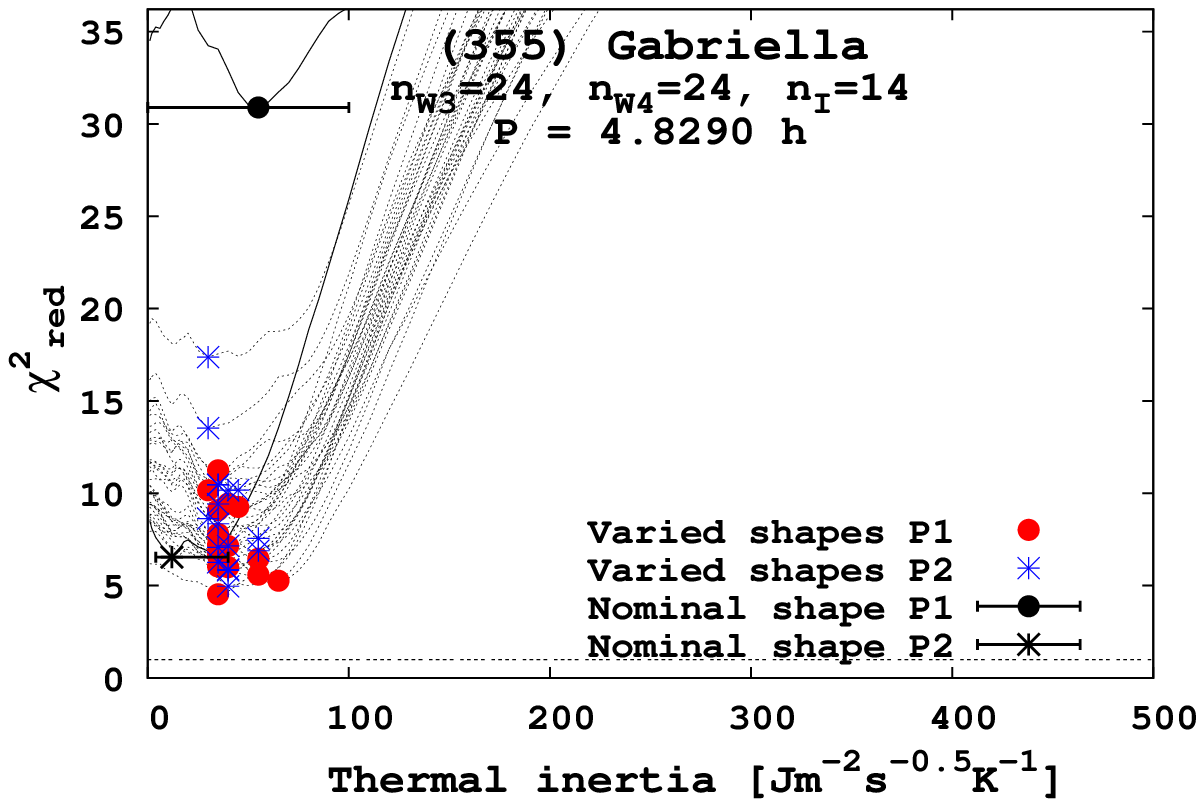}}\\
\end{center}
\caption{VS-TPM fits in the thermal inertia parameter space for eight asteroids. Each plot also contains the number of thermal infrared measurements in WISE W3 and W4 filters and in all four IRAS filters, and the rotation period.}
\end{figure*}

\begin{figure*}[!htbp]
\begin{center}
\resizebox{0.8\hsize}{!}{\includegraphics{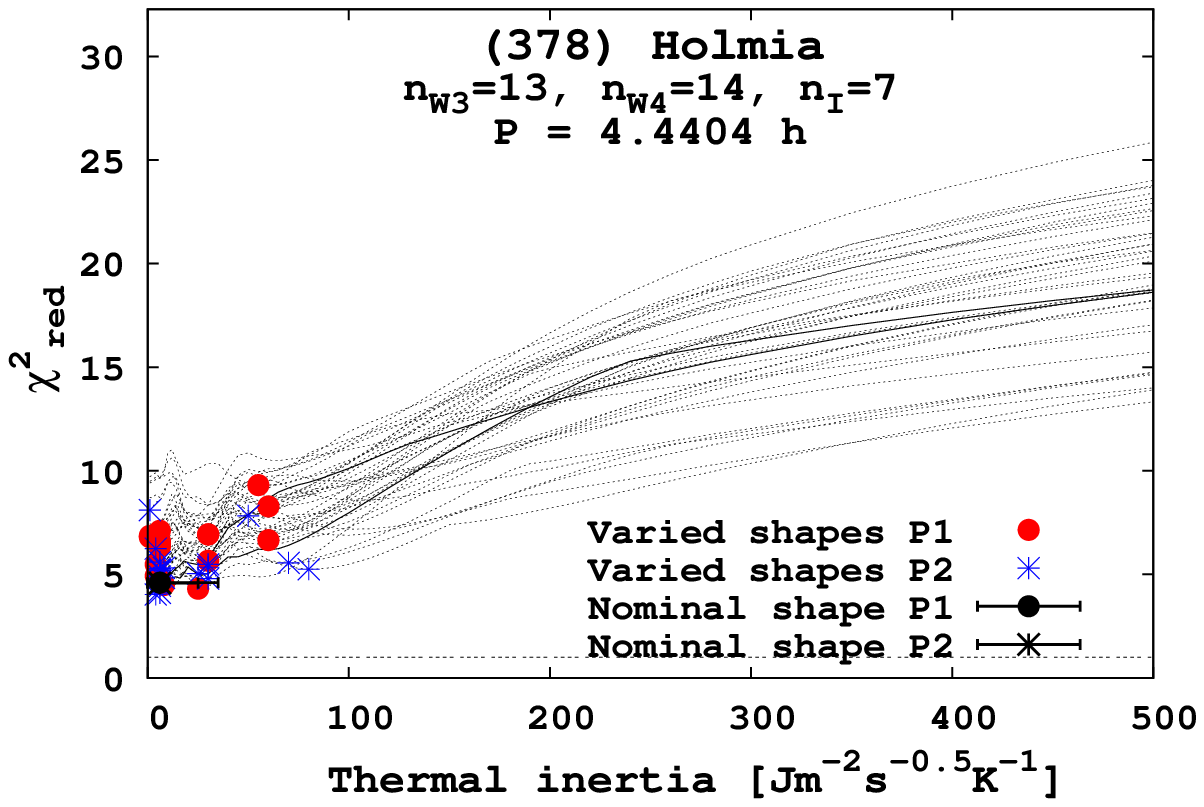}\includegraphics{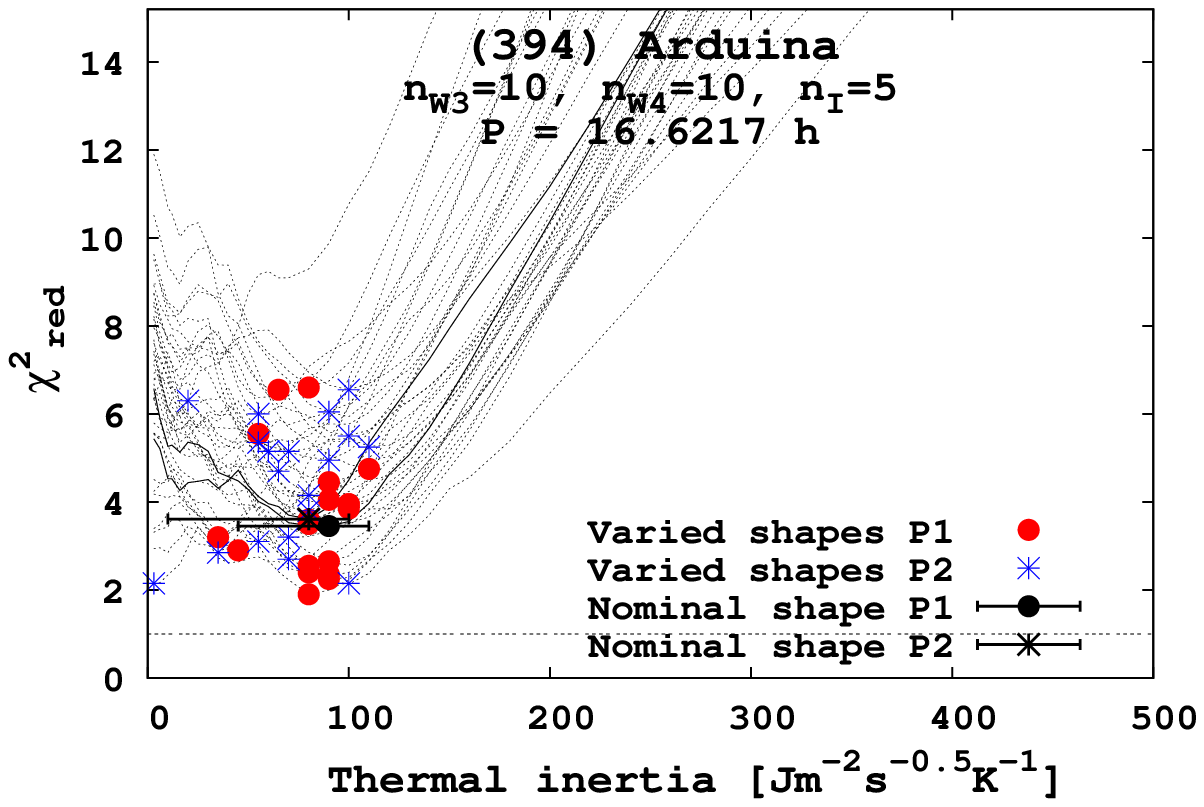}}\\
\resizebox{0.8\hsize}{!}{\includegraphics{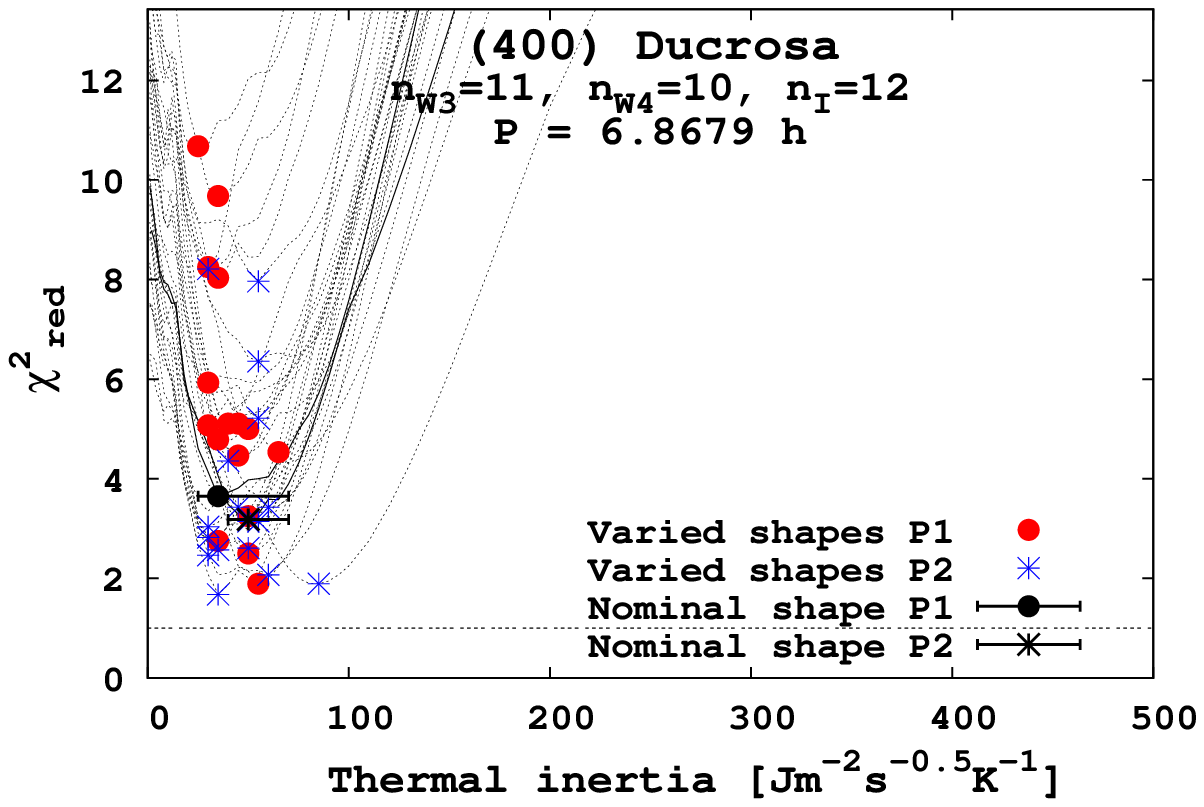}\includegraphics{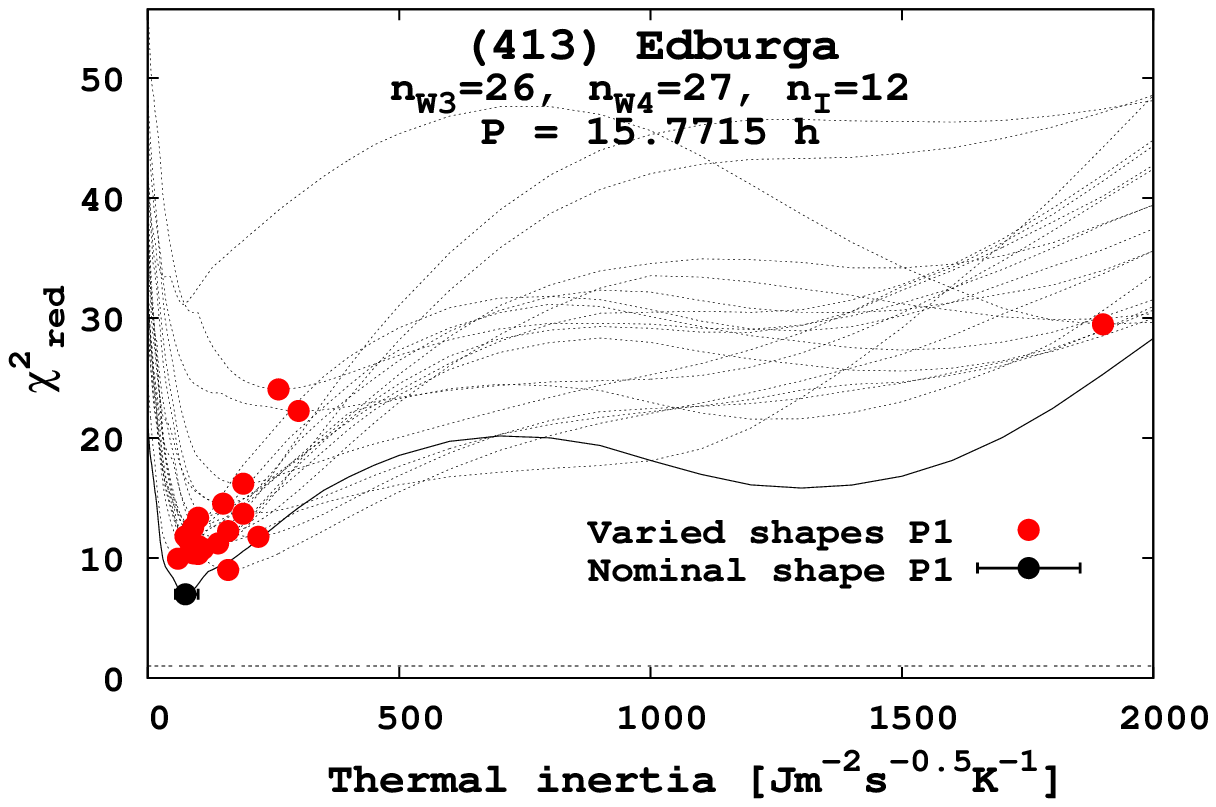}}\\
\resizebox{0.8\hsize}{!}{\includegraphics{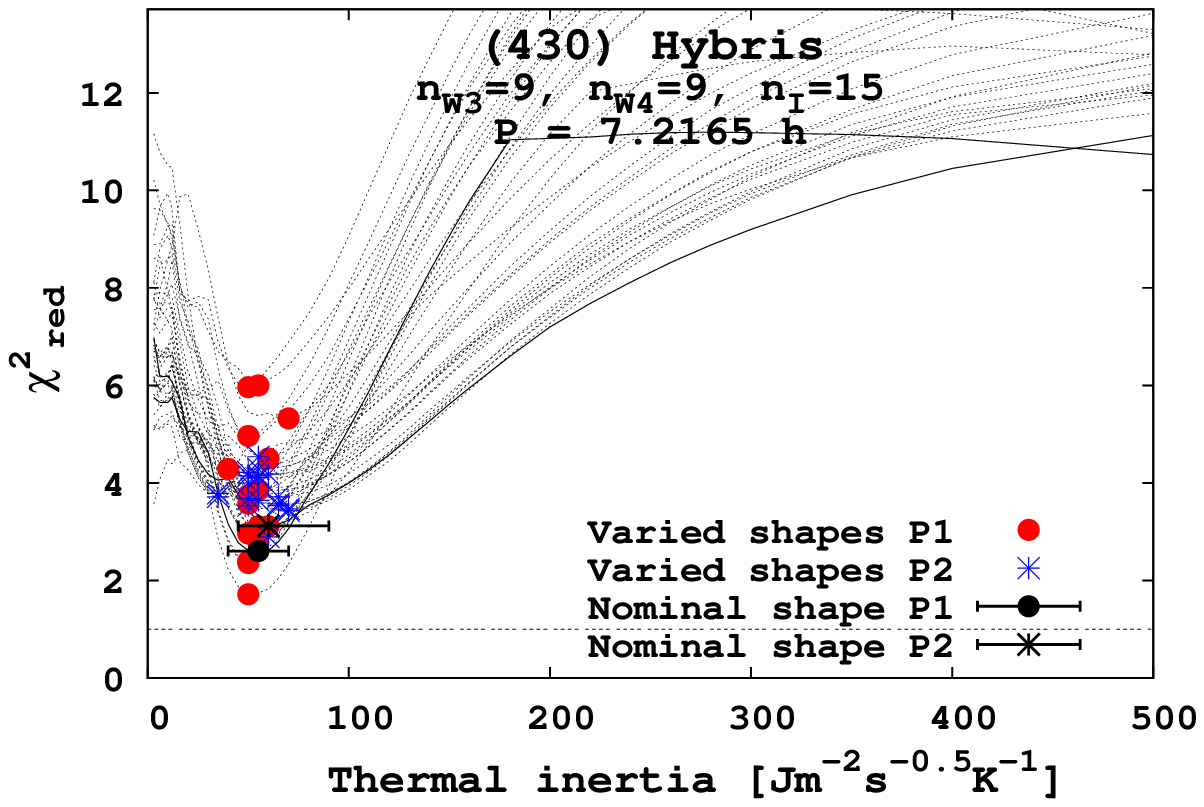}\includegraphics{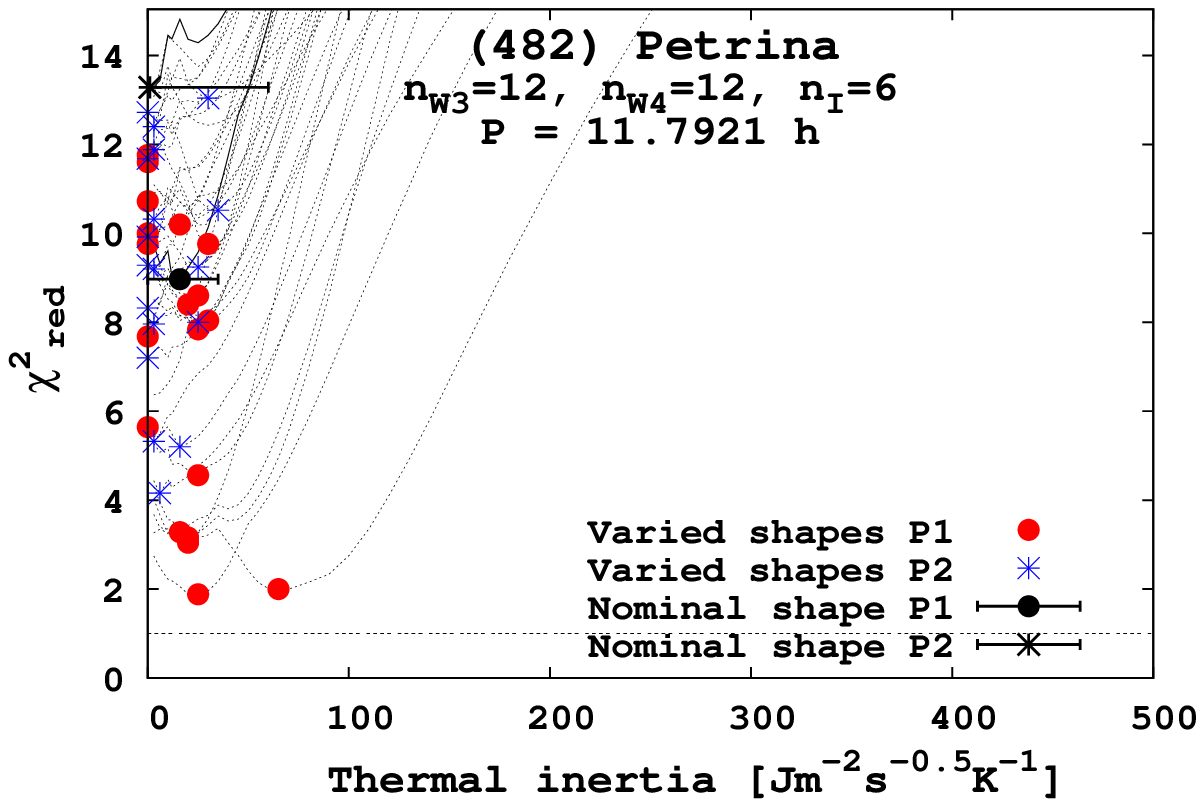}}\\
\resizebox{0.8\hsize}{!}{\includegraphics{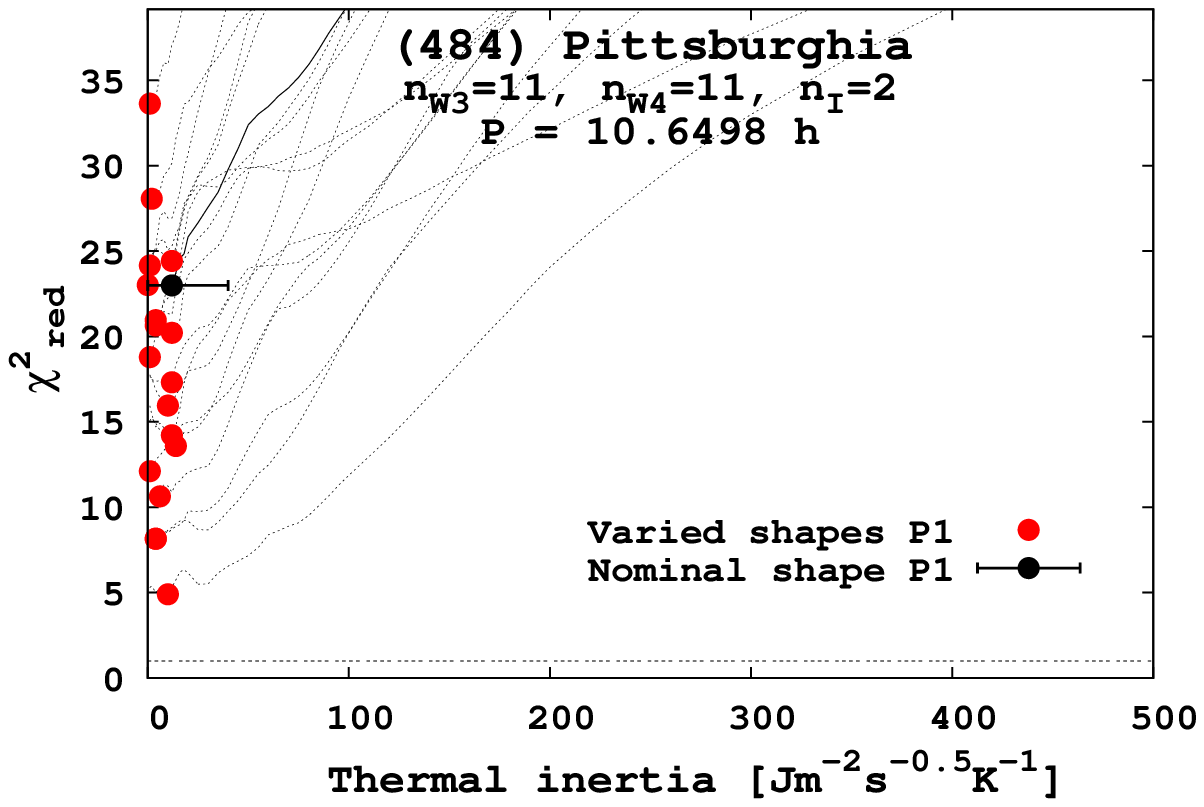}\includegraphics{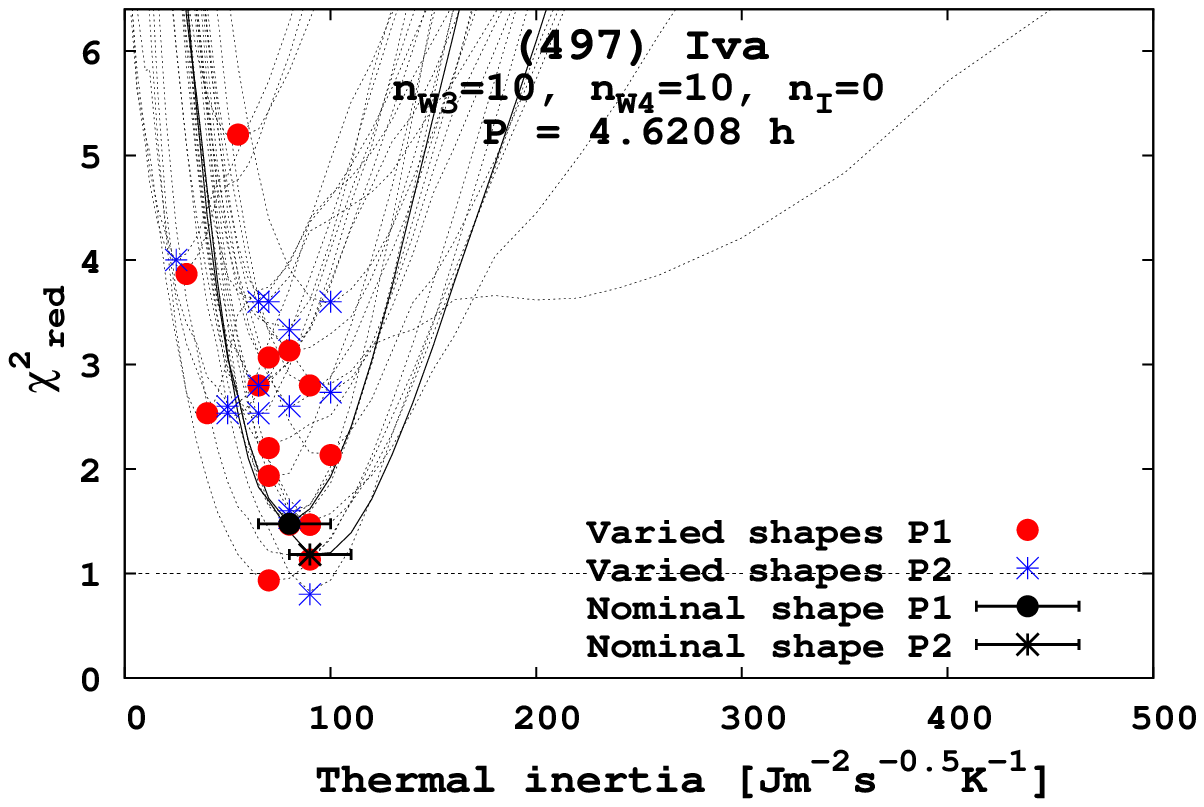}}\\
\end{center}
\caption{VS-TPM fits in the thermal inertia parameter space for eight asteroids. Each plot also contains the number of thermal infrared measurements in WISE W3 and W4 filters and in all four IRAS filters, and the rotation period.}
\end{figure*}

\begin{figure*}[!htbp]
\begin{center}
\resizebox{0.8\hsize}{!}{\includegraphics{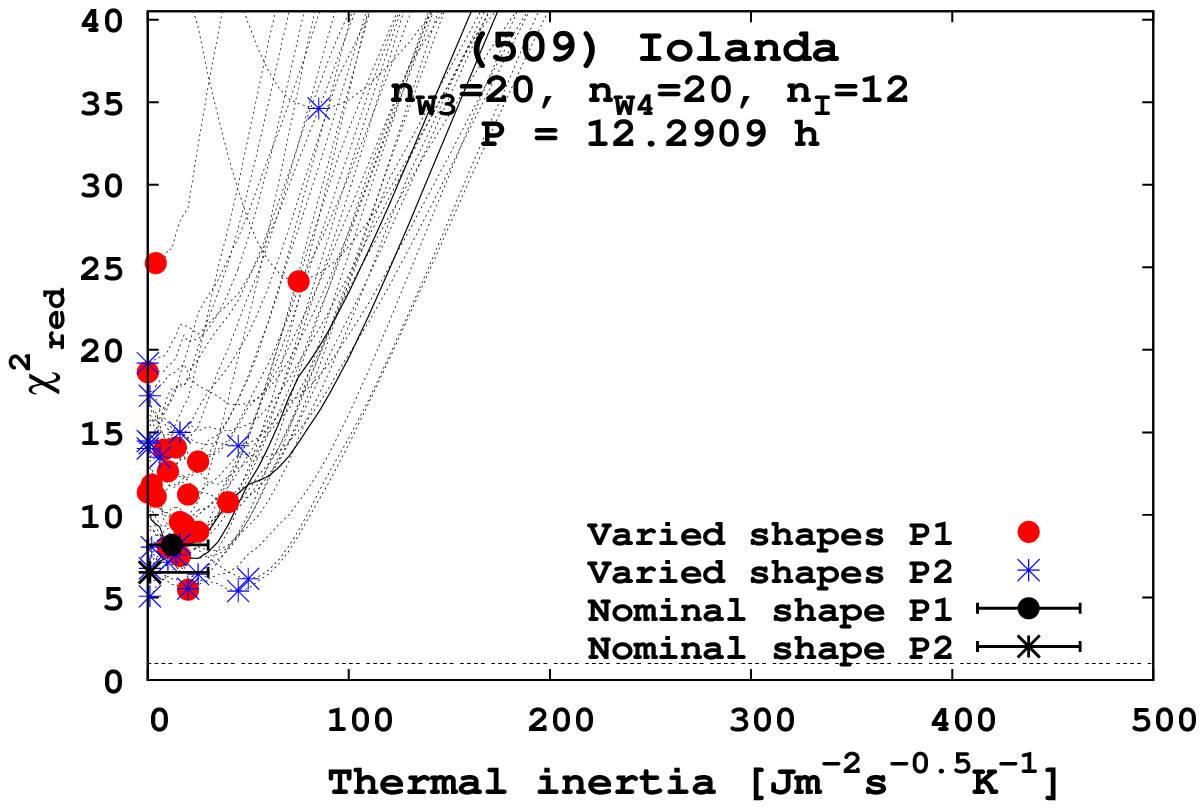}\includegraphics{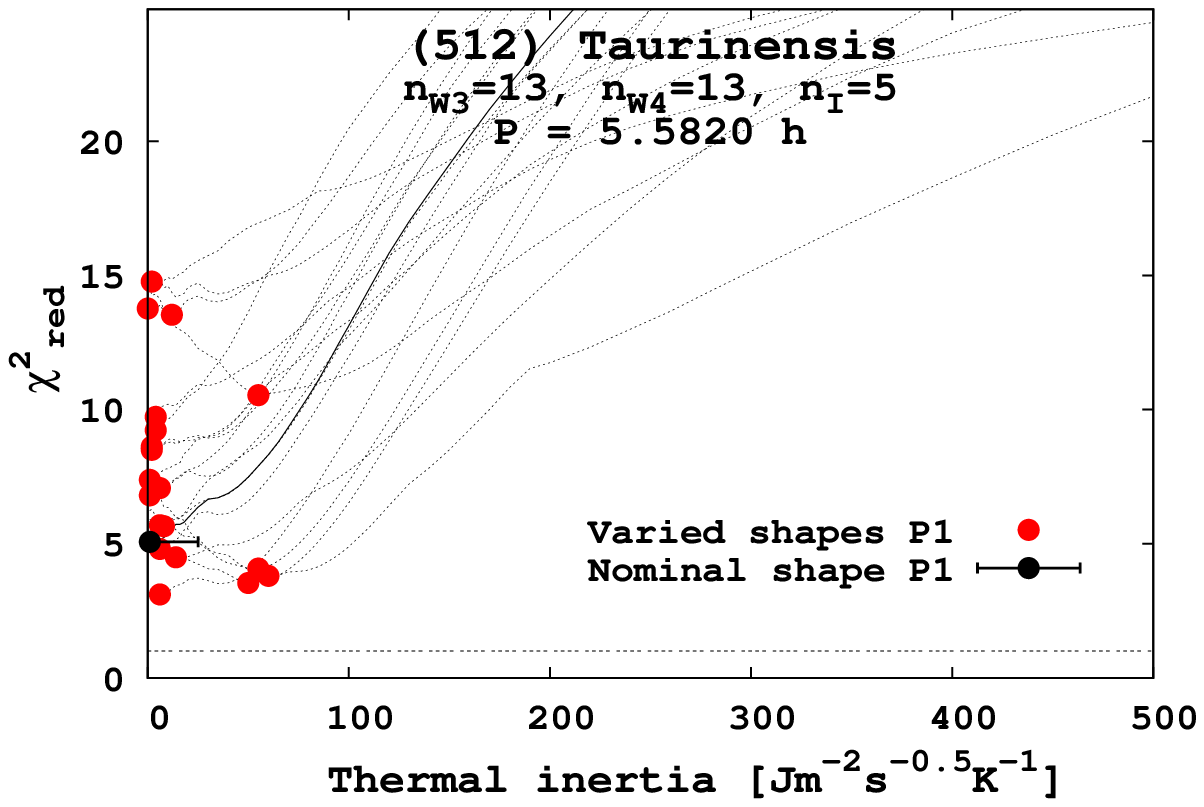}}\\
\resizebox{0.8\hsize}{!}{\includegraphics{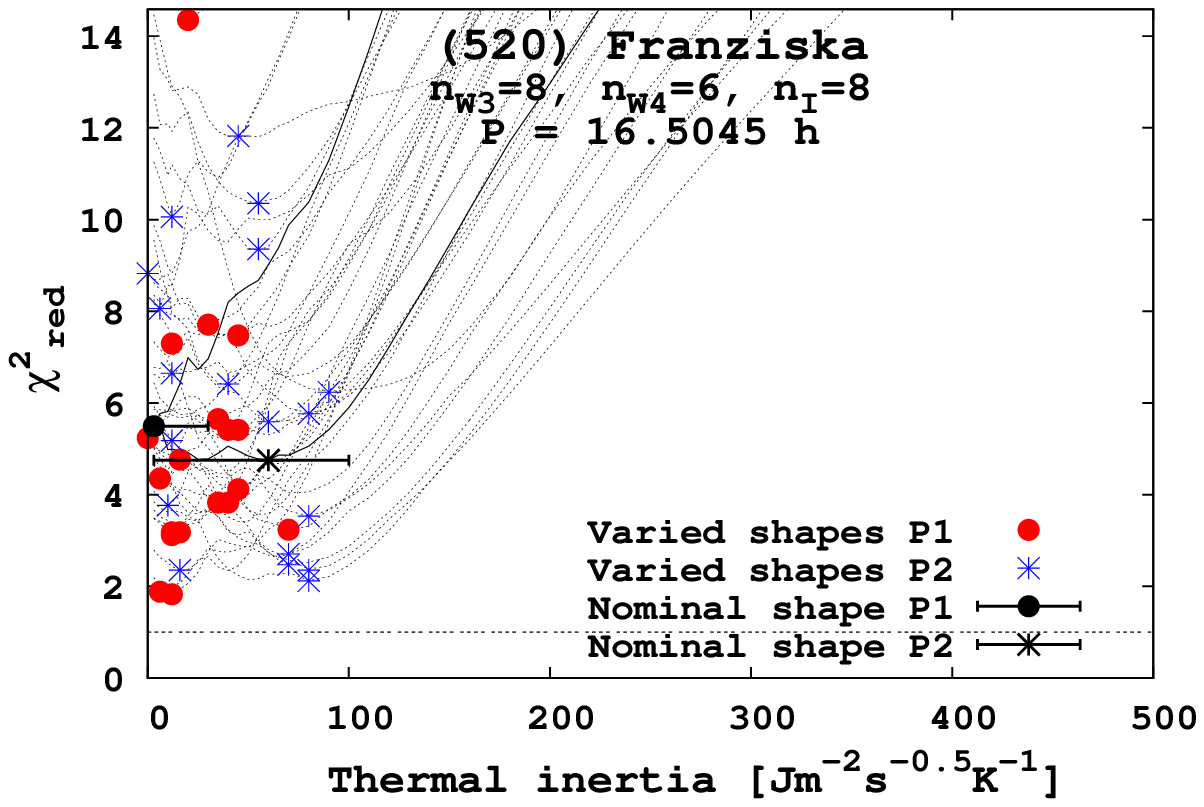}\includegraphics{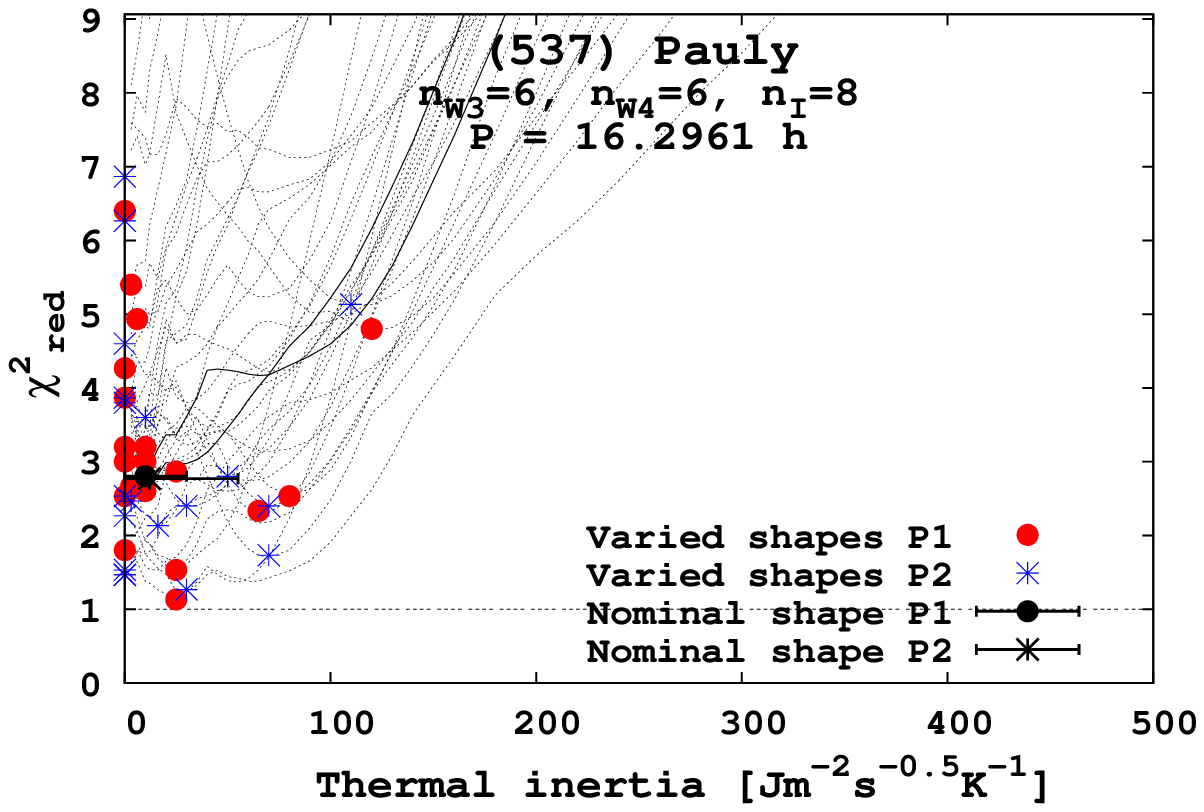}}\\
\resizebox{0.8\hsize}{!}{\includegraphics{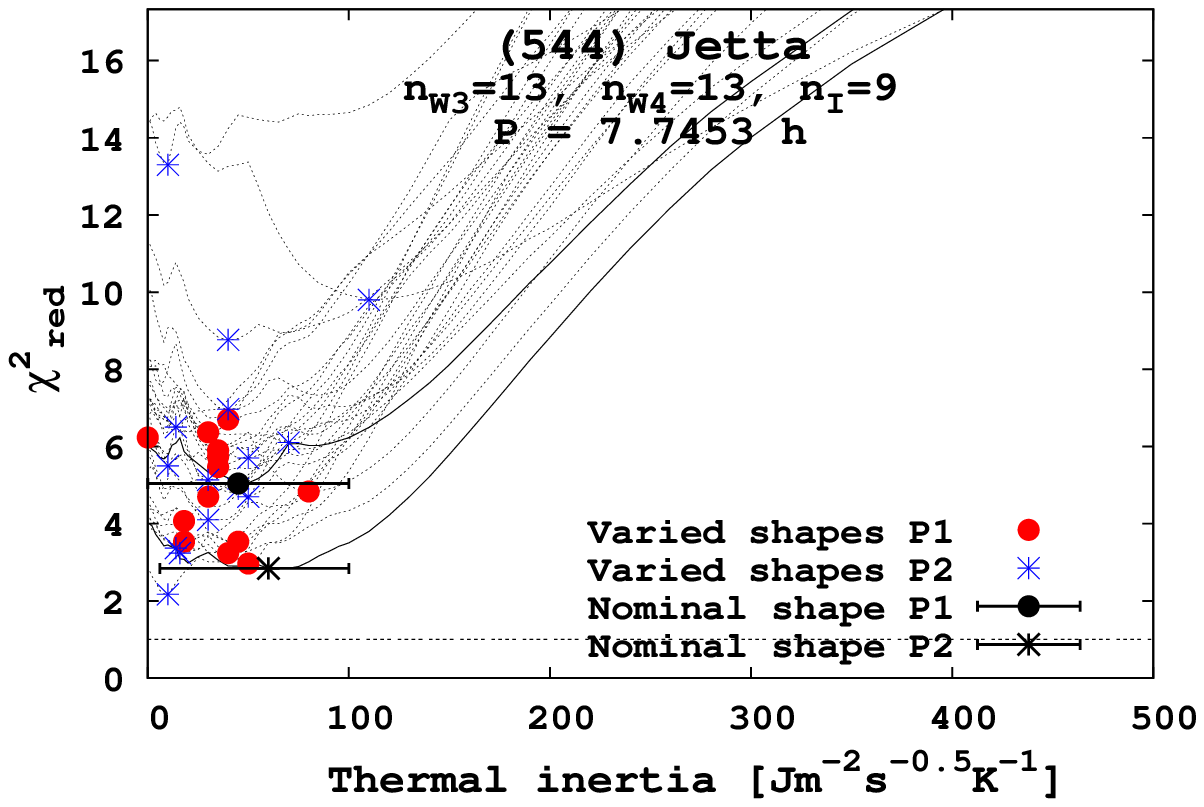}\includegraphics{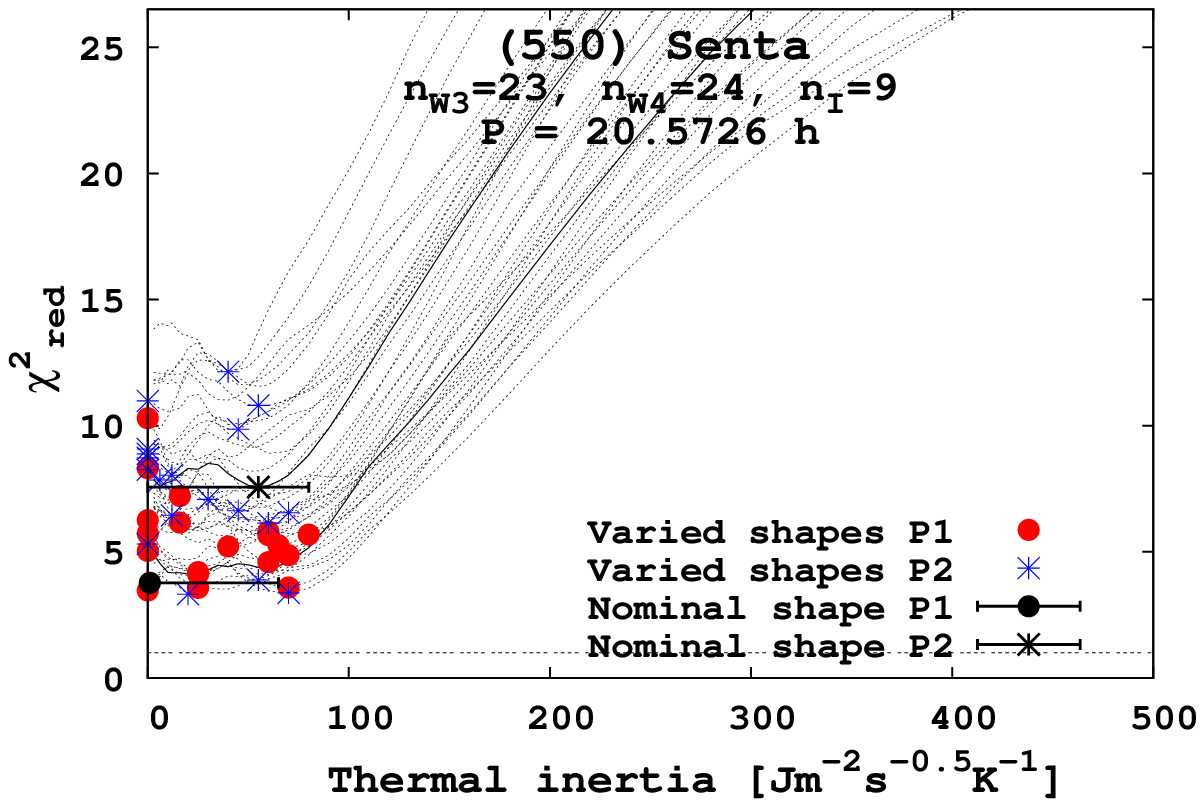}}\\
\resizebox{0.8\hsize}{!}{\includegraphics{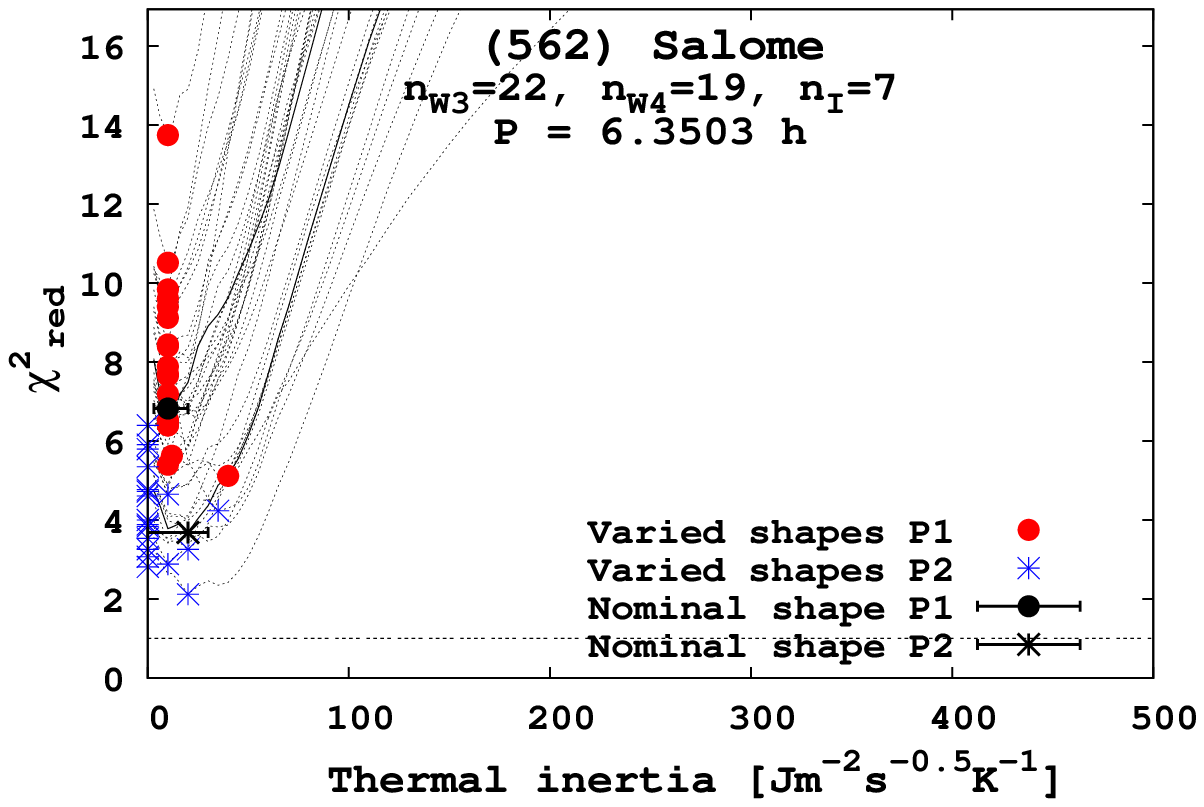}\includegraphics{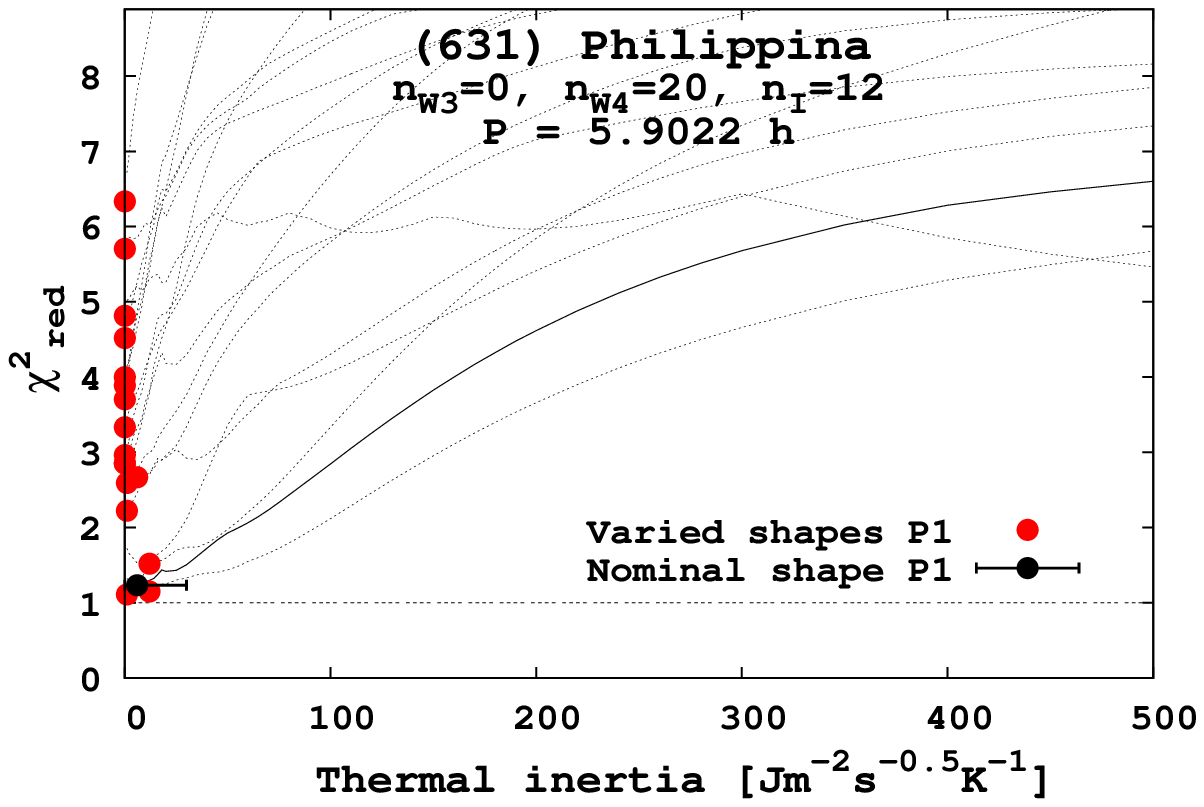}}\\
\end{center}
\caption{VS-TPM fits in the thermal inertia parameter space for eight asteroids. Each plot also contains the number of thermal infrared measurements in WISE W3 and W4 filters and in all four IRAS filters, and the rotation period.}
\end{figure*}

\begin{figure*}[!htbp]
\begin{center}
\resizebox{0.8\hsize}{!}{\includegraphics{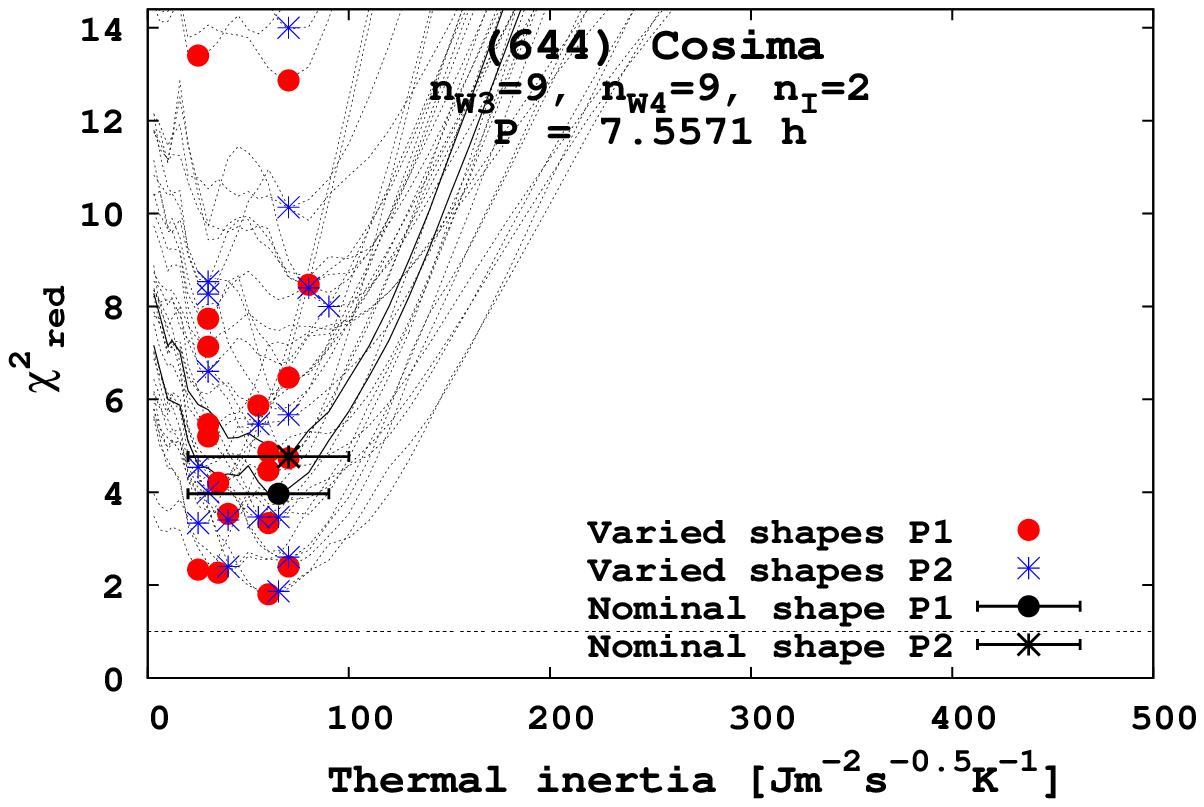}\includegraphics{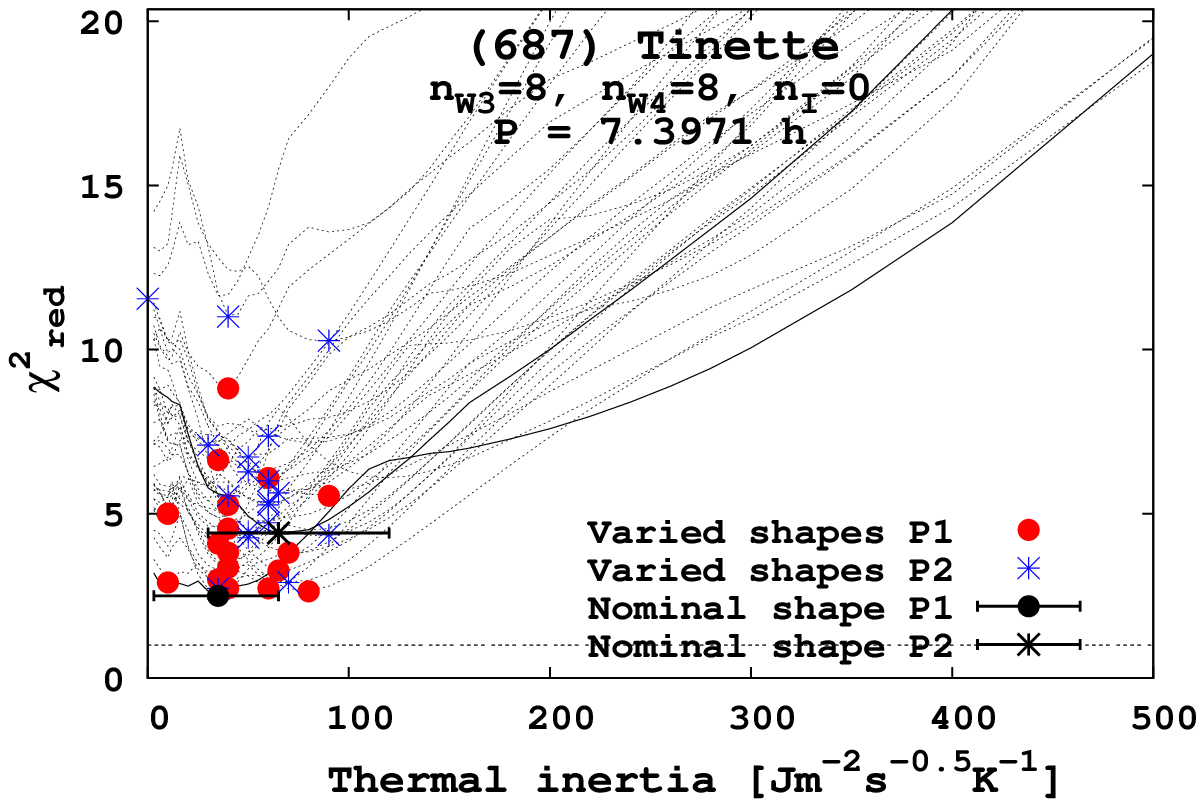}}\\
\resizebox{0.8\hsize}{!}{\includegraphics{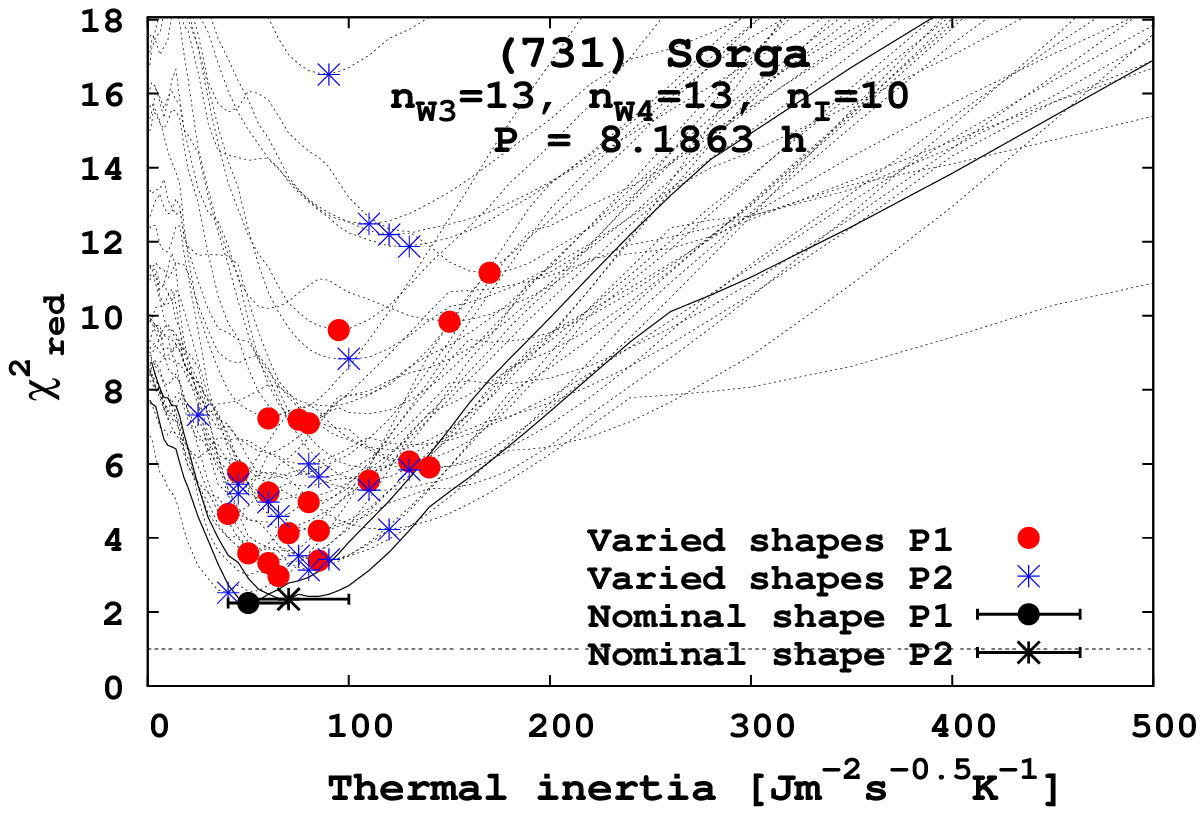}\includegraphics{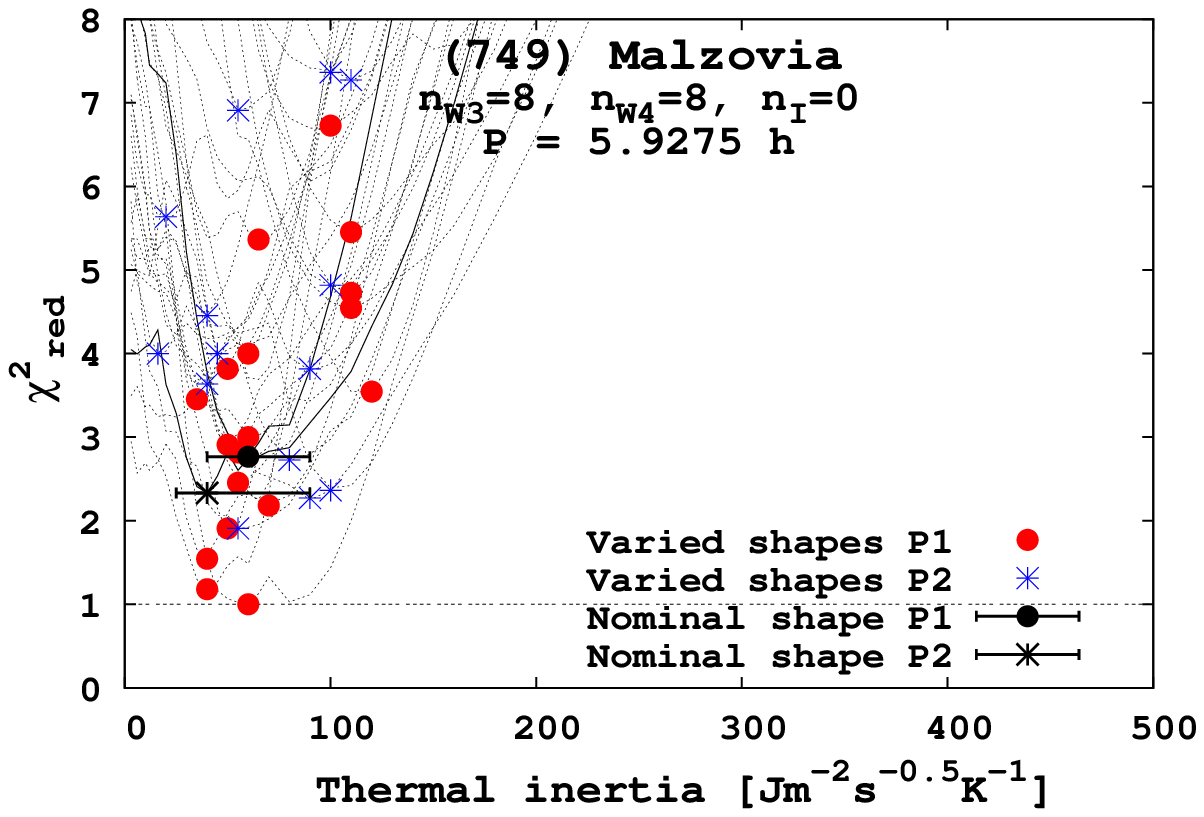}}\\
\resizebox{0.8\hsize}{!}{\includegraphics{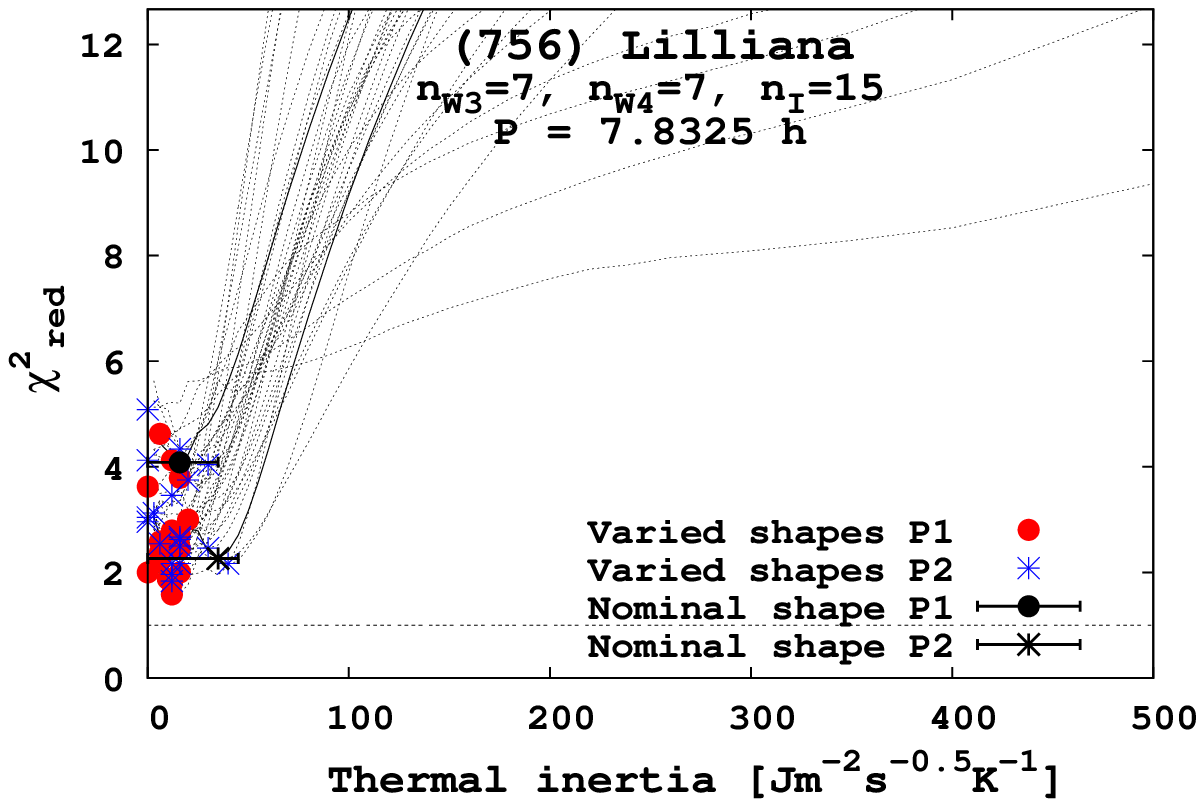}\includegraphics{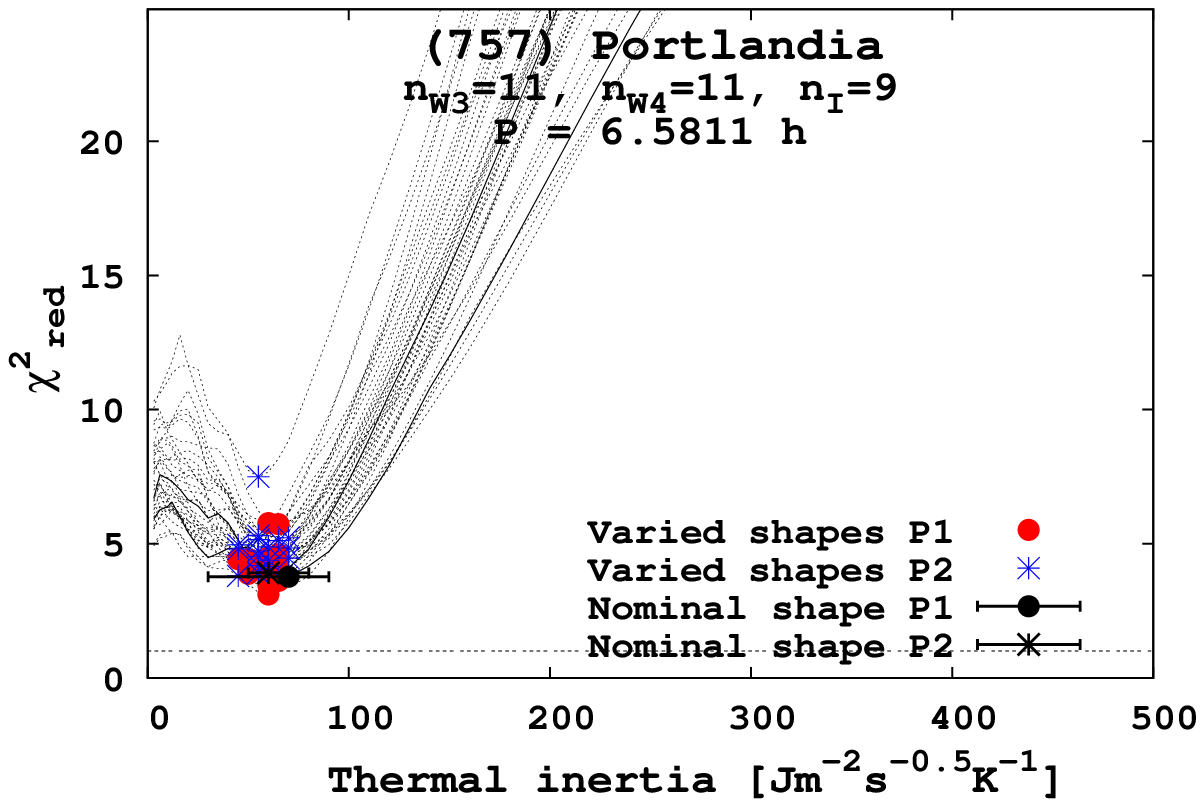}}\\
\resizebox{0.8\hsize}{!}{\includegraphics{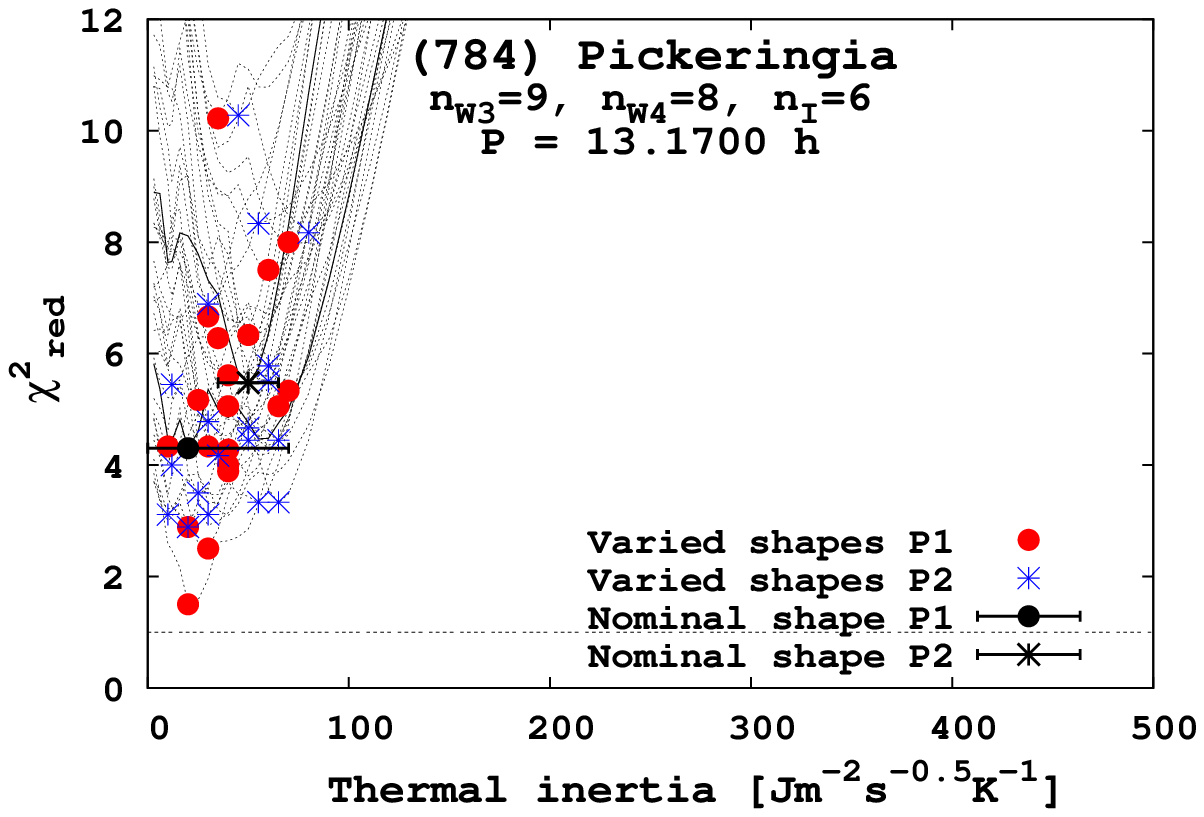}\includegraphics{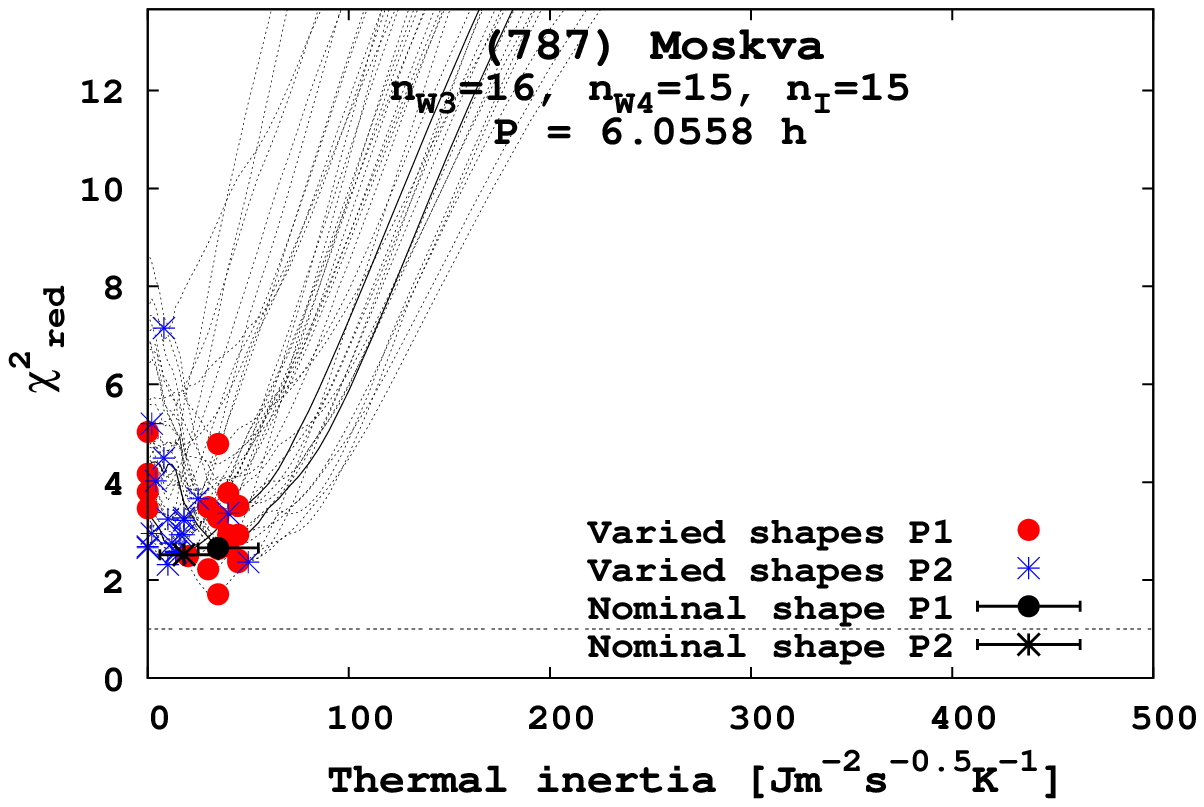}}\\
\end{center}
\caption{VS-TPM fits in the thermal inertia parameter space for eight asteroids. Each plot also contains the number of thermal infrared measurements in WISE W3 and W4 filters and in all four IRAS filters, and the rotation period.}
\end{figure*}

\begin{figure*}[!htbp]
\begin{center}
\resizebox{0.8\hsize}{!}{\includegraphics{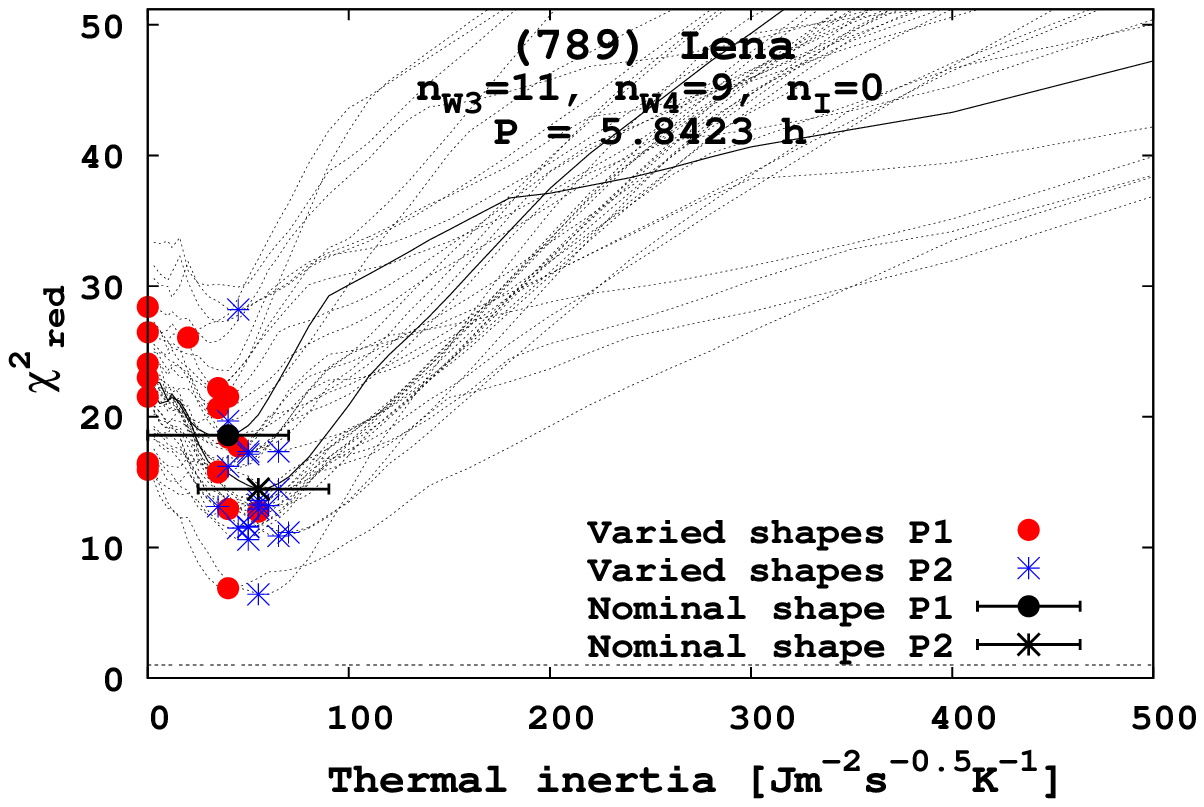}\includegraphics{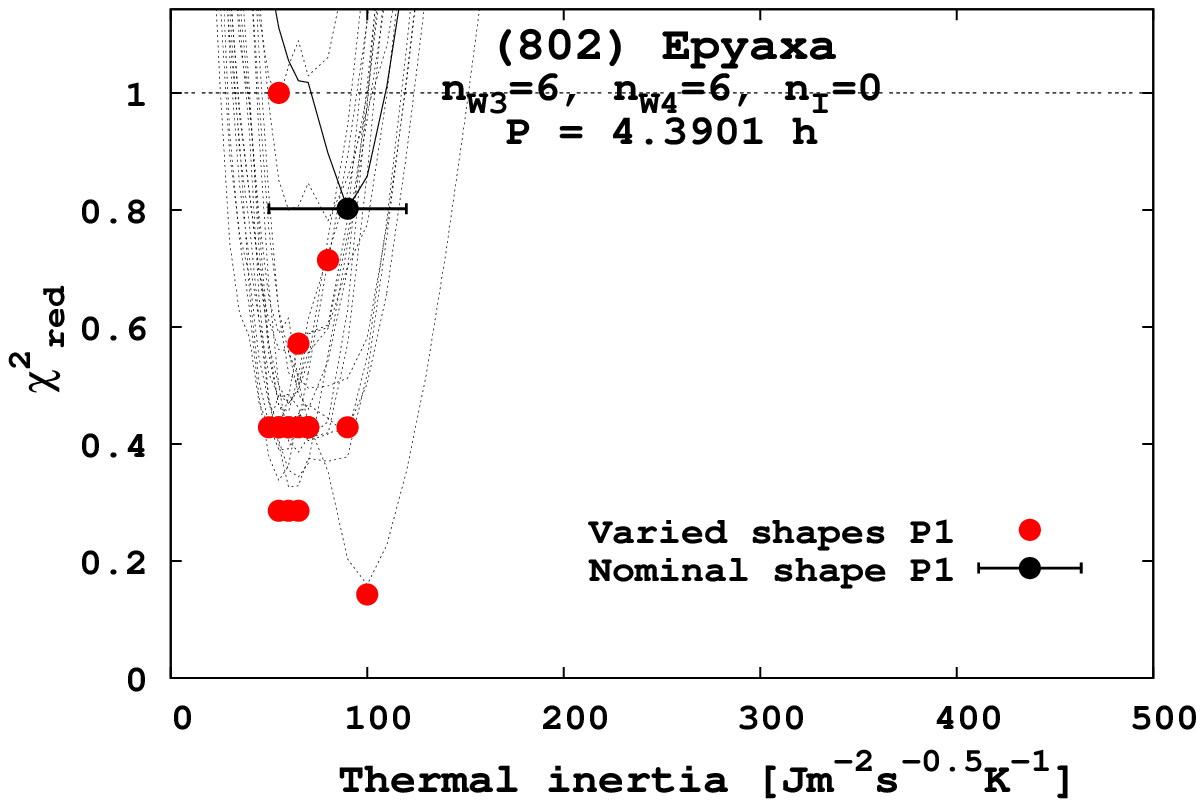}}\\
\resizebox{0.8\hsize}{!}{\includegraphics{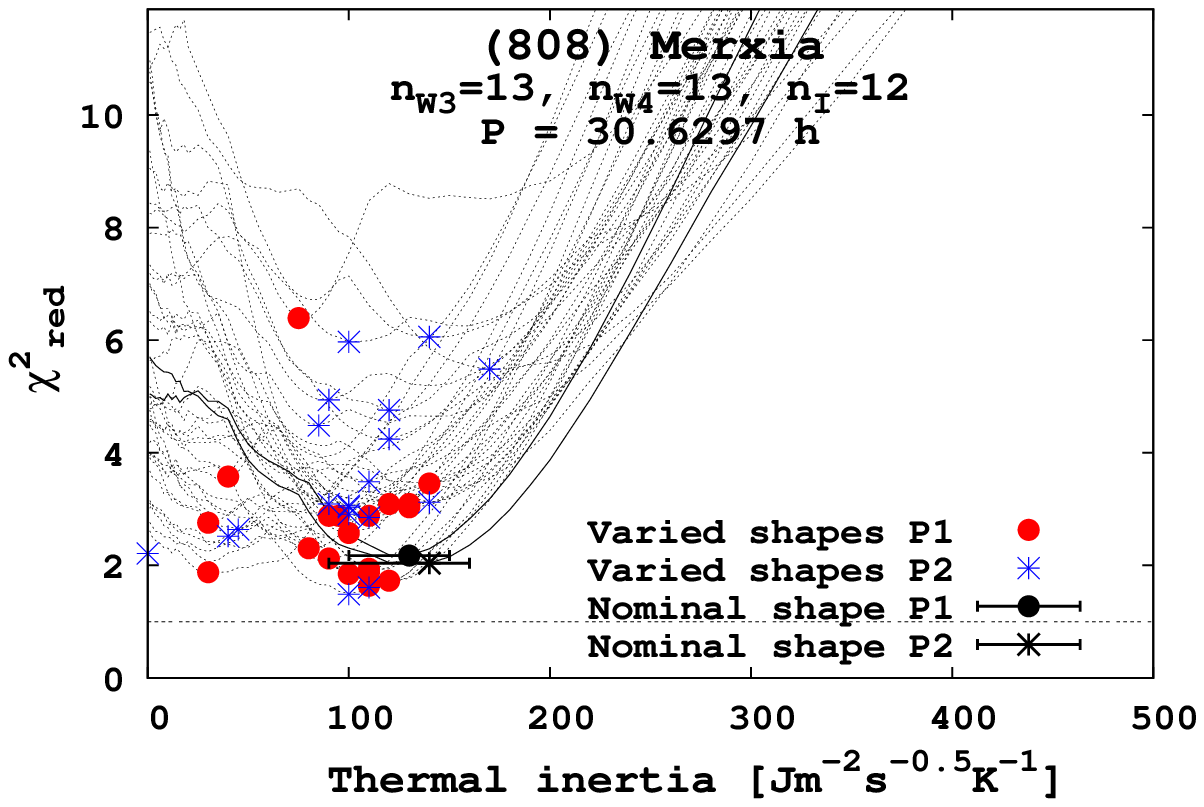}\includegraphics{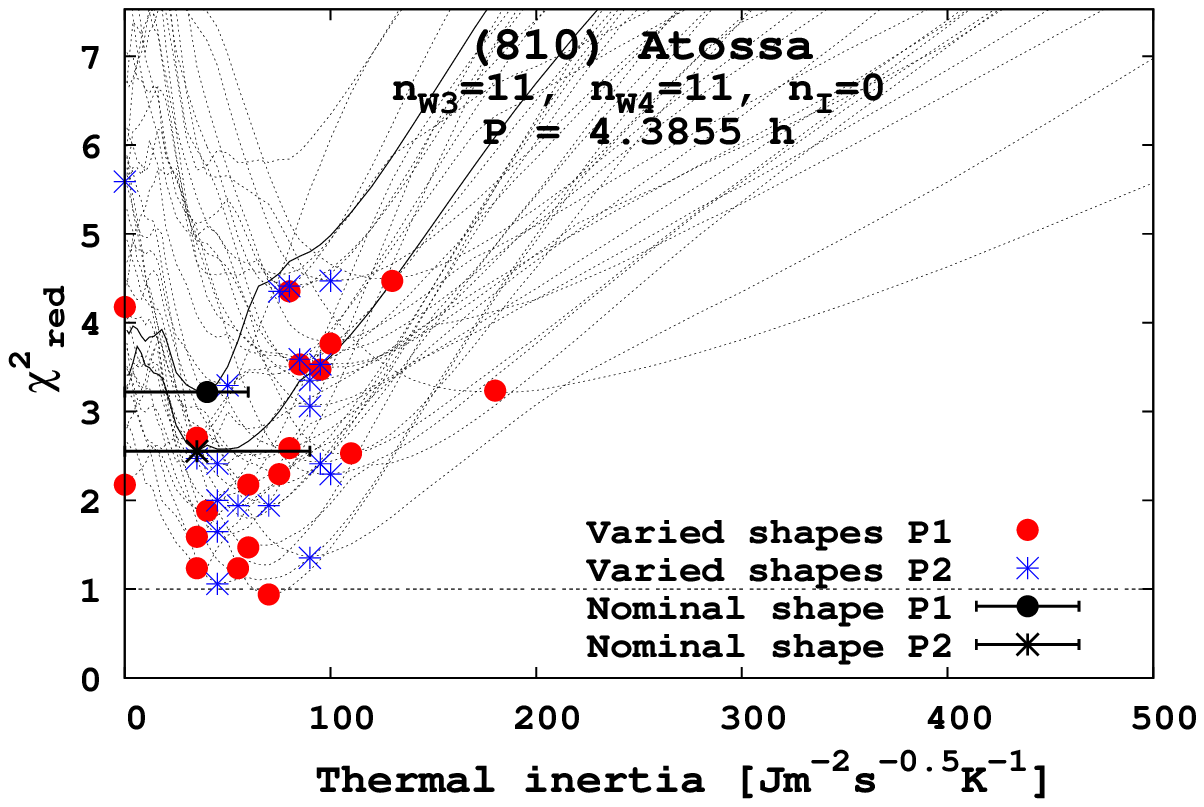}}\\
\resizebox{0.8\hsize}{!}{\includegraphics{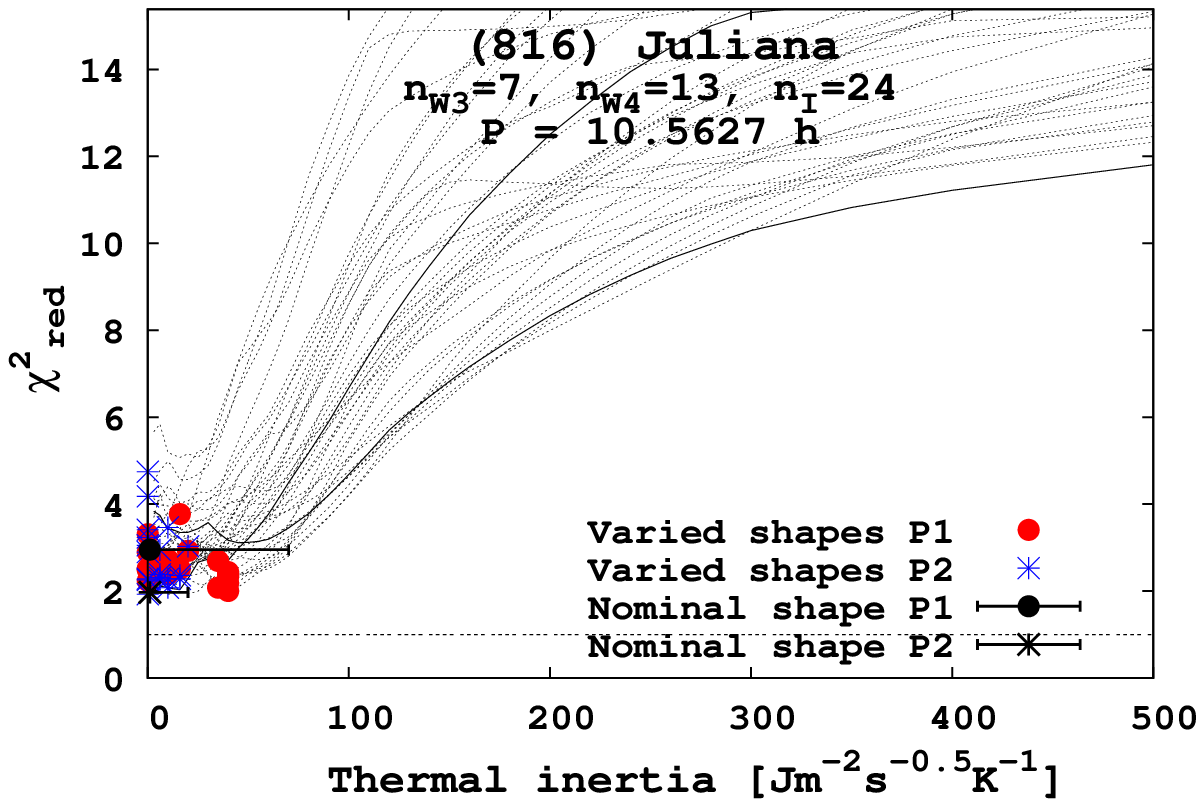}\includegraphics{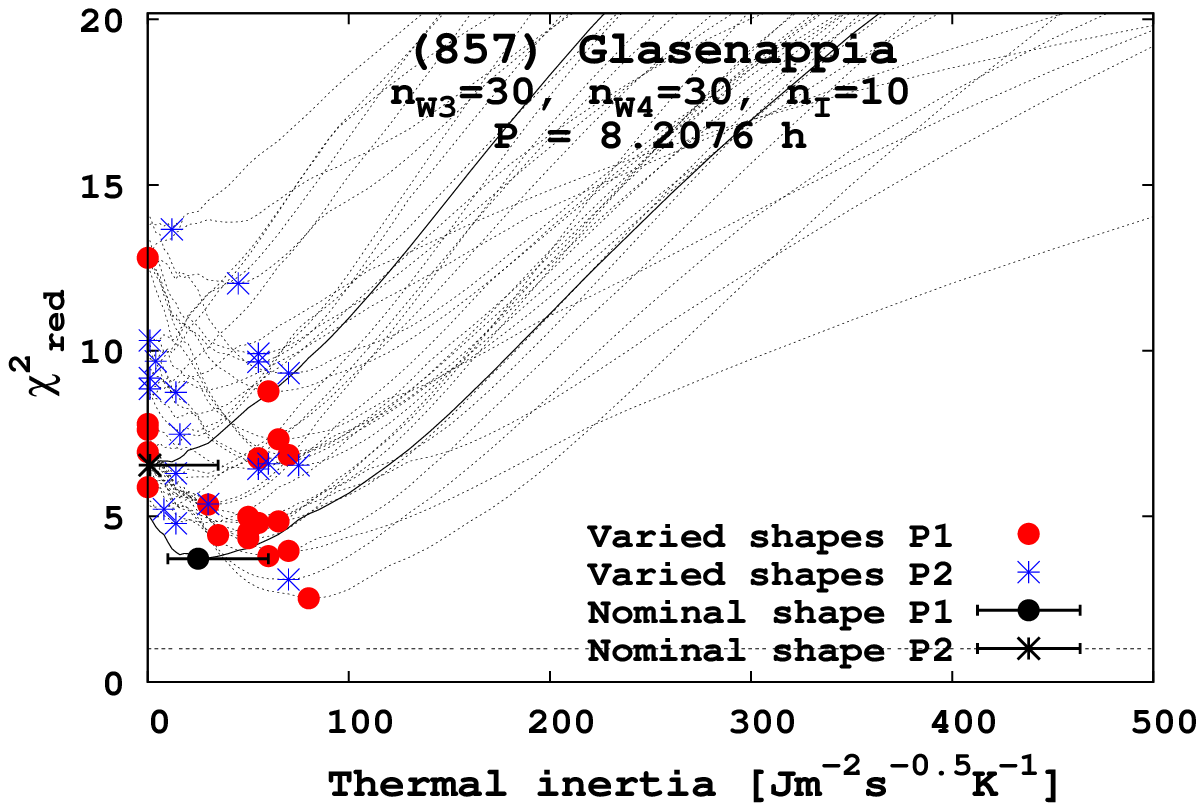}}\\
\resizebox{0.8\hsize}{!}{\includegraphics{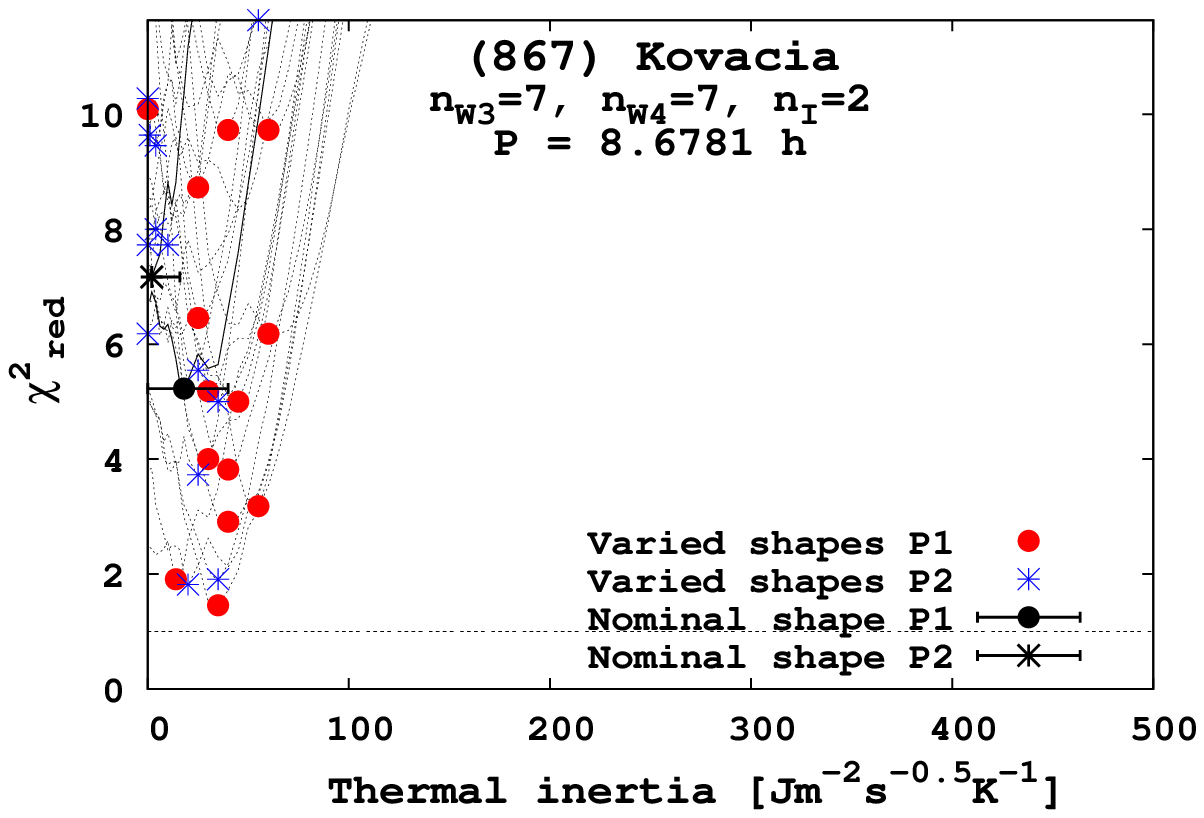}\includegraphics{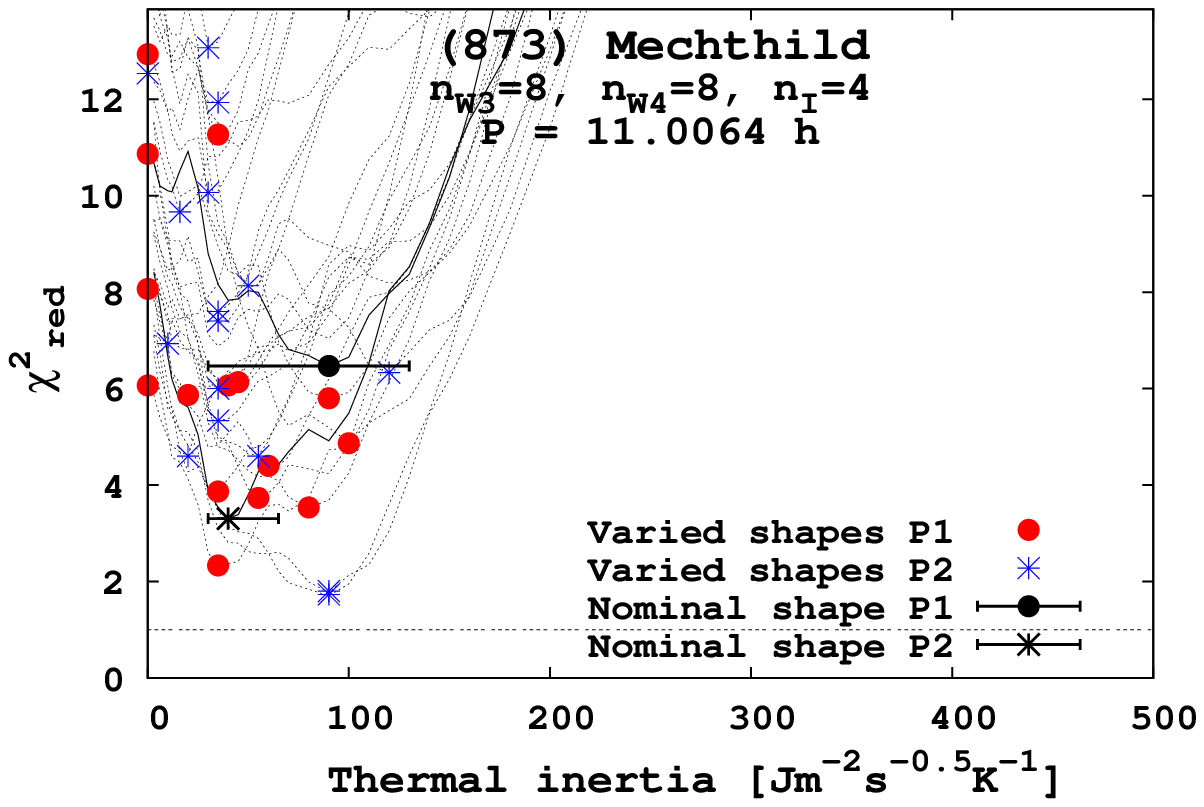}}\\
\end{center}
\caption{VS-TPM fits in the thermal inertia parameter space for eight asteroids. Each plot also contains the number of thermal infrared measurements in WISE W3 and W4 filters and in all four IRAS filters, and the rotation period.}
\end{figure*}

\begin{figure*}[!htbp]
\begin{center}
\resizebox{0.8\hsize}{!}{\includegraphics{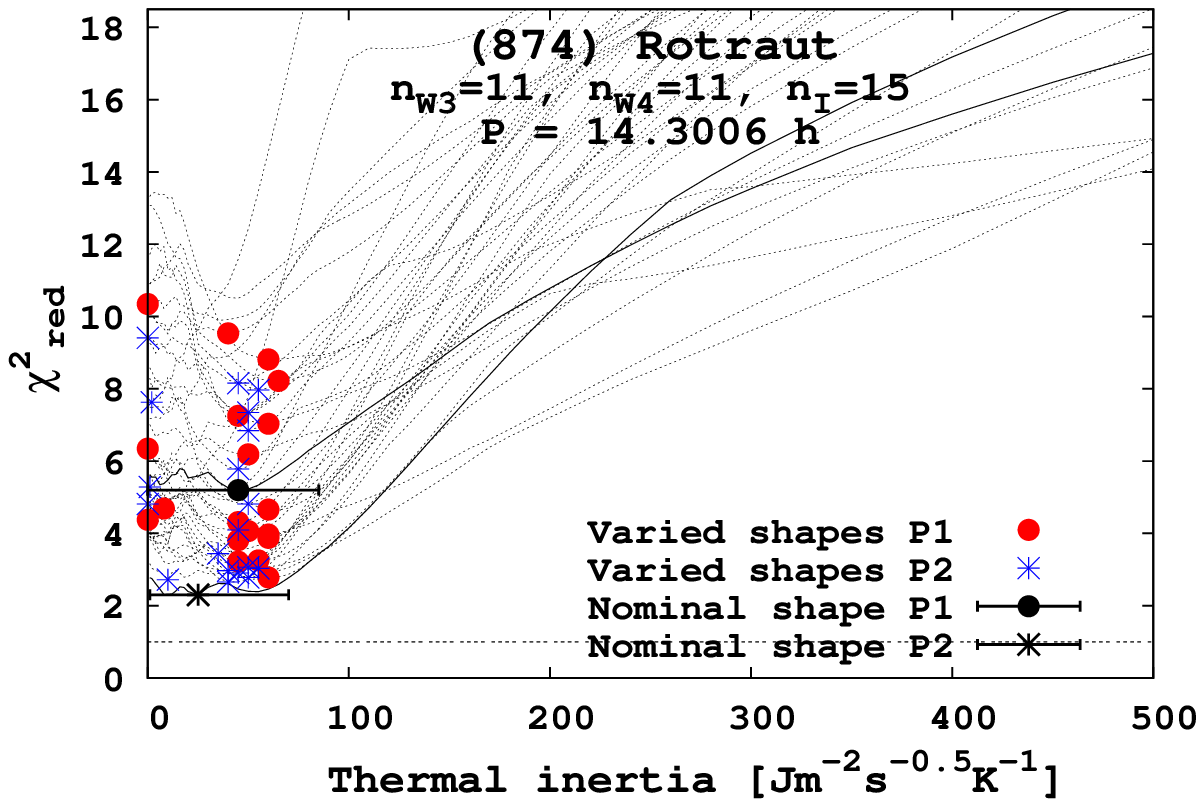}\includegraphics{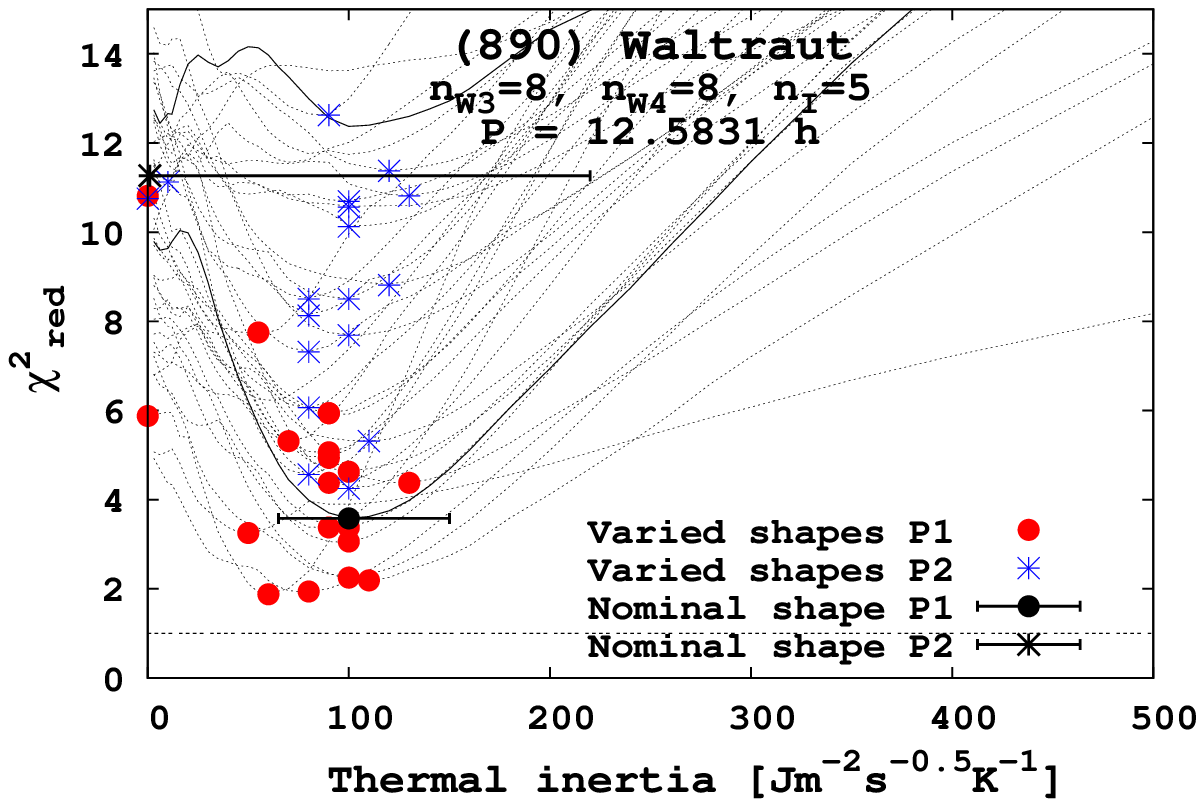}}\\
\resizebox{0.8\hsize}{!}{\includegraphics{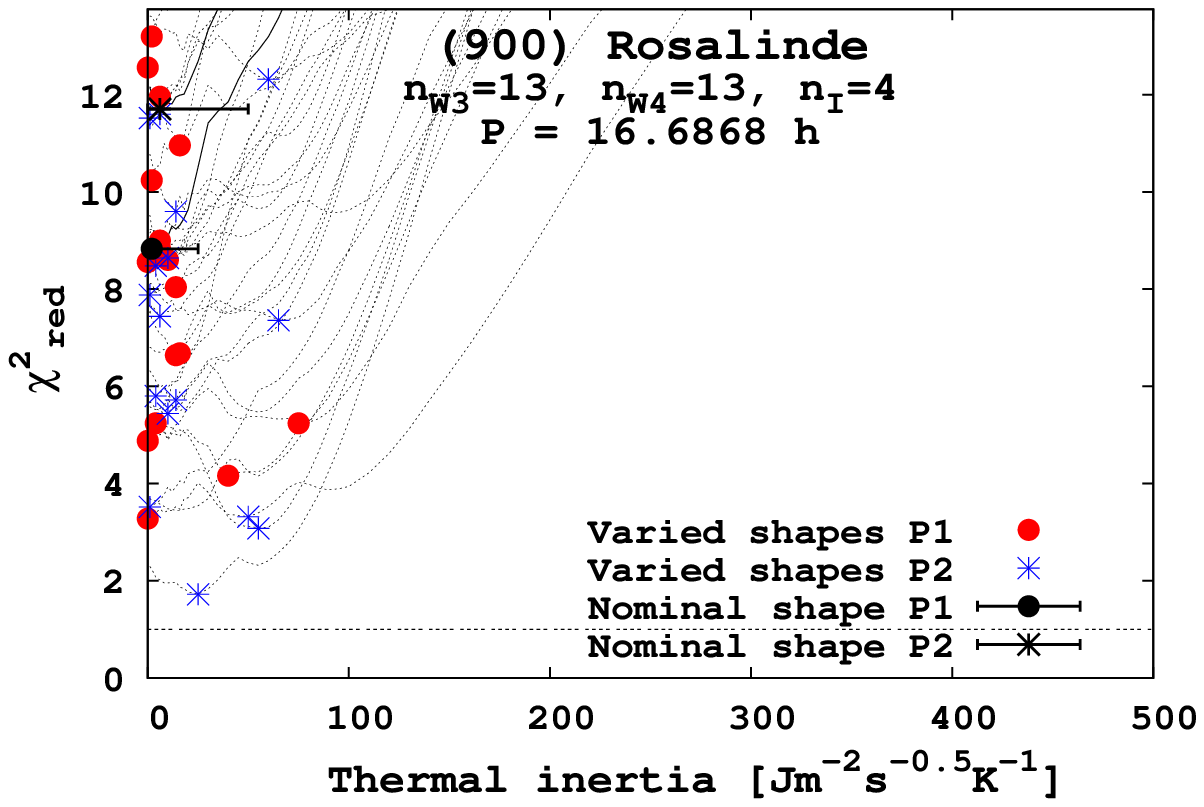}\includegraphics{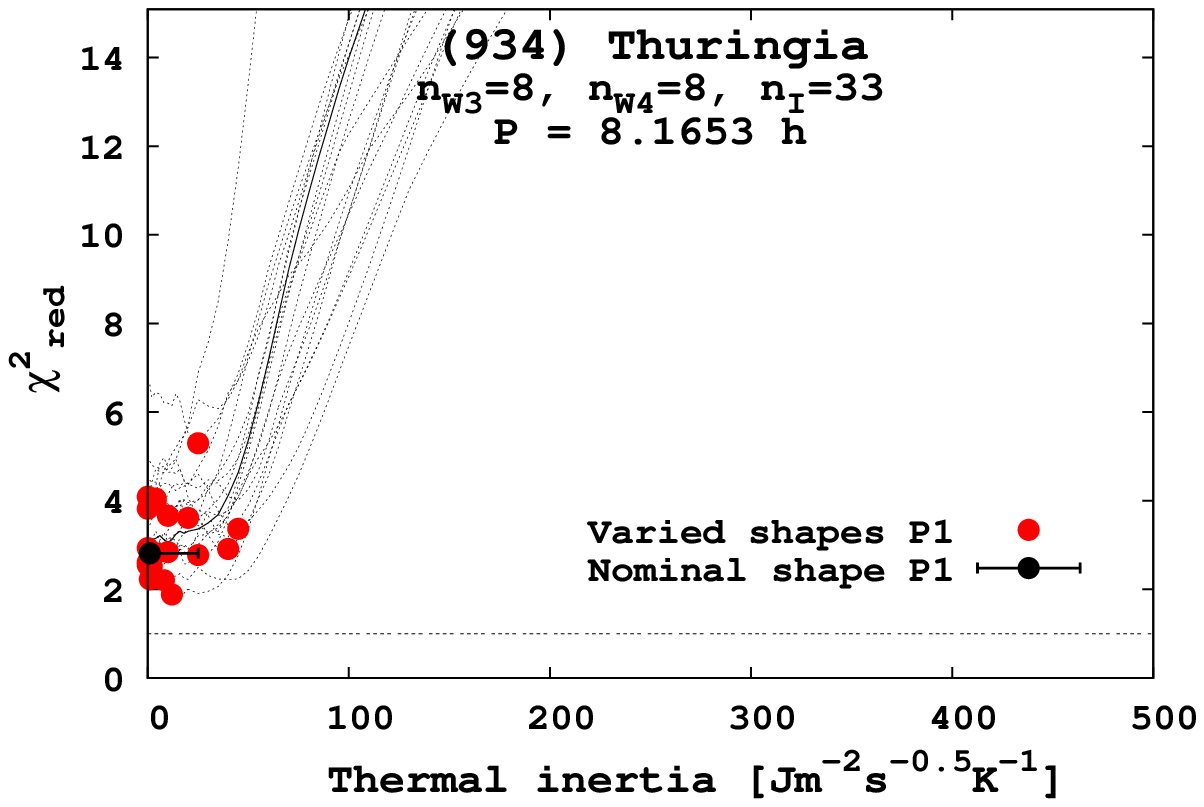}}\\
\resizebox{0.8\hsize}{!}{\includegraphics{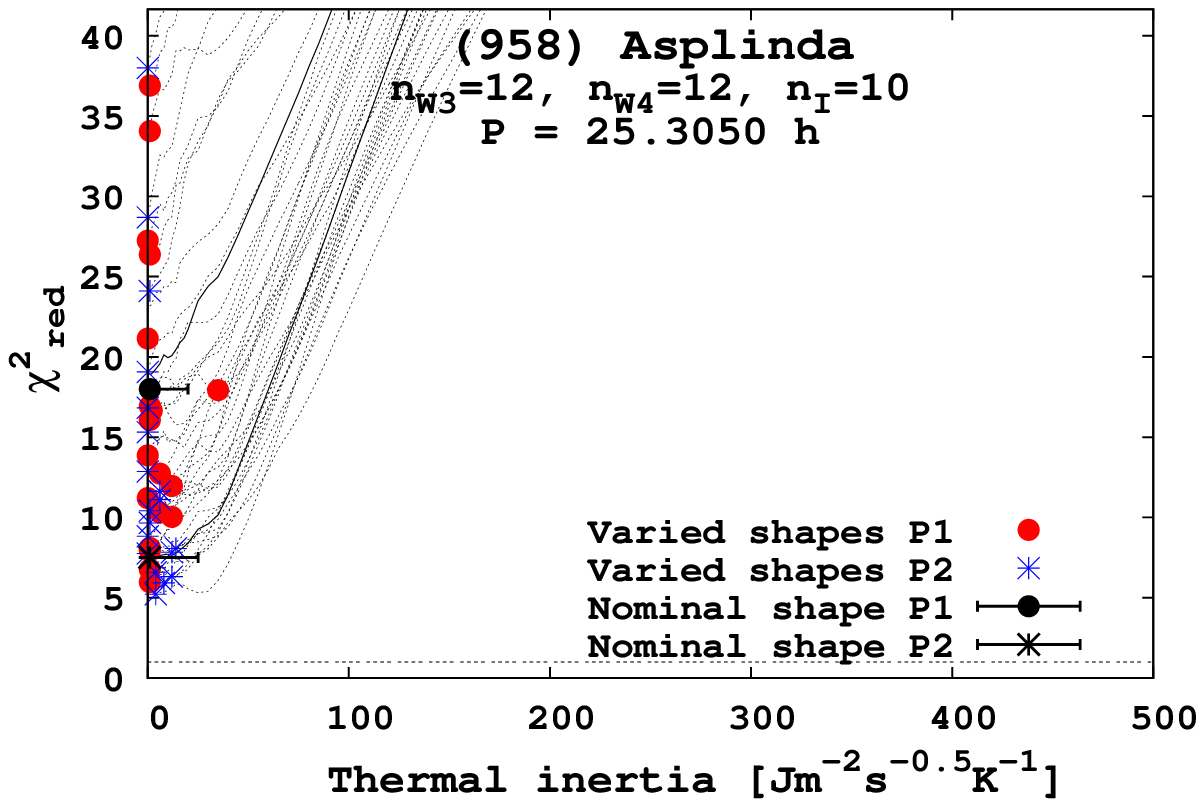}\includegraphics{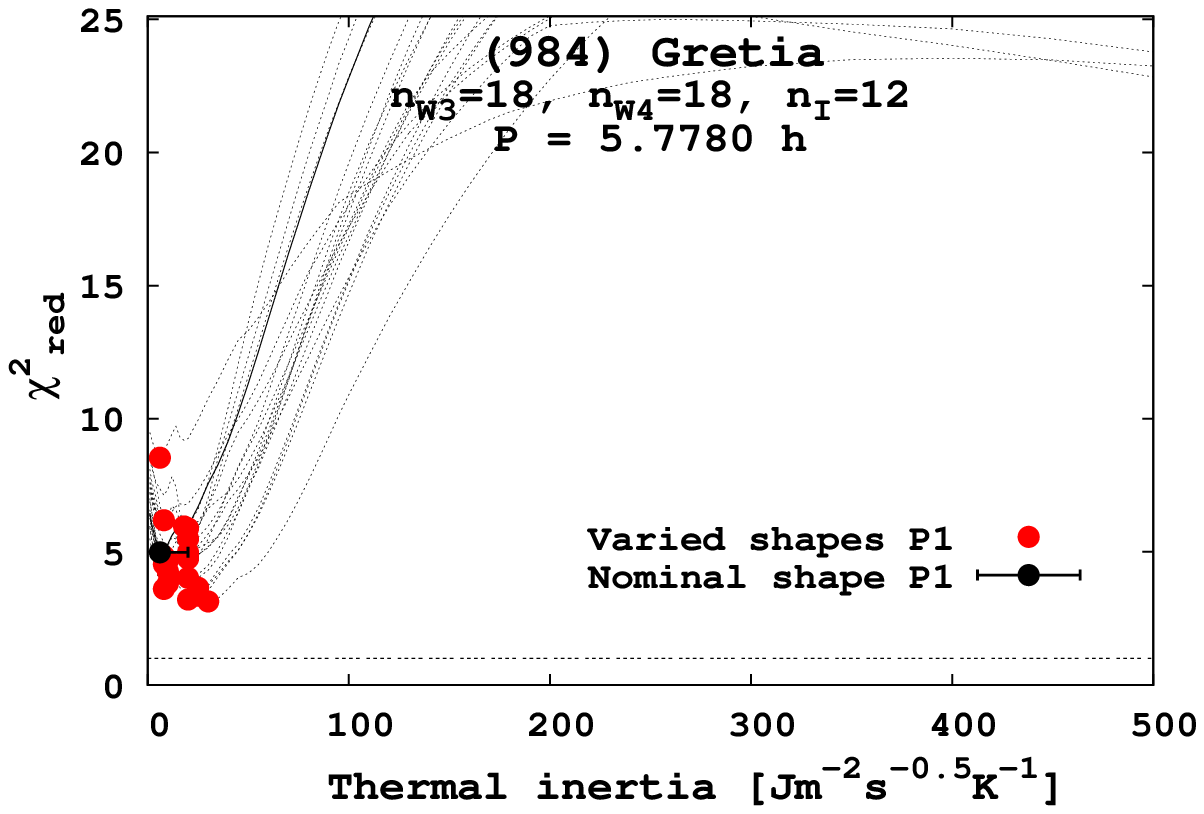}}\\
\resizebox{0.8\hsize}{!}{\includegraphics{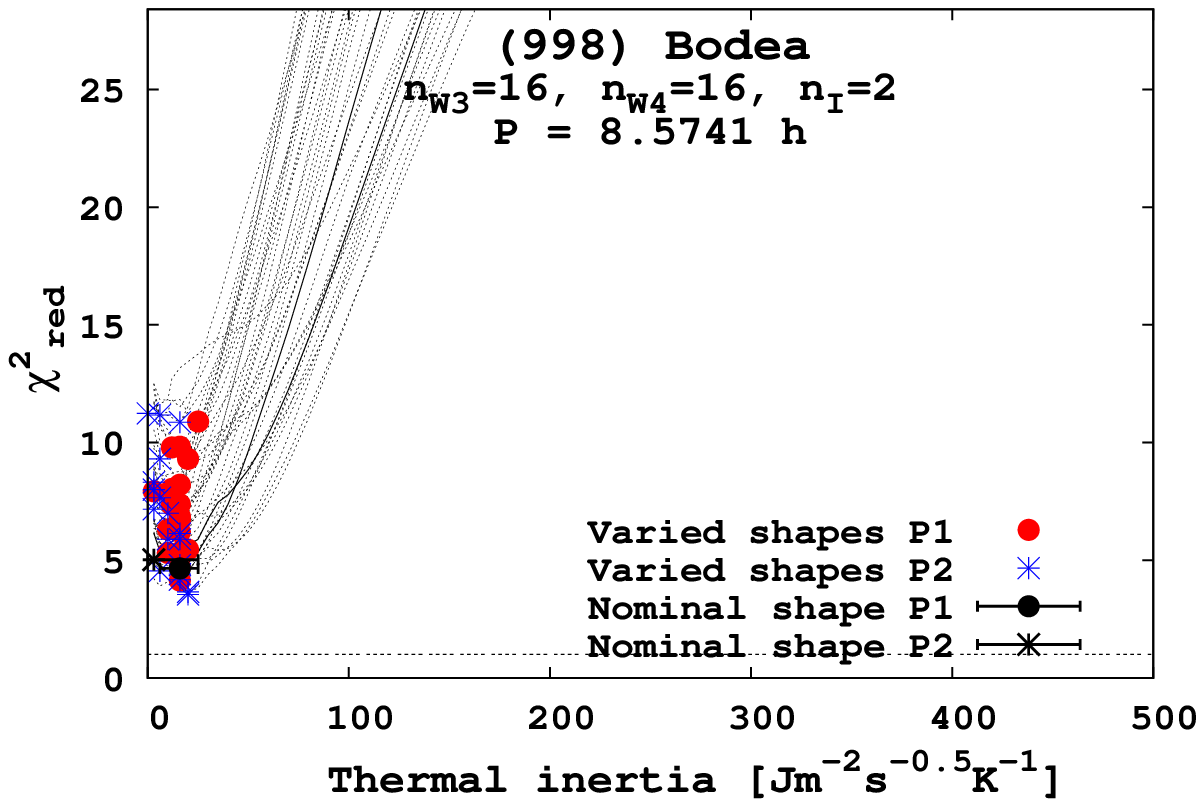}\includegraphics{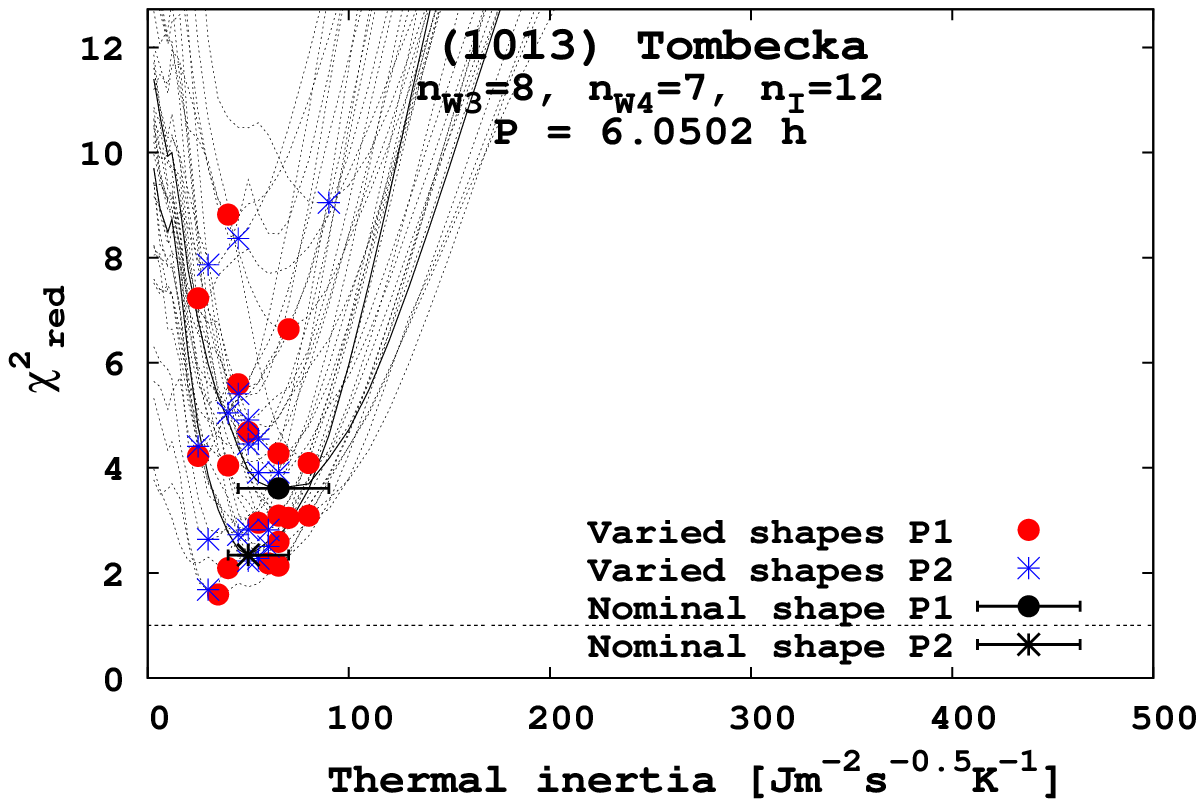}}\\
\end{center}
\caption{VS-TPM fits in the thermal inertia parameter space for eight asteroids. Each plot also contains the number of thermal infrared measurements in WISE W3 and W4 filters and in all four IRAS filters, and the rotation period.}
\end{figure*}

\begin{figure*}[!htbp]
\begin{center}
\resizebox{0.8\hsize}{!}{\includegraphics{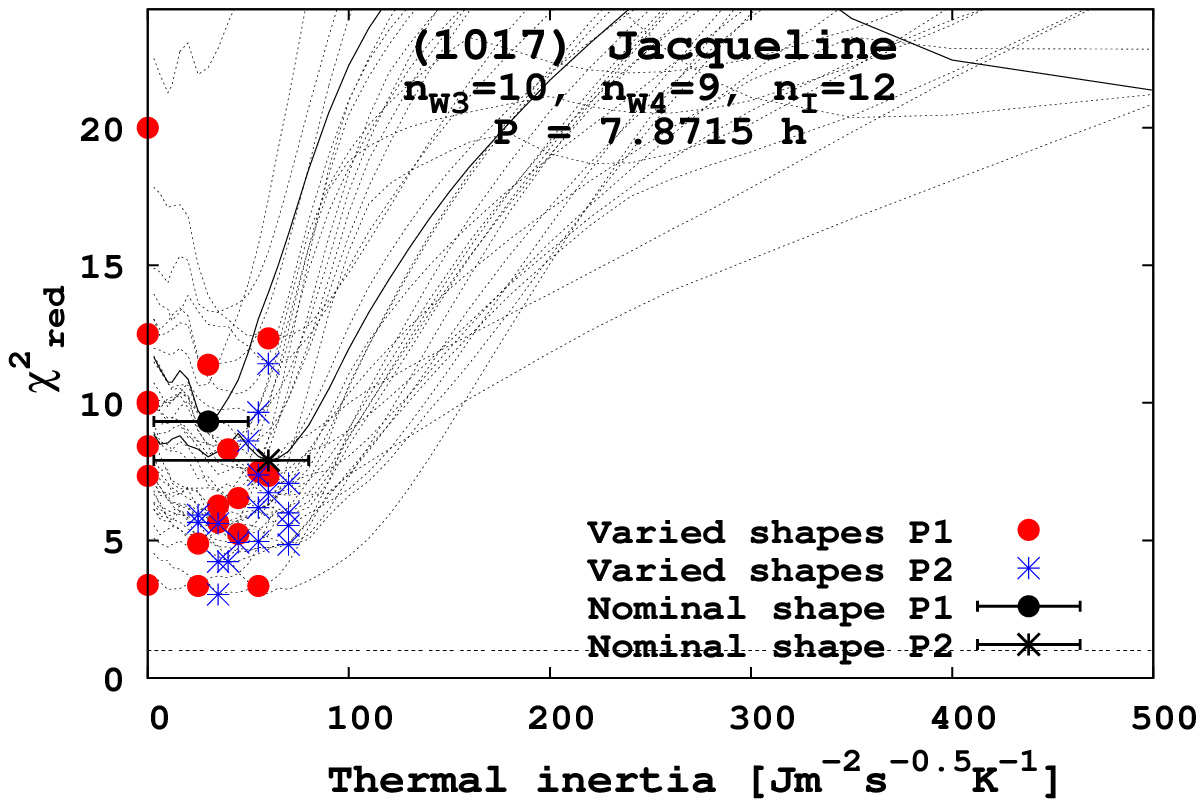}\includegraphics{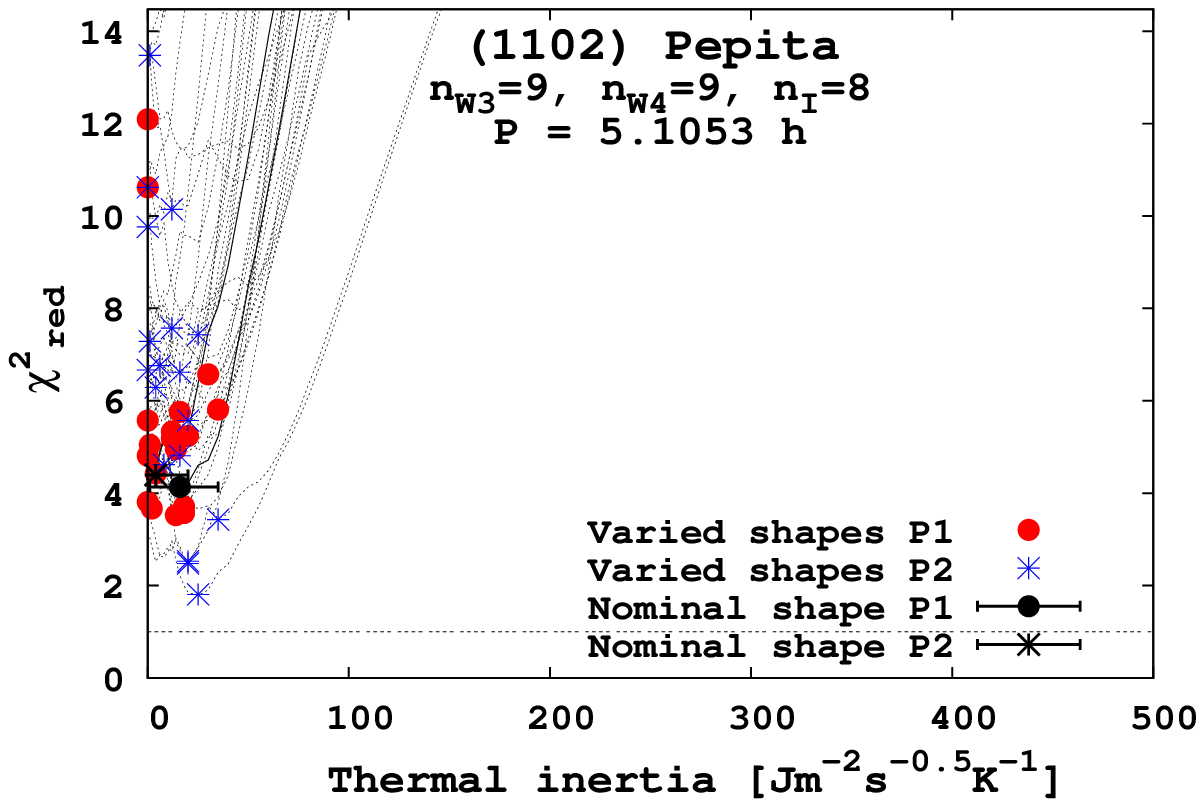}}\\
\resizebox{0.8\hsize}{!}{\includegraphics{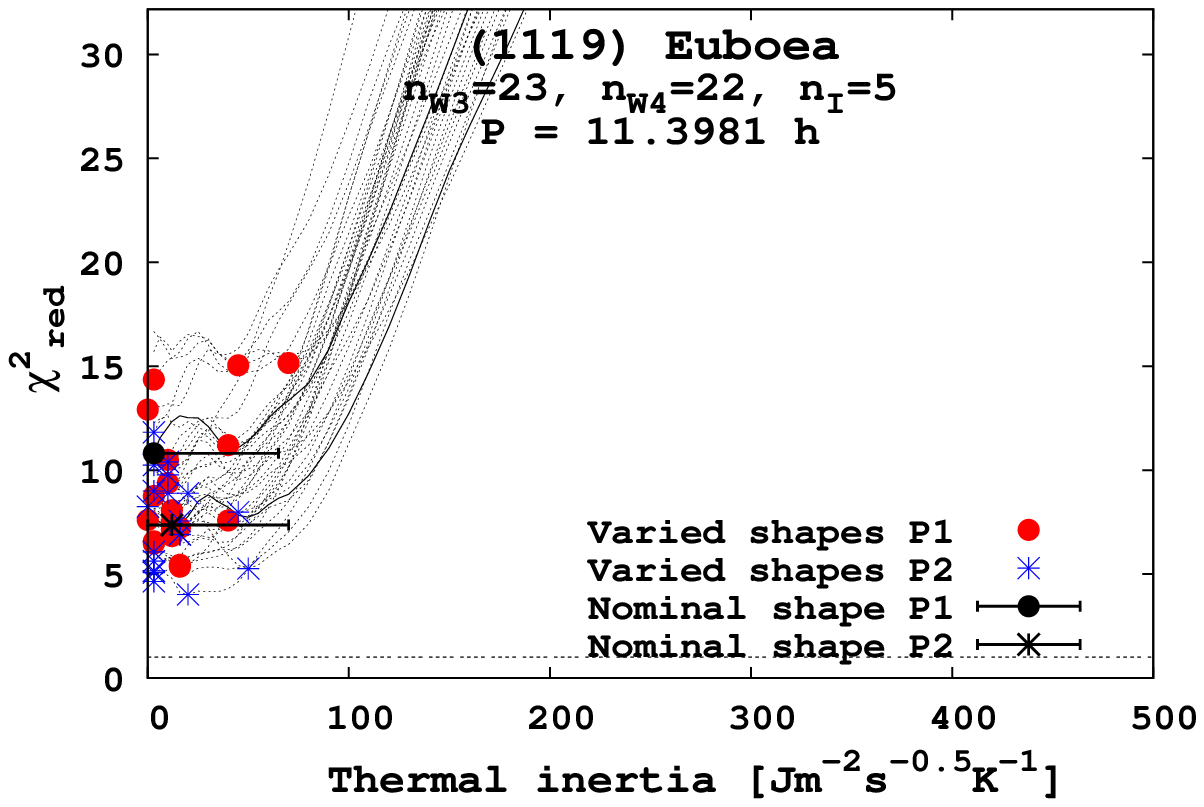}\includegraphics{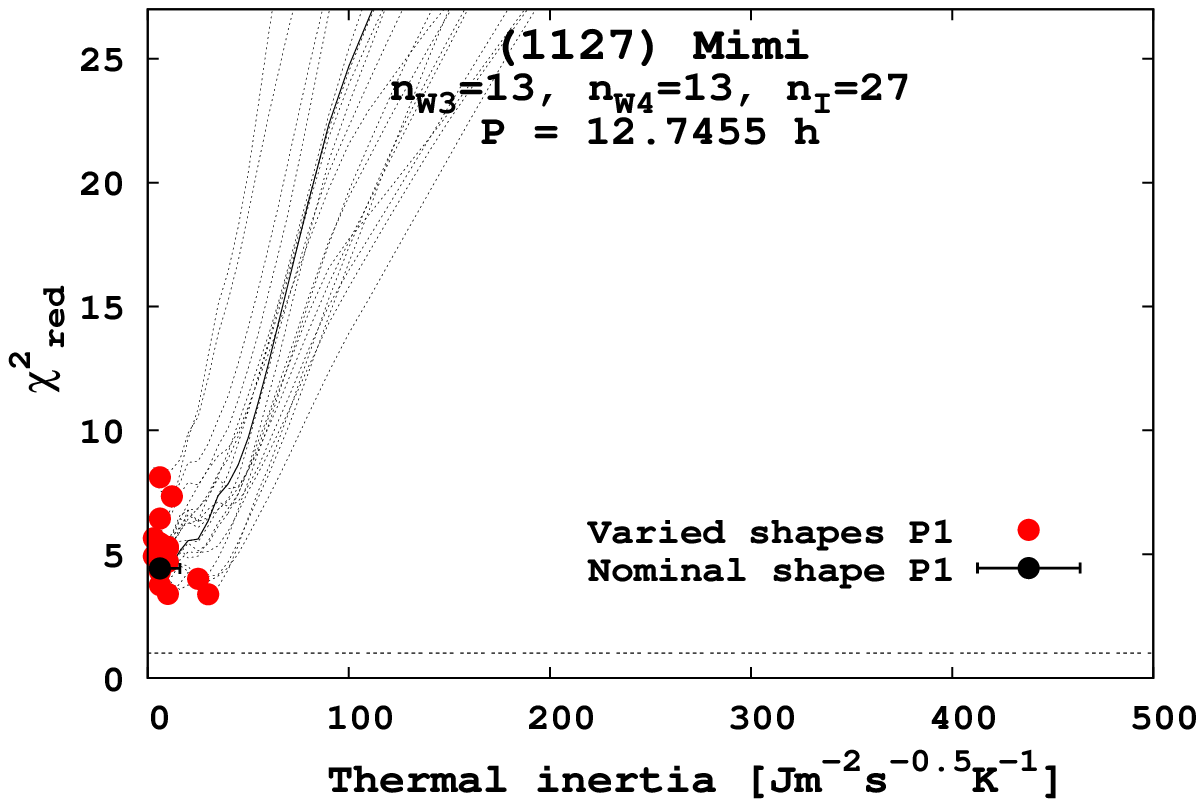}}\\
\resizebox{0.8\hsize}{!}{\includegraphics{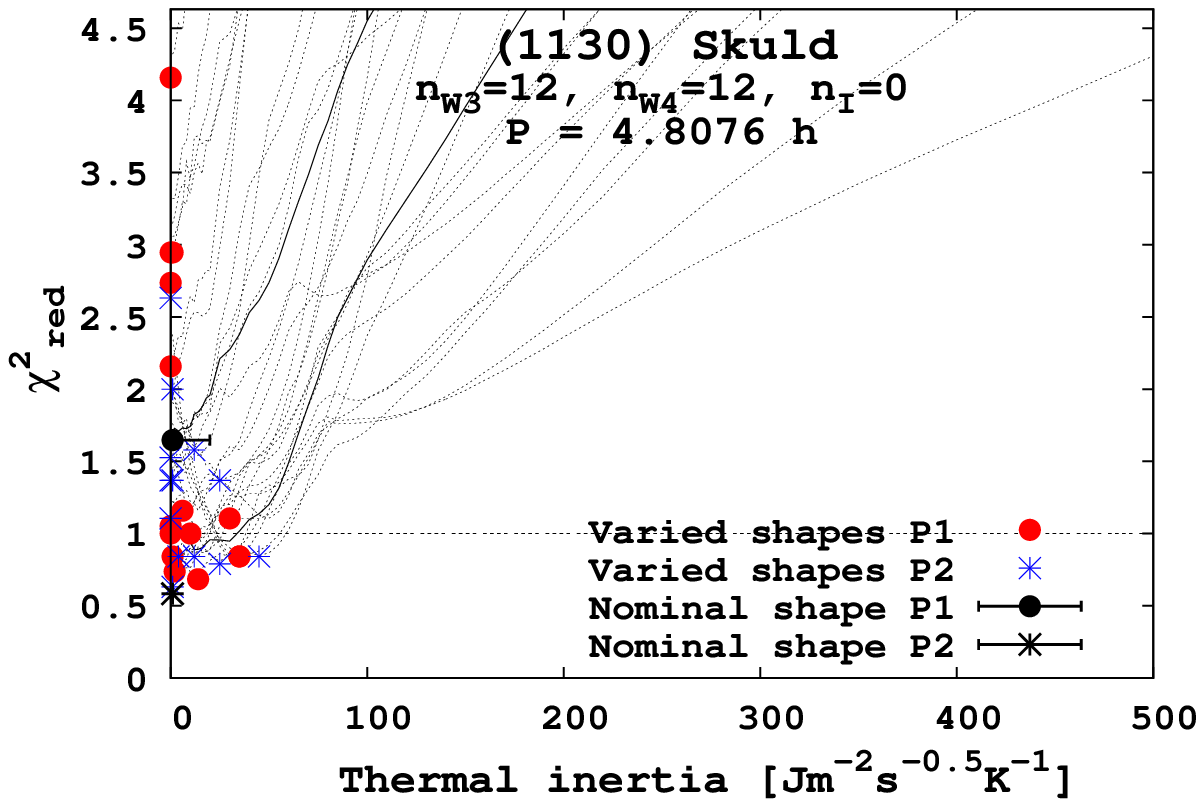}\includegraphics{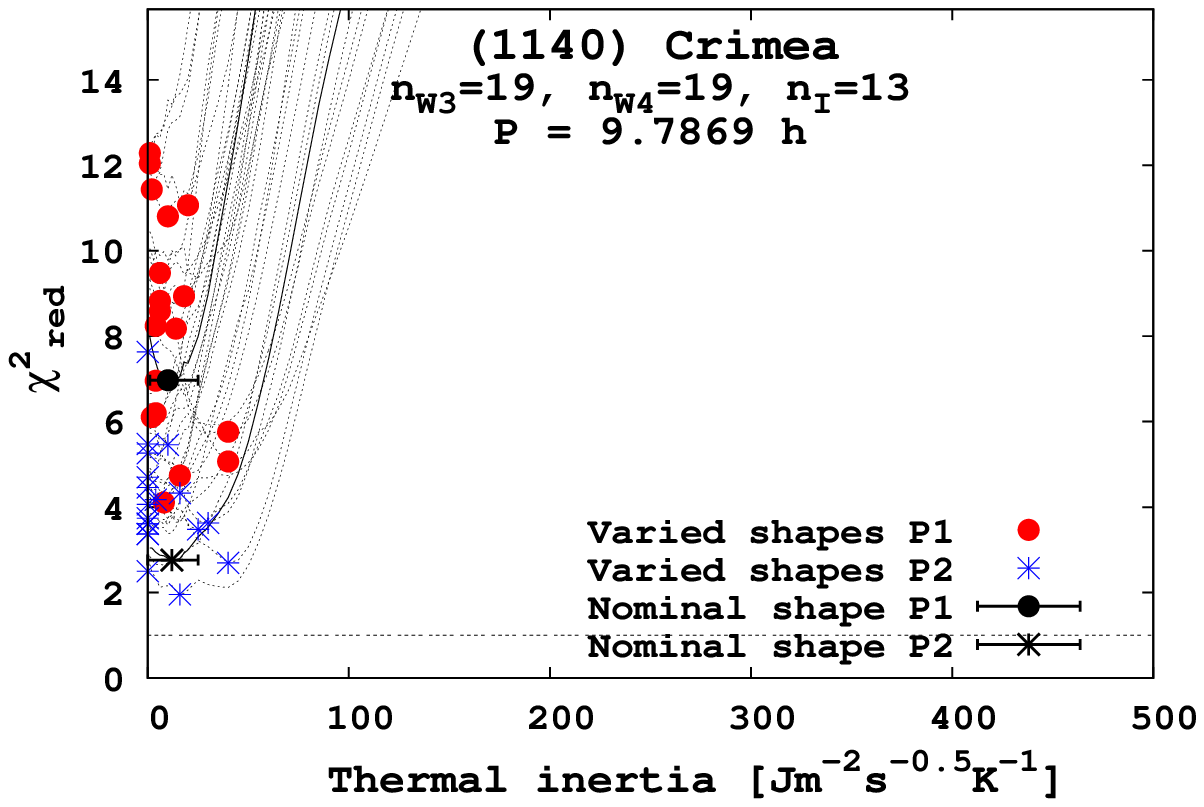}}\\
\resizebox{0.8\hsize}{!}{\includegraphics{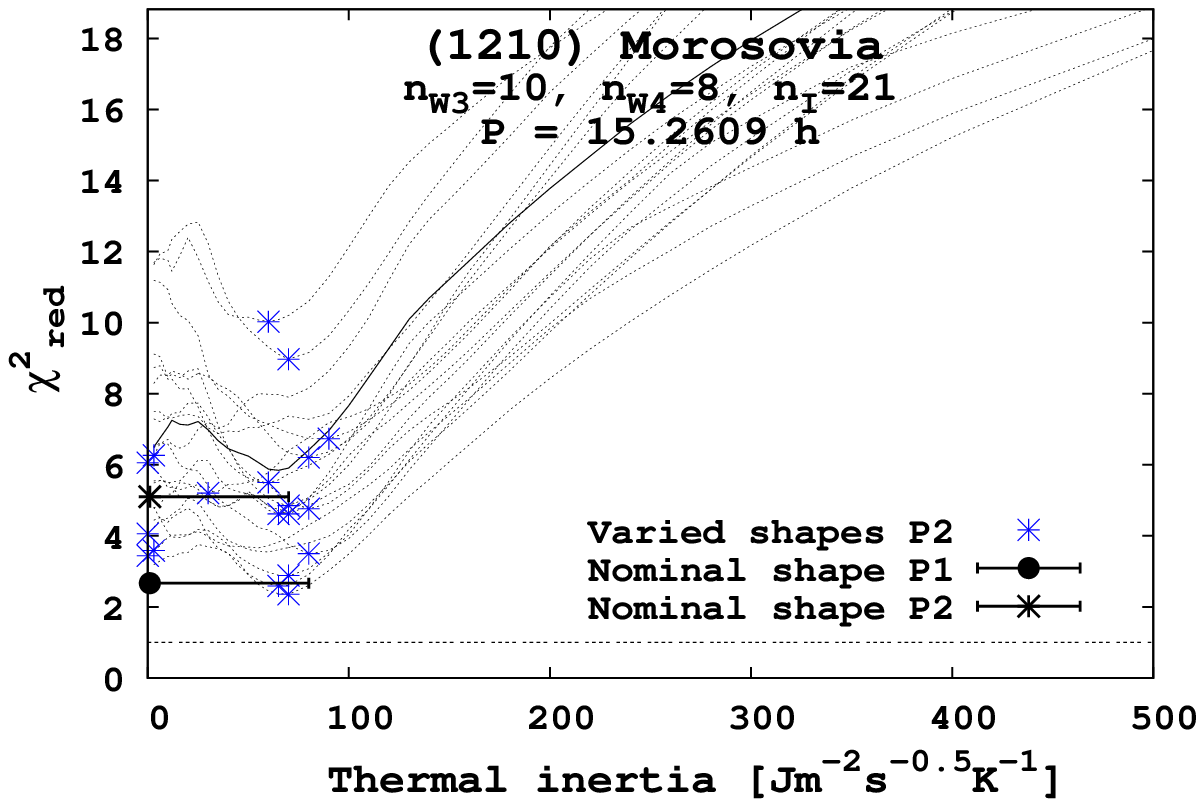}\includegraphics{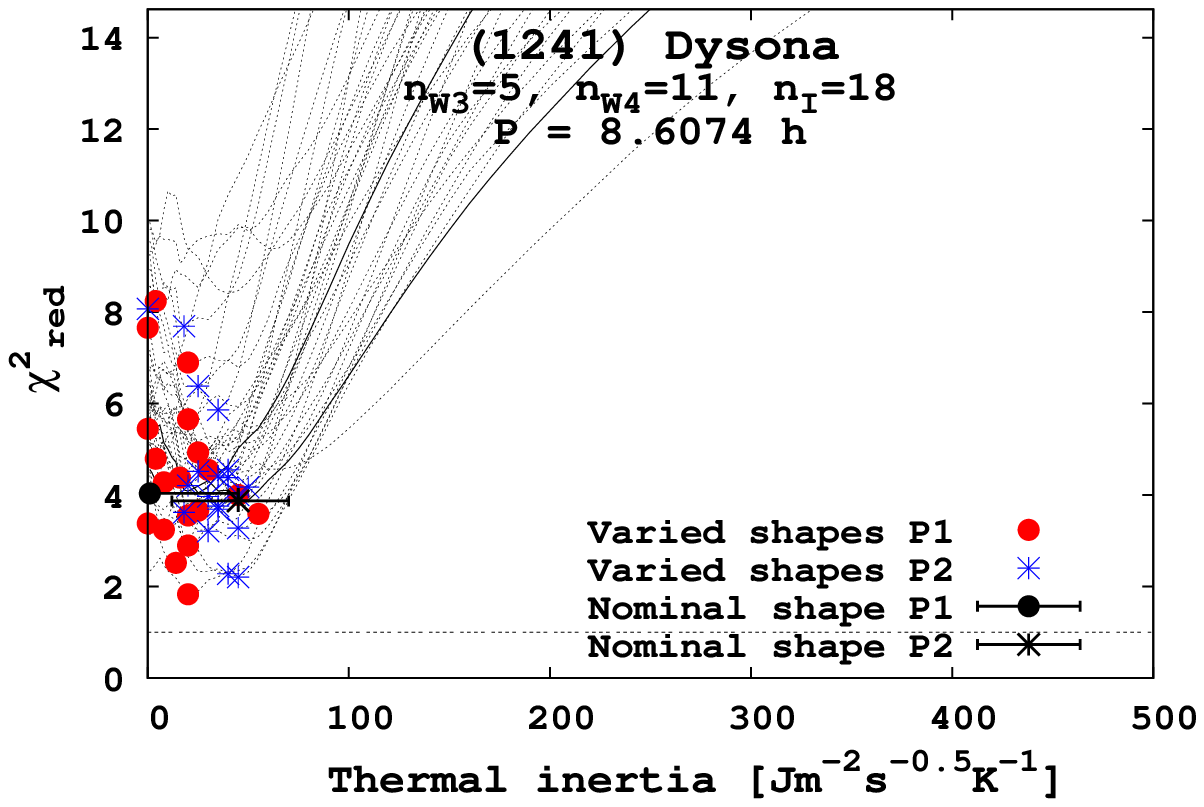}}\\
\end{center}
\caption{VS-TPM fits in the thermal inertia parameter space for eight asteroids. Each plot also contains the number of thermal infrared measurements in WISE W3 and W4 filters and in all four IRAS filters, and the rotation period.}
\end{figure*}

\begin{figure*}[!htbp]
\begin{center}
\resizebox{0.8\hsize}{!}{\includegraphics{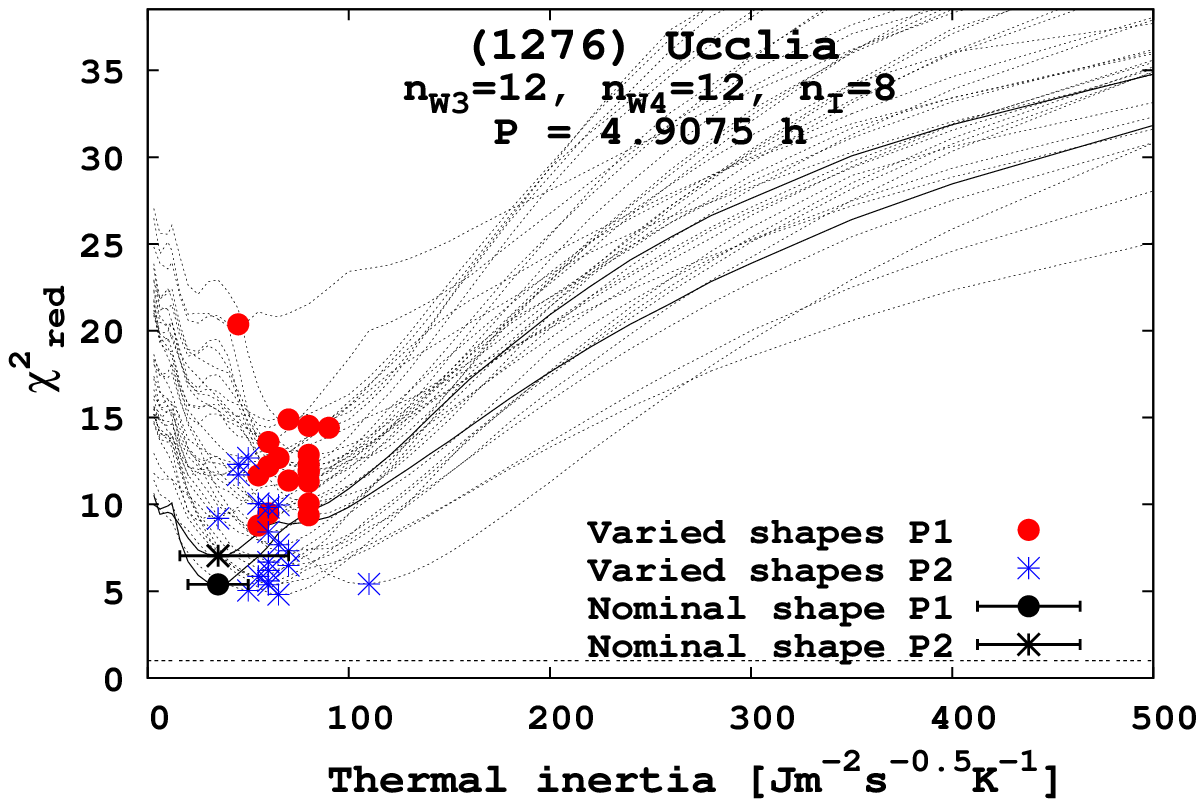}\includegraphics{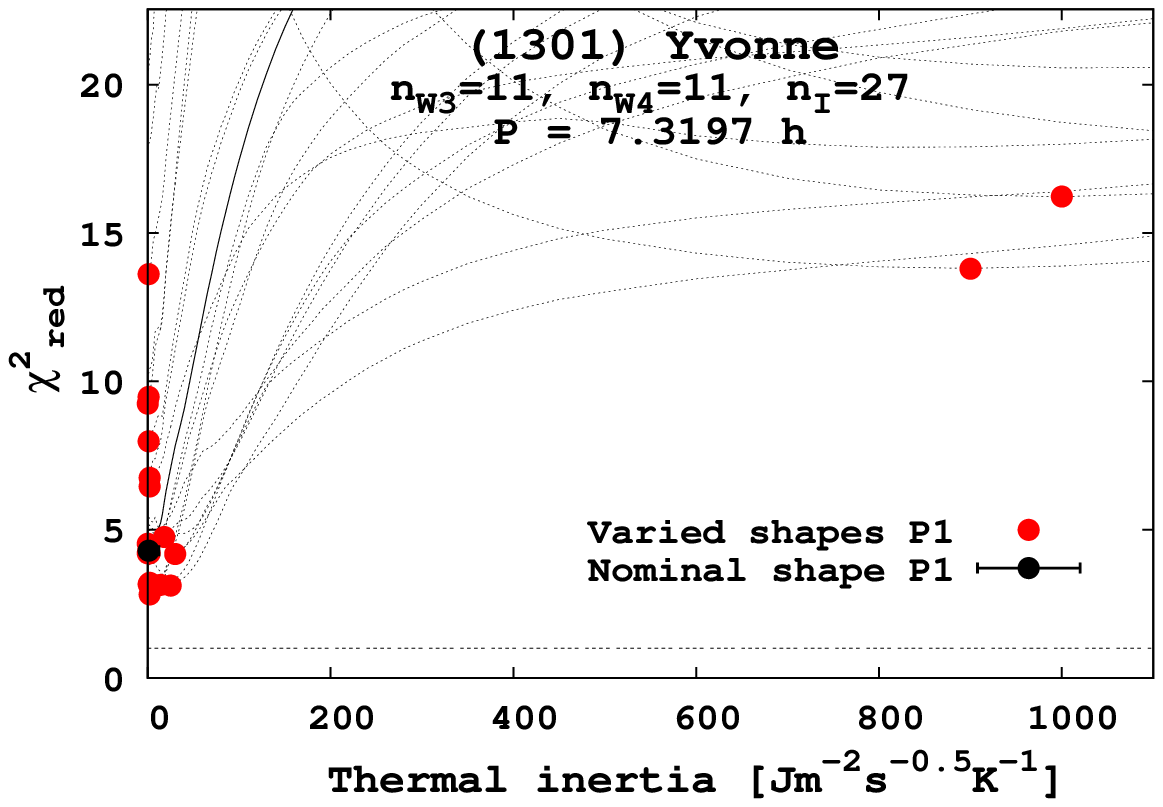}}\\
\resizebox{0.8\hsize}{!}{\includegraphics{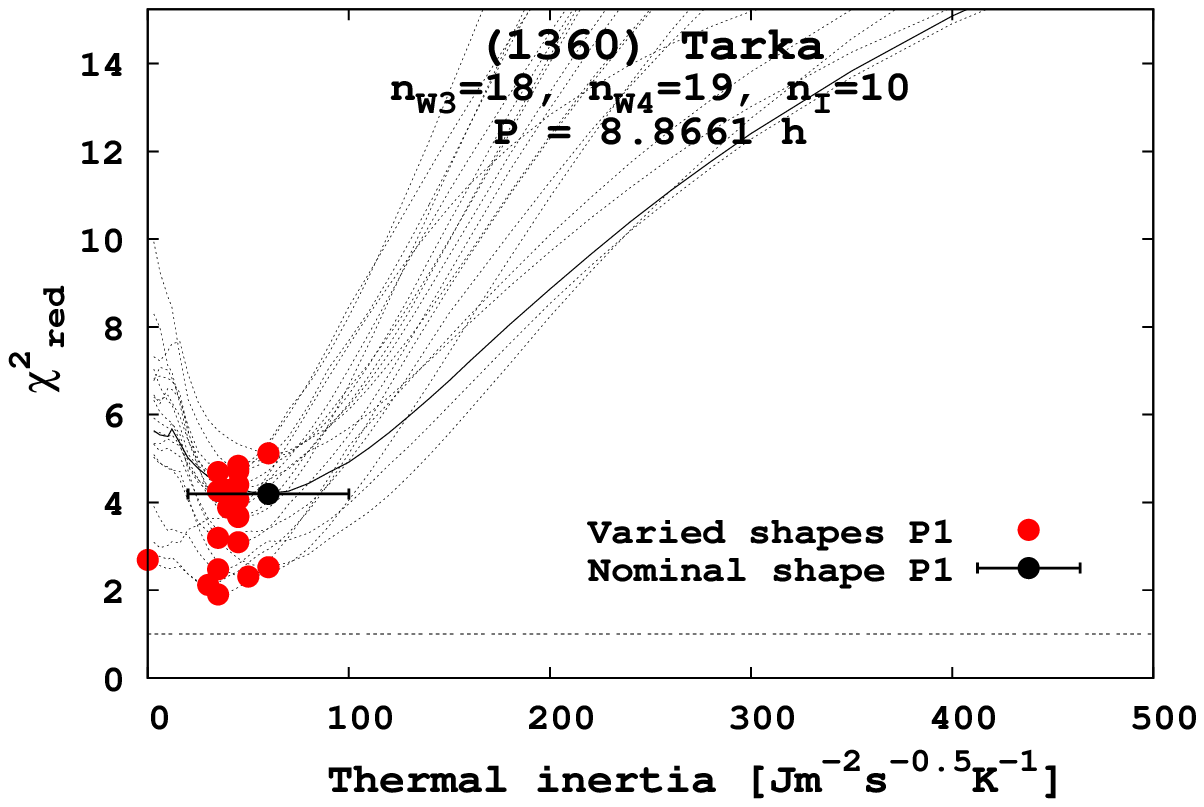}\includegraphics{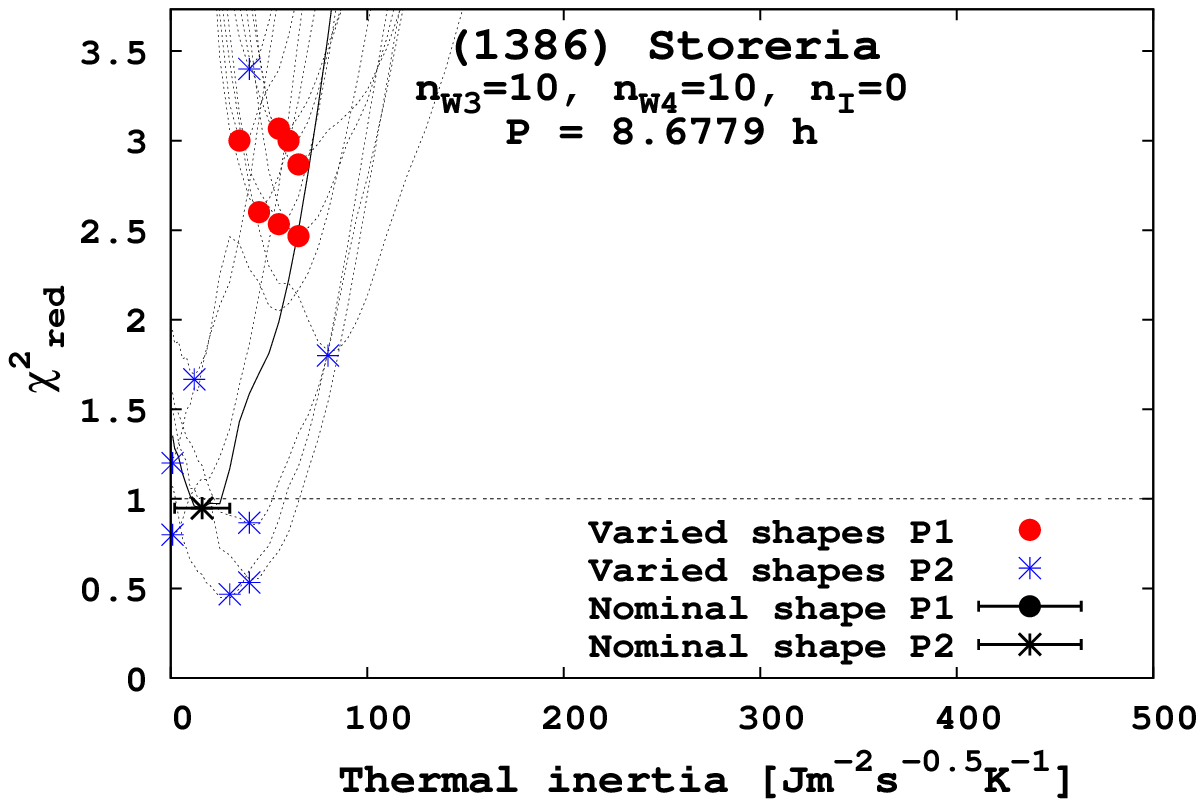}}\\
\resizebox{0.8\hsize}{!}{\includegraphics{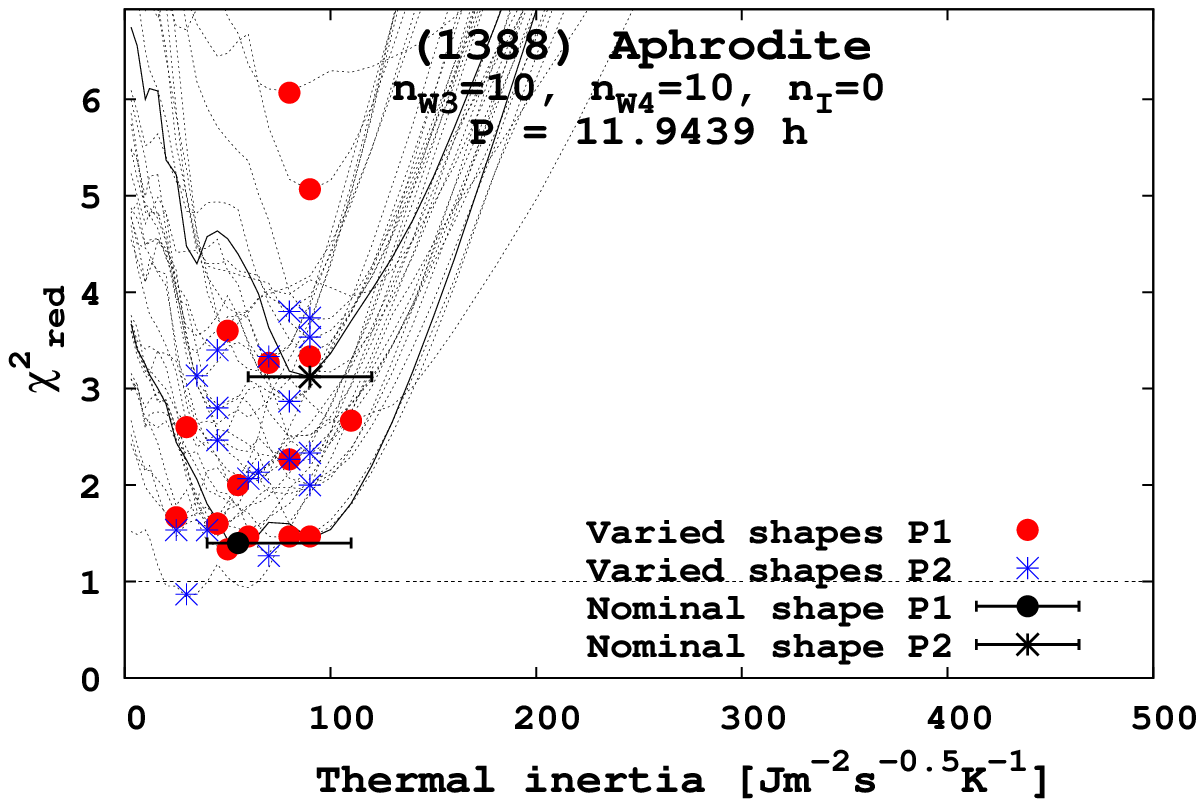}\includegraphics{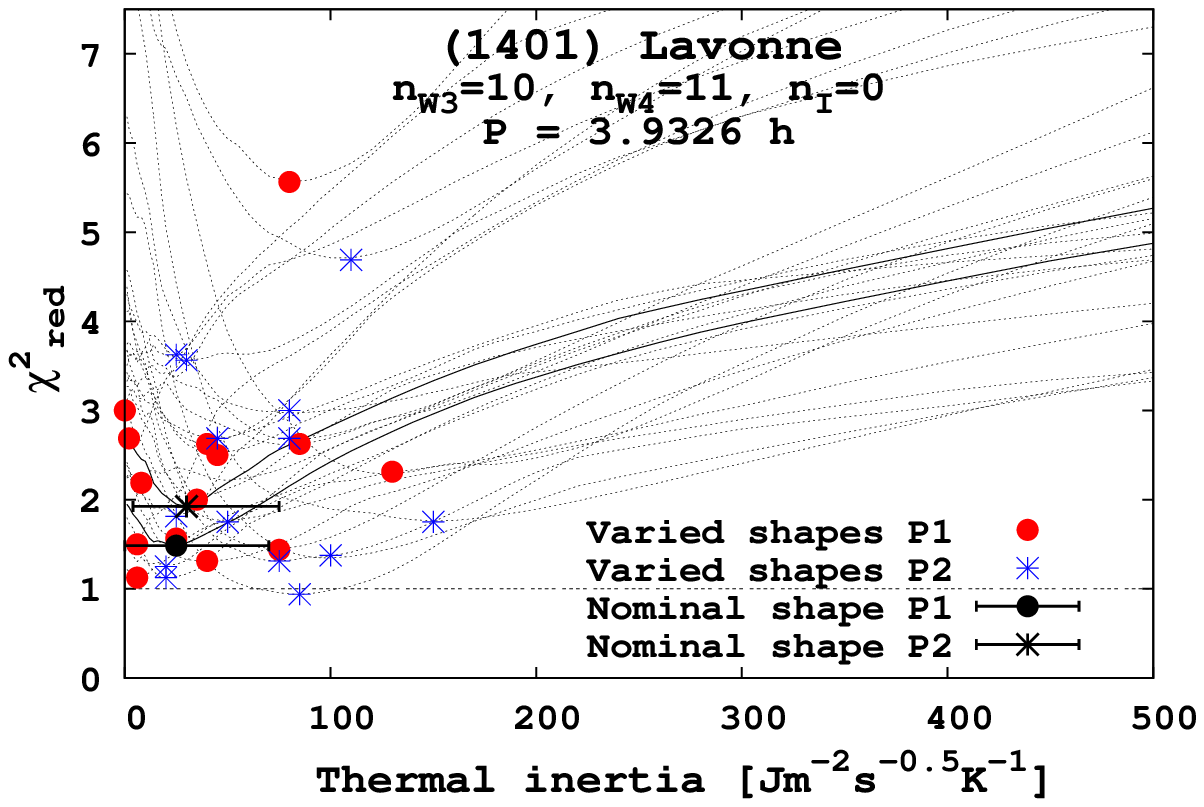}}\\
\resizebox{0.8\hsize}{!}{\includegraphics{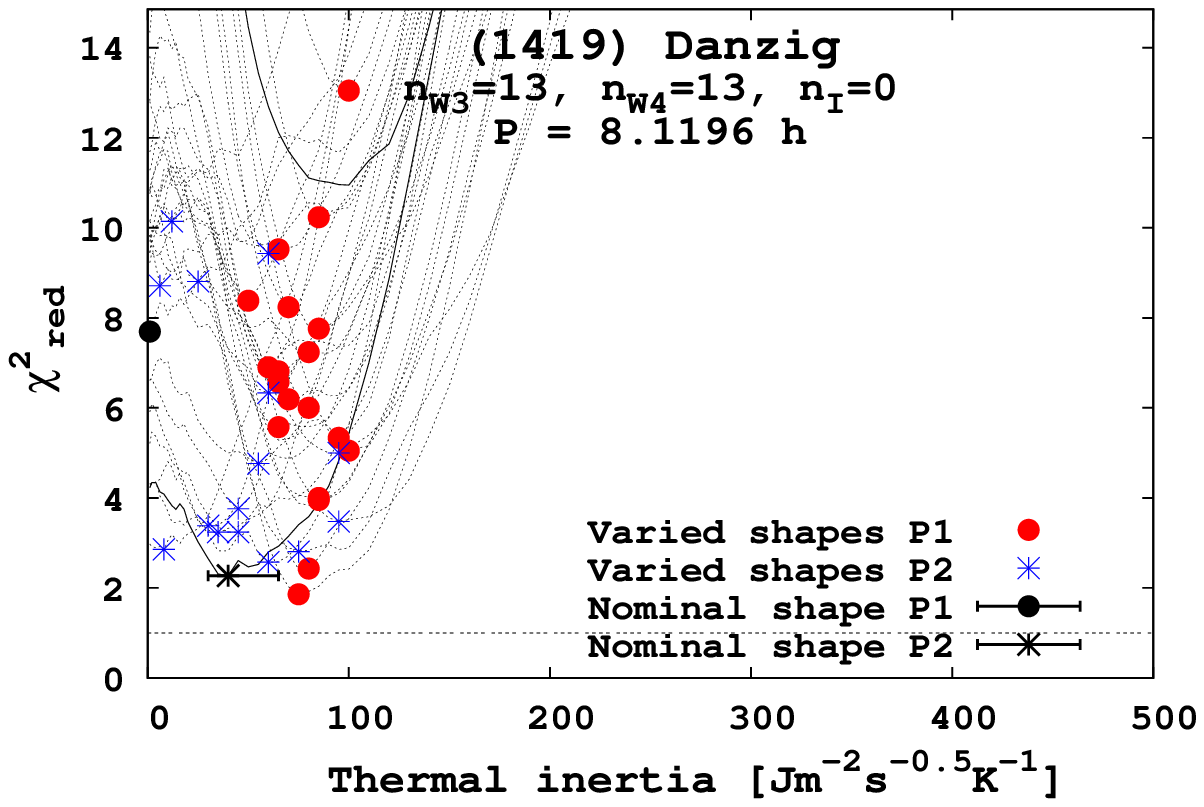}\includegraphics{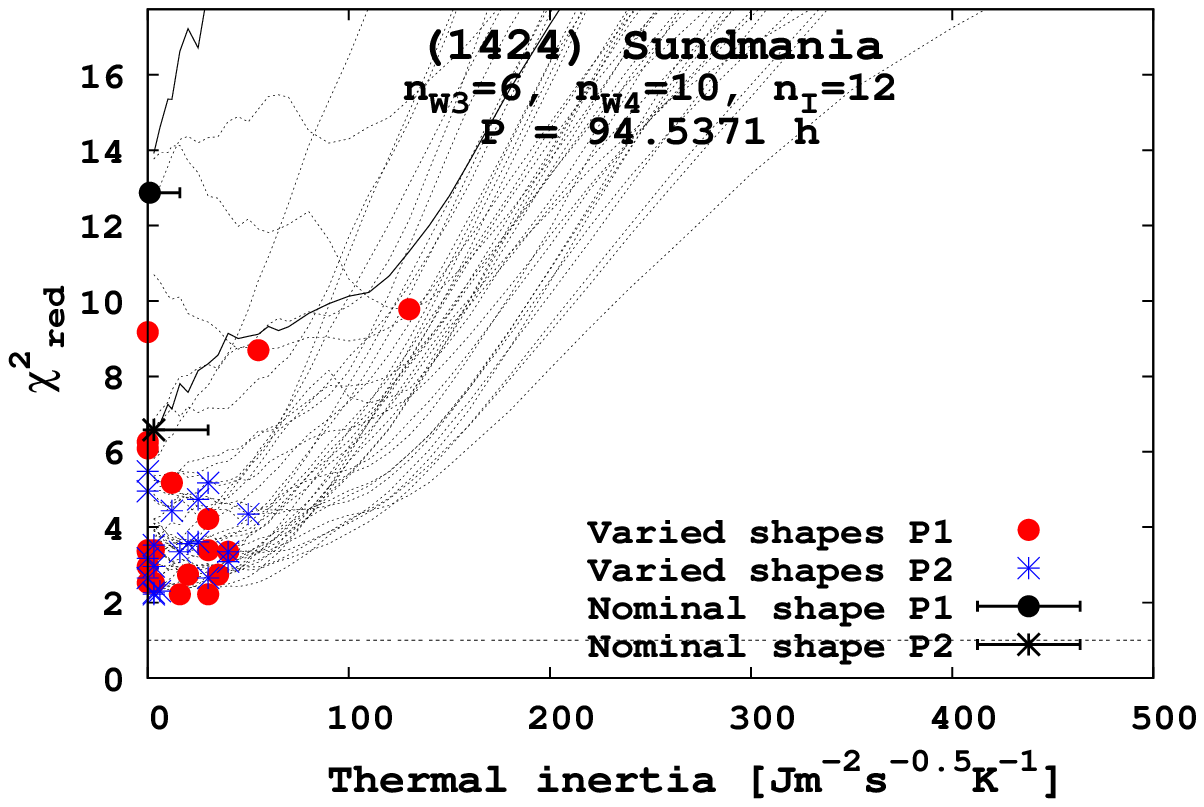}}\\
\end{center}
\caption{VS-TPM fits in the thermal inertia parameter space for eight asteroids. Each plot also contains the number of thermal infrared measurements in WISE W3 and W4 filters and in all four IRAS filters, and the rotation period.}
\end{figure*}

\begin{figure*}[!htbp]
\begin{center}
\resizebox{0.8\hsize}{!}{\includegraphics{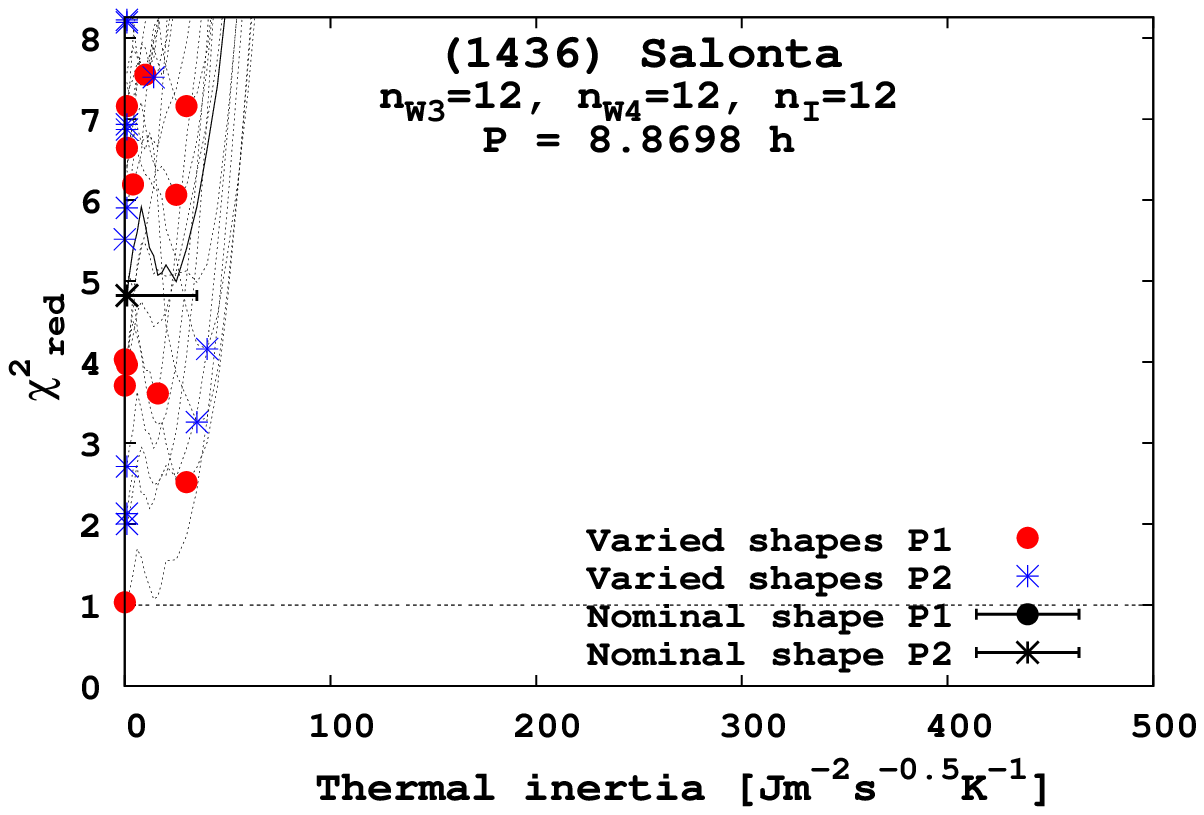}\includegraphics{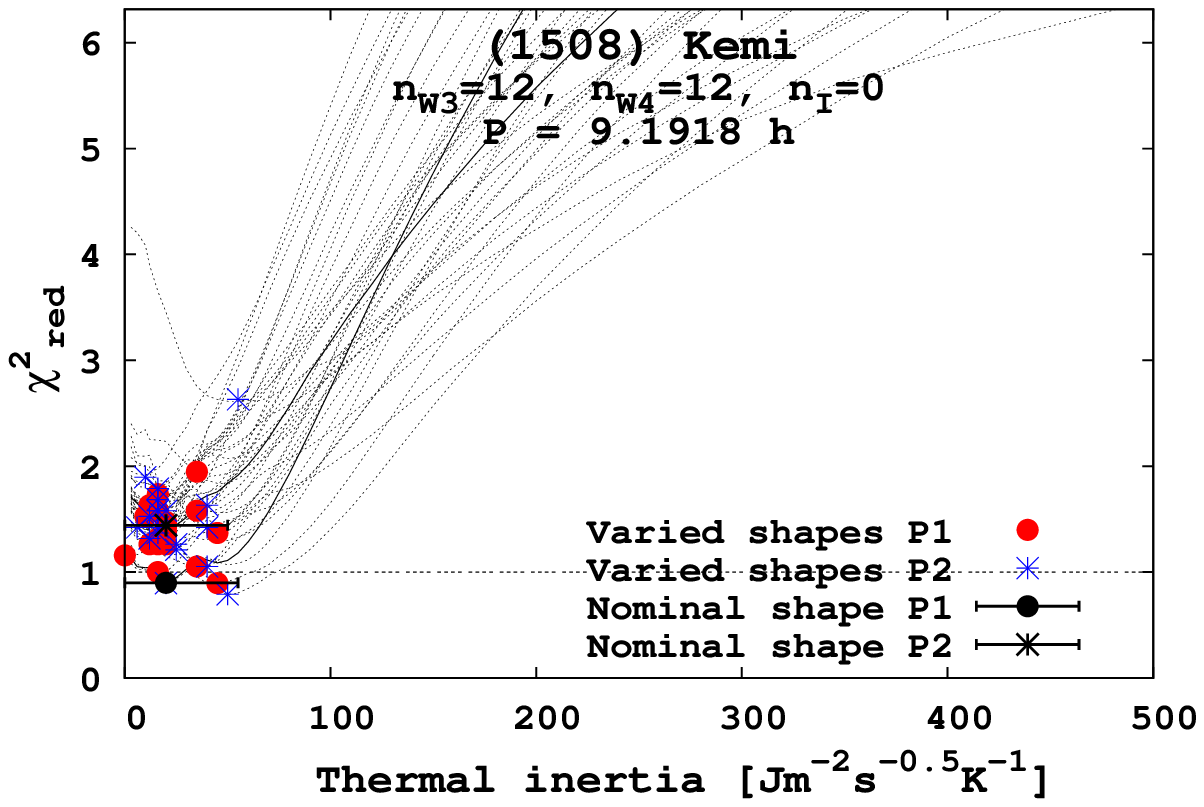}}\\
\resizebox{0.8\hsize}{!}{\includegraphics{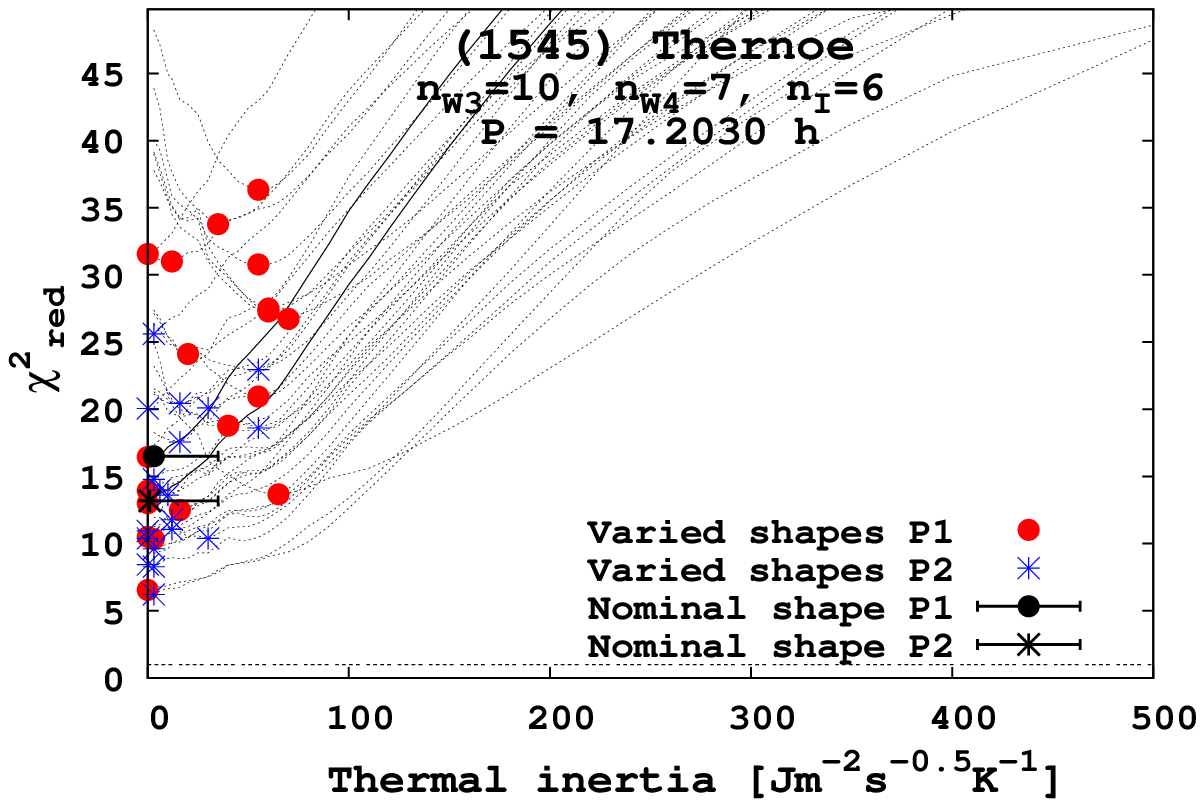}\includegraphics{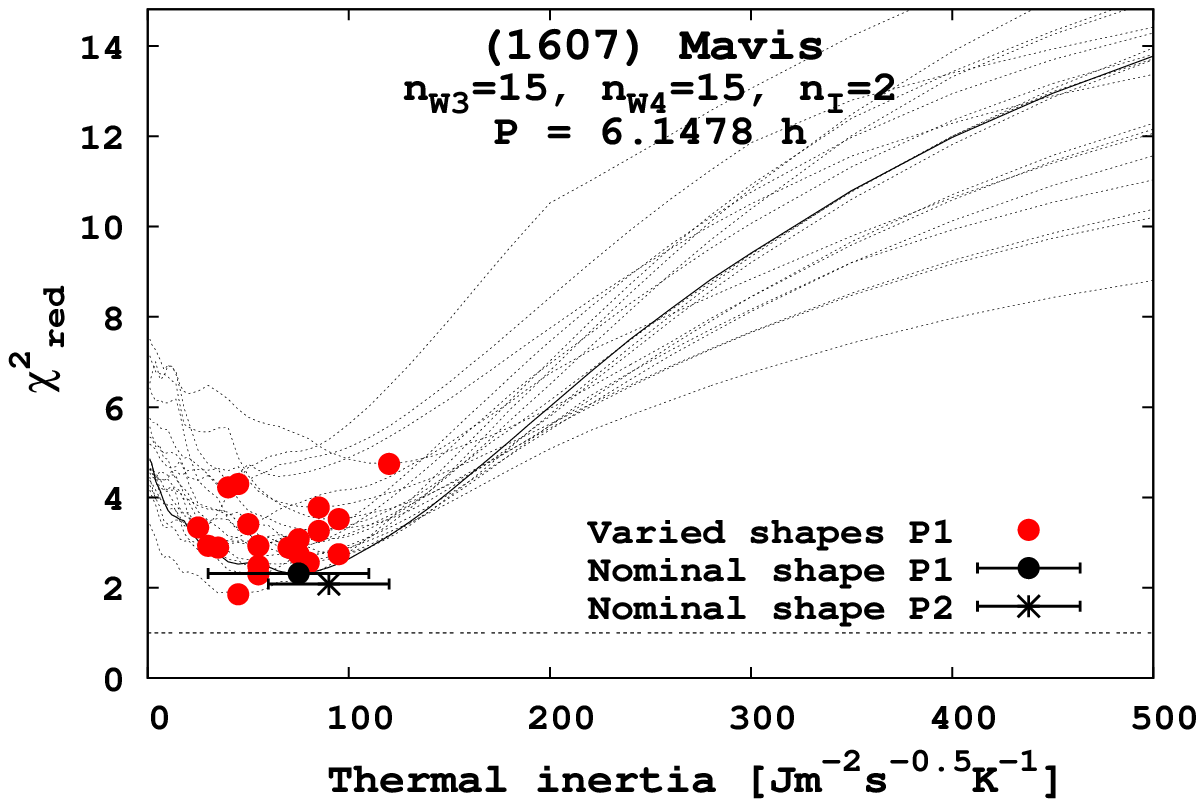}}\\
\resizebox{0.8\hsize}{!}{\includegraphics{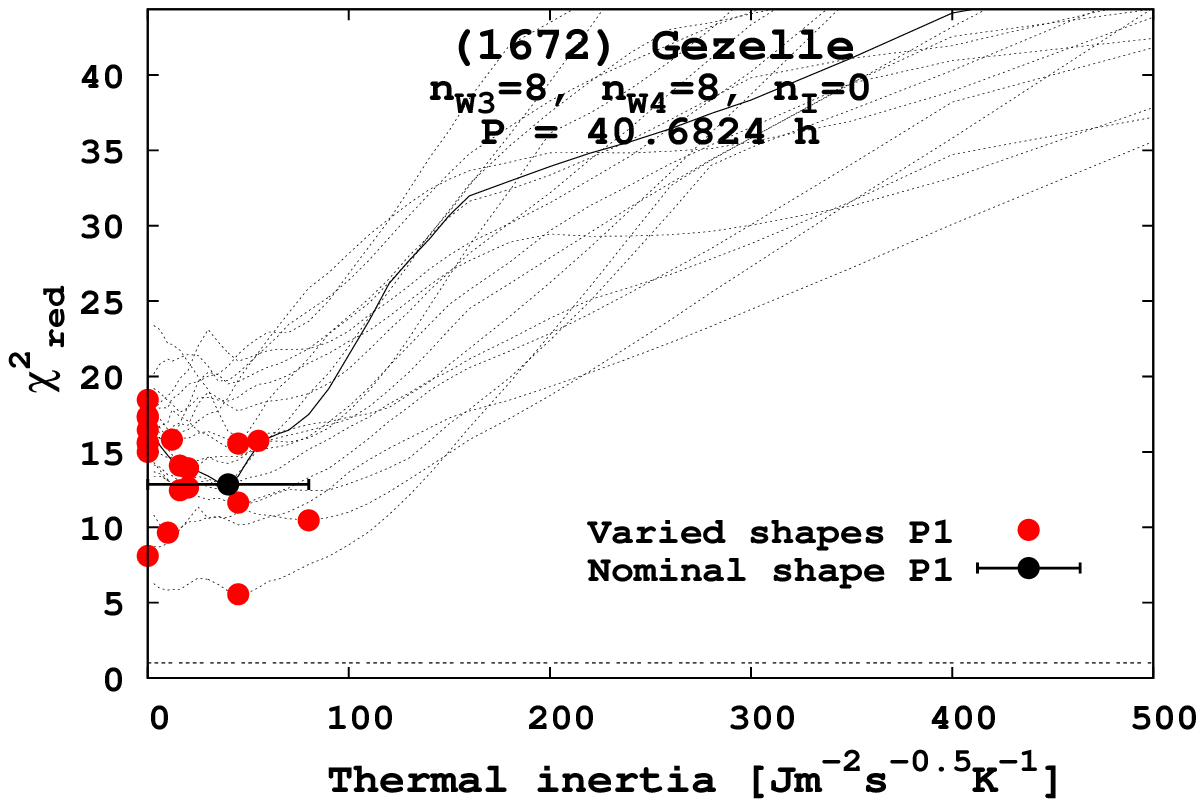}\includegraphics{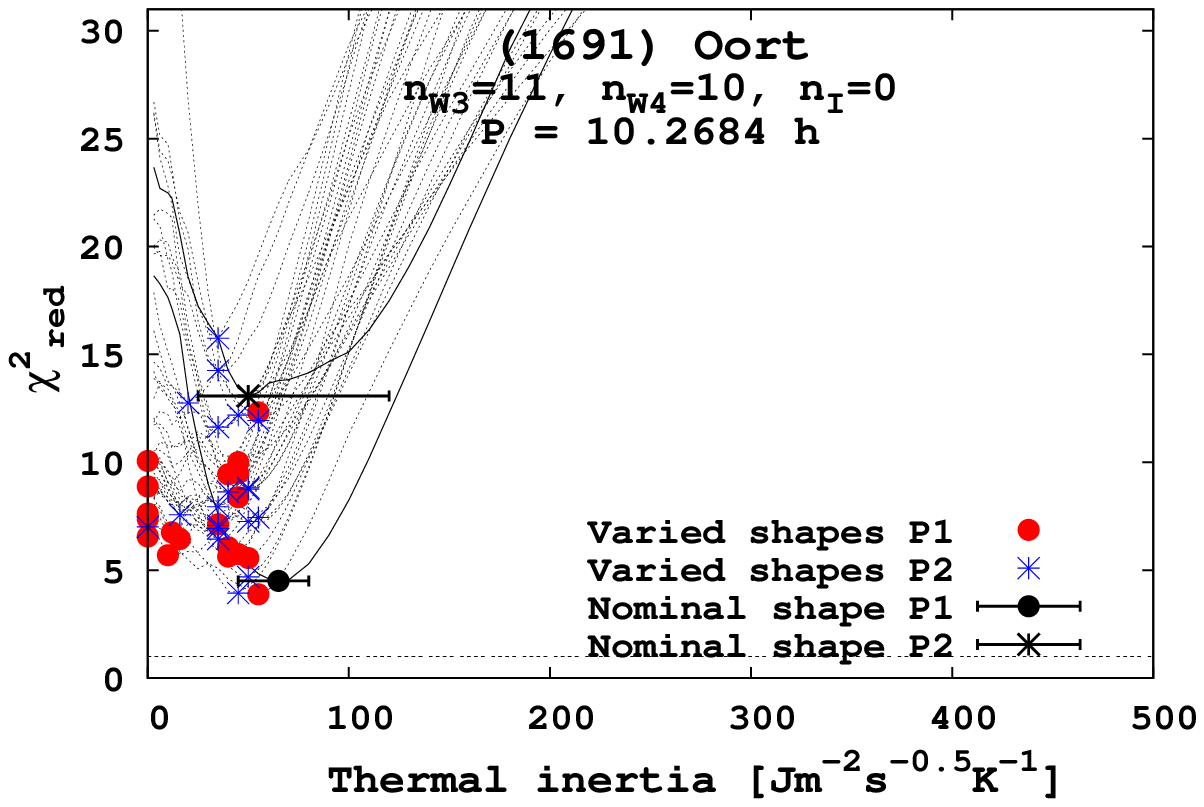}}\\
\resizebox{0.8\hsize}{!}{\includegraphics{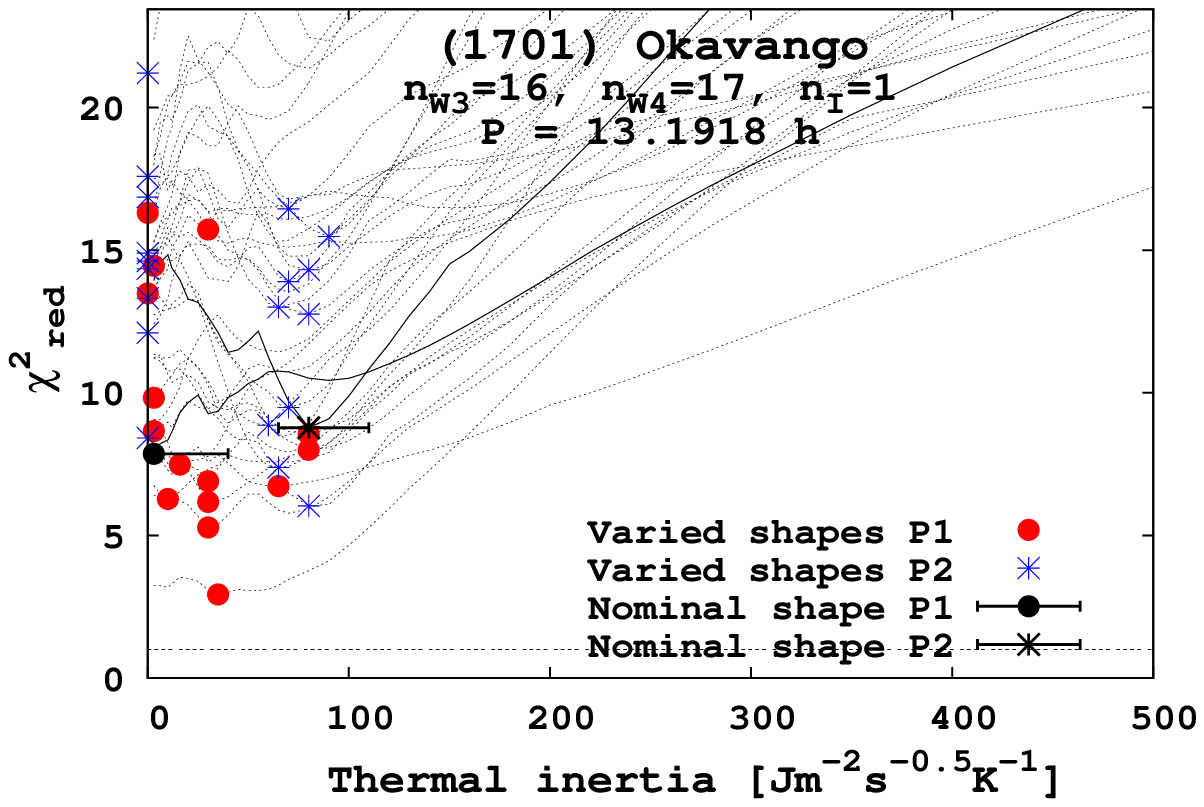}\includegraphics{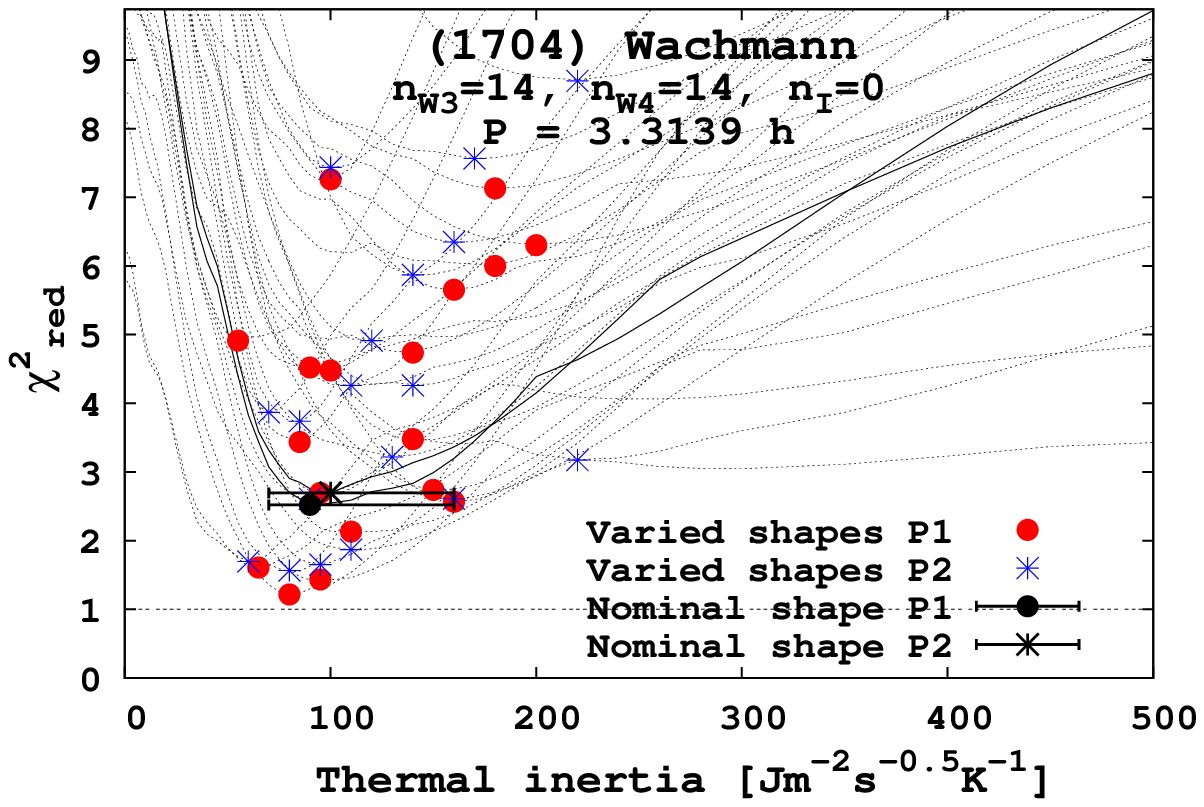}}\\
\end{center}
\caption{VS-TPM fits in the thermal inertia parameter space for eight asteroids. Each plot also contains the number of thermal infrared measurements in WISE W3 and W4 filters and in all four IRAS filters, and the rotation period.}
\end{figure*}

\begin{figure*}[!htbp]
\begin{center}
\resizebox{0.8\hsize}{!}{\includegraphics{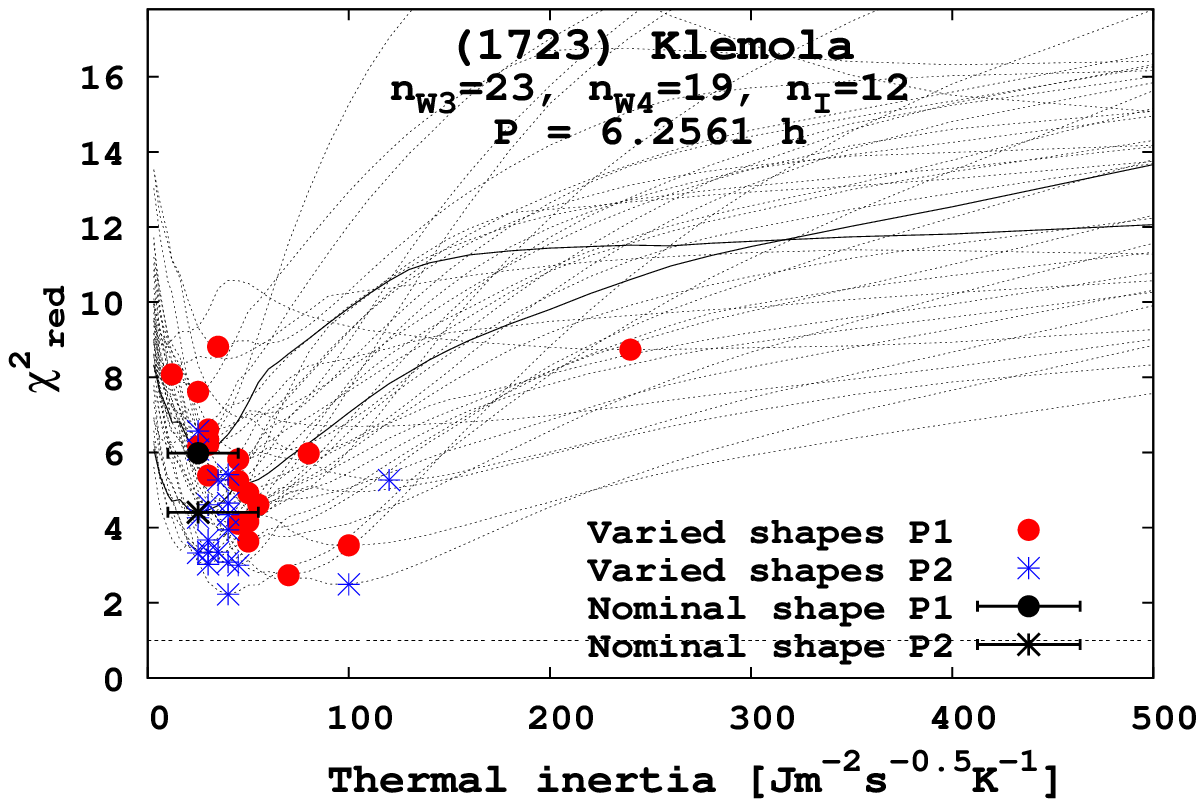}\includegraphics{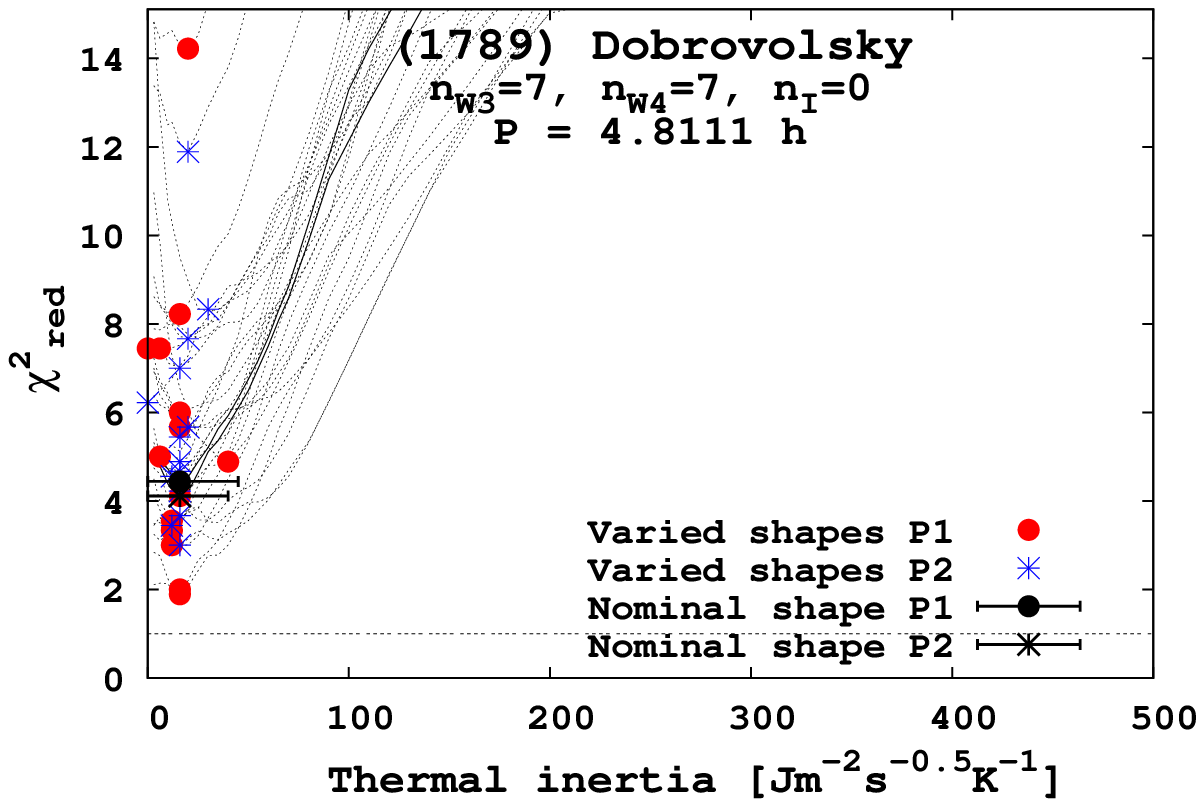}}\\
\resizebox{0.8\hsize}{!}{\includegraphics{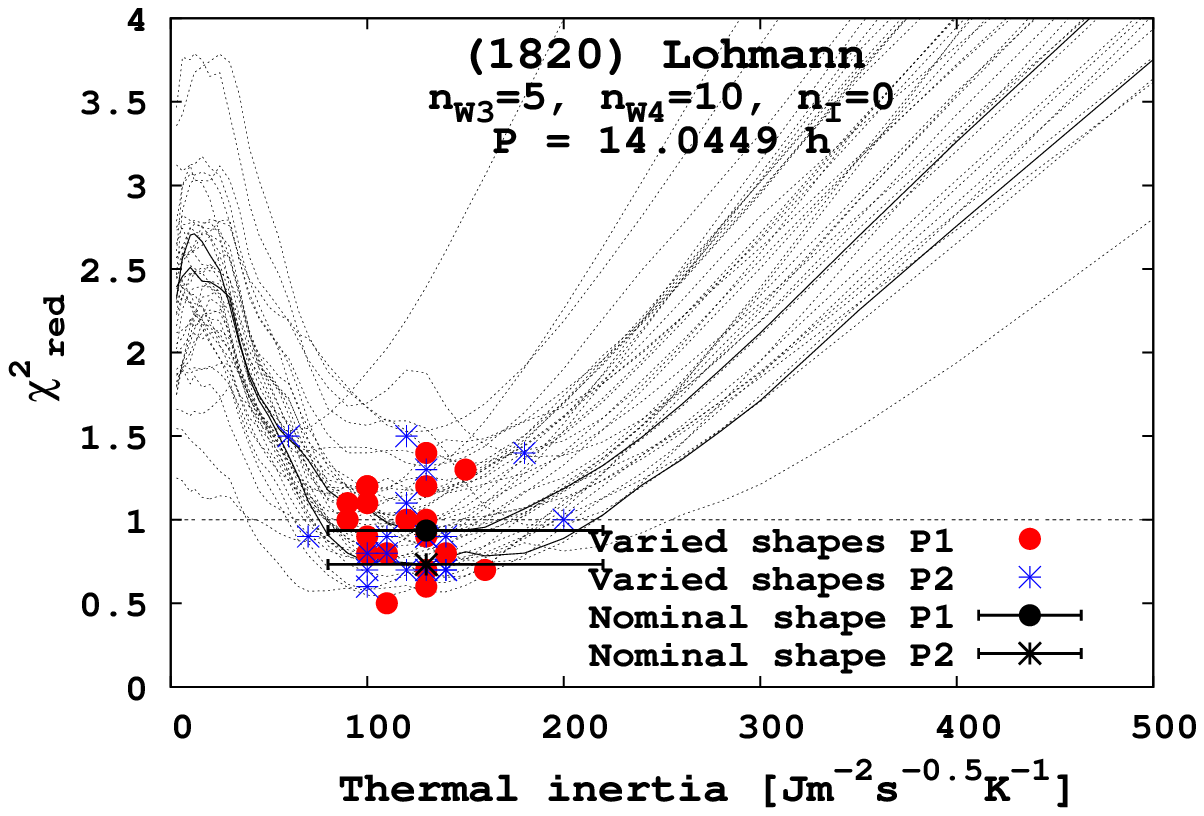}\includegraphics{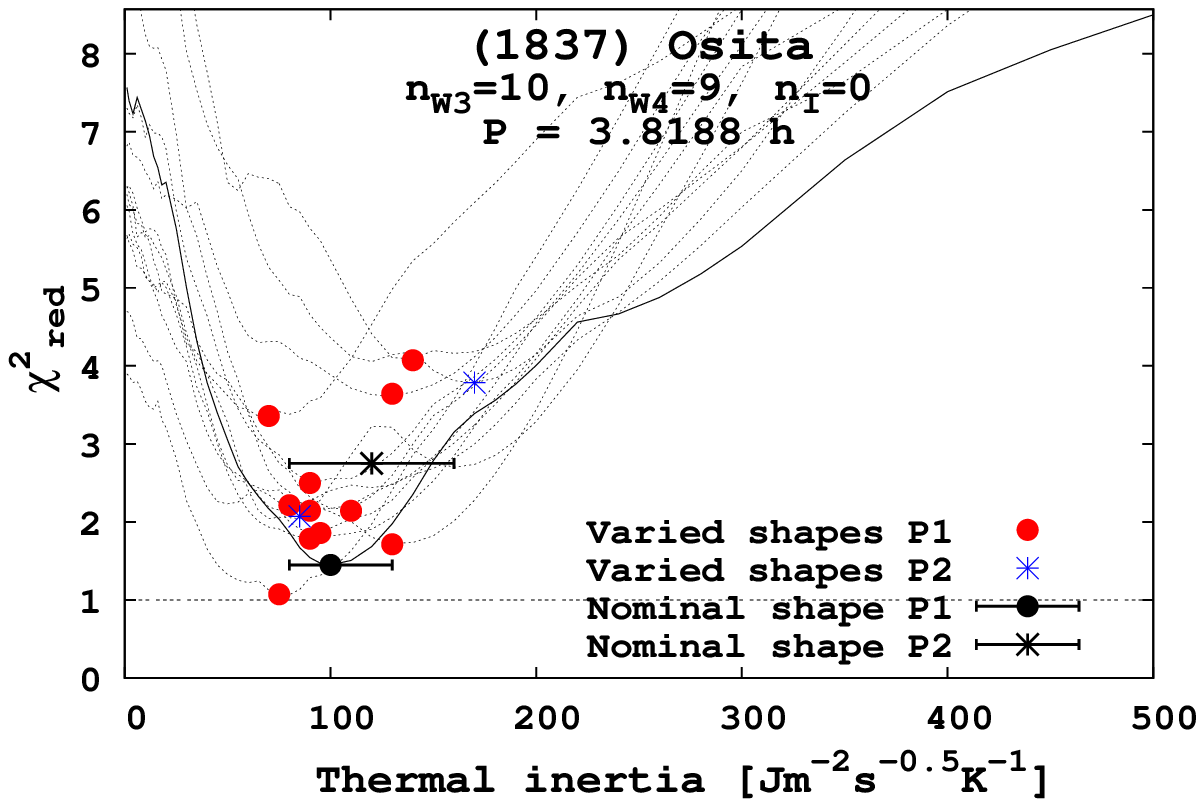}}\\
\resizebox{0.8\hsize}{!}{\includegraphics{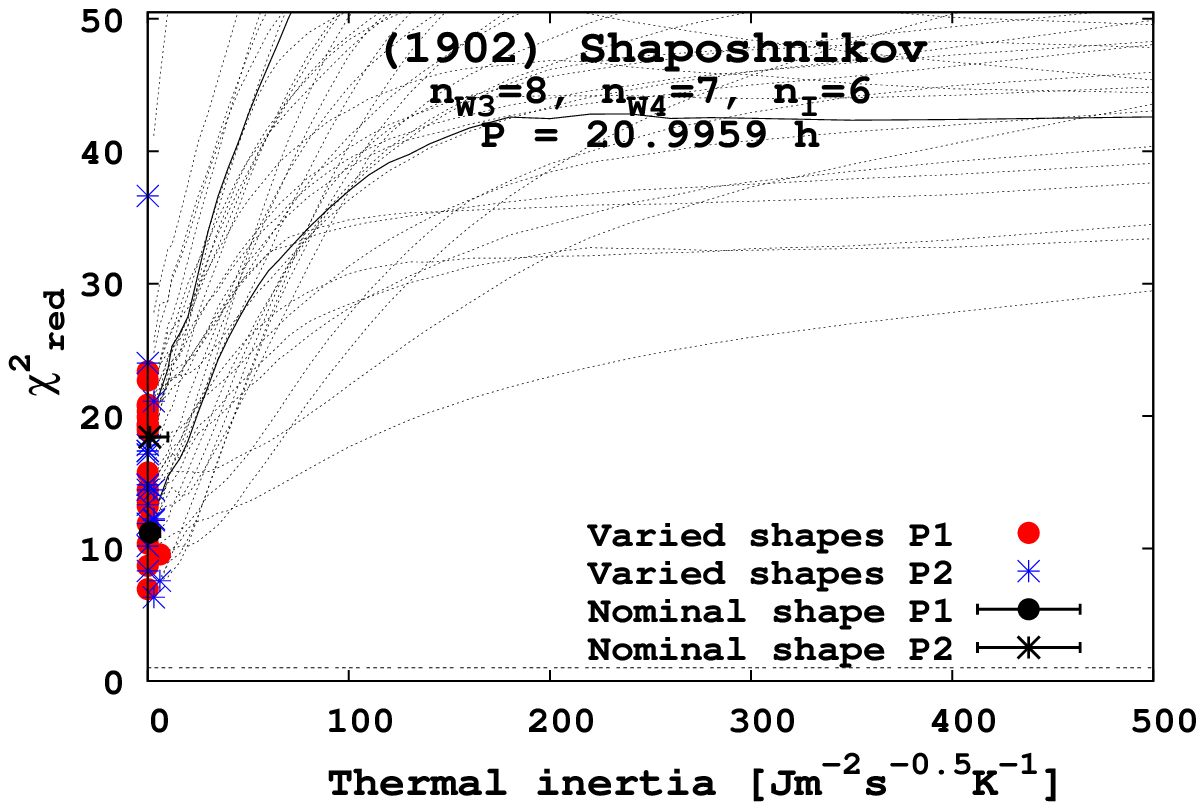}\includegraphics{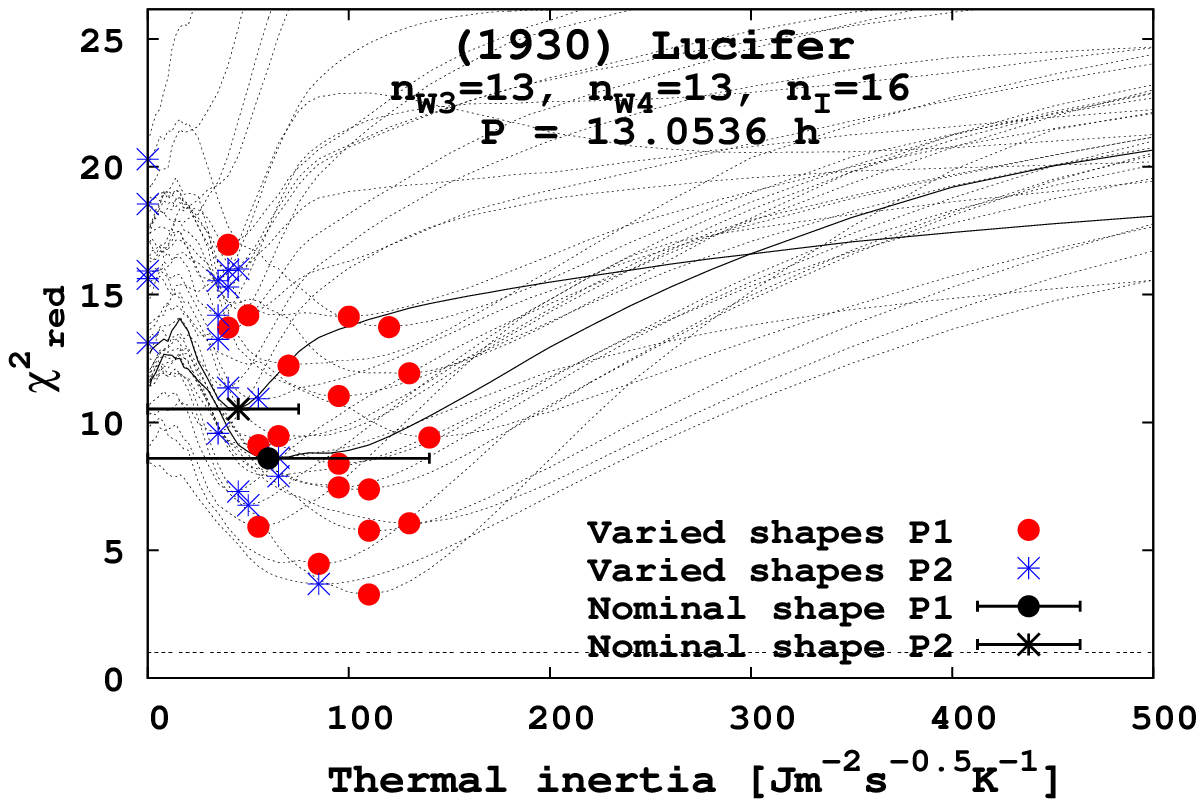}}\\
\resizebox{0.8\hsize}{!}{\includegraphics{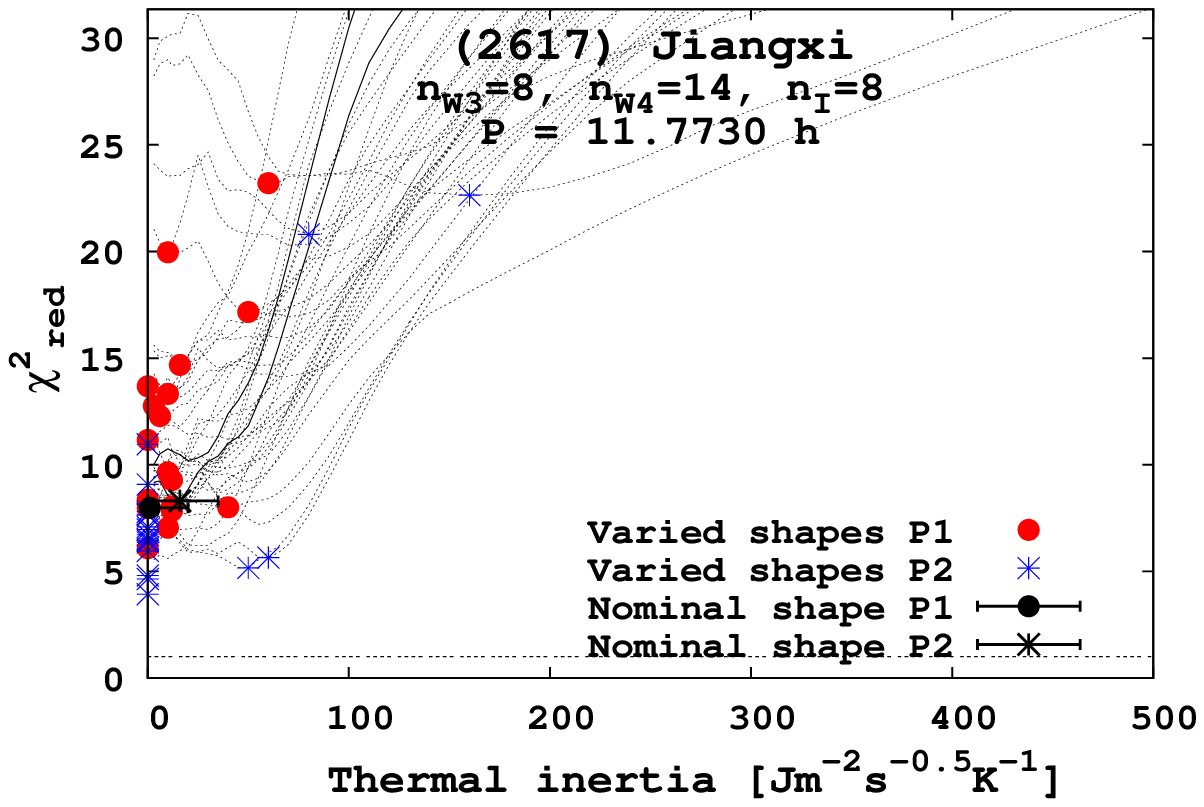}\includegraphics{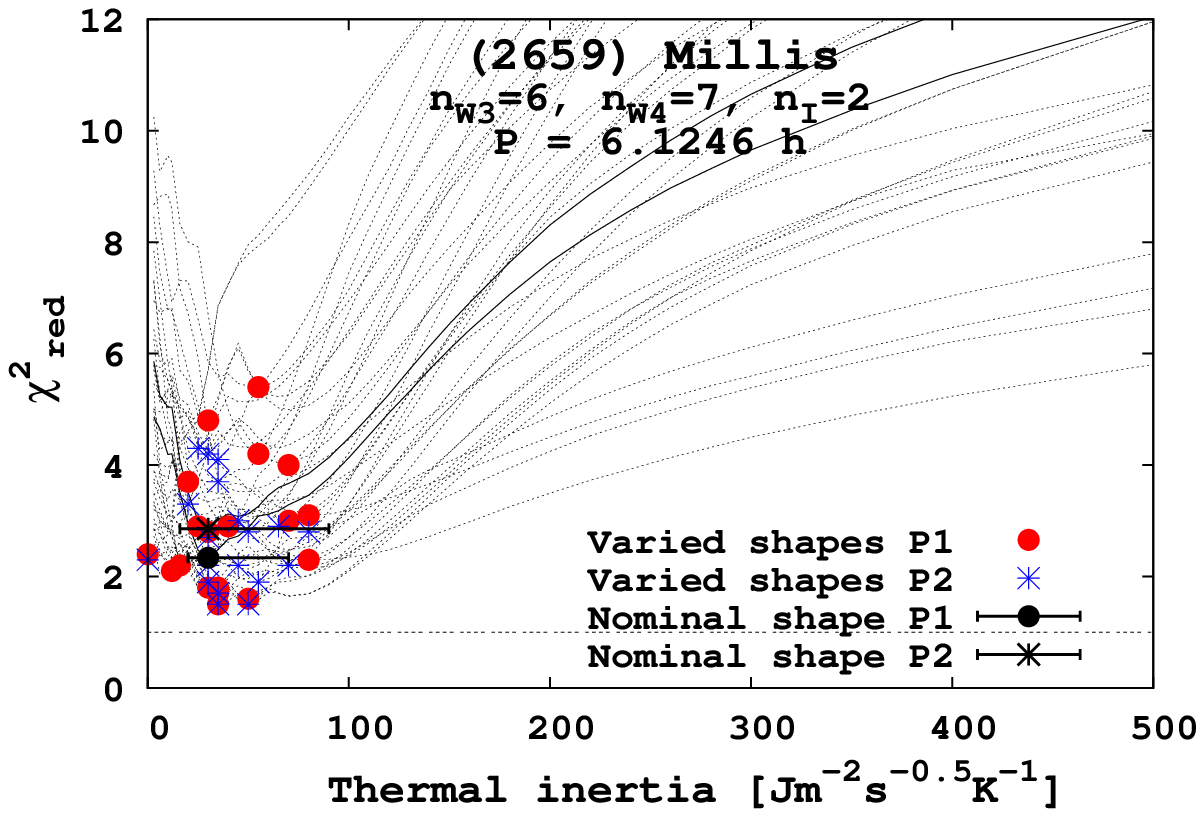}}\\
\end{center}
\caption{VS-TPM fits in the thermal inertia parameter space for eight asteroids. Each plot also contains the number of thermal infrared measurements in WISE W3 and W4 filters and in all four IRAS filters, and the rotation period.}
\end{figure*}

\begin{figure*}[!htbp]
\begin{center}
\resizebox{0.8\hsize}{!}{\includegraphics{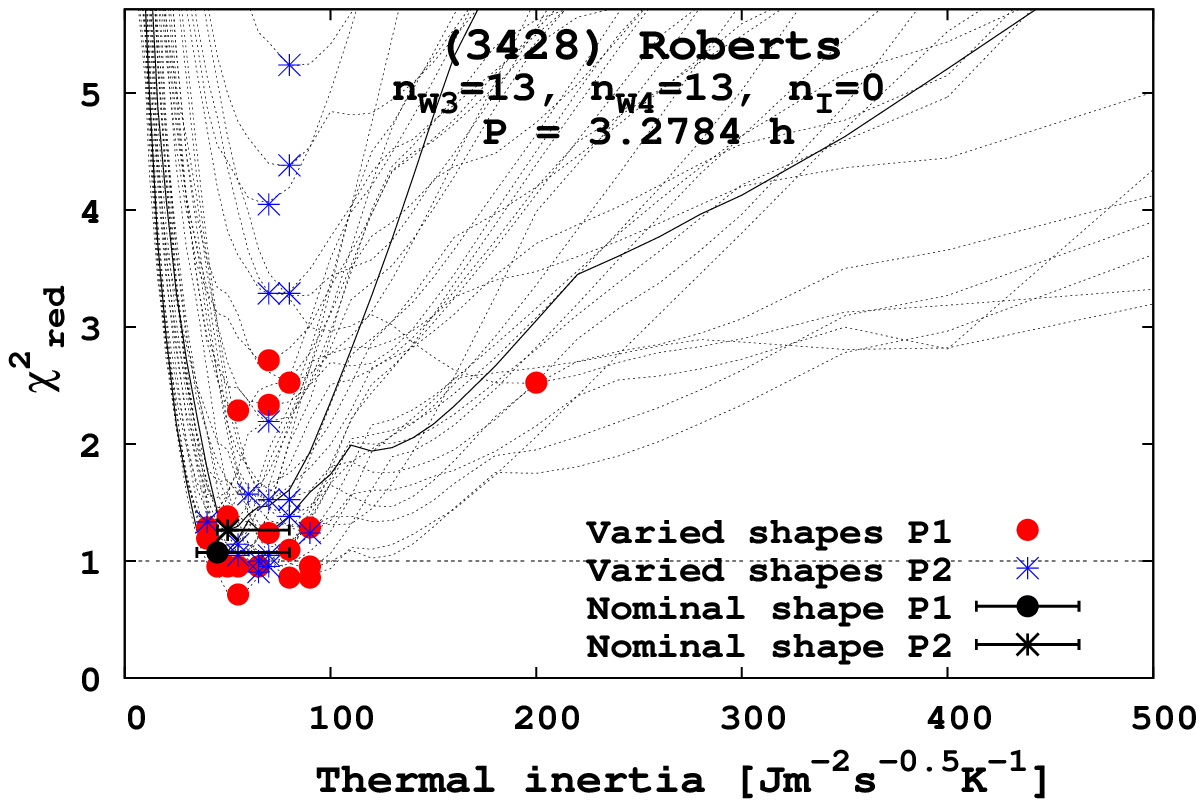}\includegraphics{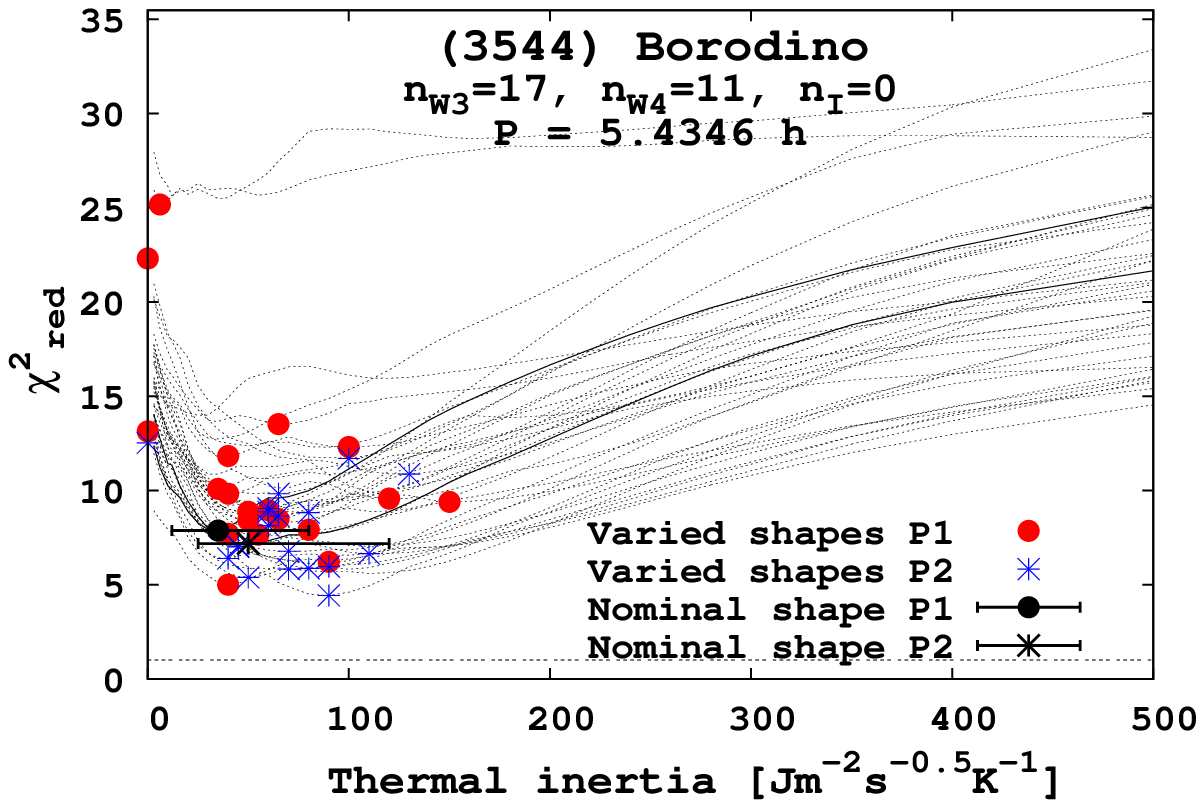}}\\
\resizebox{0.8\hsize}{!}{\includegraphics{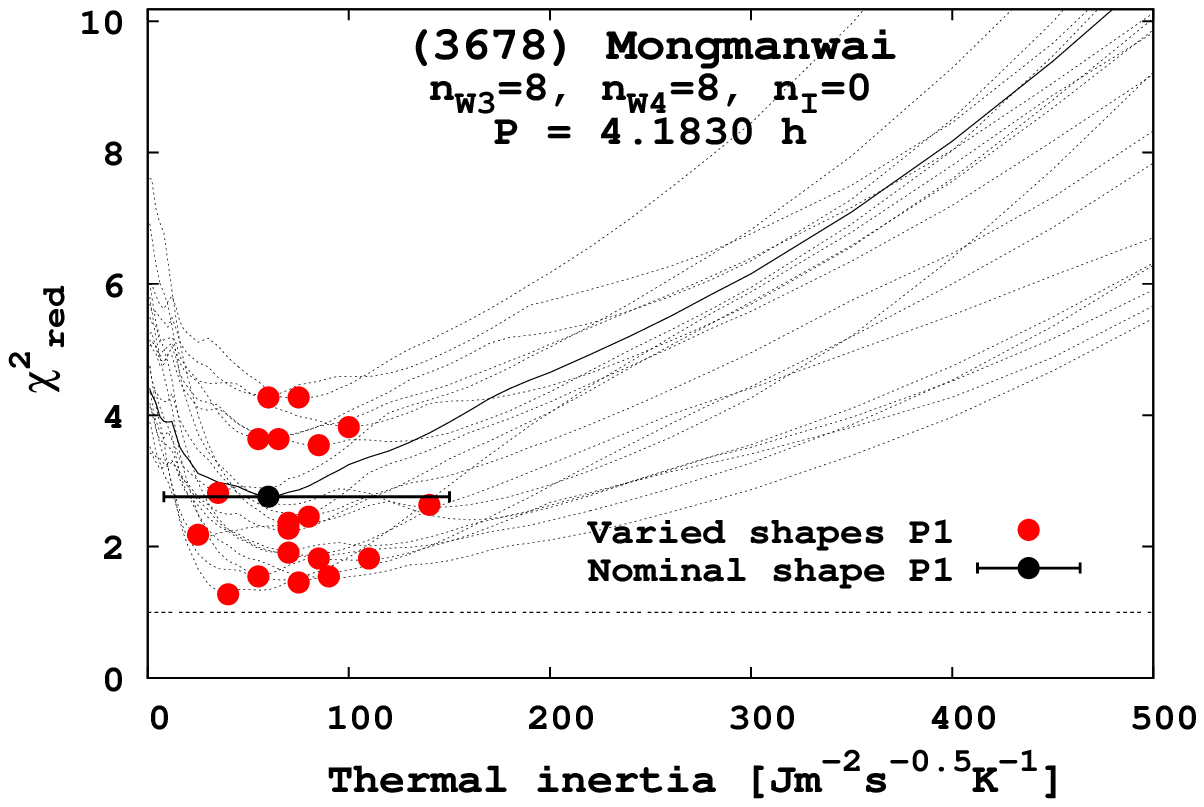}\includegraphics{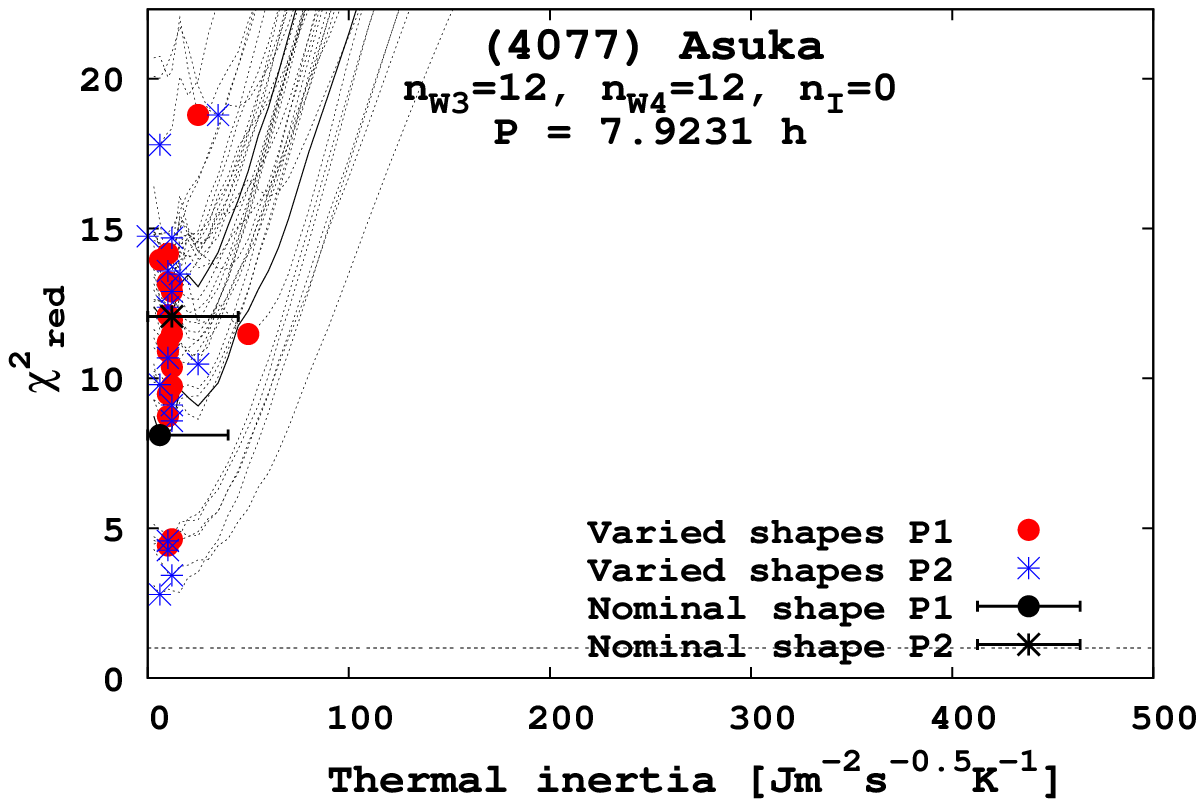}}\\
\resizebox{0.8\hsize}{!}{\includegraphics{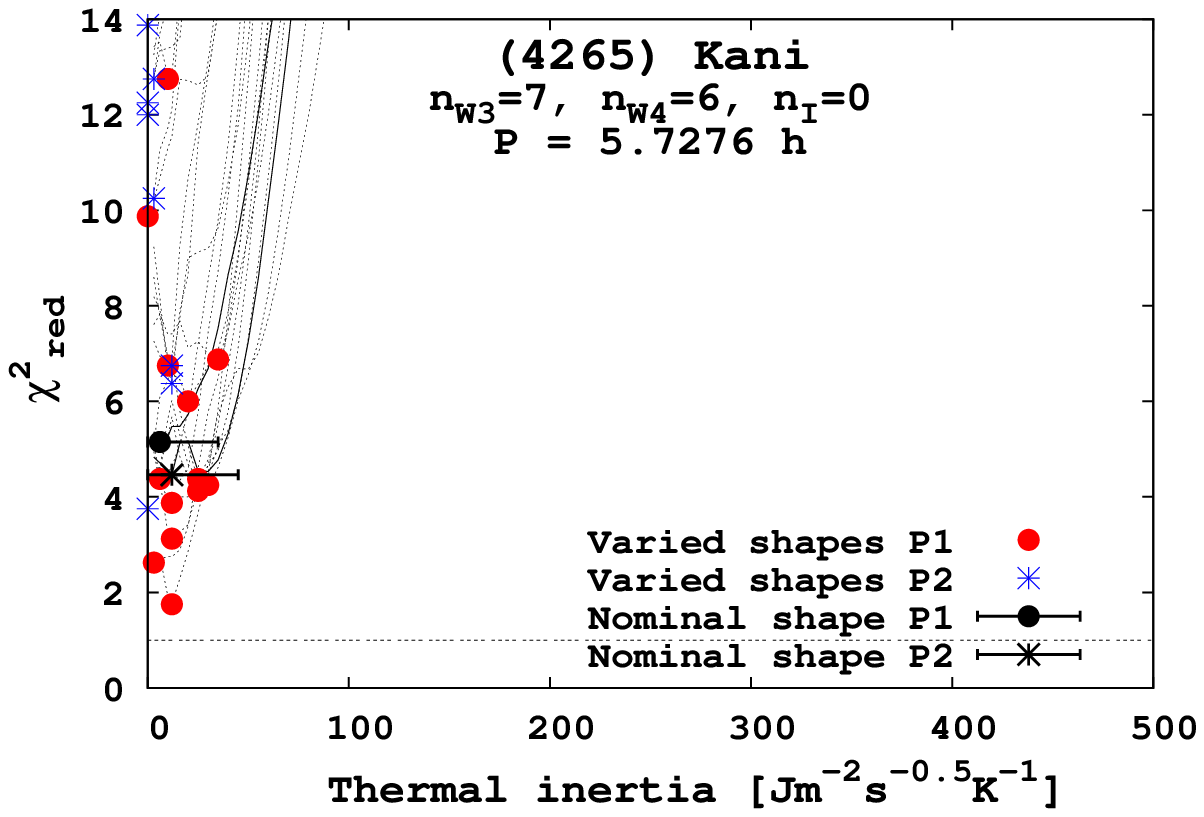}\includegraphics{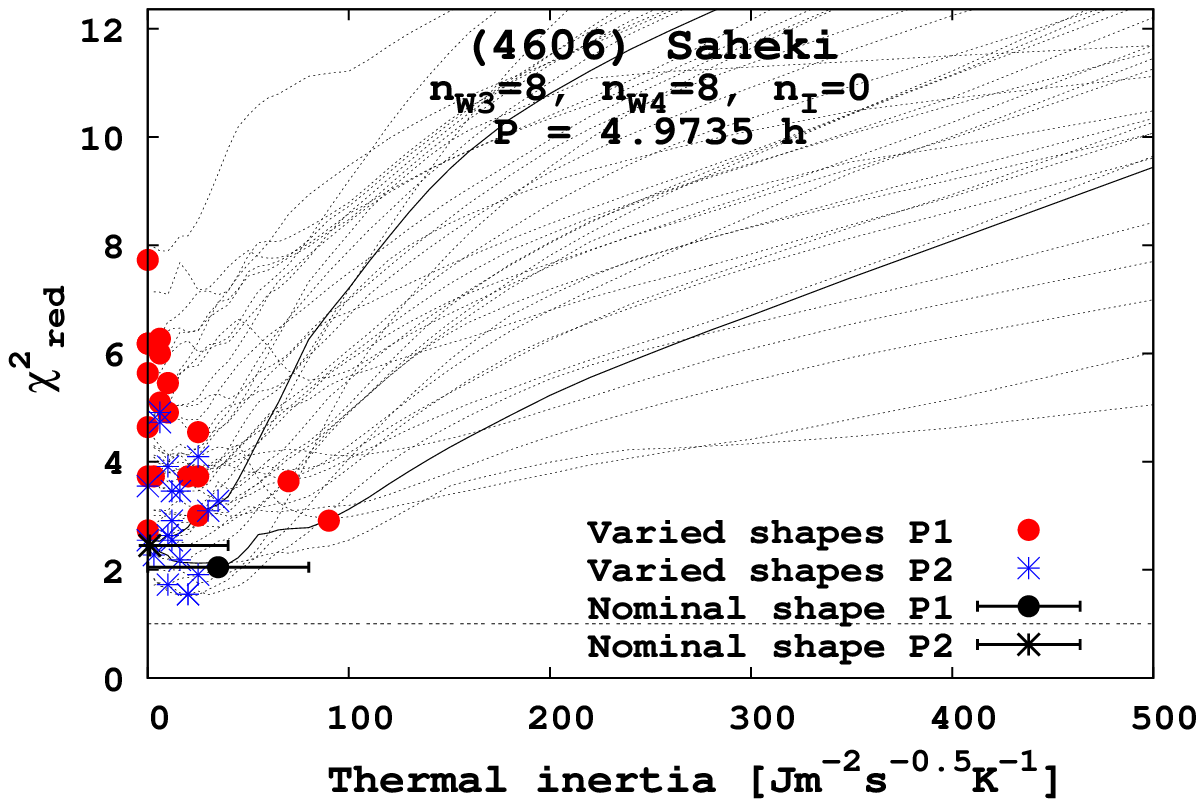}}\\
\resizebox{0.8\hsize}{!}{\includegraphics{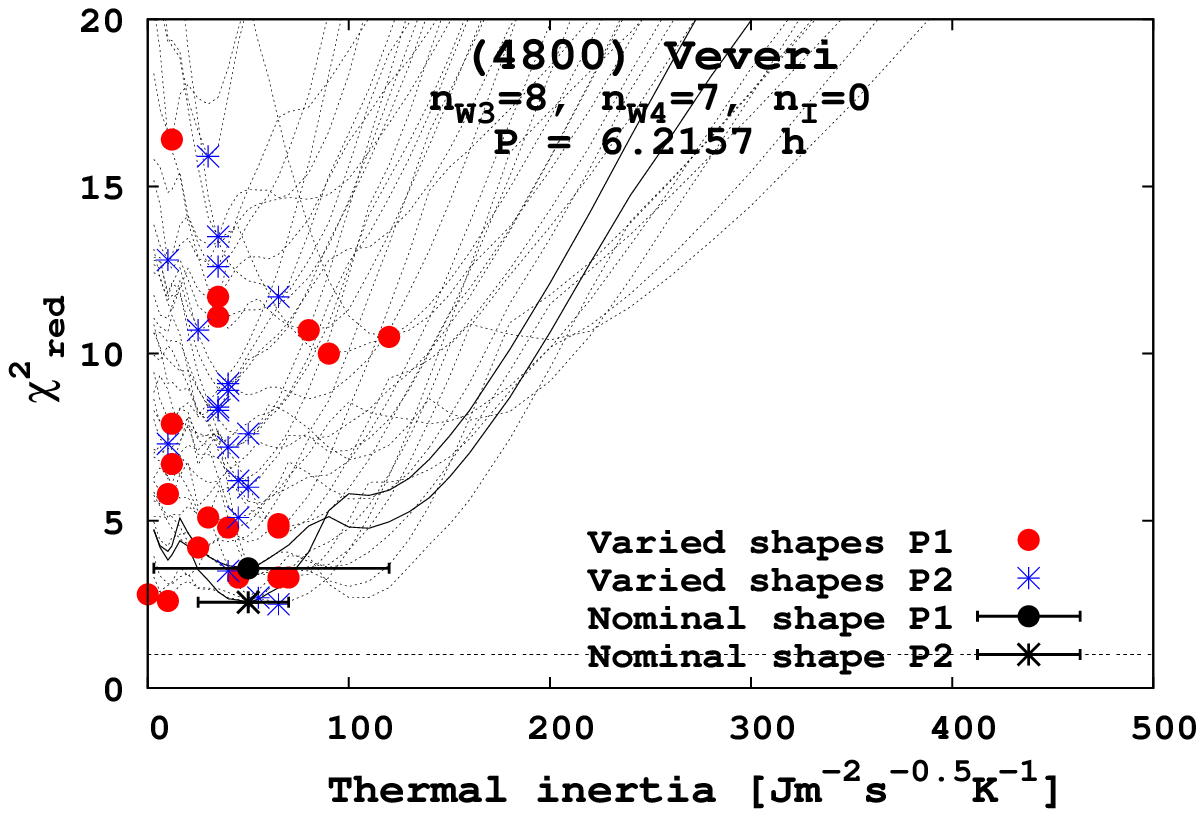}\includegraphics{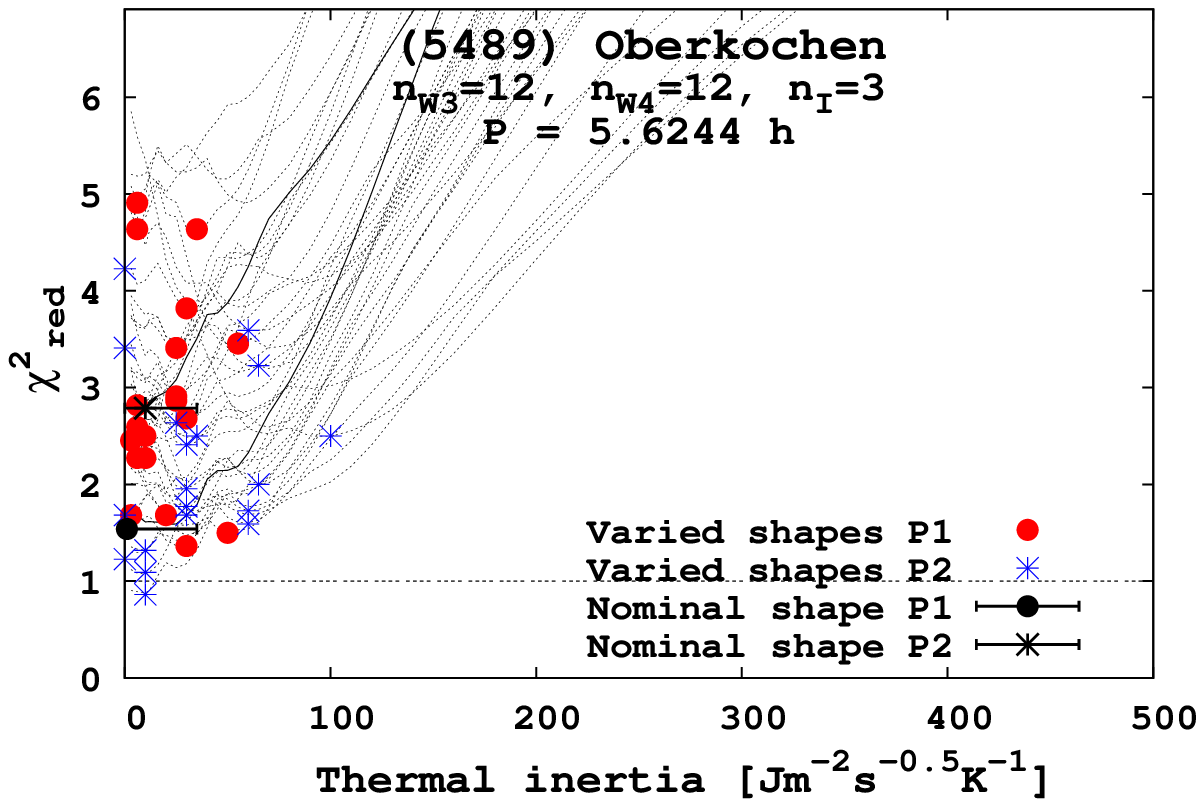}}\\
\end{center}
\caption{VS-TPM fits in the thermal inertia parameter space for eight asteroids. Each plot also contains the number of thermal infrared measurements in WISE W3 and W4 filters and in all four IRAS filters, and the rotation period.}
\end{figure*}

\begin{figure*}[!htbp]
\begin{center}
\resizebox{0.8\hsize}{!}{\includegraphics{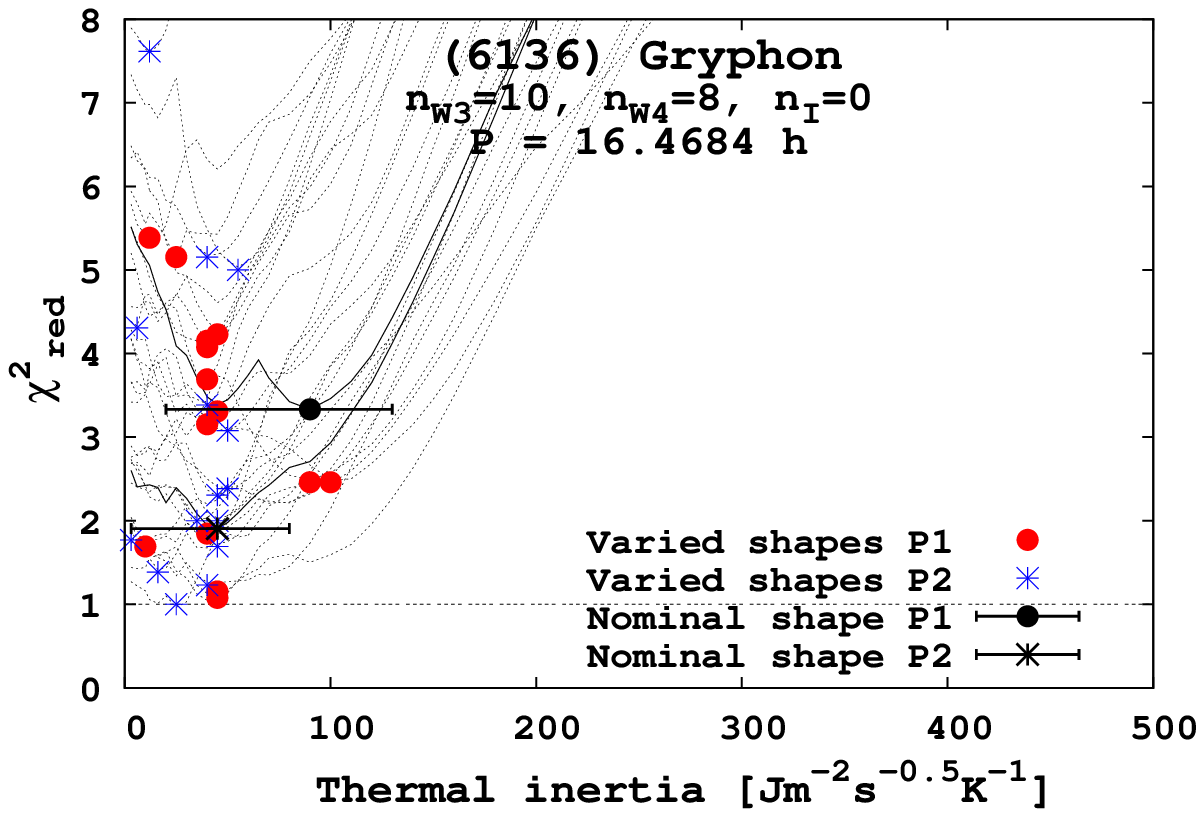}\includegraphics{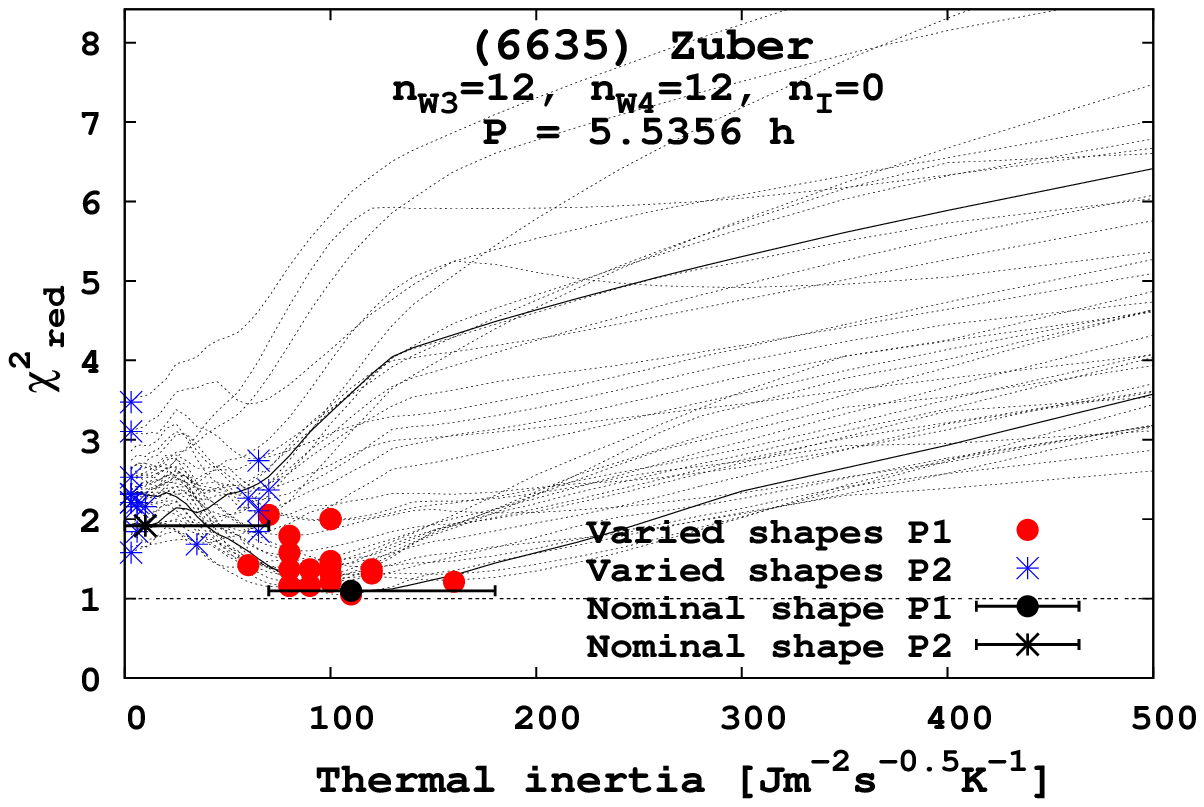}}\\
\resizebox{0.4\hsize}{!}{\includegraphics{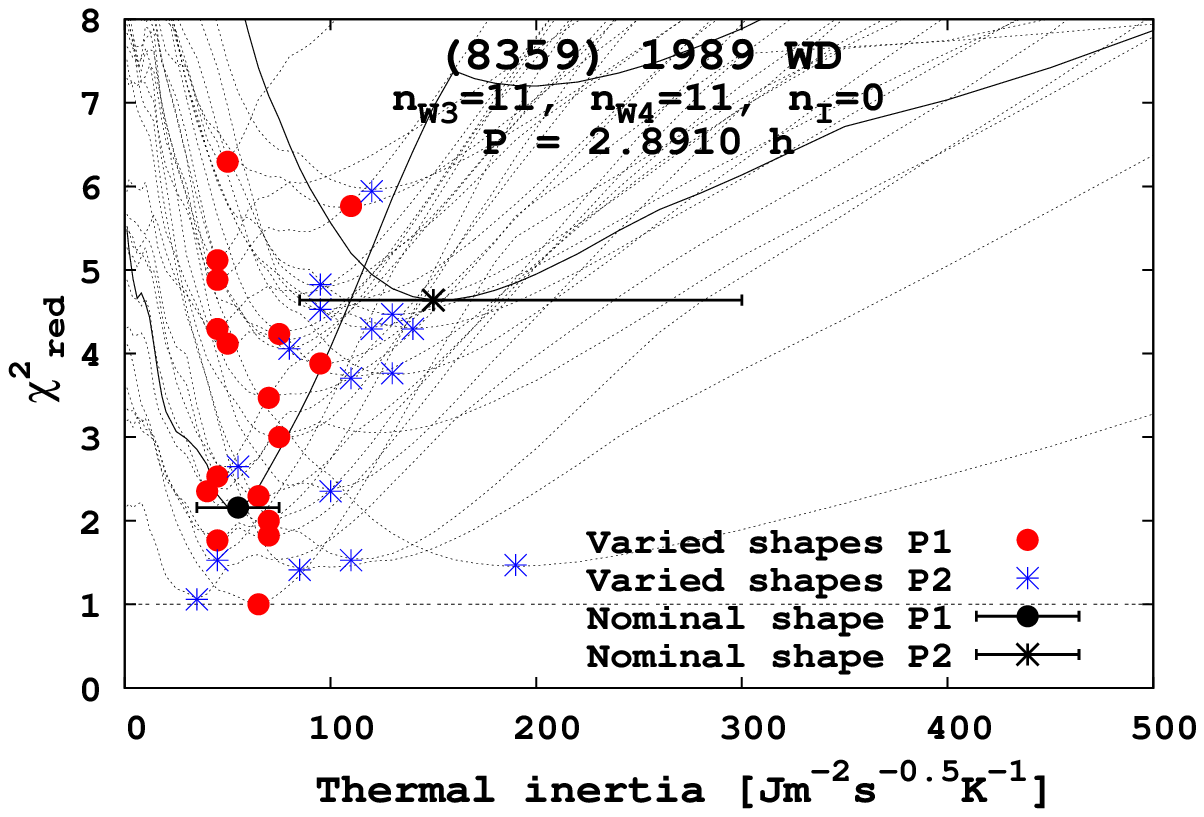}}\\
\end{center}
\caption{VS-TPM fits in the thermal inertia parameter space for eight asteroids. Each plot also contains the number of thermal infrared measurements in WISE W3 and W4 filters and in all four IRAS filters, and the rotation period.}
\end{figure*}



\end{document}